\documentclass[aps, prd, preprintnumbers]{revtex4}

\usepackage[colorlinks = true, linkcolor = blue]{hyperref}
\usepackage{xcolor}
\usepackage{ulem}
\usepackage{graphicx}
\usepackage{tabularx}
\usepackage{makecell}
\usepackage{subcaption}
\usepackage{multirow}
\usepackage{float}

\usepackage{siunitx}

\usepackage{amsmath}
\usepackage[e]{esvect}
\usepackage[capitalise]{cleveref} %% automatically format reference notations, has to be loaded after 'amsmath'
\usepackage{amssymb}
\usepackage{slashed}
\usepackage{physics}

\usepackage{pifont} %% http://ctan.org/pkg/pifont

\begin{document}

\title{Global analysis of fragmentation functions to light neutral hadrons}

\author{Jun Gao$^{1}$, ChongYang Liu$^{1}$, Mengyang Li$^{2}$, XiaoMin~Shen$^{3,1}$, Hongxi Xing$^{2,4,5}$, Yuxiang Zhao$^{3,5,6,7}$, Yiyu Zhou$^{8,9,10}$}

\affiliation{
	$^1${State Key Laboratory of Dark Matter Physics, Shanghai Key Laboratory for Particle Physics and Cosmology, Key Laboratory for Particle Astrophysics and Cosmology (MOE), School of Physics and Astronomy, Shanghai Jiao Tong University, Shanghai 200240, China}
	\\
	$^2${State Key Laboratory of Nuclear Physics and Technology, Institute of Quantum Matter, South China Normal University, Guangzhou 510006, China}
	\\
	$^3${Institute of Modern Physics, Chinese Academy of Sciences, Lanzhou, Gansu 730000, China}
	\\
	$^4${Guangdong Basic Research Center of Excellence for Structure and Fundamental Interactions of Matter, Guangdong Provincial Key Laboratory of Nuclear Science, Guangzhou 510006, China}
	\\
	$^5${Southern Center for Nuclear-Science Theory (SCNT), Institute of Modern Physics, Chinese Academy of Sciences, Huizhou 516000, China}
	\\
	$^6${University of Chinese Academy of Sciences, Beijing 100049, China}
	\\
	$^7${Key Laboratory of Quark and Lepton Physics (MOE) and Institute of Particle Physics, Central China Normal University, Wuhan 430079, China}
	\\
	$^8${Key Laboratory of Atomic and Subatomic Structure and Quantum Control (MOE), Guangdong-Hong Kong Joint Laboratory of Quantum Matter, Guangzhou 510006, China}
	\\
	$^9${Department of Physics, University of Turin, via Pietro Giuria 1, I-10125 Torino, Italy}
	\\
	$^{10}${INFN, Section of Turin, via Pietro Giuria 1, I-10125 Torino, Italy}
}

\email{\\
{jung49@sjtu.edu.cn} \\
{liucy1999@sjtu.edu.cn} \\
{limengyang@m.scnu.edu.cn} \\
{xiaominshen@impcas.ac.cn} \\
{hxing@m.scnu.edu.cn} \\
{yxzhao@impcas.ac.cn} \\
{yiyu.zhou@unito.it} \\
}

\begin{abstract}
	Fragmentation functions (FFs) are crucial non-perturbative components in quantum chromodynamics (QCD), playing a vital role in predictions and understanding of the hadronization process.
	In this paper, we present the FFs for $K_S^0$, $\eta$, $\pi^0$ mesons, and $\Lambda$ baryons in the context of global QCD analysis.
	The data included in the fit are from single inclusive $e^+ e^-$ annihilation (SIA), semi-inclusive deep-inelastic scattering (SIDIS) and proton-proton collisions, with kinematic cuts carefully applied to ensure validity of collinear factorization and perturbative QCD expansion.
	For the first time, data from SIDIS and hadron-in-jet production in SIA have been incorporated into the extraction of FFs for light-flavor neutral hadrons.
	Our analysis reveals that these data play a critical role in constraining the gluon distribution, and in distinguishing between different quark flavors.
	Pulls from different datasets are also studied by performing alternative fits with systematically subtracting groups of data from the nominal fit.
	For the quality of the fit, good $\chi^2$ values are achieved for most of the datasets, and FFs are generally well constrained within the momentum fraction region $\pqty{0.1, 0.5}$.
	The extracted $K_S^0$ fragmentation functions, together with the $K_S^0$ FFs constructed from $K^{\pm}$ FFs via isospin symmetry, are used to test isospin symmetry in kaon fragmentation.
	Although a definitive conclusion cannot be reached yet, these studies have identified several potential measurements that can be performed at existing facilities, which may ultimately help us to arrive at a conclusive answer.
	With the comprehensive species of FFs extracted within the NPC framework, we are able to perform a test on the momentum sum rule with the light-flavor charged and neutral hadrons.
	These neutral hadrons are found to carry a relatively smaller fraction of the fragmenting parton's momentum compared to the charged ones.
	The central and Hessian error sets of the fitted FFs, denoted as \texttt{NPC23}, are publicly available in the form of LHAPDF6 grids.
\end{abstract}

\maketitle

\pagebreak
\tableofcontents
\newpage

\section{Introduction}

Understanding hadronization is essential for exploring color confinement in quantum chromodynamics (QCD).
In high-energy collisions, hadrons are abundantly produced via fragmentation of primary quarks and gluons.
Fragmentation functions (FFs) are proposed to describe the hadronization process, characterizing the probability density of an outgoing parton (quark or gluon) transforming into color neutral hadrons, with respect to the light-cone momentum fraction it carries \cite{Berman:1971xz, Field:1977fa, Feynman:1978dt}.

As one of the fundamental objects of non-perturbative QCD, FFs are widely studied and have many important applications~\cite{Metz:2016swz}.
In collinear factorization, the cross sections can be factorized into perturbatively calculable short-distance matrix elements, and non-perturbative distribution functions \cite{Collins:1989gx}, including parton distribution functions (PDFs) and FFs, each characterizing the hard scattering process and universal long-distance effects, respectively.
Therefore, leveraging the universality property, FFs can be extracted from various hadron production processes, such as $e^+ e^-$ annihilation (SIA), semi-inclusive deep-inelastic scattering (SIDIS), and proton-proton collisions. These extracted FFs can then be utilized to make predictions for hadron production in different high-energy processes, providing a powerful tool for theoretical and phenomenological studies in QCD.
Apart from collinear factorization, FFs are also involved in transverse momentum dependent (TMD) factorization.
The hadronization in TMD processes is more complicated than in collinear processes due to the inclusion of an additional kinematic variable, namely the transverse momentum $\boldsymbol{k}_T$ of the hadron with respect to the fragmenting parton.
However, since TMD FFs can be matched to collinear FFs~\cite{Collins:2011zzd}, our knowledge of collinear FFs remains crucial in the study of TMD physics.
Recently, study of the transverse energy-energy correlator (TEEC) has brought a lot of insight into TMD physics \cite{Li:2020bub, Kang:2023oqj, Kang:2024otf}.
TEEC is defined by integrating the differential cross section weighted by the transverse energy of the final-state particles.
Since the final-state hadrons are summed over, the calculation of the TEEC must be approximated due to limitations in the available hadron FFs.
This approximation arises because the current set of FFs is restricted to only a few types of hadrons.
The situation is even worse when nuclear modification is studied \cite{Kang:2023oqj, Kang:2024otf} because the availability of nuclear FFs is more limited \cite{Sassot:2009sh, Zurita:2021kli, Soleymaninia:2023dds, Doradau:2024wli}.
Besides, the study of polarization phenomena, especially the $\Lambda$ baryon spontaneous transverse polarization \cite{Kang:2021kpt} also relies on precise inputs of collinear unpolarized FFs of $\Lambda$.
With the key applications of collinear FFs for various hadrons, their determination has become a more pressing subject.
Unfortunately, direct calculation of FFs has not been realized due to its non-perturbative and time-dependent nature, although initial efforts of using quantum computing have been proposed recently~\cite{Li:2024nod, Grieninger:2024axp}.
Global analysis remains the most robust method for determining the FFs.
In the past few decades, great progress has been made towards the calculation of the hard scattering cross sections at higher orders in QCD.
In particular, coefficient functions for SIA at the next-to-next-to-leading order (NNLO) have been calculated in Refs.~\cite{Rijken:1996vr, Rijken:1996ns, Mitov:2006wy, Blumlein:2006rr}.
Results of threshold resummation are available at the next-to-next-to-leading logarithmic ($\mathrm{N}^2\mathrm{LL}$) \cite{Cacciari:2001cw, Moch:2009my} and next-to-next-to-next-to-leading logarithmic ($\mathrm{N}^3\mathrm{LL}$) \cite{Xu:2024rbt} accuracy, which are important at large momentum fractions.
For SIDIS, the next-to-leading order (NLO) coefficient functions are available in Refs.~\cite{Altarelli:1979kv, Nason:1993xx, Furmanski:1981cw, Graudenz:1994dq, deFlorian:1997zj, deFlorian:2012wk}.
The NNLO coefficient functions are recently computed in Refs.~\cite{Goyal:2023zdi, Bonino:2024qbh}.
On the other hand, approximate NNLO and $\mathrm{N^3LO}$ results have been obtained via threshold resummation~\cite{Abele:2021nyo, Abele:2022wuy}.
The hard cross sections for single inclusive hadron production in proton-proton collisions are calculated up to NLO in \cite{Aversa:1988vb, deFlorian:2002az, Jager:2002xm} and NNLO in \cite{Czakon:2025yti}.
Besides, hadron production inside jets can provide direct access to the shape of FFs \cite{Kang:2016ehg, Kaufmann:2019ksh, Kang:2019ahe, Wang:2020kar} and result in a strong constraint to the gluon FFs, as demonstrated in Ref. \cite{Anderle:2017cgl} for the FFs of $D^*$-meson and in Ref. \cite{Gao:2024nkz,Gao:2024dbv} for the FFs of light charged hadrons.
Corresponding jet fragmentations have been calculated at NLO in Refs. \cite{Arleo:2013tya, Liu:2023fsq, Zidi:2024lid, Caletti:2024xaw, Bonino:2024adk}.
Finally, the scale dependence of FFs, which is evolved by the Dokshitzer-Gribov-Lipatov-Altarelli-Parisi (DGLAP) evolution equations with time-like splitting kernels, can be perturbatively calculated.
The time-like splitting kernels are available at $\order{\alpha_s^3}$ in Refs.~\cite{Mitov:2006ic, Moch:2007tx, Almasy:2011eq, Chen:2020uvt, Ebert:2020qef, Luo:2020epw}, where $\alpha_s$ is the strong coupling constant.
With the results from perturbative QCD calculations and precise experimental measurements at hand, there are several groups providing phenomenological determinations of FFs.
For light charged hadrons, including $\pi^\pm$, $K^\pm$, and $p/\overline{p}$,
representative efforts can be found in the works of BKK \cite{Binnewies:1994ju, Binnewies:1995pt}, KKP \cite{Kniehl:2000fe}, BFGW \cite{Bourhis:2000gs}, S. Kretzer \cite{Kretzer:2000yf}, KLC \cite{Kretzer:2001pz}, DSS~\cite{deFlorian:2007ekg,deFlorian:2014xna,deFlorian:2017lwf,Borsa:2021ran}, HKNS~\cite{Hirai:2007cx}, AKK~\cite{Albino:2008fy}, NNFF~\cite{Bertone:2018ecm}, MAPFF~\cite{Khalek:2021gxf}, JAM~\cite{Moffat:2021dji} and NPC23~\cite{Gao:2024nkz,Gao:2024dbv, Gao:2025hlm}.
They are carried out at NLO in QCD with global data sets and different theoretical prescriptions.
There also exist determinations of FFs at NNLO with SIA data only~\cite{Anderle:2015lqa, Bertone:2017tyb,Soleymaninia:2018uiv,Soleymaninia:2020bsq, Abdolmaleki:2021yjf}, and at approximate NNLO with SIA and SIDIS data~\cite{Borsa:2022vvp,AbdulKhalek:2022laj}.
The full NNLO analysis of light charged hadron FFs using SIA and SIDIS is recently completed in \cite{Gao:2025hlm}.
For light neutral hadrons, for instance, $K_S^0$, $\Lambda$, and $\eta$, their FFs are less known due to the lack of precision data.
The $K_S^0$ meson FFs have been extracted in the past by BKK96 \cite{Binnewies:1995kg}, AKK05 \cite{Albino:2005mv}, AKK08 \cite{Albino:2008fy}, SAK20 \cite{Soleymaninia:2020ahn}, LAXZ \cite{Li:2024etc} and FF24 \cite{Soleymaninia:2024jam} using SIA data only (except AKK08 that used $pp$ data as well).
The $\Lambda$ baryon FFs have been extracted by DSV \cite{deFlorian:1997zj}, AKK05 \cite{Albino:2005mv}, AKK08 \cite{Albino:2008fy} and SAK20 \cite{Soleymaninia:2020ahn}, also using only SIA data (with the exception of AKK08 which also used $pp$ data).
The $\eta$ meson FFs have been fitted in the past by AESSS \cite{Aidala:2010bn} with SIA and $pp$ data, and by LAXZ \cite{Li:2024etc} with SIA data using NNLO formula and higher twist effects.

In this work, we present a global analysis at NLO in QCD for the FFs of light neutral hadrons, namely $K_S^0$, $\Lambda$, $\eta$, and $\pi^0$.
We include for the first time the SIDIS data from ZEUS \cite{ZEUS:2011cdi}, and for the first time hadron-in-jet data from SIA three jet production.
The inclusion of the latter dataset has been made feasible by the development of the FMNLO program \cite{Liu:2023fsq, Zhou:2024cyk}.
This advancement has resulted in a noticeable constraint on the gluon FFs.
This analysis extends our previous work on FFs of light charged hadrons (NPC23) \cite{Gao:2024nkz, Gao:2024dbv, Gao:2025hlm} and allows a further test on the sum rules of FFs.
We are also able to study the isospin symmetry in the FFs of light hadrons, which is found to be valid for pions.
On the other hand, our analysis reveals that the current experimental data are insufficient to draw definitive conclusions regarding isospin symmetry in kaon FFs. To address this limitation, we propose several decisive measurements that could be carried out in existing experimental facilities.

The remainder of the paper is organized as follows.
\Cref{s.data} lists the experimental data included in the fit.
In \cref{s.theory}, we discuss the theoretical aspect of our global QCD analysis, including the parameterization of FFs and the goodness-of-fit criterion.
\Cref{s.quality-of-fit} presents the quality of fit to the data by listing the $\chi^2$ values and showing detailed comparisons between theory and data.
In \cref{s.NPC23-FFs}, we present the NPC23 delivered FFs.
In \cref{s.prediction}, we provide the physical quantities of interest, predicted by the fitted FFs of the neutral hadrons.
Finally, in \cref{s.conclusion}, we present a summary of our findings and draw conclusions based on the results and discussions presented throughout the paper.

\section{Experimental data sets fitted}
\label{s.data}

In this section, we summarize the experimental data used for the extraction of the $K_S^0$ meson, $\eta$ meson and $\Lambda$ baryon FFs.
The $\pi^0$ FFs, on the other hand, are derived from $\pi^{\pm}$ FFs via isospin symmetry.
A list of $\pi^0$ datasets is also provided to benchmark the resulting $\pi^0$ FFs.

\subsection{$K_S^0$ data}
\label{ss.K0S-data}

For the extraction of $K_S^0$ FFs, data from SIA, SIDIS, and $pp$ collisions are incorporated.
The SIA datasets for $K_S^0$ are summarized in \cref{t.SIA-data-below-Z-pole,t.SIA-data-at-Z-pole}, with information on collaboration, year of paper publication, center-of-mass energy, identified hadrons (parton flavor tags and in-jet production are also specified), measured observable, and number of data points after applying the cuts.
For SIA experiments at $Z$-pole, we include data from OPAL \cite{OPAL:1991ixp, OPAL:1995ebr, OPAL:2000dkf}, ALEPH \cite{ALEPH:1996oqp}, DELPHI \cite{DELPHI:1994qgk} and SLD \cite{SLD:1998coh}.
We also include data below $Z$-pole from TPC \cite{TPCTwoGamma:1984eoj}, MARK II \cite{Schellman:1984yz}, TASSO \cite{TASSO:1984nda, TASSO:1989jyt}, HRS \cite{Derrick:1985wd}, CELLO \cite{CELLO:1989adw}, TOPAZ \cite{TOPAZ:1994voc} and Belle \cite{Belle:2024vua}.
Additionally, we include $c$-tagged and $b$-tagged data from SLD \cite{SLD:1998coh} and three-jet data from ALEPH \cite{ALEPH:1999udi}.
The SLD $c$ and $b$-tagged $K_S^0$ production data are supposed to provide sensitivity to the heavy quark distributions.
On the other hand, since the third most energetic jet in three-jet events is typically a gluon jet, the ALEPH dataset \cite{ALEPH:1999udi} is expected to be sensitive to gluon FFs.
The DELPHI data at the center-of-mass energies $\sqrt{s} = 183~\mathrm{GeV}$ and 189 GeV \cite{DELPHI:2000ahn} are not included in our fit, since they are known to be in tension with the other world data \cite{Albino:2005mv, Albino:2008fy, Soleymaninia:2020ahn}.
We have also excluded the OPAL light flavor tagged data \cite{OPAL:1999zfe} from our fits as those data are based on leading-order analysis.
The BESIII measurement on $K_S^0$ production \cite{BESIII:2022zit} is not included in this fit as the treatment of mass correction at low hadron momentum is beyond the scope of this work.
Moreover, we observe that the NPC23 NNLO FFs \cite{Gao:2025hlm} yield predictions that are consistent with the data for both charged and neutral light hadrons in the large hadron momentum region \cite{BESIII:2025mbc}.

\begin{table}[]
	\setcellgapes{2.5 pt}
	\makegapedcells
	\begin{tabular}{|c|c|c|c|c|c|c|c|}
		\hline
		collaboration                    & year & $\sqrt{s}~[\mathrm{GeV}]$    & observable                                              & $N_{\mathrm{pt}} (K^0_S)$ & $N_{\mathrm{pt}} (\Lambda)$ & $N_{\mathrm{pt}} (\eta)$ & $N_{\mathrm{pt}} (\pi^0)$ \\
		\hline
		ARGUS~\cite{ARGUS:1989orf}       & 1990 & 9.46                         & $\frac{s}{\beta} \frac{\dd{\sigma}}{\dd{x_h}}$          & --                        & --                          & 6                        & 14                        \\
		\hline
		Belle~\cite{Belle:2024vua}       & 2024 & 10.52                        & $\frac{\dd{\sigma}}{\dd{x_p}}$                          & 35                        & --                          & --                       & --                        \\
		\hline
		TASSO~\cite{TASSO:1984nda}       & 1985 & 14, 22 \& 34                 & $\frac{s}{\beta} \frac{\dd{\sigma}}{\dd{x_h}}$          & 9, 6 \& 13                & 3, 4 \& 7                   & --                       & --                        \\
		\hline
		TASSO~\cite{TASSO:1989jyt}       & 1990 & 14.8, 21.5, 34.5, 35 \& 42.6 & $\frac{s}{\beta} \frac{\dd{\sigma}}{\dd{x_h}}$          & 9, 6, 13, 13 \& 13        & --                          & --                       & --                        \\
		\hline
		TPC~\cite{TPCTwoGamma:1984eoj}   & 1984 & 29                           & $\frac{1}{\beta \sigma_h} \frac{\dd{\sigma}}{\dd{x_h}}$ & 8                         & --                          & --                       & --                        \\
		\hline
		MARK II~\cite{Schellman:1984yz}  & 1985 & 29                           & $\frac{1}{\beta \sigma_h} \frac{\dd{\sigma}}{\dd{x_h}}$ & 17                        & --                          & --                       & --                        \\
		\hline
		HRS~\cite{Derrick:1985wd}        & 1987 & 29                           & $\frac{s}{\beta} \frac{\dd{\sigma}}{\dd{x_h}}$          & 12                        & 8                           & --                       & --                        \\
		\hline
		HRS~\cite{HRS:1987aky}           & 1988 & 29                           & $\frac{s}{\beta} \frac{\dd{\sigma}}{\dd{x_h}}$          & --                        & --                          & 13                       & --                        \\
		\hline
		SLAC~\cite{delaVaissiere:1984xg} & 1985 & 29                           & $\frac{s}{\beta} \frac{\dd{\sigma}}{\dd{x_h}}$          & --                        & 15                          & --                       & --                        \\
		\hline
		TASSO~\cite{TASSO:1981uqa}       & 1981 & 33                           & $\frac{s}{\beta}\frac{\dd{\sigma}}{\dd{x_h}}$           & --                        & 5                           & --                       & --                        \\
		\hline
		JADE~\cite{JADE:1985bzp}         & 1985 & 34.4                         & $\frac{s}{\beta} \frac{\dd{\sigma}}{\dd{x_h}}$          & --                        & --                          & 2                        & 10                        \\
		\hline
		TASSO~\cite{TASSO:1988qlu}       & 1989 & 34.8 \& 42.1                 & $\frac{s}{\beta} \frac{\dd{\sigma}}{\dd{x_h}}$          & --                        & 10 \& 5                     & --                       & --                        \\
		\hline
		JADE~\cite{JADE:1989ewf}         & 1990 & 35                           & $\frac{s}{\beta} \frac{\dd{\sigma}}{\dd{x_h}}$          & --                        & --                          & 3                        & 9                         \\
		\hline
		CELLO~\cite{CELLO:1989adw}       & 1990 & 35                           & $\frac{1}{\beta \sigma_h} \frac{\dd{\sigma}}{\dd{x_h}}$ & 9                         & 7                           & --                       & --                        \\
		\hline
		CELLO~\cite{CELLO:1989byk}       & 1990 & 35                           & $\frac{s}{\beta} \frac{\dd{\sigma}}{\dd{x_h}}$          & --                        & --                          & 5                        & 15                        \\
		\hline
		TASSO~\cite{TASSO:1988jma}       & 1989 & 44                           & $\frac{1}{\sigma_h}\frac{\dd{\sigma}}{\dd{x_p}}$        & --                        & --                          & --                       & 6                         \\
		\hline
		JADE~\cite{JADE:1989ewf}         & 1990 & 44                           & $\frac{s}{\beta} \frac{\dd{\sigma}}{\dd{x_h}}$          & --                        & --                          & --                       & 6                         \\
		\hline
		TOPAZ~\cite{TOPAZ:1994voc}       & 1995 & 58                           & $\frac{1}{\sigma_h} \frac{\dd{\sigma}}{\dd{\xi}}$       & 4                         & --                          & --                       & --                        \\
		\hline
	\end{tabular}
	\caption{
		Summary of SIA datasets at center-of-mass energies below $Z$-boson mass that are used in $K_S^0$, $\Lambda$, $\eta$ and $\pi^0$ FFs analyses.
		In the header line, $\sqrt{s}$ indicates the center-of-mass energy and $N_{\mathrm{pt}}$ denotes the number of points (after applying the kinematic cuts).
		The type of observed hadron is labeled in parentheses after $N_{\mathrm{pt}}$.
		The column ``year'' shows the year of paper publication.
		The kinematic variables in the observables are defined in \cref{e.K0S-SIA-dsigma-d_xE,e.K0S-SIA-dsigma-d_xp-and-d_xi} and corresponding texts.
	}
	\label{t.SIA-data-below-Z-pole}
\end{table}

\begin{table}[]
	\setcellgapes{2.5 pt}
	\makegapedcells
	\begin{tabular}{|c|c|c|c|c|c|c|c|}
		\hline
		collaboration                             & year & $\sqrt{s}~[\mathrm{GeV}]$ & observable                                                     & $N_{\mathrm{pt}} (K^0_S)$ & $N_{\mathrm{pt}} (\Lambda)$ & $N_{\mathrm{pt}} (\eta)$ & $N_{\mathrm{pt}} (\pi^0)$ \\
		\hline
		OPAL~\cite{OPAL:1991ixp}                  & 1991 & 91.2                      & $\frac{1}{\beta \sigma_h} \frac{\dd{\sigma}}{\dd{x_h}}$        & 7                         & --                          & --                       & --                        \\
		\hline
		OPAL~\cite{OPAL:1995ebr}                  & 1995 & 91.2                      & $\frac{1}{\sigma_h} \frac{\dd{\sigma}}{\dd{x_h}}$              & 16                        & --                          & --                       & --                        \\
		\hline
		OPAL~\cite{OPAL:1996gsw}                  & 1997 & 91.2                      & $\frac{1}{\beta \sigma_h} \frac{\dd{\sigma}}{\dd{x_h}}$        & --                        & 12                          & --                       & --                        \\
		\hline
		OPAL~\cite{OPAL:1998enc}                  & 1998 & 91.2                      & $\frac{1}{\sigma_h} \frac{\dd{\sigma}}{\dd{x_h}}$              & --                        & --                          & 11                       & 10                        \\
		\hline
		OPAL~\cite{OPAL:2000dkf}                  & 2000 & 91.2                      & $\frac{1}{\sigma_h} \frac{\dd{\sigma}}{\dd{x_h}}$              & 16                        & --                          & --                       & --                        \\
		\hline
		ALEPH~\cite{ALEPH:1992zhm}                & 1992 & 91.2                      & $\frac{1}{\sigma_h} \frac{\dd{\sigma}}{\dd{x_h}}$              & --                        & --                          & 8                        & --                        \\
		\hline
		ALEPH~\cite{ALEPH:1994fts}                & 1994 & 91.2                      & $\frac{1}{\sigma_h} \frac{\dd{\sigma}}{\dd{\xi}}$              & --                        & 14                          & --                       & --                        \\
		\hline
		ALEPH~\cite{ALEPH:1996pxg}                & 1997 & 91.2                      & $\frac{1}{\sigma_h}\frac{\dd{\sigma}}{\dd{x_p}}$               & --                        & --                          & --                       & 20                        \\
		\hline
		ALEPH~\cite{ALEPH:1996oqp}                & 1998 & 91.2                      & $\frac{1}{\sigma_h} \frac{\dd{\sigma}}{\dd{x_p}}$              & 16                        & 16                          & --                       & --                        \\
		\hline
		ALEPH~\cite{ALEPH:1999udi}                & 2000 & 91.2                      & $\frac{1}{\sigma_h} \frac{\dd{\sigma}}{\dd{\xi}}$              & 14                        & --                          & --                       & --                        \\
		\hline
		ALEPH~\cite{ALEPH:1999udi}                & 2000 & 91.2                      & $\frac{1}{\sigma_h} \frac{\dd{\sigma}}{\dd{x_h}}$              & --                        & --                          & 18                       & --                        \\
		\hline
		ALEPH (jet)~\cite{ALEPH:1999udi}          & 2000 & 91.2                      & $\frac{1}{\sigma_{\mathrm{jet}}} \frac{\dd{\sigma}}{\dd{\xi}}$ & 12, 13 \& 11              & 13, 12 \& 9                 & --                       & --                        \\
		\hline
		ALEPH (jet)~\cite{ALEPH:1999udi}          & 2000 & 91.2                      & $\frac{1}{\sigma_{\mathrm{jet}}} \frac{\dd{\sigma}}{\dd{x_h}}$ & --                        & --                          & 7, 6 \& 4                & 8, 8 \& 6                 \\
		\hline
		ALEPH~\cite{ALEPH:2001tfk}                & 2002 & 91.2                      & $\frac{1}{\sigma_h} \frac{\dd{\sigma}}{\dd{x_p}}$              & --                        & --                          & 5                        & --                        \\
		\hline
		DELPHI~\cite{DELPHI:1993vpj}              & 1993 & 91.2                      & $\frac{1}{\sigma_h} \frac{\dd{\sigma}}{\dd{x_p}}$              & --                        & 7                           & --                       & --                        \\
		\hline
		DELPHI~\cite{DELPHI:1994qgk}              & 1995 & 91.2                      & $\frac{1}{\sigma_h} \frac{\dd{\sigma}}{\dd{x_p}}$              & 13                        & --                          & --                       & --                        \\
		\hline
		DELPHI~\cite{DELPHI:1995ase}              & 1996 & 91.2                      & $\frac{1}{\sigma_h} \frac{\dd{\sigma}}{\dd{x_p}}$              & --                        & --                          & --                       & 17                        \\
		\hline
		DELPHI ($b$-tagged)~\cite{DELPHI:1995ase} & 1996 & 91.2                      & $\frac{1}{\sigma_h} \frac{\dd{\sigma}}{\dd{x_p}}$              & --                        & --                          & --                       & 15                        \\
		\hline
		SLD~\cite{SLD:1998coh}                    & 1999 & 91.2                      & $\frac{1}{\sigma_h} \frac{\dd{\sigma}}{\dd{x_p}}$              & 9                         & 9                           & --                       & --                        \\
		\hline
		SLD (tagged)~\cite{SLD:1998coh}           & 1999 & 91.2                      & $\frac{1}{\sigma_h} \frac{\dd{\sigma}}{\dd{x_p}}$              & 9 \& 9                    & 5 \& 5                      & --                       & --                        \\
		\hline
		L3~\cite{L3:1992pbe}                      & 1992 & 91.2                      & $\frac{1}{\sigma_h} \frac{\dd{\sigma}}{\dd{x_p}}$              & --                        & --                          & 4                        & --                        \\
		\hline
		L3~\cite{L3:1994gkb}                      & 1994 & 91.2                      & $\frac{1}{\sigma_h} \frac{\dd{\sigma}}{\dd{x_h}}$              & --                        & --                          & 10                       & --                        \\
		\hline
		L3~\cite{L3:1994gkb}                      & 1994 & 91.2                      & $\frac{1}{\sigma_h} \frac{\dd{\sigma}}{\dd{x_p}}$              & --                        & --                          & --                       & 12                        \\
		\hline
	\end{tabular}
	\caption{
		Same as \cref{t.SIA-data-below-Z-pole}, but for SIA datasets with $\sqrt{s}$ at $Z$-boson mass.
		For SLD (tagged) \cite{SLD:1998coh}, the numbers of data points are provided respectively for $c$-tagged and $b$-tagged hadron production.
		For ALEPH three-jet events measurement \cite{ALEPH:1999udi}, the numbers of data points are listed respectively for hadron in three jets, ordered by jet energies.
	}
	\label{t.SIA-data-at-Z-pole}
\end{table}

\begin{table}[]
	\setcellgapes{2.5 pt}
	\makegapedcells
	\begin{tabular}{|c|c|c|c|c|c|c|c|c|c|}
		\hline
		collaboration                 & year & $\sqrt{s}~\bqty{\mathrm{GeV}}$ & observable                                                                                           & $N_{\mathrm{pt}} (K_0^S)$ & $N_{\mathrm{pt}} (\Lambda)$ & $N_{\mathrm{pt}} (\eta)$ & $N_{\mathrm{pt}} (\pi^0)$ \\
		\hline
		ZEUS \cite{ZEUS:2011cdi}      & 2012 & 318                            & $\frac{1}{N_{\mathrm{DIS}}} \frac{\dd{N_{h}}}{\dd{z_p}}$                                             & 5, 5 \& 2                 & 5, 3 \& 1                   & --                       & --                        \\
		\hline
		CMS \cite{CMS:2011jlm}        & 2011 & 900                            & $\frac{\dd{N_{h}}}{\dd{p_T}} \bigg/ \frac{\dd{N_{K_S^0}}}{\dd{p_T}}$                                 & --                        & 4                           & --                       & --                        \\
		\hline
		STAR \cite{STAR:2009qzv}      & 2010 & 200                            & $\frac{E \dd[3]{\sigma_h}}{\dd[3]{\vv{p}}}$                                                          & --                        & --                          & --                       & 8                         \\
		\hline
		PHENIX \cite{PHENIX:2010hvs}  & 2011 & 200                            & $\frac{E \dd[3]{\sigma_{h}}}{\dd[3]{\vv{p}}} \bigg/ \frac{E \dd[3]{\sigma_{\pi^0}}}{\dd[3]{\vv{p}}}$ & --                        & --                          & 14                       & --                        \\
		\hline
		PHENIX  \cite{PHENIX:2007kqm} & 2007 & 200                            & $\frac{E \dd[3]{\sigma_h}}{\dd[3]{\vv{p}}}$                                                          & --                        & --                          & --                       & 17                        \\
		\hline
		PHENIX \cite{PHENIX:2015fxo}  & 2016 & 510                            & $\frac{E \dd[3]{\sigma_h}}{\dd[3]{\vv{p}}}$                                                          & --                        & --                          & --                       & 22                        \\
		\hline
		ALICE \cite{ALICE:2017nce}    & 2017 & 2760                           & $\frac{E \dd[3]{\sigma_{h}}}{\dd[3]{\vv{p}}} \bigg/ \frac{E \dd[3]{\sigma_{\pi^0}}}{\dd[3]{\vv{p}}}$ & --                        & --                          & 6                        & --                        \\
		\hline
		ALICE \cite{ALICE:2017nce}    & 2017 & 2760                           & $\frac{E \dd[3]{\sigma_h}}{\dd[3]{\vv{p}}}$                                                          & --                        & --                          & --                       & 16                        \\
		\hline
		ALICE \cite{ALICE:2012wos}    & 2012 & 7000                           & $\frac{E \dd[3]{\sigma_{h}}}{\dd[3]{\vv{p}}} \bigg/ \frac{E \dd[3]{\sigma_{\pi^0}}}{\dd[3]{\vv{p}}}$ & --                        & --                          & 4                        & --                        \\
		\hline
		ALICE \cite{ALICE:2012wos}    & 2012 & 7000                           & $\frac{E \dd[3]{\sigma_h}}{\dd[3]{\vv{p}}}$                                                          & --                        & --                          & --                       & 13                        \\
		\hline
		ALICE \cite{ALICE:2017ryd}    & 2018 & 8000                           & $\frac{E \dd[3]{\sigma_{h}}}{\dd[3]{\vv{p}}} \bigg/ \frac{E \dd[3]{\sigma_{\pi^0}}}{\dd[3]{\vv{p}}}$ & --                        & --                          & 13                       & --                        \\
		\hline
		ALICE \cite{ALICE:2017ryd}    & 2018 & 8000                           & $\frac{E \dd[3]{\sigma_h}}{\dd[3]{\vv{p}}}$                                                          & --                        & --                          & --                       & 24                        \\
		\hline
		ALICE \cite{ALICE:2020jsh}    & 2021 & 13000 \& 7000                  & $\frac{\dd{N_{h}^{13~\mathrm{TeV}}}}{\dd{p_T}} \Bigg/ \frac{\dd{N_{h}^{7~\mathrm{TeV}}}}{\dd{p_T}}$  & 10                        & 7                           & --                       & --                        \\
		\hline
		ALICE \cite{ALICE:2020jsh}    & 2021 & 13000                          & $\frac{\dd{N_{h}}}{\dd{p_T}} \bigg/ \frac{\dd{N_{\pi^{\pm}}}}{\dd{p_T}}$                             & 15                        & --                          & --                       & --                        \\
		\hline
		ALICE \cite{ALICE:2024vgi}    & 2024 & 13000                          & $\frac{E \dd[3]{\sigma_{h}}}{\dd[3]{\vv{p}}} \bigg/ \frac{E \dd[3]{\sigma_{\pi^0}}}{\dd[3]{\vv{p}}}$ & --                        & --                          & 14                       & --                        \\
		\hline
	\end{tabular}
	\caption{
		Same as \cref{t.SIA-data-below-Z-pole}, but for SIDIS and $pp$ collisions.
		The kinematic variables are defined in \cref{e.K0S-SIDIS-dsigma-d_zp,e.K0S-pp-dsigma-d_pt,e.eta-pp-dsigma-d_p} and surrounding texts.
		For ZEUS \cite{ZEUS:2011cdi}, the numbers of data points are given respectively for three $Q^2$ bins: $\pqty{160, 640}~\mathrm{GeV}^2$, $\pqty{640, 2560}~\mathrm{GeV}^2$ and $\pqty{2560, 20140}~\mathrm{GeV}^2$.
	}
	\label{t.SIDIS-and-pp-data}
\end{table}

In \cref{t.SIA-data-below-Z-pole,t.SIA-data-at-Z-pole}, the kinematic variables in
\begin{equation}
	\frac{1}{\beta \sigma_h} \frac{\dd{\sigma}}{\dd{x_h}}
	, \quad
	\frac{s}{\beta} \frac{\dd{\sigma}}{\dd{x_h}}
	, \quad
	\frac{1}{\sigma_h} \frac{\dd{\sigma}}{\dd{x_h}}
	, \label{e.K0S-SIA-dsigma-d_xE}
\end{equation}
are defined as $x_h \equiv 2 E_{K_S^0} / \sqrt{s}$, $\beta \equiv p_{K_S^0} / E_{K_S^0}$, and $\sigma_h$ is the total hadronic cross section.
Here $p_{K_S^0}$ and $E_{K_S^0}$ are the momentum and energy of the identified $K_S^0$, $\sqrt{s}$ is the center-of-mass energy of the $e^+ e^-$ collision.

On the other hand, the observables that are differential in $x_p$ or $\xi$:
\begin{equation}
	\frac{1}{\sigma_h} \frac{\dd{\sigma}}{\dd{\xi}}
	, \quad
	\frac{1}{\sigma_h} \frac{\dd{\sigma}}{\dd{x_p}}
	, \quad
	\frac{1}{\sigma_{\mathrm{jet}}} \frac{\dd{\sigma}}{\dd{x_p}}
	, \label{e.K0S-SIA-dsigma-d_xp-and-d_xi}
\end{equation}
are converted to distributions in $x_h$ such that they are directly computed by FMNLO without further hadron mass corrections.
Here $x_p \equiv 2 p_{K_S^0} / \sqrt{s}$, $\xi \equiv - \ln(x_p)$ and $\sigma_{\mathrm{jet}}$ is the total cross section for the three-jet production.
For SIA we also need the kinematic variable $Q^2$ which is defined as the square of invariant mass of the exchanged photon.

The SIDIS and $pp$ datasets for $K_S^0$ are summarized in \cref{t.SIDIS-and-pp-data}.
We include experimental data for SIDIS from ZEUS \cite{ZEUS:2011cdi} for the observable:
\begin{equation}
	\frac{1}{N_{\mathrm{DIS}}} \frac{\dd{N_{K_S^0}}}{\dd{z_p}}
	, \label{e.K0S-SIDIS-dsigma-d_zp}
\end{equation}
where $z_p \equiv 2 p_{K_S^0} / Q$ with $p_{K_S^0}$ being the $K_S^0$ momentum and $Q^2 \equiv -q^2$ being the virtuality.
$N_{\mathrm{DIS}}$ is the number of deep inelastic scattering (DIS) events in the corresponding $Q^2$ and $z_p$ bins.
Finally, $q^2$ is given by $q^2 = \pqty{\ell - \ell'}^2$ in the following process:
\begin{equation}
	e \pqty{\ell} + p \pqty{P}
	\to
	e' \pqty{\ell'} + K_S^0 \pqty{p_{K_S^0}} + X
	,
\end{equation}
where $e$ and $e'$ denote the incoming and scattered electron, respectively, while $p$, $K_S^0$ and $X$ represent the incoming proton beam, outgoing $K_S^0$ and unobserved particles, respectively.
Their momenta are given in the parentheses.
In addition, we also define the Bjorken variable $x_B \equiv Q^2 / \pqty{2 P \cdot q}$ and event inelasticity $y \equiv \pqty{P \cdot q} / \pqty{P \cdot \ell}$.
In the calculation, we used the relation:
\begin{equation}
	\frac{1}{N_{\mathrm{DIS}}} \frac{\dd{N_{K_S^0}}}{\dd{z_p}}
	=
	\frac{1}{\sigma_{\mathrm{DIS}}} \frac{\dd{\sigma_{K_S^0}}}{\dd{z_p}}
	,
\end{equation}
where $\sigma_{\mathrm{DIS}}$ is the DIS cross section.
The SIDIS data are given in 5 different $Q^2$ bins.
To ensure validity of perturbative QCD formulae at NLO, we exclude from our study to the two lowest $Q^2$ bins: $\pqty{10, 40}~\mathrm{GeV}^2$ and $\pqty{40, 160}~\mathrm{GeV}^2$.
This is also consistent with the choice made in the previous NPC analysis for charged hadrons \cite{Gao:2024dbv}.
Furthermore, a separate H1 measurement \cite{H1:1996kfw} of $K_S^0$ production at relatively low $Q$ reveals significant discrepancies between predictions and various MC predictions, indicating the possible influence of mechanisms beyond fragmentation processes.

For the data of $pp$ collisions from ALICE~\cite{ALICE:2020jsh}, in order to avoid dealing with various normalization, we choose to use the ratios of $K_S^0$ production between different center-of-mass energies, or ratios of $K_S^0$ production over $\pi^{\pm}$ production at the same center-of-mass energy:
\begin{equation}
	\frac{\dd{N_{K_S^0}^{13~\mathrm{TeV}}}}{\dd{p_T}} \Bigg/ \frac{\dd{N_{K_S^0}^{7~\mathrm{TeV}}}}{\dd{p_T}}
	, \quad
	\frac{\dd{N_{K_S^0}}}{\dd{p_T}} \bigg/ \frac{\dd{N_{\pi^{\pm}}}}{\dd{p_T}}
	. \label{e.K0S-pp-dsigma-d_pt}
\end{equation}
Here $p_T$ is the transverse momentum of the identified hadron ($K_S^0$ or $\pi^{\pm}$ in this case).
In order to maintain the validity of leading-twist factorization and perturbative convergence, we apply kinematic cuts $x_h > 0.05$ for SIA, $z_p > 0.05$ for SIDIS, and $p_T > 4~\mathrm{GeV}$ for $pp$ collisions, respectively.

\subsection{$\Lambda$ data}

For $\Lambda$ production only the sum of $\Lambda$ and $\overline{\Lambda}$ baryons have been measured.
Thus, our fitted $\Lambda$ FFs should be interpreted as an average of FFs for both $\Lambda$ and $\overline{\Lambda}$.
The summaries of $\Lambda$ production in SIA, SIDIS and $pp$ collisions are given in \cref{t.SIA-data-below-Z-pole,t.SIA-data-at-Z-pole,t.SIDIS-and-pp-data}, respectively.
The kinetic variables for $\Lambda$ production in \cref{t.SIA-data-below-Z-pole,t.SIA-data-at-Z-pole,t.SIDIS-and-pp-data} are similar to those for $K_S^0$ productions, as given in \cref{ss.K0S-data}.
We also apply the same kinematic cut as in the $K^0_S$ case.

For SIA data at $Z$-pole, we include data from DELPHI \cite{DELPHI:1993vpj}, ALEPH \cite{ALEPH:1994fts, ALEPH:1996oqp, ALEPH:1999udi}, OPAL \cite{OPAL:1996gsw} and SLD \cite{SLD:1998coh}.
While for data below $Z$-pole, we include data from SLAC \cite{delaVaissiere:1984xg}, TASSO \cite{TASSO:1981uqa, TASSO:1984nda, TASSO:1988qlu}, HRS \cite{Derrick:1985wd} and CELLO \cite{CELLO:1989adw}.
In particular, the SLD $c$-tagged and $b$-tagged $\Lambda$ production data \cite{SLD:1998coh} are crucial in constraining $c$ and $b$ FFs,
while the ALEPH $\Lambda$-in-jet production data (especially $\Lambda$ produced in the third most energetic jet) \cite{ALEPH:1999udi} can help in pinning down the gluon FFs.
The single-inclusive $\Lambda$ production data from the same ALEPH paper \cite{ALEPH:1999udi}, however, is not included in our fit due to its tension with the ALEPH 1998 measurement \cite{ALEPH:1996oqp}.
Also, the OPAL 1992 data \cite{OPAL:1992asw} is not included in our fit as it is superseded by the OPAL 1997 measurement \cite{OPAL:1996gsw}.
We note that including both the inclusive and heavy-quark tagged data from SLD \cite{SLD:1998coh} is equivalent to including the light-flavor tagged data since the latter is simply difference of the two former measurements.
The DELPHI data at center-of-mass energies $\sqrt{s} = 183~\mathrm{GeV}$ and 189 GeV \cite{DELPHI:2000ahn} are not included in our fit also because they are known to have tension with the other world data \cite{Albino:2005mv, Albino:2008fy, Soleymaninia:2020ahn}.

For SIDIS data from ZEUS \cite{ZEUS:2011cdi}, we exclude the two lowest $Q^2$ bins due to the same reason discussed in \cref{ss.K0S-data}.
In the case of $pp$ collision data, we include the ratio data of $\Lambda$ productions at different center-of-mass energies from ALICE \cite{ALICE:2020jsh}, and the yield ratio of $\Lambda$ over $K_S^0$ at the same center-of-mass energy from CMS \cite{CMS:2011jlm}.
For the data on the ratio of type $\Lambda / K_S^0$, special treatment has to be given, since we are not fitting the $K_S^0$ and $\Lambda$ FFs simultaneously.
In particular, we included a fully correlated uncertainty of 5\% in the ratio data of the type $\Lambda / K_S^0$ so as to properly propagate the uncertainty from $K_S^0$ FFs.
Due to the substantial discrepancy between measurements and event generator simulations observed in \cite{ALICE:2020jsh, CMS:2011jlm}, we have chosen to exclude the $\Lambda$ to $K_S^0$ production data from ALICE 2021 at 13 TeV \cite{ALICE:2020jsh}, as well as the CMS 2011 data at 7 TeV \cite{CMS:2011jlm}, from our fit.

\subsection{$\eta$ data}
\label{ss.eta-data}

We have gathered a comprehensive list of datasets from both SIA and $pp$ measurements for the global analysis of $\eta$ FFs, which are summarized in \cref{t.SIA-data-below-Z-pole,t.SIA-data-at-Z-pole} and \cref{t.SIDIS-and-pp-data}, respectively.
The kinetic variables presented in \cref{t.SIA-data-below-Z-pole,t.SIA-data-at-Z-pole} for $\eta$ productions are similar to those for the $K_S^0$ productions, as given in \cref{ss.K0S-data}.
We also apply the same kinematic cut as in the $K^0_S$ case.

In the context of SIA measurements, most of the data was collected at the $Z$-pole energy, with significant contributions from the ALEPH \cite{ALEPH:1992zhm, ALEPH:1999udi, ALEPH:2001tfk}, L3 \cite{L3:1992pbe, L3:1994gkb} and OPAL \cite{OPAL:1998enc} collaborations.
As noted in the previous sections, $\eta$-in-jet measurements from ALEPH \cite{ALEPH:1999udi} can provide additional constraints on the gluon fragmentation functions.
To complement the high-energy measurements, we have included lower-energy data from the JADE \cite{JADE:1985bzp, JADE:1989ewf}, CELLO \cite{CELLO:1989byk}, HRS \cite{HRS:1987aky} and ARGUS \cite{ARGUS:1989orf} experiments.
These low-energy datasets, particularly the ARGUS data (with a center-of-mass energy of 9.46 GeV), are significant for flavor separation, as ARGUS operates below the threshold for $b$ quark production.
The $\eta$ production measurements reported by Belle \cite{Belle:2024vua} have been excluded from our analysis due to the observed discrepancies between Belle's results and other global datasets, as discussed in \cref{ss.Belle-impact}.

For $pp$ collisions, high-energy data are collected primarily by the ALICE collaboration \cite{ALICE:2012wos, ALICE:2017nce, ALICE:2017ryd} at center-of-mass energies of 2.76, 7, and 8 TeV.
Additionally, the PHENIX collaboration \cite{PHENIX:2010hvs} provides lower-energy measurements at 200 GeV, which are also included in our fits.
The observables from PHENIX \cite{PHENIX:2010hvs} and ALICE \cite{ALICE:2012wos, ALICE:2017nce, ALICE:2017ryd} are all given in ratios
\begin{equation}
	\frac{E \dd[3]{\sigma_{\eta}}}{\dd[3]{\vec{p}}} \bigg/ \frac{E \dd[3]{\sigma_{\pi^0}}}{\dd[3]{\vec{p}}}
	, \label{e.eta-pp-dsigma-d_p}
\end{equation}
where $E$ and $\vv{p}$ are the energy and three momentum of the identified hadron, respectively.
The predictions for $\pi^0$ production are calculated using the $\pi^0$ FFs constructed from $\pi^{\pm}$ FFs determined in the previous NPC analysis \cite{Gao:2024dbv} via isospin symmetry (see \cref{e.pi_0-construction-u-and-d,e.pi_0-construction-unfavored}).
The Hessian uncertainties of the constructed $\pi^0$ FFs are very small and can be omitted.

\subsection{$\pi^0$ data}

We have gathered a comprehensive dataset on $\pi^0$ production to validate the $\pi^0$ FFs derived from NPC23 $\pi^{\pm}$ FFs~\cite{Gao:2024dbv}.
These data are summarized in \cref{t.SIA-data-below-Z-pole,t.SIA-data-at-Z-pole} and \cref{t.SIDIS-and-pp-data}.
In the context of SIA, we have incorporated measurements at the $Z$-pole from ALEPH \cite{ALEPH:1996pxg, ALEPH:1999udi}, DELPHI \cite{DELPHI:1995ase} L3 \cite{L3:1994gkb}, and OPAL \cite{OPAL:1998enc}.
Additionally, we included jet measurements from ALEPH \cite{ALEPH:1999udi} and $b$-tagged data from DELPHI \cite{DELPHI:1995ase}, which contribute to flavor separation as mentioned previously.
The observables are normalized by the total hadronic cross section, expressed as $1 / \sigma_h \times \dd{\sigma} / \dd{x_p}$, or by the jet cross section in the case of jet measurements, represented as $1 / \sigma_{\mathrm{jet}} \times \dd{\sigma} / \dd{x_p}$.
We also included low-energy SIA measurements from JADE \cite{JADE:1985bzp,JADE:1989ewf}, CELLO \cite{Elze:1989gm}, TASSO \cite{TASSO:1988jma}, and ARGUS \cite{ARGUS:1989orf}, spanning center-of-mass energies from 9.46 GeV to 44 GeV.
The observables in this range are mostly given as $s / \beta \times \dd{\sigma} / \dd{x_h}$, except for TASSO, which provides a normalized cross section.
The kinematic variables $\beta$, $x_h$ and $x_p$ the similar to those in the $K_S^0$ case, defined in \cref{ss.K0S-data}.

For $pp$ collisions, we have included data from the ALICE collaboration \cite{ALICE:2017nce, ALICE:2012wos, ALICE:2017ryd} at center-of-mass energies of 2.76, 7 and 8 TeV, as well as low-energy measurements from STAR \cite{STAR:2009qzv} and PHENIX \cite{PHENIX:2007kqm} at 200 GeV, and PHENIX \cite{PHENIX:2015fxo} at 510 GeV.
The relevant observable for $pp$ collisions is the invariant cross section, $E \dd[3]{\sigma} / \dd[3]{\vv{p}}$, with $E$ and $\vv{p}$ being the energy and three momentum of the identified $\pi^0$, respectively.

\clearpage
\section{Theoretical inputs to the NPC23 analysis}
\label{s.theory}

In this section, we outline the parameterization of the NPC23 fit and summarize the details of theoretical computations.

\subsection{Parametrization}
\label{ss.parameterization}

The master formula for the parameterization for FFs is given by
\begin{equation}
	z D_i^h \pqty{z, Q_0}
	=
	z^{\alpha_i^h} \pqty{1-z}^{\beta_i^h}
	\exp(\sum_{n=0}^m a^h_{i,n} z^{n/2})
	, \label{e.parameterization}
\end{equation}
where the subscript $i$ labels different parton flavors, the superscript $h$ labels different hadron species. By varying $m = 0, 1, 2 \cdots$, one can freely choose different parameterizations.
In practice, we increase the value of $m$ until no discernible improvement in fit quality can be obtained.
$\Bqty{\alpha_i^h, \beta_i^h, a^h_{i,n}}$ are the fitted parameters, and the initial scale is chosen as $Q_0 = 5~\mathrm{GeV}$.
The initial scale is chosen to satisfy two key requirements: (1) being above the bottom quark mass threshold so that the $b$ distribution can be parameterized under fixed flavor number scheme with $n_f = 5$, and (2) remaining below most experimental hard scales used in our global fit to avoid backward evolution of FFs.
We use a zero-mass scheme for heavy quarks and a number of flavors $n_f = 5$.

The parameterization choice for $K_S^0$ is summarized in \cref{t.K0S-parameterization}.
In total, there are 20 free parameters.
The parameters for quark FFs are made symmetric between quark and antiquark due to charge conjugation symmetry, which requires $D_q^{K_S^0} = D_{\overline{q}}^{K_S^0}$ (notice that the antiparticle of $K_S^0$ is itself).
The same choice is made in BKK96 \cite{Binnewies:1995kg}, AKK05 \cite{Albino:2005mv}, AKK08 \cite{Albino:2008fy} and SAK20 \cite{Soleymaninia:2020ahn}.
SIA data sets alone cannot differentiate fragmentation from $d$ and $s$ quarks.
The SIDIS data on $K_S^0$ production have poor quality and also show a slight preference for a larger distribution of the $d$ quark than that of the $s$ quark.
This is counter-intuitive since the fragmentation of a $d$ quark to kaons requires creation of a pair of $s \overline{s}$ from vacuum, which is more difficult due to the larger mass of the strange quark.
We therefore included an empirical penalty term for $K_S^0$ parameters:
\begin{equation}
	\pqty{D_{s}^{K_S^0} / D_{d}^{K_S^0} - 3}^2
	, \label{e.chi_2-penalty-K0S}
\end{equation}
into the $\chi^2$ to enforce the ratio of fragmentation probabilities from $s$ and $d$ quarks to be centered around 3.
This is based on the finding that the normalization of $s$ fragmenting into $K^{\pm}$ is approximately three time larger than that for $u$, as studied in \cite{Gao:2024nkz, Moffat:2021dji, Khalek:2021gxf}.
Furthermore, due to insufficient constraints provided by the data, we have imposed identical shape parameters ($\alpha$, $\beta$ and $a_1$) between $s$ and $d$, while setting $\alpha$ parameter of $d$ to be the same as that of $u$.

\begin{table}[!h]
	\setcellgapes{2.5 pt}
	\makegapedcells
	\begin{tabular}{|c|c|c|c|c|c|}
		\hline
		flavor             & favored   & $a_0$     & $\alpha$     & $\beta$     & $a_1$       \\
		\hline
		$u = \overline{u}$ & \ding{55} & \ding{51} & \ding{51}    & \ding{51}   & \ding{51}   \\
		\hline
		$d = \overline{d}$ & \ding{51} & \ding{51} & $=\alpha_u$  & \ding{51}   & \ding{51}   \\
		\hline
		$s = \overline{s}$ & \ding{51} & \ding{51} & $= \alpha_u$ & $= \beta_d$ & $= a_{1,d}$ \\
		\hline
		$c = \overline{c}$ & \ding{55} & \ding{51} & \ding{51}    & \ding{51}   & \ding{51}   \\
		\hline
		$b = \overline{b}$ & \ding{55} & \ding{51} & \ding{51}    & \ding{51}   & \ding{51}   \\
		\hline
		$g$                & \ding{55} & \ding{51} & \ding{51}    & \ding{51}   & \ding{51}   \\
		\hline
	\end{tabular}
	\caption{
		Summary of the parameterization for $K_S^0$ FFs.
		Check marks under ``favored'' represent the flavor is favored, while cross marks denote unfavored.
		Check marks under each parameter mean the parameter is free to vary.
		There are in total 20 free parameters for $K_S^0$ FFs.
	}
	\label{t.K0S-parameterization}
\end{table}

In the context of $\Lambda$ baryon, since the data are reconstructed from averaging between $\Lambda$ and its antiparticle $\overline{\Lambda}$, the parameters are also made symmetric between quarks and antiquarks, with the gluon parameters still being independent.
Moreover, the quark constituents of $\Lambda$ is $u d s$, and the masses of $u$ and $d$ are much smaller than that of $\Lambda$ which is 1.115 GeV, we assume that the $u$ and $d$ FFs are identical.
Finally, in the absence of sufficient constraints from existing world data, we fix the parameter $a_1$ to zero for the FFs of $c$ and $b$ quarks.
The parameterization of $\Lambda$ FFs is summarized in \cref{t.Lambda-parameterization}.

\begin{table}[!h]
	\setcellgapes{2.5 pt}
	\makegapedcells
	\begin{tabular}{|c|c|c|c|c|c|}
		\hline
		flavor                                & favored   & $a_0$     & $\alpha$  & $\beta$   & $a_1$     \\
		\hline
		$u = \overline{u} = d = \overline{d}$ & \ding{51} & \ding{51} & \ding{51} & \ding{51} & \ding{51} \\
		\hline
		$s = \overline{s}$                    & \ding{51} & \ding{51} & \ding{51} & \ding{51} & \ding{51} \\
		\hline
		$c = \overline{c}$                    & \ding{55} & \ding{51} & \ding{51} & \ding{51} & \ding{55} \\
		\hline
		$b = \overline{b}$                    & \ding{55} & \ding{51} & \ding{51} & \ding{51} & \ding{55} \\
		\hline
		$g$                                   & \ding{55} & \ding{51} & \ding{51} & \ding{51} & \ding{51} \\
		\hline
	\end{tabular}
	\caption{
		Same table as \cref{t.K0S-parameterization}, but for $\Lambda$ FFs.
		There are in total 18 free parameters for $\Lambda$.
	}
	\label{t.Lambda-parameterization}
\end{table}

For the $\eta$ meson, which is its own antiparticle, we set the antiquark FFs to be equal to the corresponding quark FFs in accordance with the charge conjugation symmetry, \textit{i.e.}, $D_q^{\eta} = D_{\overline{q}}^{\eta}$.
Since $\eta$ FFs are only weakly constrained, we set the shape parameters ($\alpha$, $\beta$ and $a_1$) of $s$ to be equal to those of $u$, leaving the normalization parameter $a_0$ to vary.
Again, the gluon parameters are independent from those of the quarks.
Additionally, since no flavor tagged data is available for $\eta$, we also set $D_c^{\eta} = D_b^{\eta}$.
The parameterization for $\eta$ FFs is summarized in \cref{t.eta-parameterization}.

\begin{table}[hbtp]
	\setcellgapes{2.5 pt}
	\makegapedcells
	\begin{tabular}{|c|c|c|c|c|c|}
		\hline
		flavor                                & favored   & $a_0$     & $\alpha$     & $\beta$     & $a_1$       \\
		\hline
		$u = \overline{u} = d = \overline{d}$ & \ding{51} & \ding{51} & \ding{51}    & \ding{51}   & \ding{51}   \\
		\hline
		$s = \overline{s}$                    & \ding{51} & \ding{51} & $= \alpha_u$ & $= \beta_u$ & $= a_{1,u}$ \\ \hline
		$c = \overline{c} = b = \overline{b}$ & \ding{55} & \ding{51} & \ding{51}    & \ding{51}   & \ding{55}   \\
		\hline
		$g$                                   & \ding{55} & \ding{51} & \ding{51}    & \ding{51}   & \ding{51}   \\
		\hline
	\end{tabular}
	\caption{Same as \cref{t.K0S-parameterization}, but for $\eta$ FFs.
		The parameters $\alpha$, $\beta$ and $a_1$ of $s$ are set to be equal to those of $u$, while $a_0$ is free to vary.
		There are in total 12 free parameters for $\eta$.
	}
	\label{t.eta-parameterization}
\end{table}

\subsection{Theoretical computation method}
\label{ss.theory-computation-method}

The theoretical computation techniques have been summarized in the previous NPC23 work \cite{Gao:2024dbv}, here we only recall the key elements.
In the NPC framework, the two-loop timelike splitting kernels are used to evolve the fragmentation functions to higher scales, matching the NLO accuracy.
The splitting functions are given in \cite{Stratmann:1996hn} and are implemented with HOPPET \cite{Salam:2008qg, Salam:2008sz}.
The hard kernels are computed with FMNLOv2.0 package \cite{Liu:2023fsq, Gao:2024dbv, FMNLO} which implements processes including SIA, SIDIS and $pp$ collisions.
We adopt a zero-mass scheme for heavy quarks, with number of flavors $n_f = 5$ and $\alpha_s \pqty{m_Z} = 0.118$ throughout the fit.
Here $m_Z = 91.18~\mathrm{GeV}$ is the mass of $Z$-boson, and $\alpha_s$ denotes the strong coupling.

The CT14 NLO parton distribution functions (PDFs) \cite{Dulat:2015mca} are used in the generation of hard kernels for processes involving initial-state hadrons, \textit{i.e.}, SIDIS and $pp$ collisions.
The central factorization, renormalization and fragmentation scales ($\mu_{F,0}$, $\mu_{R,0}$ and $\mu_{D,0}$) are chosen to be the transferred momentum $Q$ for SIA and SIDIS, and to be transverse momentum of the identified final-state hadron for $pp$ collisions, when it is applicable as summarized in \cref{t.scales}.
In the case of SIA jet production the central renormalization and fragmentation scales are set to the transferred momentum $Q$ and the energy $E_j$ of individual jet respectively.
The scale variations are estimated by the half width of the envelope of theoretical predictions based on 9 scale combinations: $\mu_F/\mu_{F,0}=\mu_R/\mu_{R,0}=\{1/2,~1,~2\}$ and $\mu_D/\mu_{D,0}=\{1/2,~1,~2\}$.
The covariance matrix of $\chi^2$ calculations incorporates scale variations as theoretical uncertainties, assumed to be fully correlated among points within each subset of the data.

\begin{table}[!h]
	\setcellgapes{2.5 pt}
	\makegapedcells
	\begin{tabular}{|c|c|c|c|c|}
		\hline
		reaction    & SIA & SIDIS & $pp$  & SIA jet \\
		\hline
		$\mu_{F,0}$ & --  & $Q$   & $p_T$ & --      \\
		\hline
		$\mu_{R,0}$ & $Q$ & $Q$   & $p_T$ & $Q$     \\
		\hline
		$\mu_{D,0}$ & $Q$ & $Q$   & $p_T$ & $E_j$   \\
		\hline
	\end{tabular}
	\caption{
		Central scale choices for single inclusive hadron production in SIA, SIDIS, and $pp$ collisions, and hadron-in-jet production in SIA.
	}
	\label{t.scales}
\end{table}

\subsection{Goodness of fit function and the covariance matrix}

We follow~\cite{Stump:2001gu, Pumplin:2002vw} and quantify the goodness of fit with the likelihood function:
\begin{equation}
	\chi^2 \pqty{\Bqty{\alpha, \beta, a_i}, \Bqty{\lambda}}
	=
	\sum_{e,j} \frac{1}{s_{e,j}^2} \pqty{D_{e,j} - T_{e,j} - \sum_{k = 1}^{N_e^{\pqty{\beta}}} \beta_{e,j,k} \lambda_{e,k}}^2
	+
	\sum_e \sum_{k = 1}^{N_e^{\pqty{\beta}}} \lambda_{e,k}^2
	,
\end{equation}
where $\Bqty{\alpha, \beta, a_i}$ and $\Bqty{\lambda}$ are the set of fitted parameters and nuisance parameters, respectively.
The indices $e$, $j$ and $k$ represent the experiment, data point, and various sources of correlated uncertainties, respectively.
We also denote the number of correlated uncertainties in experiment $e$ as $N_e^{\pqty{\beta}}$.
The nuisance parameter $\lambda_{e,k}$ combined with $\beta_{e,j,k}$ controls the amount of additive shift coming from the $k$th correlated uncertainty at the $j$th data point in the experiment $e$.
Furthermore, $D_{e,j}$ denotes the $j$th data value of the experiment $e$, and $T_{e,j}$ and $s_{e,j}$ are its corresponding theoretical prediction and uncorrelated uncertainty.

In practice, the values of $\Bqty{\lambda}$ are determined by minimizing the likelihood function $\chi^2 \pqty{\Bqty{\alpha, \beta, a_i}, \Bqty{\lambda}}$ with respect to $\Bqty{\lambda}$, which leads to the profiled $\chi^2$ function \cite{Stump:2001gu}:
\begin{equation}
	\chi^2 \pqty{\Bqty{\alpha, \beta, a_i}, \Bqty{\hat{\lambda}}}
	=
	\sum_e \sum_{j,j'} \pqty{D_{e,j} - T_{e,j}} \mathcal{C}_{e,jj'}^{-1} \pqty{D_{e,j'} - T_{e,j'}}
\end{equation}
where $\hat{\lambda}$ are the best fit nuisance parameters and $\mathcal{C}_e^{-1}$ is the inverse of the covariance matrix $\mathcal{C}_e$:
\begin{equation}
	\mathcal{C}_{e,jj'}
	\equiv
	s_{e,j}^2 \delta_{jj'}
	+
	\sum_{k = 1}^{N_e^{\pqty{\beta}}} \beta_{e,j,k} \beta_{e,j',k}
	, \quad
	\mathcal{C}_{e,jj'}^{-1}
	=
	\frac{\delta_{jj'}}{s_{e,j}^2}
	-
	\sum_{k, k' = 1}^{N_e^{\pqty{\beta}}} \frac{\beta_{e,j,k}}{s_{e,j}^2} A_{e,kk'}^{-1} \frac{\beta_{e,j',k'}}{s_{e,j'}^2}
	,
\end{equation}
and $A_{kk'}$ is defined as \cite{Stump:2001gu}:
\begin{equation}
	A_{e,kk'}
	\equiv
	\delta_{kk'}
	+
	\sum_j \frac{\beta_{e,j,k} \beta_{e,j,k'}}{s_j^2}
	.
\end{equation}
Note that in this analysis we have included correlated uncertainties due to normalizations in experimental measurements as well as scale variations of the NLO calculations.

\clearpage
\section{Quality of fit to data}
\label{s.quality-of-fit}

In this section, we present the predictions of FMNLO-nominated FFs compared to the experimental data.
The quality of fit to $K_S^0$, $\Lambda$ and $\eta$ data is discussed in dedicated sections.
Furthermore, the quality of predictions for the $\pi^0$ data using the FFs constructed from the $\pi^{\pm}$ FFs via isospin symmetry is also discussed.
Overall, a good description is found for all hadron productions, with a few exceptions that will be pointed out.

\subsection{$K_S^0$ production}
\label{ss.K0S-fit-quality}

\begin{figure}[htbp]
	\centering
	\includegraphics[width = 0.9 \textwidth]{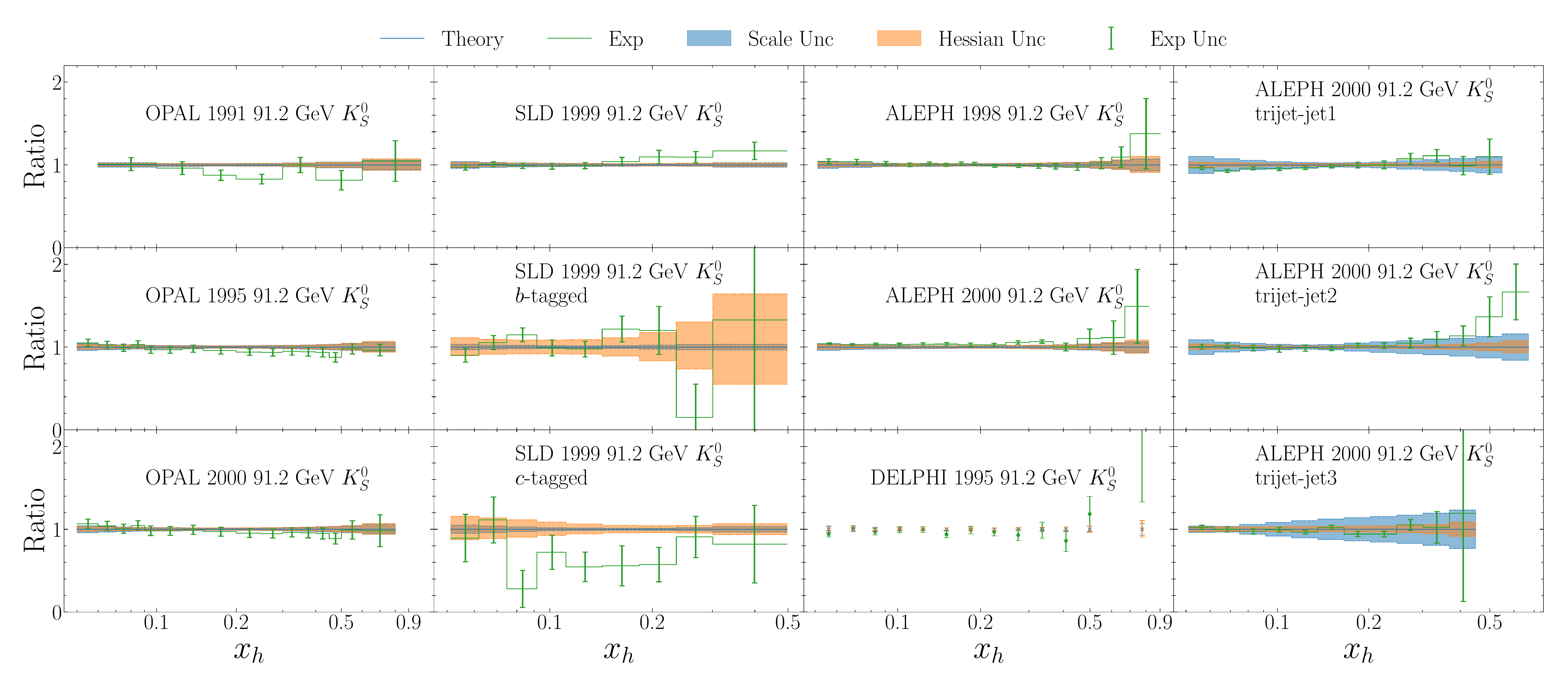}
	\caption{
		The data over theory plot for $K_S^0$ production in SIA at $Z$-pole energy.
		The $x$-axis is plotted in logarithmic scale for $x_h$ defined in \cref{ss.K0S-data}.
		The data and experimental uncertainties are shown in green error bars, while the scale and Hessian uncertainties are shown in blue and orange bands, respectively.
		The data, scale and Hessian uncertainties are normalized to our best-fit theory values.
	}
	\label{f.SIA-Z_pole-K0S}
\end{figure}

In \cref{t.chi_2-K0S}, we present the $\chi^2$ values together with the corresponding $\chi^2/N_{\mathrm{pt}}$ values for each experimental dataset, comparing our best-fit predictions with the experimental measurements.
Here $N_{\mathrm{pt}}$ denotes the number of data points.
Furthermore, we provide the sum of the $\chi^2$ values for the SIA, SIDIS, $pp$ collisions, and the total fitted datasets to give an overall perspective on the fit quality across these different categories.
The global fit to all datasets yields a total $\chi^2$ of 327.93 for 349 data points, giving a $\chi^2/N_{\mathrm{pt}}$ value of 0.94, signifying solid agreement between the theoretical predictions and the measured data.
Additionally, the $\chi^2 / \mathrm{d.o.f.}$ for the $K_S^0$ fit is 1.00 with the degree of freedom $\mathrm{d.o.f.} \equiv N_{\mathrm{pt}} - N_{\mathrm{par}}$ and $N_{\mathrm{par}} = 20$ is the number of free parameters for the $K_S^0$ FFs.
For the SIA datasets, the overall $\chi^2/N_{\mathrm{pt}}$ is 0.99 for 312 data points, showing a generally good match between theory and experiment.
However, some datasets show larger discrepancies, such as the TASSO 1990 dataset \cite{TASSO:1989jyt} at 42.6 GeV with $\chi^2/N_{\mathrm{pt}}$ of 2.46 and HRS 1987 dataset \cite{Derrick:1985wd} at 29 GeV with $\chi^2/N_{\mathrm{pt}}$ of 3.03.
While the HRS data was generally not well fitted in previous analyses \cite{Albino:2008fy, Soleymaninia:2020ahn, Li:2024etc}, the TASSO 42.6 GeV data was well fitted in AKK05 \cite{Albino:2005mv}, AKK08 \cite{Albino:2008fy}, SAK20 \cite{Soleymaninia:2020ahn} and FF24 \cite{Soleymaninia:2024jam}.
Additionally, the SLD $c$-tagged data \cite{SLD:1998coh} has a relatively large $\chi^2 / N_{\mathrm{pt}} = 2.58$ and this is due to the few outliers of the data (see \cref{f.SIA-Z_pole-K0S}).
For the SIDIS data, the agreement between theory and data is also very good, showing a $\chi^2/N_{\mathrm{pt}}$ of 0.86 over 12 data points.
The $pp$ collision data also aligns well between theory and experiment, with a $\chi^2/N_{\mathrm{pt}}$ of 0.37 over 25 data points.
In general, most datasets are well described by theoretical predictions, with the majority of $\chi^2/N_{\mathrm{pt}}$ values close to 1, indicating a successful fit, despite a few cases where larger deviations arise.

\begin{table}[]
	\setcellgapes{2.5 pt}
	\makegapedcells
	\begin{tabular}{|c|c|c|c|c|c|}
		\hline
		collaboration                                                                                & year & $\sqrt{s}~\bqty{\mathrm{GeV}}$ & $\chi^2$        & $N_{\mathrm{pt}}$ & $\chi^2 / N_{\mathrm{pt}}$ \\
		\hline
		Belle \cite{Belle:2024vua}                                                                   & 2024 & 10.58                          & 14.9            & 35                & 0.43                       \\
		TASSO \cite{TASSO:1984nda}                                                                   & 1985 & 14                             & 5.12            & 9                 & 0.57                       \\
		TASSO \cite{TASSO:1984nda}                                                                   & 1985 & 22                             & 5.54            & 6                 & 0.92                       \\
		TASSO \cite{TASSO:1984nda}                                                                   & 1985 & 34                             & 14.86           & 13                & 1.14                       \\
		TASSO \cite{TASSO:1989jyt}                                                                   & 1990 & 14.8                           & 10.89           & 9                 & 1.21                       \\
		TASSO \cite{TASSO:1989jyt}                                                                   & 1990 & 21.5                           & 3.40            & 6                 & 0.57                       \\
		TASSO \cite{TASSO:1989jyt}                                                                   & 1990 & 34.5                           & 19.54           & 13                & 1.50                       \\
		TASSO \cite{TASSO:1989jyt}                                                                   & 1990 & 35                             & 13.52           & 13                & 1.04                       \\
		TASSO \cite{TASSO:1989jyt}                                                                   & 1990 & 42.6                           & 32.01           & 13                & 2.46                       \\
		TPC \cite{TPCTwoGamma:1984eoj}                                                               & 1984 & 29                             & 3.02            & 8                 & 0.38                       \\
		MARK II \cite{Schellman:1984yz}                                                              & 1985 & 29                             & 11.06           & 17                & 0.65                       \\
		HRS \cite{Derrick:1985wd}                                                                    & 1987 & 29                             & 36.33           & 12                & 3.03                       \\
		CELLO \cite{CELLO:1989adw}                                                                   & 1990 & 35                             & 3.48            & 9                 & 0.39                       \\
		TOPAZ \cite{TOPAZ:1994voc}                                                                   & 1995 & 58                             & 0.18            & 4                 & 0.05                       \\
		OPAL \cite{OPAL:1991ixp}                                                                     & 1991 & 91.2                           & 8.03            & 7                 & 1.15                       \\
		OPAL \cite{OPAL:1995ebr}                                                                     & 1995 & 91.2                           & 12.97           & 16                & 0.81                       \\
		OPAL \cite{OPAL:2000dkf}                                                                     & 2000 & 91.2                           & 8.98            & 16                & 0.56                       \\
		ALEPH \cite{ALEPH:1996oqp}                                                                   & 1998 & 91.2                           & 7.22            & 16                & 0.45                       \\
		ALEPH \cite{ALEPH:1999udi}                                                                   & 2000 & 91.2                           & 13.24           & 14                & 0.95                       \\
		ALEPH jet 1 \cite{ALEPH:1999udi}                                                             & 2000 & 91.2                           & 14.85           & 12                & 1.24                       \\
		ALEPH jet 2 \cite{ALEPH:1999udi}                                                             & 2000 & 91.2                           & 9.34            & 13                & 0.72                       \\
		ALEPH jet 3 \cite{ALEPH:1999udi}                                                             & 2000 & 91.2                           & 9.28            & 11                & 0.84                       \\
		DELPHI \cite{DELPHI:1994qgk}                                                                 & 1995 & 91.2                           & 8.12            & 13                & 0.62                       \\
		SLD \cite{SLD:1998coh}                                                                       & 1999 & 91.2                           & 7.61            & 9                 & 0.85                       \\
		SLD $c$-tagged \cite{SLD:1998coh}                                                            & 1999 & 91.2                           & 23.23           & 9                 & 2.58                       \\
		SLD $b$-tagged \cite{SLD:1998coh}                                                            & 1999 & 91.2                           & 11.27           & 9                 & 1.30                       \\
		\hline
		SIA sum                                                                                      &      &                                & 308.43          & 312               & 0.99                       \\
		\hline
		ZEUS $Q^2 \in \pqty{160, 640} \, \mathrm{GeV}^2$ \cite{ZEUS:2011cdi}                         & 2012 & 318                            & 4.38            & 5                 & 0.88                       \\
		ZEUS $Q^2 \in \pqty{640, 2560} \, \mathrm{GeV}^2$ \cite{ZEUS:2011cdi}                        & 2012 & 318                            & 2.97            & 5                 & 0.59                       \\
		ZEUS $Q^2 \in \pqty{2560, 10240} \, \mathrm{GeV}^2$ \cite{ZEUS:2011cdi}                      & 2012 & 318                            & 3.00            & 2                 & 1.50                       \\
		\hline
		SIDIS sum                                                                                    &      &                                & 10.35           & 12                & 0.86                       \\
		\hline
		ALICE $\pqty{N_{K_S^0}^{13~\mathrm{TeV}} / N_{K_S^0}^{7~\mathrm{TeV}}}$ \cite{ALICE:2020jsh} & 2021 & 13000 \& 7000                  & 3.19            & 10                & 0.32                       \\
		ALICE $\pqty{N_{K_S^0} / N_{\pi^{\pm}}}$ \cite{ALICE:2020jsh}                                & 2021 & 13000                          & 5.96            & 15                & 0.40                       \\
		\hline
		$pp$ sum                                                                                     &      &                                & 9.15            & 25                & 0.37                       \\
		\hline
		\textbf{total sum}                                                                           &      &                                & \textbf{327.93} & \textbf{349}      & \textbf{0.94}              \\
		\hline
	\end{tabular}
	\caption{
		The $\chi^2$, number of data points ($N_{\mathrm{pt}}$) and $\chi^2 / N_{\mathrm{pt}}$ for the global datasets of $K_S^0$ production.
		Additional details, including information on collaboration, the year of publication, and center-of-mass energy, are also provided.
		Sum of $\chi^2$ is also given for SIA, SIDIS and $pp$ collision subgroups, as well as for grand total of all datasets.
	}
	\label{t.chi_2-K0S}
\end{table}

\cref{f.SIA-Z_pole-K0S} presents comparisons between theoretical predictions and experimental data for $K_S^0$ production in SIA at the $Z$-pole, including measurements from OPAL \cite{OPAL:1991ixp, OPAL:1995ebr, OPAL:2000dkf}, SLD \cite{SLD:1998coh}, ALEPH \cite{ALEPH:1996oqp, ALEPH:1999udi} and DELPHI \cite{DELPHI:1994qgk} collaborations.
The figure presents twelve panels with all results normalized to our best-fit predictions. Scale uncertainties and Hessian uncertainties are represented by blue and orange bands, respectively.
The OPAL measurements from 1991 \cite{OPAL:1991ixp}, 1995 \cite{OPAL:1995ebr} and 2000 \cite{OPAL:2000dkf} show agreement with theoretical predictions throughout the $x_h$ range, with some fluctuations in the intermediate $x_h$ region, but generally within experimental uncertainties.
For SLD measurements \cite{SLD:1998coh}, three different datasets are presented: inclusive, $c$-tagged and $b$-tagged events.
The $c$-tagged and $b$-tagged measurements show larger Hessian uncertainties compared to the inclusive case, reflecting the relatively weak constraint of the data caused by the large experimental uncertainties.
In the low $x_h$ region, the SLD $c$-tagged data exhibit larger uncertainties than their $b$-tagged counterparts, resulting in correspondingly greater Hessian uncertainties in theory values of the fit.
This is further supported by the observation that the $c$-quark FF has a $\sim$30\% relative uncertainty at $x_h \approx 0.06$, compared to $\sim$20\% for the $b$-quark FF (see \cref{f.FFs-K0S}).
On the other hand, in the intermediate-$x_h$ region, the $c$-quark FF shows smaller Hessian uncertainties since the SLD $c$-tagged data there are more precise than the $b$-tagged data, as well as because of the stronger indirect constraints from inclusive data.
The ALEPH data from 1998 \cite{ALEPH:1996oqp} and 2000 \cite{ALEPH:1999udi} are well described by the theory.
Although a slightly upward trend in the high-$x_h$ region is observed, it remains within the experimental uncertainties.
The DELPHI 1995 \cite{DELPHI:1994qgk} data points are notably consistent with theoretical predictions, except for the last point which has large experimental uncertainty.
The last column of \cref{f.SIA-Z_pole-K0S} shows the prediction of ALEPH three-jet data \cite{ALEPH:1999udi}, respectively, for jet 1, 2, and 3.
It is noticed that the scale uncertainties now become discernible. This is attributed to the fact that in the jet production process, $\alpha_s$ enters at tree level, and thus scale variation via running of the strong coupling constant can contribute at leading order (LO).

\begin{figure}[htbp]
	\centering
	\includegraphics[width = 0.9 \textwidth]{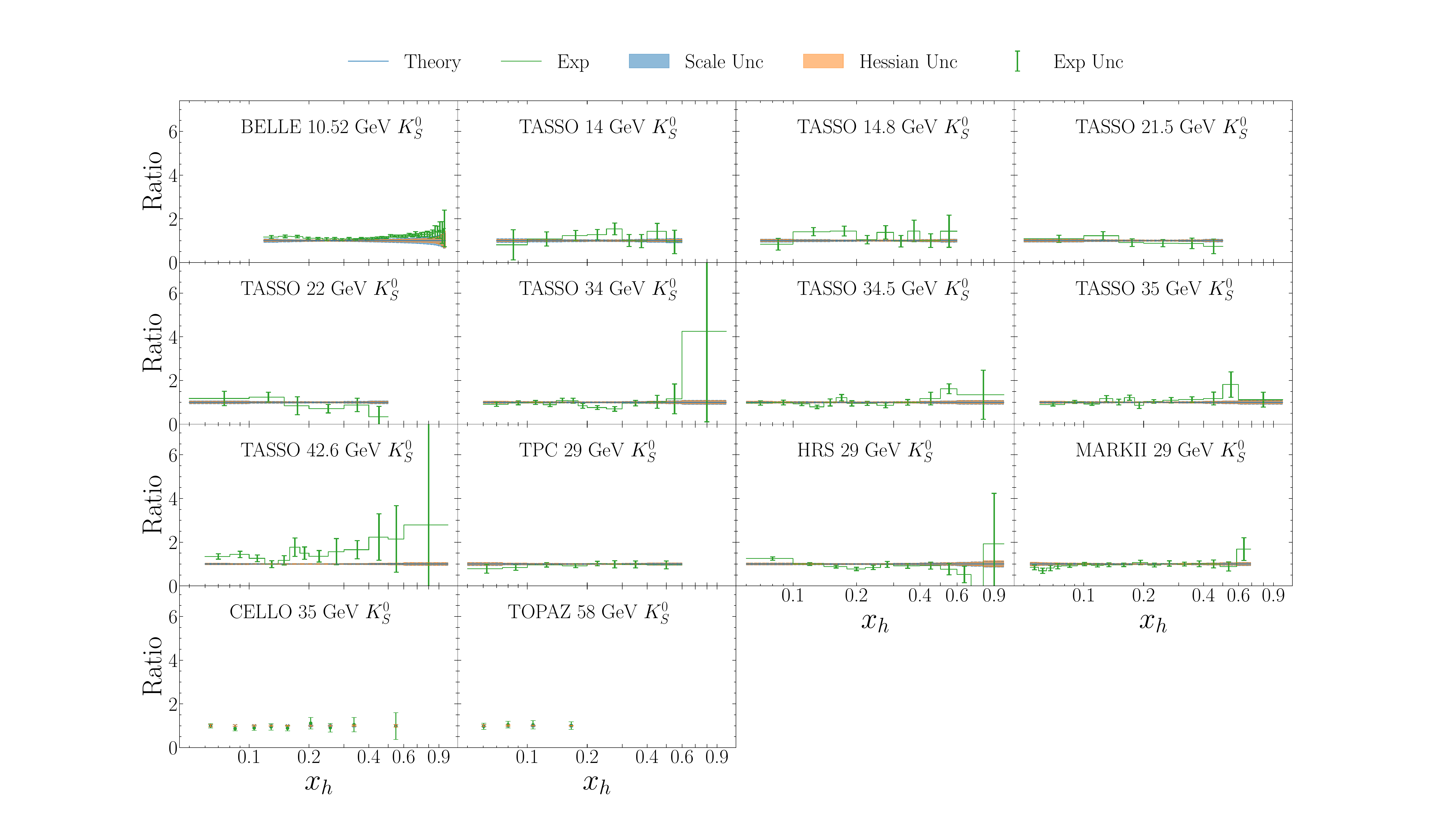}
	\caption{
		Same as \cref{f.SIA-Z_pole-K0S} but for $K_S^0$ production in SIA below $Z$-pole energy.
	}
	\label{f.SIA-below-Z_pole-K0S}
\end{figure}

\cref{f.SIA-below-Z_pole-K0S} shows a comparison between theoretical predictions and experimental data for $K_S^0$ production in SIA with a center-of-mass energy below $Z$ mass, with all results normalized to our best-fit predictions.
The figure consists of fourteen panels, among them the first two rows (and the first panel in the third row) show Belle \cite{Belle:2024vua} and TASSO \cite{TASSO:1984nda, TASSO:1989jyt} data-theory agreement with center-of-mass energies ranging from 10.52 to 42.6 GeV, while the remaining panels present TPC \cite{TPCTwoGamma:1984eoj}, HRS \cite{Derrick:1985wd}, MARK II \cite{Schellman:1984yz}, CELLO \cite{CELLO:1989adw} and TOPAZ \cite{TOPAZ:1994voc} measurements.
The values and uncertainties are displayed the same way as in \cref{f.SIA-Z_pole-K0S}.
The theoretical predictions align well with the Belle measurements across most of the kinematic range, except for the large $x_h$ region where data has substantial experimental uncertainties.
For most TASSO measurements, the data points show relatively larger fluctuations around the theoretical predictions, particularly in the high-$x_h$ region, but the theory is in general within experimental uncertainties.
The measurements from TPC \cite{TPCTwoGamma:1984eoj} at 29 GeV are well described by our theoretical predictions.
For HRS \cite{Derrick:1985wd} and MARK II \cite{Schellman:1984yz} measurements at 29 GeV, the data exhibits some fluctuations around the theory predictions, but are generally within the experimental uncertainties.
The CELLO measurement \cite{CELLO:1989adw} at 35 GeV has larger experimental uncertainties, particularly at high $x_h$, yet a good description is still achieved.
The TOPAZ data \cite{TOPAZ:1994voc} is measured at 58 GeV, with four points in the low $x_h$ region, and they show good agreement with the theoretical predictions.

\begin{figure}[htbp]
	\centering
	\includegraphics[width = 0.9 \textwidth]{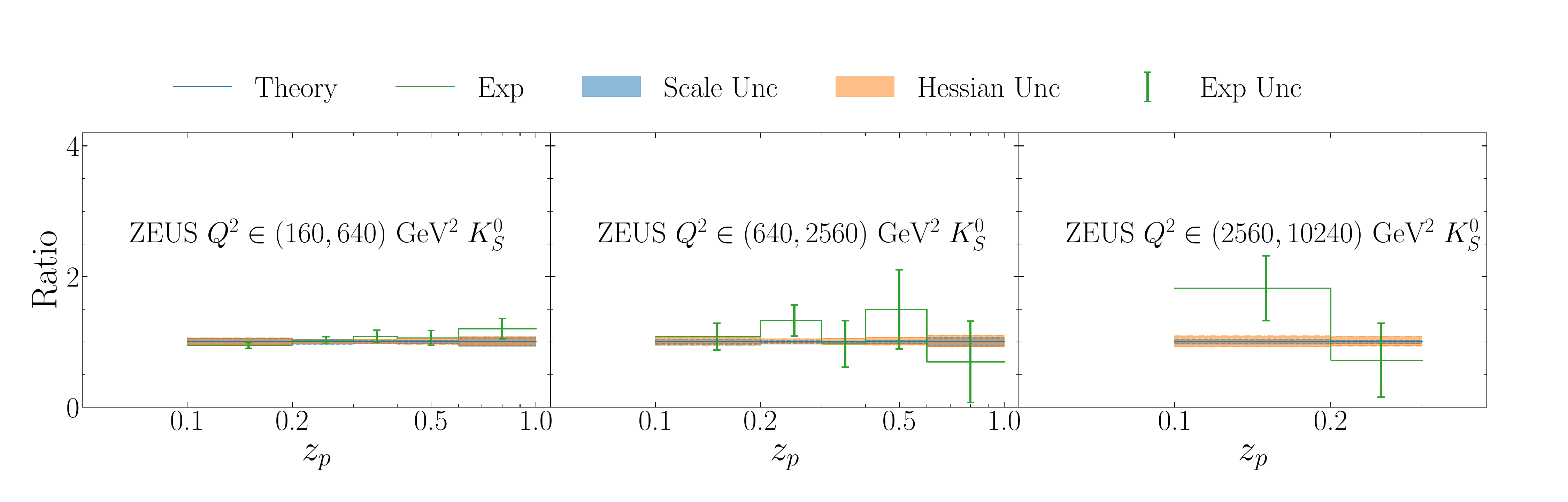}
	\caption{
		Same as \cref{f.SIA-Z_pole-K0S} but for $K_S^0$ production in SIDIS process.
		The ranges of photon virtuality, $Q^2$, are also labeled in each panel.
	}
	\label{f.SIDIS-K0S}
\end{figure}

\cref{f.SIDIS-K0S} presents comparisons between theoretical predictions and SIDIS measurements from ZEUS \cite{ZEUS:2011cdi} for $K_S^0$ production in three different $Q^2$ ranges: $\pqty{160, 640} \, \mathrm{GeV}^2$, $\pqty{640, 2560} \, \mathrm{GeV}^2$ and $\pqty{2560, 10240} \, \mathrm{GeV}^2$.
For the lowest $Q^2$ range, $\pqty{160, 640} \, \mathrm{GeV}^2$, the theoretical predictions show excellent agreement with data across the entire $z_p$ range, with experimental uncertainties generally comparable to or smaller than the theoretical uncertainties.
In the intermediate $Q^2$ range, $\pqty{640, 2560} \, \mathrm{GeV}^2$, the data exhibits larger fluctuations around the theory predictions, particularly in the high-$z_p$ region, though the theory values generally remain within the experimental uncertainties.
For the highest $Q^2$ range, $\pqty{2560, 10240} \, \mathrm{GeV}^2$, the prediction and data show larger deviations.
However, due to the relatively larger experimental uncertainties, the theoretical values remain roughly within the 1$\sigma$ region.
Both scale and Hessian uncertainties are relatively stable across all three $Q^2$ ranges and the entire $z_p$ range.
Finally, we elaborate on the results related to the lowest $Q^2$-bin data.
During iteration of the fits, we observe that without the penalty term described in \cref{e.chi_2-penalty-K0S}, this particular dataset favors a $d$ distribution compared to the $s$ distribution, which is unphysical.
In accordance with the physics considerations discussed in \cref{ss.parameterization}, we included the aforementioned penalty term in order to correct this behavior.
As a result, the $s$ distribution becomes larger than that of $d$ and the increase in $\chi^2$ is only marginal (total $\Delta \chi^2 = 3.77$ for a total of 349 data points, and $\Delta \chi^2 = 1.70, 0.82$ for ZEUS at $Q^2 \in \pqty{160, 640}\, \mathrm{GeV}^2$ and ZEUS at $Q^2 \in \pqty{640, 2560}\, \mathrm{GeV}^2$, respectively).
Although in the NPC23 nominal results, the $d$ and $s$ distributions follow the physical picture, we emphasize that to achieve flavor separation between the $d$ and $s$ distributions purely based on data, more precise SIDIS measurements are still needed.

\begin{figure}[htbp]
	\centering
	\includegraphics[width = 0.9 \textwidth]{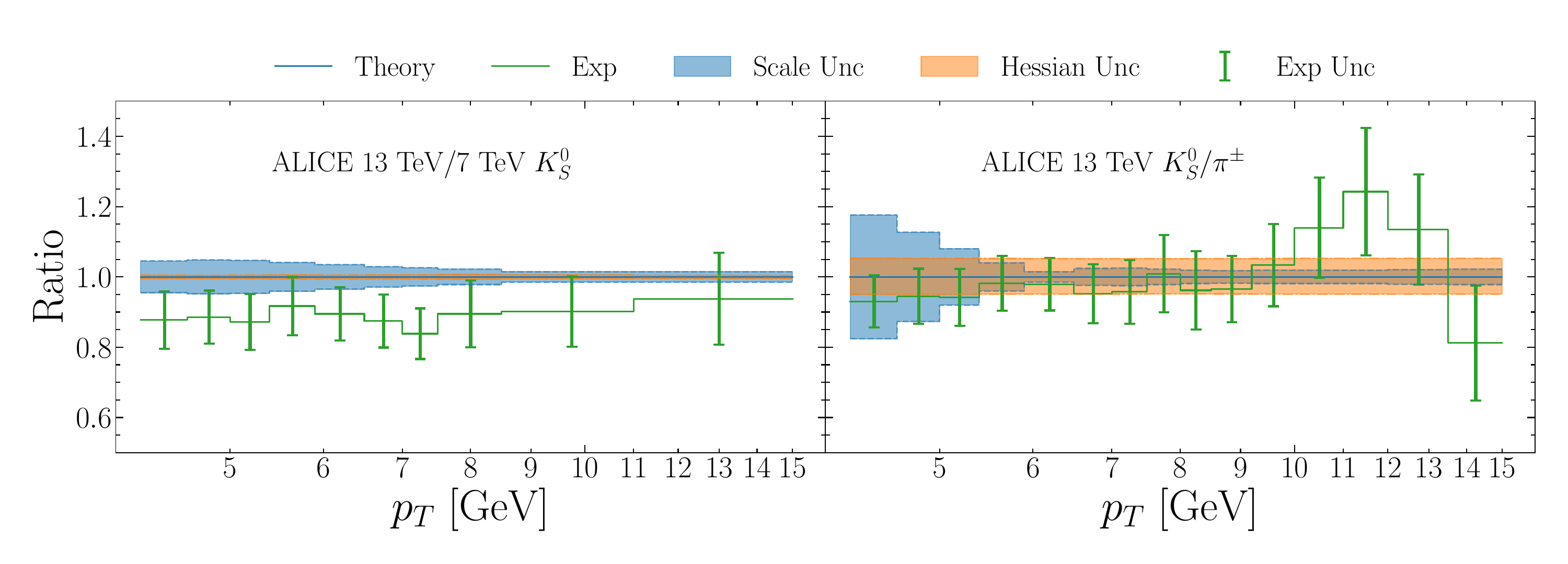}
	\caption{
		Same as \cref{f.SIA-Z_pole-K0S} but for $K_S^0$ production in $pp$ collisions.
		The first panel shows the data-theory comparison for the ALICE data of $K_S^0$ production ratio between $\sqrt{s} = 13~\mathrm{TeV}$ and $\sqrt{s} = 7~\mathrm{TeV}$.
		The second panel shows the data-theory comparison for the data of $K_S^0$ production over $\pi^{\pm}$ production at $\sqrt{s} = 13~\mathrm{TeV}$.
	}
	\label{f.pp-K0S}
\end{figure}

In \cref{f.pp-K0S} the comparisons between the theoretical predictions and the ALICE measurements \cite{ALICE:2020jsh} are shown.
The data in the left panel is the ratio of $K_S^0$ production cross sections between $\sqrt{s} = 13~\mathrm{TeV}$ and $\sqrt{s} = 7~\mathrm{TeV}$, while the data in the right panel is the ratio of $K_S^0$ production to $\pi^{\pm}$ production at $\sqrt{s} = 13~\mathrm{TeV}$.
In the left panel, the theoretical predictions are slightly larger than the data across the entire $p_T$ range, though the theoretical values are generally within the experimental uncertainties.
This indicates a tension between the ALICE $K_S^0$ production ratio data (13 TeV / 7 TeV) and the other world data, and further measurements could help us better understand it.
The scale uncertainties are larger than the Hessian uncertainties in the low-$p_T$ region and decrease with increasing $p_T$. This is caused by the fact that the running of strong coupling constant is more steep at lower scales.
On the other hand, the Hessian uncertainties remain relatively stable across the entire $p_T$ range.
For the $K_S^0/\pi^{\pm}$ ratio data at $\sqrt{s} = 13~\mathrm{TeV}$, which is compared to the theoretical prediction in the right panel, good agreement between theory and data is observed at low and intermediate $p_T$ values.
However, some fluctuations are observed in the high-$p_T$ region and this is understood as a result of the lower statistics when $p_T$ gets larger.
Both scale and Hessian uncertainties are relatively small for this observable, with scale uncertainties slightly more prominent in the low-$p_T$ region.

In conclusion, the fit to the $K_S^0$ global datasets is successful, with a total $\chi^2 / N_{\mathrm{pt}} = 0.94$.
The individual $\chi^2 / N_{\mathrm{pt}}$ values for the SIA, SIDIS and $pp$ collision datasets are 0.99, 0.86 and 0.37, respectively.
Apart from TASSO at 42.6 GeV \cite{TASSO:1989jyt}, HRS at 29 GeV \cite{Derrick:1985wd} and the SLD $c$-tagged data \cite{SLD:1998coh}, the global fit achieved a satisfactory $\chi^2 / N_{\mathrm{pt}}$ for all experimental datasets.
Notably, we found that the SIDIS dataset \cite{ZEUS:2011cdi} in $Q^2$ range $\pqty{160, 640} \, \mathrm{GeV}^2$ favors a larger $d$-to-$K_S^0$ distribution than that of $s$.
To ensure consistency with the physics arguments discussed in \cref{ss.parameterization}, we apply the penalty term \cref{e.chi_2-penalty-K0S} to correct the behaviour.

\clearpage
\subsection{$\Lambda$ production}
\label{ss.Lambda-fit-quality}

In \cref{t.chi_2-Lambda}, we present the $\chi^2$, $\chi^2/N_{\mathrm{pt}}$ values and summary for different categories similar to that in the $K_S^0$ fit.
In summary, we have obtained a good fit with total $\chi^2/N_{\mathrm{pt}} = 0.80$.
In addition, the $\chi^2 / \mathrm{d.o.f.}$ for the $\Lambda$ fit is 0.88.
For individual processes, a good description is found for SIA and $pp$ collisions, each giving a $\chi^2/N_{\mathrm{pt}}$ of 0.78 and 0.43, respectively.
However, for SIDIS data, the description is slightly worse with $\chi^2/N_{\mathrm{pt}} = 1.60$.
This is caused by the large $\chi^2/N_{\mathrm{pt}} = 2.66$ from the ZEUS measurement \cite{ZEUS:2011cdi} in the $Q^2$ bin of $\pqty{160, 640} \, \mathrm{GeV}^2$, and it will be discussed later in this section.

\begin{table}[]
	\setcellgapes{2.5 pt}
	\makegapedcells
	\begin{tabular}{|c|c|c|c|c|c|}
		\hline
		collaboration                                                                                    & year & $\sqrt{s}~\bqty{\mathrm{GeV}}$ & $\chi^2$        & $N_{\mathrm{pt}}$ & $\chi^2/N_{\mathrm{pt}}$ \\
		\hline
		TASSO \cite{TASSO:1984nda}                                                                       & 1985 & 14                             & 1.25            & 3                 & 0.42                     \\
		TASSO \cite{TASSO:1984nda}                                                                       & 1985 & 22                             & 2.81            & 4                 & 0.70                     \\
		TASSO \cite{TASSO:1984nda}                                                                       & 1985 & 34                             & 4.37            & 7                 & 0.62                     \\
		SLAC \cite{delaVaissiere:1984xg}                                                                 & 1985 & 29                             & 22.46           & 15                & 1.50                     \\
		HRS \cite{Derrick:1985wd}                                                                        & 1987 & 29                             & 6.05            & 8                 & 0.76                     \\
		TASSO \cite{TASSO:1981uqa}                                                                       & 1981 & 33                             & 4.48            & 5                 & 0.90                     \\
		TASSO \cite{TASSO:1988qlu}                                                                       & 1989 & 34.8                           & 16.23           & 10                & 1.62                     \\
		TASSO \cite{TASSO:1988qlu}                                                                       & 1989 & 42.1                           & 1.49            & 5                 & 0.30                     \\
		CELLO \cite{CELLO:1989adw}                                                                       & 1990 & 35                             & 3.43            & 7                 & 0.49                     \\
		DELPHI \cite{DELPHI:1993vpj}                                                                     & 1993 & 91.2                           & 10.49           & 7                 & 1.50                     \\
		ALEPH \cite{ALEPH:1994fts}                                                                       & 1994 & 91.2                           & 6.36            & 14                & 0.45                     \\
		ALEPH \cite{ALEPH:1996oqp}                                                                       & 1998 & 91.2                           & 3.60            & 16                & 0.22                     \\
		ALEPH jet 1 \cite{ALEPH:1999udi}                                                                 & 2000 & 91.2                           & 16.58           & 13                & 1.28                     \\
		ALEPH jet 2 \cite{ALEPH:1999udi}                                                                 & 2000 & 91.2                           & 3.53            & 12                & 0.29                     \\
		ALEPH jet 3 \cite{ALEPH:1999udi}                                                                 & 2000 & 91.2                           & 3.38            & 9                 & 0.38                     \\
		OPAL \cite{OPAL:1996gsw}                                                                         & 1997 & 91.2                           & 6.23            & 12                & 0.52                     \\
		SLD \cite{SLD:1998coh}                                                                           & 1999 & 91.2                           & 4.19            & 9                 & 0.47                     \\
		SLD $c$-tagged \cite{SLD:1998coh}                                                                & 1999 & 91.2                           & 11.41           & 5                 & 2.28                     \\
		SLD $b$-tagged \cite{SLD:1998coh}                                                                & 1999 & 91.2                           & 1.17            & 5                 & 0.23                     \\
		\hline
		SIA sum                                                                                          &      &                                & 129.52          & 166               & 0.78                     \\
		\hline
		ZEUS $Q^2 \in \pqty{160, 640} \, \mathrm{GeV}^2$ \cite{ZEUS:2011cdi}                             & 2012 & 318                            & 13.30           & 5                 & 2.66                     \\
		ZEUS $Q^2 \in \pqty{640, 2560} \, \mathrm{GeV}^2$ \cite{ZEUS:2011cdi}                            & 2012 & 318                            & 1.13            & 3                 & 0.38                     \\
		ZEUS $Q^2 \in \pqty{2560, 10240} \, \mathrm{GeV}^2$ \cite{ZEUS:2011cdi}                          & 2012 & 318                            & 0.004           & 1                 & 0.004                    \\
		\hline
		SIDIS sum                                                                                        &      &                                & 14.43           & 9                 & 1.60                     \\
		\hline
		CMS $\pqty{N_{\Lambda} / N_{K_S^0}}$ \cite{CMS:2011jlm}                                          & 2011 & 900                            & 2.97            & 4                 & 0.74                     \\
		ALICE $\pqty{N_{\Lambda}^{13~\mathrm{TeV}} / N_{\Lambda}^{7~\mathrm{TeV}}}$ \cite{ALICE:2020jsh} & 2021 & 13000 \& 7000                  & 1.71            & 7                 & 0.24                     \\
		\hline
		$pp$ sum                                                                                         &      &                                & 4.68            & 11                & 0.43                     \\
		\hline
		\textbf{total sum}                                                                               &      &                                & \textbf{148.63} & \textbf{186}      & \textbf{0.80}            \\
		\hline
	\end{tabular}
	\caption{Same as \cref{t.chi_2-K0S} but for $\Lambda$ production data.
	}
	\label{t.chi_2-Lambda}
\end{table}

\begin{figure}
	\centering
	\includegraphics[width = 0.9 \textwidth]{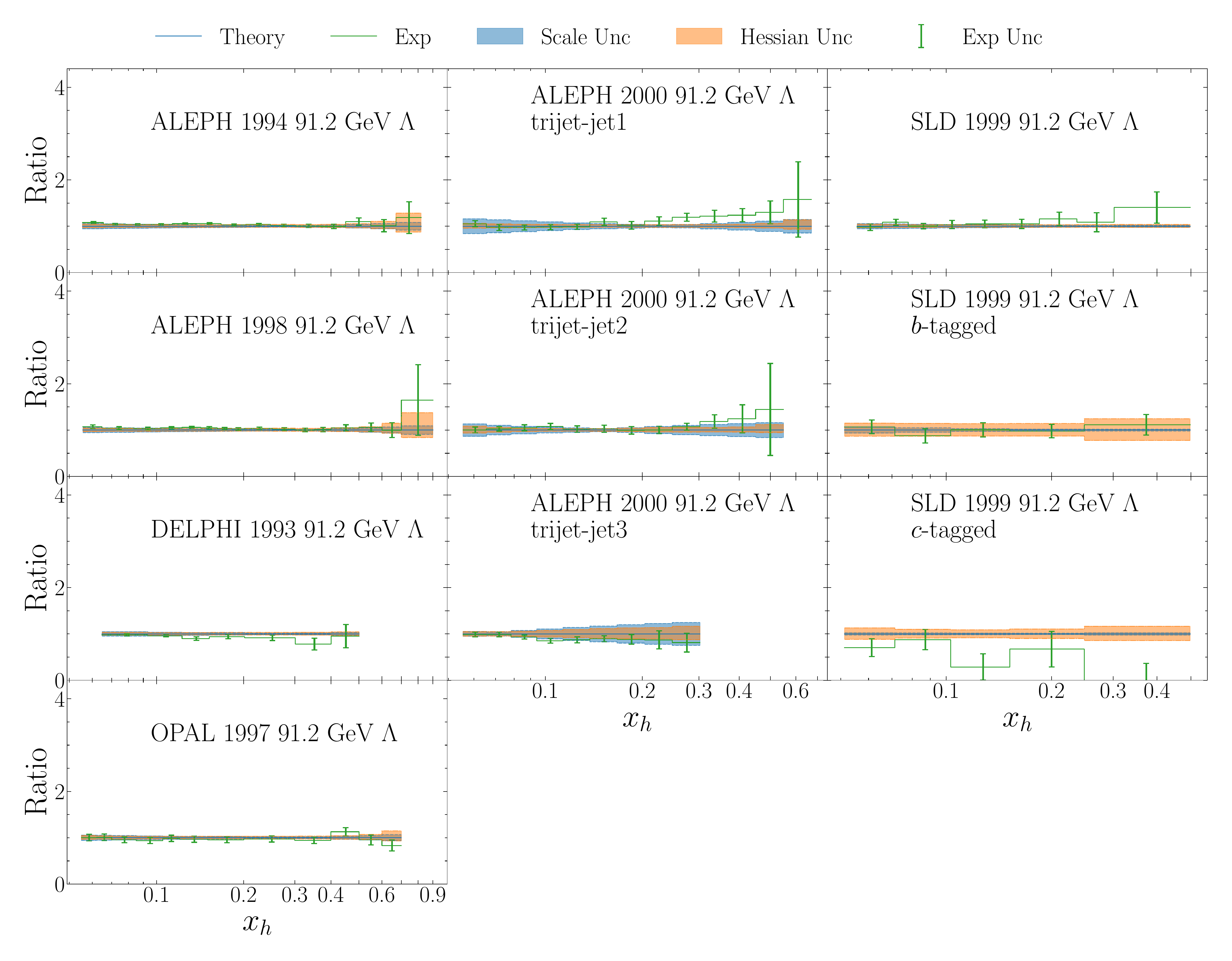}
	\caption{
		Same as \cref{f.SIA-Z_pole-K0S} but for $\Lambda$ production in SIA at $Z$-pole energy.
	}
	\label{f.SIA-Z_pole-Lambda}
\end{figure}

In \cref{f.SIA-Z_pole-Lambda} we show the data-theory comparison for $\Lambda$ production in the SIA process at the $Z$-pole, including measurements from DELPHI \cite{DELPHI:1993vpj}, ALEPH \cite{ALEPH:1994fts, ALEPH:1996oqp, ALEPH:1999udi}, OPAL \cite{OPAL:1996gsw} and SLD \cite{SLD:1998coh} collaboration.
The experimental data and their uncertainties, as well as the theoretical uncertainties (scale and Hessian), are normalized to our best-fit theoretical results.
The comparison with ALEPH 1994 \cite{ALEPH:1994fts} and ALEPH 1998 \cite{ALEPH:1996oqp} data is given in the first two panels in column 1.
A good agreement between theory and data is found, with reasonable scale and Hessian uncertainty bands.
The Hessian uncertainty becomes larger at the high-$x_h$ end, reflecting the growth of experimental uncertainties.
The comparison to ALEPH 2000 $\Lambda$-in-jet data \cite{ALEPH:1999udi} is shown in column 2 of \cref{f.SIA-Z_pole-Lambda}, for jet 1, 2 and 3 from the top to bottom panels, respectively.
A good description is also found for all three cases, with some fluctuations in the high-$x_h$ region.
This can be understood as a result of the decreasing experimental precision with lower statistics in the high $x_h$ region.
Similarly to the case of the ALEPH $K_S^0$-in-jet production data shown in \cref{f.SIA-Z_pole-K0S}, the scale uncertainties for the $\Lambda$-in-jet production are discernibly larger compared to other scenarios.
Again, this is caused by the fact that $\alpha_s$ enters at LO for jet production (see comments on \cref{f.SIA-Z_pole-K0S}).
As for the SLD 1999 data \cite{SLD:1998coh}, shown in the third column of \cref{f.SIA-Z_pole-Lambda}, we find a good description for both inclusive and $b$-tagged $\Lambda$ productions.
The exception is for the SLD $c$-tagged data, which has a $\chi^2/N_{\mathrm{pt}}$ of 2.28.
This is caused by the third point which is an outlier.
Larger Hessian uncertainties are also observed in $c$/$b$-tagged scenarios,  this reflects the relatively larger experimental uncertainties again.
The DELPHI \cite{DELPHI:1993vpj} and OPAL \cite{OPAL:1996gsw} data are compared to our theory in the last two panels of column 1. Except for a few fluctuations at large $x_h$, a good alignment is found.
Finally, we would like to mention that the inclusive $\Lambda$ production from ALEPH 2000 paper \cite{ALEPH:1999udi} is not included in our fit, because we observe that the ALEPH 2000 \cite{ALEPH:1999udi} data has some tension with the ALEPH 1998 \cite{ALEPH:1996oqp} measurements.

\begin{figure}
	\centering
	\includegraphics[width = 0.9 \textwidth]{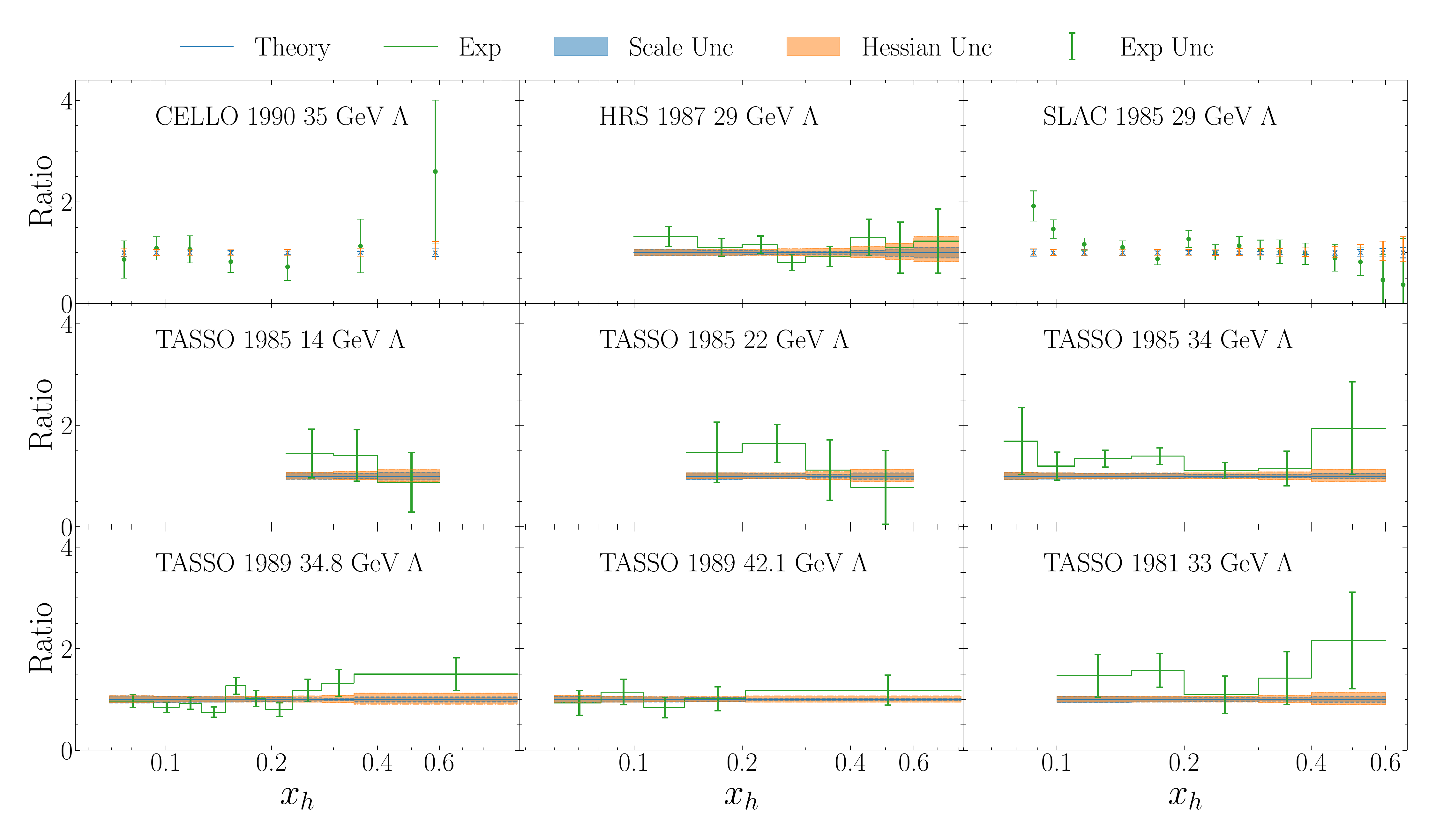}
	\caption{
		Same as \cref{f.SIA-Z_pole-K0S} but for $\Lambda$ production in SIA below $Z$-pole energy.
		Notice the data points and theory values are only connected horizontally if the measurement is provided with $x_h$ bins.
	}
	\label{f.SIA-below-Z_pole-Lambda}
\end{figure}

In \cref{f.SIA-below-Z_pole-Lambda}, we show the data-theory comparison for $\Lambda$ production in SIA at center-of-mass energies below the $Z$-boson mass.
In particular, data from the TASSO \cite{TASSO:1981uqa, TASSO:1984nda, TASSO:1988qlu}, SLAC \cite{delaVaissiere:1984xg}, HRS \cite{Derrick:1985wd}, and CELLO \cite{CELLO:1989adw} collaborations are included.
Starting with the first panel, which compares the CELLO 1990 data \cite{CELLO:1989adw} to the theoretical predictions, we find that the experimental results are well described, except for the last point, where the data and theory lie outside the 1$\sigma$ error bands.
The second panel presents the comparison with HRS 1987 data \cite{Derrick:1985wd}. Except for some fluctuations at low $x_h$, the theory is consistent with the data.
We also observe that the Hessian uncertainty is noticeably larger for HRS, which reflects the larger experimental uncertainties.
The third panel in the first row compares the data and theory for the SLAC 1985 data \cite{delaVaissiere:1984xg}, showing good agreement, with only a few differences at low $x_h$.
In the second row, the TASSO 1985 data \cite{TASSO:1984nda} at center-of-mass energies of 14, 22, and 34 GeV are compared to the theoretical predictions in the second panels, from left to right, respectively.
A good description is also achieved in these cases.
In the third row, the TASSO 1989 data \cite{TASSO:1988qlu} at center-of-mass energies of 34.8 and 42.1 GeV, as well as the TASSO 1981 data \cite{TASSO:1981uqa} at 33 GeV are presented, where a good description is achieved in general.

\begin{figure}
	\centering
	\includegraphics[width = 0.9 \textwidth]{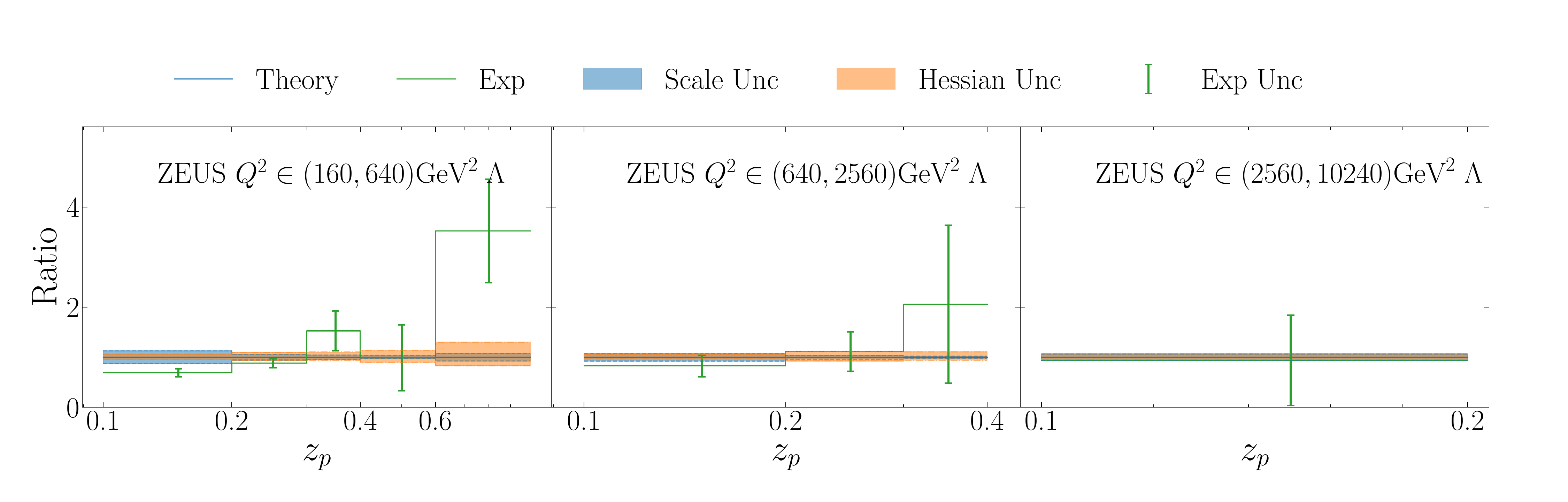}
	\caption{
		Same as \cref{f.SIA-Z_pole-K0S} but for $\Lambda$ production in SIDIS process.
		The ranges of photon virtuality, $Q^2$, are also labeled in each panel.
	}
	\label{f.SIDIS-Lambda}
\end{figure}

In \cref{f.SIDIS-Lambda}, we show the data-theory comparison for the SIDIS process measured by the ZEUS collaboration \cite{ZEUS:2011cdi}.
The data is grouped into three different $Q^2$ ranges: $\pqty{160, 640} \, \mathrm{GeV}^2$, $\pqty{640, 2560} \, \mathrm{GeV}^2$ and $\pqty{2560, 10240} \, \mathrm{GeV}^2$.
A good description is achieved for datasets in the intermediate and highest $Q^2$ ranges. However, we find that the data in the lowest $Q^2$ range can not be well described due to the first and last data points, which deviate significantly from theoretical predictions.
For the first point in the lowest $Q^2$ bin, it is realized that the momentum of the $\Lambda$ baryon is around $3 \sim 4.5~\mathrm{GeV}$, approaching the mass of $\Lambda$ which is 1.115 GeV.
At this point, the mass corrections might become more important, which is not implemented in this work.
As for the last data point in the lowest $Q^2$ bin, it has much smaller uncertainty compared to the second-to-last point and it might be an outlier.
(The experimental uncertainty of the last data point appears to be larger because it is normalized by the central theoretical value which is approximately 3.5 times smaller than the experimental value.)
Thus, further precision measurements from SIDIS remain important, not only for resolving the tension between datasets, but also for better flavor separation in the extraction of $\Lambda$ FFs.

\begin{figure}
	\centering
	\includegraphics[width = 0.9 \textwidth]{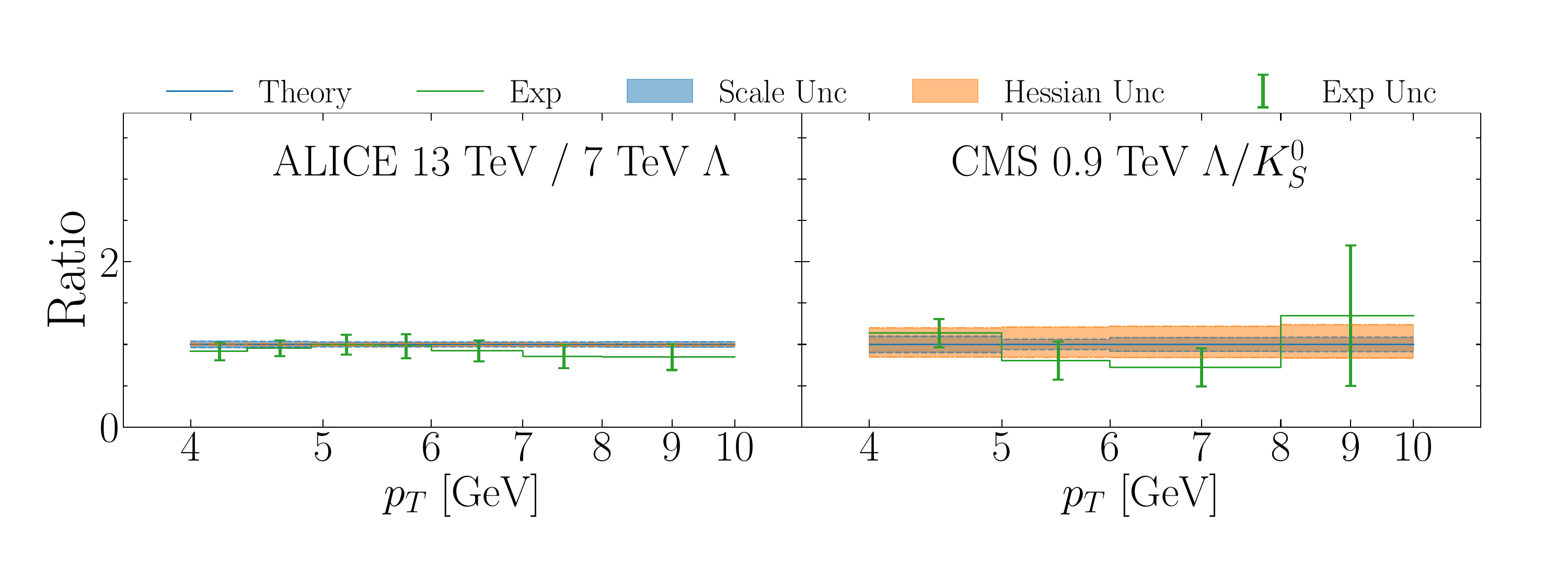}
	\caption{
		Same as \cref{f.SIA-Z_pole-K0S} but for $K_S^0$ production in $pp$ collisions.
		The first panel shows the data-theory comparison for the ALICE data of $\Lambda$ production ratio between $\sqrt{s} = 13~\mathrm{TeV}$ and $\sqrt{s} = 7~\mathrm{TeV}$.
		The second panel shows the data-theory comparison for the data of $\Lambda$ production over $K_S^0$ production at $\sqrt{s} = 900~\mathrm{GeV}$.
	}
	\label{f.pp-Lambda}
\end{figure}

In \cref{f.pp-Lambda} the data-theory comparison is shown for CMS \cite{CMS:2011jlm} and ALICE \cite{ALICE:2020jsh} measurements.
The data in the left panel is the ratio of $\Lambda$ production cross sections between $\sqrt{s} = 13~\mathrm{TeV}$ and $\sqrt{s} = 7~\mathrm{TeV}$ from ALICE, while the data in the right panel is the ratio of $\Lambda$ production over $K_S^0$ production at $\sqrt{s} = 900~\mathrm{GeV}$ from CMS.
In the left panel, the theoretical predictions are slightly above the data in the low and high $p_T$ regions, but are always within experimental uncertainties.
The scale uncertainties in this case are larger than the Hessian uncertainties in the entire $p_T$ range, while the Hessian uncertainties remain relatively stable across the $p_T$ range.
Again, this is due to the difference of slope in running of $\alpha_s$ at different scales (see comments on \cref{f.pp-K0S}).
For the $\Lambda/K_S^0$ ratio data at $\sqrt{s} = 13~\mathrm{TeV}$, which is compared to the theory in the right panel, a good agreement between theory and data is achieved in general.
However, some fluctuations are observed in the middle-$p_T$ region but the theory remains within the experimental uncertainty.
The relatively larger Hessian uncertainty also reflects the weaker constraints of the experimental data.

In conclusion, the fit to the $\Lambda$ global datasets is successful, with a total $\chi^2 / N_{\mathrm{pt}} = 0.80$.
The individual $\chi^2 / N_{\mathrm{pt}}$ values for SIA, SIDIS and $pp$ collision datasets are 0.78, 1.60 and 0.43, respectively.
The data-theory agreement for the SLD $c$-tagged data \cite{SLD:1998coh} and the SIDIS dataset \cite{ZEUS:2011cdi} in $Q^2$ range $\pqty{160, 640} \, \mathrm{GeV}^2$ are deteriorated, potentially by outliers.
This further underscores the critical need for future high-precision SIDIS measurements.
Finally, the inclusive $\Lambda$ production data from ALEPH 2000 paper \cite{ALEPH:1999udi} is not included in our fit due to its tension with the ALEPH 1998 \cite{ALEPH:1996oqp} measurements.

\clearpage
\subsection{$\eta$ production}

In \cref{t.chi_2-eta}, we present the $\chi^2$, $\chi^2/N_{\mathrm{pt}}$ values and summary for different categories similar to that in the $K_S^0$ fit.
For the global dataset, we obtain a total $\chi^2$ value of 151.21 for 153 data points, resulting in a $\chi^2/N_{\mathrm{pt}}$ of 0.99, which indicates good overall agreement between the theoretical predictions and the experimental data.
In addition, the $\chi^2 / \mathrm{d.o.f.}$ for the $\eta$ fit is 1.07.
Examining different experimental categories, we find that the SIA measurements are well described with $\chi^2/N_{\mathrm{pt}} = 1.15$ for 102 data points.
Moreover, a good agreement with the $pp$ collision data is also achieved with $\chi^2/N_{\mathrm{pt}} = 0.67$ across 51 data points.
For individual experiments, most datasets are well described with $\chi^2/N_{\mathrm{pt}} \approx 1$.
However, the $\eta$-in-jet production (the third jet in particular) from the ALEPH 2000 data \cite{ALEPH:1999udi}, and the inclusive $\eta$ production from the ALEPH 2002 data \cite{ALEPH:2001tfk} can not be described very well.
The fit quality to these two datasets will be discussed in more detail later.

\begin{table}[]
	\setcellgapes{2.5 pt}
	\makegapedcells
	\begin{tabular}{|c|c|c|c|c|c|}
		\hline
		collaboration                    & year & $\sqrt{s}~\bqty{\mathrm{GeV}}$ & $\chi^2$        & $N_{\mathrm{pt}}$ & $\chi^2/N_{\mathrm{pt}}$ \\
		\hline
		ARGUS \cite{ARGUS:1989orf}       & 1990 & 9.46                           & 5.94            & 6                 & 0.99                     \\
		HRS \cite{HRS:1987aky}           & 1988 & 29                             & 18.07           & 13                & 1.39                     \\
		JADE \cite{JADE:1985bzp}         & 1985 & 34.4                           & 2.29            & 2                 & 1.14                     \\
		JADE \cite{JADE:1989ewf}         & 1990 & 35                             & 3.29            & 3                 & 1.09                     \\
		CELLO \cite{CELLO:1989byk}       & 1990 & 35                             & 3.47            & 5                 & 0.69                     \\
		L3 \cite{L3:1992pbe}             & 1992 & 91.2                           & 5.83            & 4                 & 1.46                     \\
		L3 \cite{L3:1994gkb}             & 1994 & 91.2                           & 10.46           & 10                & 1.05                     \\
		ALEPH \cite{ALEPH:1992zhm}       & 1992 & 91.2                           & 1.48            & 8                 & 0.18                     \\
		ALEPH \cite{ALEPH:1999udi}       & 2000 & 91.2                           & 18.39           & 18                & 1.02                     \\
		ALEPH jet 1 \cite{ALEPH:1999udi} & 2000 & 91.2                           & 11.26           & 7                 & 1.61                     \\
		ALEPH jet 2 \cite{ALEPH:1999udi} & 2000 & 91.2                           & 1.95            & 6                 & 0.33                     \\
		ALEPH jet 3 \cite{ALEPH:1999udi} & 2000 & 91.2                           & 10.49           & 4                 & 2.62                     \\
		ALEPH \cite{ALEPH:2001tfk}       & 2002 & 91.2                           & 17.18           & 5                 & 3.44                     \\
		OPAL \cite{OPAL:1998enc}         & 1998 & 91.2                           & 7.12            & 11                & 0.65                     \\
		\hline
		SIA sum                          &      &                                & 117.20          & 102               & 1.15                     \\
		\hline
		PHENIX \cite{PHENIX:2010hvs}     & 2011 & 200                            & 7.61            & 14                & 0.51                     \\
		ALICE \cite{ALICE:2017nce}       & 2017 & 2760                           & 5.37            & 6                 & 0.90                     \\
		ALICE \cite{ALICE:2012wos}       & 2012 & 7000                           & 1.26            & 4                 & 0.32                     \\
		ALICE \cite{ALICE:2017ryd}       & 2018 & 8000                           & 12.64           & 13                & 0.97                     \\
		ALICE \cite{ALICE:2024vgi}       & 2024 & 13000                          & 7.11            & 14                & 0.51                     \\
		\hline
		$pp$ sum                         &      &                                & 33.99           & 51                & 0.67                     \\
		\hline
		\textbf{total sum}               &      &                                & \textbf{151.21} & \textbf{153}      & \textbf{0.99}            \\
		\hline
	\end{tabular}
	\caption{
		Same as \cref{t.chi_2-K0S} but for $\eta$ production data.
	}
	\label{t.chi_2-eta}
\end{table}

\begin{figure}[htbp]
	\centering
	\includegraphics[width = 0.9 \textwidth]{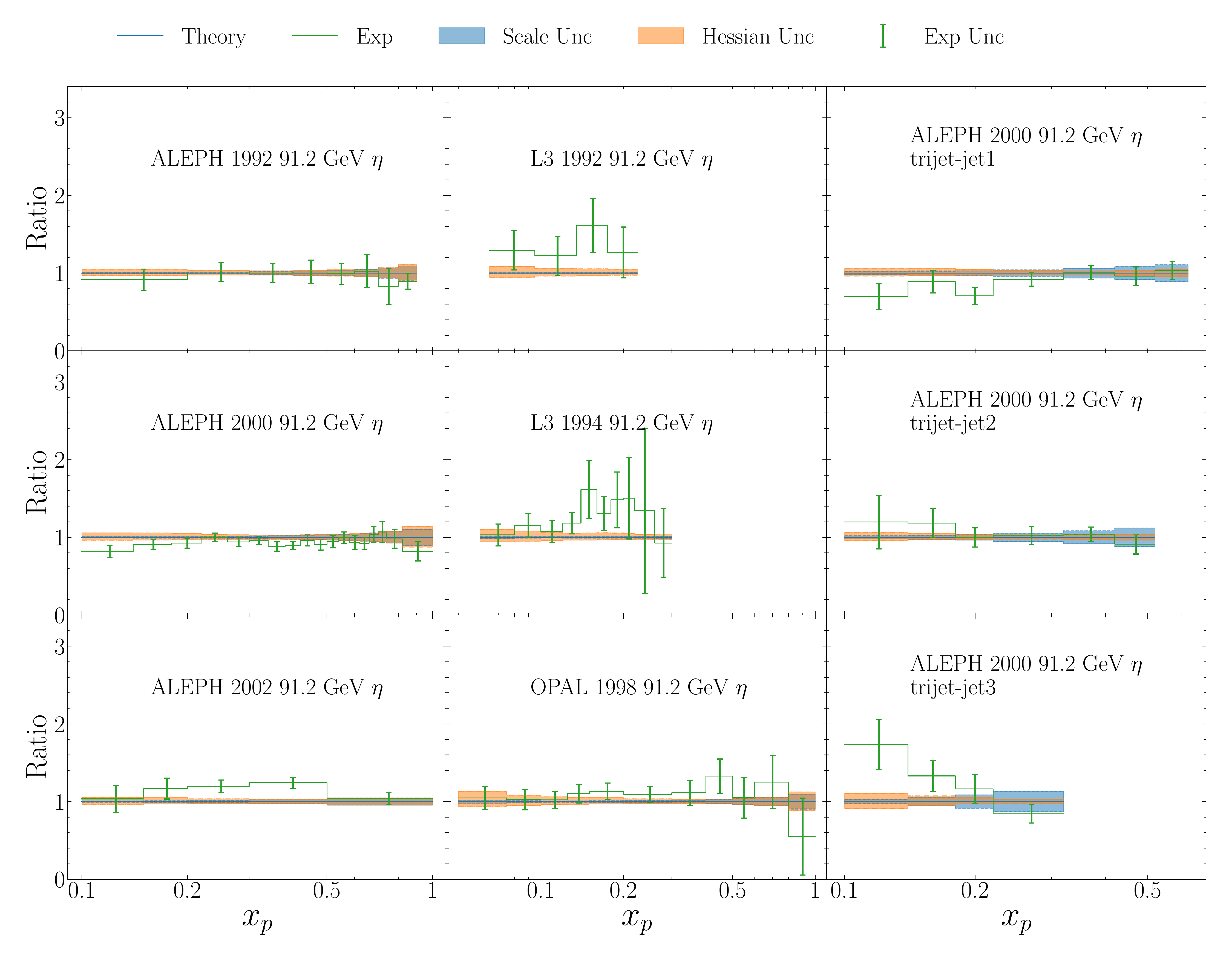}
	\caption{
		Same as \cref{f.SIA-Z_pole-K0S} but for $\eta$ production in SIA at $Z$-pole energy.
	}
	\label{f.SIA-Z_pole-eta}
\end{figure}

In \cref{f.SIA-Z_pole-eta}, the ratio of experimental data to theoretical predictions for $\eta$ meson production at the $Z$-pole is presented, including data from ALEPH \cite{ALEPH:1992zhm, ALEPH:1999udi, ALEPH:2001tfk}, L3 \cite{L3:1992pbe, L3:1994gkb} and OPAL \cite{OPAL:1998enc} collaborations.
Notice that for $\eta$ production in SIA at $Z$-pole, the cross sections are differential in $x_p$ instead of $x_h$.
The predictions to ALEPH 2002 \cite{ALEPH:2001tfk} underestimate the data, while the predictions to ALEPH 2000 \cite{ALEPH:1999udi} overshoot the data, indicating a possible inconsistency between these two measurements.
Here we note that neither AESSS \cite{Aidala:2010bn} nor LAXZ \cite{Li:2024etc} can describe the ALEPH 2002 well, which further confirms the tension between the ALEPH 2002 data and the other world data.
The ALEPH 1992 data \cite{ALEPH:1992zhm}, on the other hand, is very well described with $\chi^2/N_{\mathrm{pt}} = 0.18$.
The comparisons to L3 1992 \cite{L3:1992pbe} and L3 1994 \cite{L3:1994gkb} data are shown in the first two panels of the second column in \cref{f.SIA-Z_pole-eta}, in which a generally good description is achieved, with some fluctuations that are mostly within the experimental uncertainties.
For OPAL data \cite{OPAL:1998enc} a good description is found with $\chi^2/N_{\mathrm{pt}} = 0.65$.

\cref{f.SIA-Z_pole-eta} also shows the comparison between the experimental data and the theoretical predictions for $\eta$ meson production in tri-jet events at the $Z$-pole energy, with measurement conducted by ALEPH \cite{ALEPH:1999udi}.
With $\eta$ produced in the first, second, and third jet shown from top to bottom of the last column in \cref{f.SIA-Z_pole-eta}, respectively.
For jet 1, the theoretical predictions are consistent with the data in the range $x_p > 0.3$, however, the theory values are found to be higher than the data in the low-$x_p$ region.
For jet 2, in the low-$x_p$ region, the data is underestimated by the theory, but both are within the uncertainties.
Jet 3 shows the worst agreement, with the data significantly exceeding the theoretical prediction in the low-$x_p$ region.
The Hessian uncertainties in this case are also significantly larger than those in other experiments, indicating a weak constraining power from the data.
We also notice that the scale uncertainties in predictions for the tri-jet experiment are again more pronounced than those in other experiments, as previously discussed for the case of $K_S^0$ and $\Lambda$ in \cref{ss.K0S-fit-quality,ss.Lambda-fit-quality}.

\begin{figure}[htbp]
	\centering
	\includegraphics[width = 0.9 \textwidth]{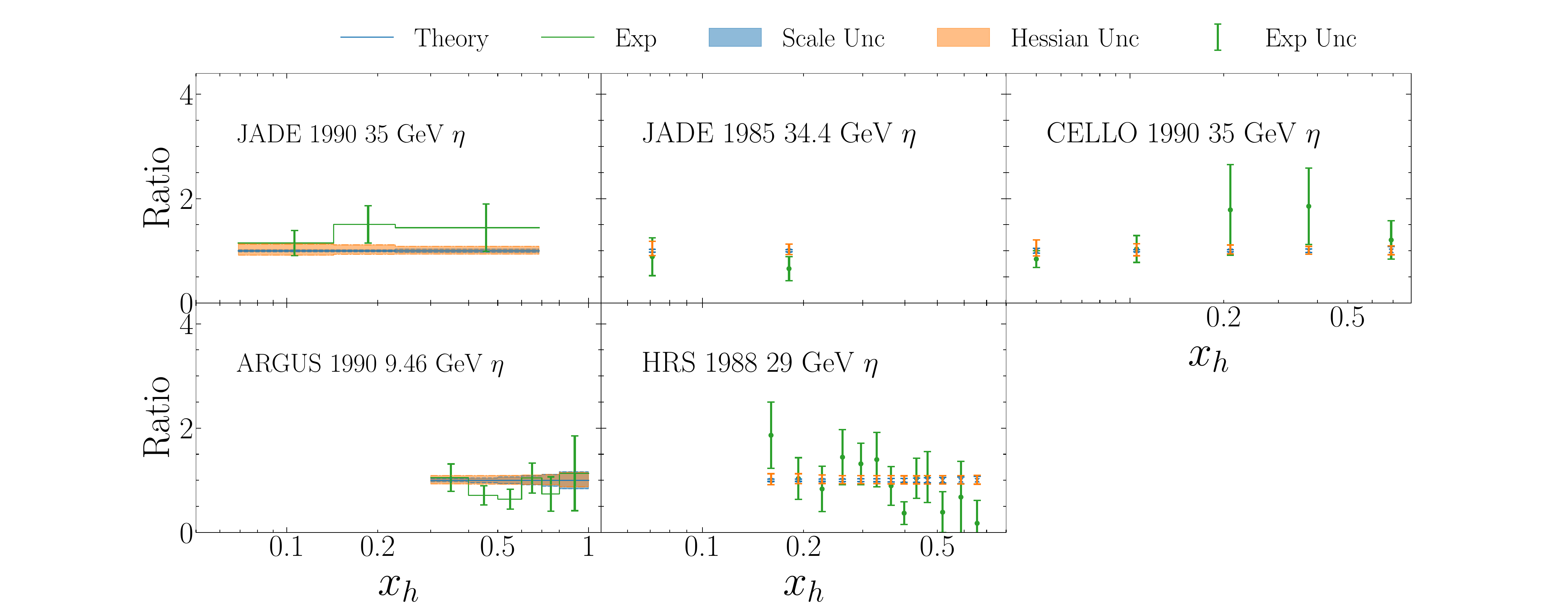}
	\caption{
		Same as \cref{f.SIA-Z_pole-K0S} but for $\eta$ production in SIA below $Z$-pole energy.
	}
	\label{f.SIA-below-Z_pole-eta}
\end{figure}

In \cref{f.SIA-below-Z_pole-eta}, the theory and data for SIA are compared with center-of-mass energies below the $Z$-boson mass.
These include ARGUS at 9.46 GeV \cite{ARGUS:1989orf}, HRS at 29 GeV \cite{HRS:1987aky}, JADE at 34.4 GeV \cite{JADE:1985bzp} and 35 GeV \cite{JADE:1989ewf}, as well as CELLO at 35 GeV \cite{CELLO:1989byk}.
For the JADE and CELLO datasets, the experimental data are generally well described by the theoretical predictions, with slight deviations at high $x_p$ values.
For the ARGUS and HRS datasets, most of the data falls within the theory predicted bounds, except for minor deviations in the intermediate-$x_p$ regions.

\begin{figure}[htbp]
	\centering
	\includegraphics[width = 0.9 \textwidth]{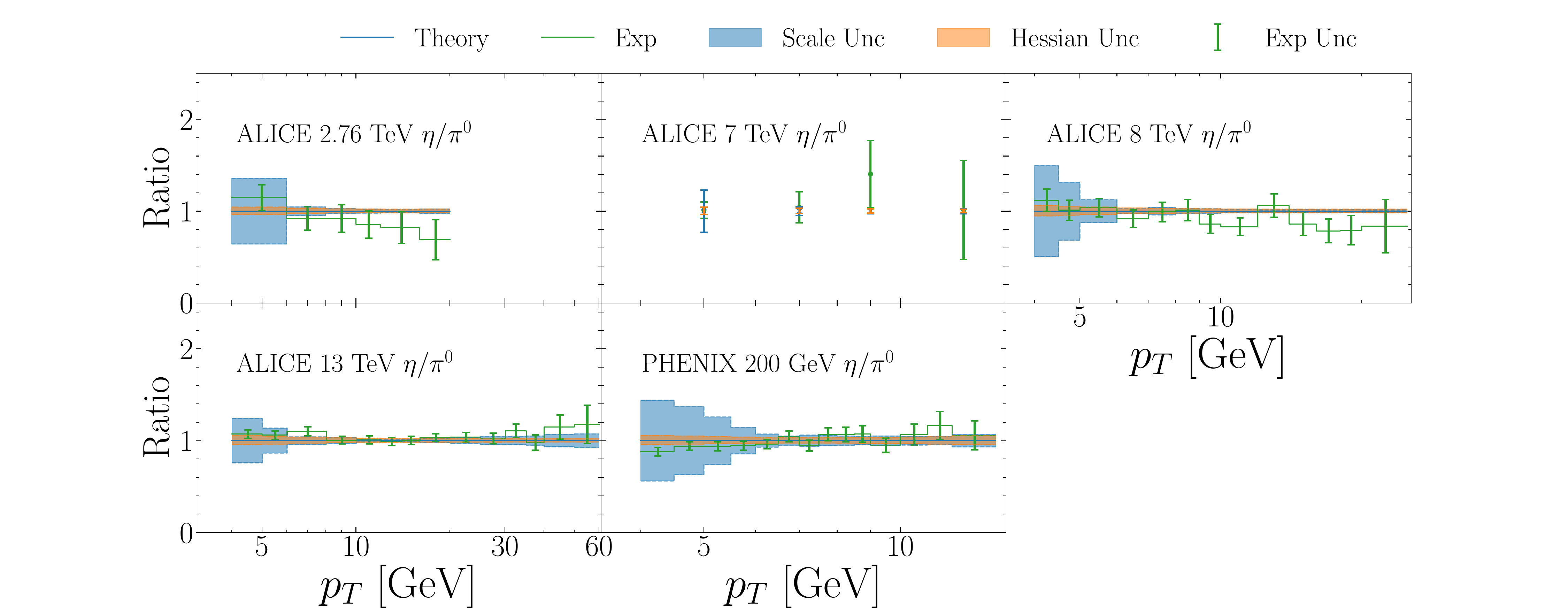}
	\caption{
		Same as \cref{f.SIA-Z_pole-K0S} but for data of $\eta$ production over $\pi^0$ production at various center-of-mass energies in $pp$ collisions.
	}
	\label{f.pp-eta}
\end{figure}

In \cref{f.pp-eta}, a comparison between the best-fit predictions and the data is shown.
Results from ALICE at 2.76 TeV \cite{ALICE:2017nce}, 7 TeV \cite{ALICE:2012wos}, and 8 TeV \cite{ALICE:2017ryd}, as well as from PHENIX at 200 GeV \cite{PHENIX:2010hvs}, are presented.
From the figure, it can be observed that most data points are very well described by the theory, with only a few exceptions that are generally within experimental uncertainty.
In the low-$p_T$ regions, we observe that the scale uncertainties can become significant, which has been discussed for the case of $K_S^0$ and $\Lambda$ in \cref{ss.K0S-fit-quality,ss.Lambda-fit-quality}.

In conclusion, the fit to the $\eta$ global datasets is successful, with a total $\chi^2 / N_{\mathrm{pt}} = 0.99$.
The individual $\chi^2 / N_{\mathrm{pt}}$ values for SIA and $pp$ collision datasets are 1.15 and 0.67, respectively.
Despite exceptions in ALEPH jet 3 \cite{ALEPH:1999udi} and some tension between ALEPH 2002 data \cite{ALEPH:2001tfk} and other SIA datasets, we achieve good overall agreement with all datasets.

\clearpage
\subsection{$\pi^0$ production}
\label{ss.pi_0-cross-section}

In \cref{t.chi_2-pi_0}, we show the $\chi^2$ values for $\pi^0$ production in SIA and $pp$ collisions, with $\pi^0$ FFs constructed from the NPC fitted $\pi^{\pm}$ FFs \cite{Gao:2024dbv} as described in \cref{e.pi_0-construction-u-and-d,e.pi_0-construction-unfavored}.
This table does not represent a direct fit to the $\pi^0$ data; instead, it evaluates the compatibility of the $\pi^{\pm}$-based predictions with the observed $\pi^0$ distributions across various experimental setups.
The total $\chi^2/N_{\mathrm{pt}}$ is 1.54 for $\pi^0$, while the $\chi^2 / \mathrm{d.o.f.}$ is 1.70.
Here the d.o.f. for $\pi^0$ FFs is 25, consistent with that used for the $\pi^{\pm}$ FFs fitted in \cite{Gao:2024dbv}.

\begin{table}[]
	\setcellgapes{2.5 pt}
	\makegapedcells
	\begin{tabular}{|c|c|c|c|c|c|}
		\hline
		collaboration                           & year & $\sqrt{s}~\bqty{\mathrm{GeV}}$ & $\chi^2$        & $N_{\mathrm{pt}}$ & $\chi^2/N_{\mathrm{pt}}$ \\
		\hline
		ARGUS \cite{ARGUS:1989orf}              & 1990 & 9.46                           & 20.12           & 14                & 1.44                     \\
		JADE \cite{JADE:1985bzp}                & 1985 & 34.4                           & 7.92            & 10                & 0.79                     \\
		JADE \cite{JADE:1989ewf}                & 1990 & 35                             & 8.66            & 9                 & 0.96                     \\
		JADE \cite{JADE:1989ewf}                & 1990 & 44                             & 7.30            & 6                 & 1.22                     \\
		CELLO \cite{Elze:1989gm}                & 1990 & 35                             & 31.31           & 15                & 2.09                     \\
		TASSO \cite{TASSO:1988jma}              & 1989 & 44                             & 1.66            & 6                 & 0.28                     \\
		ALEPH \cite{ALEPH:1996pxg}              & 1997 & 91.2                           & 14.85           & 20                & 0.74                     \\
		ALEPH jet 1 \cite{ALEPH:1999udi}        & 2000 & 91.2                           & 2.43            & 8                 & 0.30                     \\
		ALEPH jet 2 \cite{ALEPH:1999udi}        & 2000 & 91.2                           & 11.69           & 8                 & 1.46                     \\
		ALEPH jet 3 \cite{ALEPH:1999udi}        & 2000 & 91.2                           & 34.65           & 6                 & 5.77                     \\
		DELPHI \cite{DELPHI:1995ase}            & 1996 & 91.2                           & 4.86            & 17                & 0.29                     \\
		DELPHI $b$-tagged \cite{DELPHI:1995ase} & 1996 & 91.2                           & 9.12            & 15                & 0.61                     \\
		L3 \cite{L3:1994gkb}                    & 1994 & 91.2                           & 27.34           & 12                & 2.28                     \\
		OPAL \cite{OPAL:1998enc}                & 1998 & 91.2                           & 2.73            & 10                & 0.27                     \\
		\hline
		SIA sum                                 &      &                                & 184.64          & 156               & 1.18                     \\
		\hline
		STAR \cite{STAR:2009qzv}                & 2009 & 200                            & 6.09            & 8                 & 0.76                     \\
		PHENIX  \cite{PHENIX:2007kqm}           & 2007 & 200                            & 45.76           & 17                & 2.69                     \\
		PHENIX \cite{PHENIX:2015fxo}            & 2016 & 510                            & 66.95           & 22                & 3.04                     \\
		ALICE \cite{ALICE:2017nce}              & 2017 & 2760                           & 13.98           & 16                & 0.87                     \\
		ALICE \cite{ALICE:2012wos}              & 2012 & 7000                           & 29.27           & 13                & 2.25                     \\
		ALICE  \cite{ALICE:2017ryd}             & 2017 & 8000                           & 46.42           & 24                & 1.93                     \\
		\hline
		$pp$ sum                                &      &                                & 208.47          & 100               & 2.08                     \\
		\hline
		\textbf{total sum}                      &      &                                & \textbf{393.11} & \textbf{256}      & \textbf{1.54}            \\
		\hline
	\end{tabular}
	\caption{
		Same as \cref{t.chi_2-K0S} but for $\pi^0$ production data.
		Notice in this situation we only predict the data using the $\pi^0$ FFs that are constructed from $\pi^{\pm}$ FFs, \textit{i.e.}, a global analysis is \textit{not} performed for $\pi^0$.
	}
	\label{t.chi_2-pi_0}
\end{table}

For the SIA datasets, the total $\chi^2/N_{\mathrm{pt}}$ is 1.18 over 156 data points, reflecting a generally acceptable match between the $\pi^{\pm}$-based predictions and $\pi^0$ measurements.
However, individual experiments exhibit different levels of agreement.
Notably, the ALEPH 2000 (jet number 3 in three-jet events) dataset \cite{ALEPH:1999udi} displays a relatively high $\chi^2/N_{\mathrm{pt}}$ of 5.77 due to the high precision of the data.
This suggests potentially different constraints on gluon FFs from hadron-in-jet measurements in SIA versus $pp$ collisions, both of which contribute to the determination of charged pion FFs.
Other datasets, such as those from JADE 1985 \cite{JADE:1985bzp}, TASSO 1989 \cite{TASSO:1988jma}, DELPHI 1996 \cite{DELPHI:1995ase} and OPAL 1998 \cite{OPAL:1998enc}, yield $\chi^2/N_{\mathrm{pt}}$ values well below 1, indicating better alignment.

For the $pp$ collision datasets, unlike the $\pi^{\pm}$ fit where ratios of different hadrons at the same energy or the same hadron at different energies are used (resulting in reduced relative errors), here we rely on invariant cross sections, $E \dd[3]{\sigma}/\dd[3]{\vv{p}}$, and they generally have larger experimental uncertainties.
Despite this, several $pp$ datasets demonstrate a strong alignment with the theoretical predictions.
Specifically, the STAR 200 GeV data \cite{STAR:2009qzv} and ALICE 2.76 TeV data \cite{ALICE:2017nce} exhibit $\chi^2/N_{\mathrm{pt}}$ values of 0.76 and 0.87, respectively.
However, some datasets, such as PHENIX 2007 \cite{PHENIX:2007kqm} and 2016 \cite{PHENIX:2015fxo}, show higher $\chi^2/N_{\mathrm{pt}}$ values of 2.69 and 3.04, respectively, suggesting greater deviations.
The combined $\chi^2/N_{\mathrm{pt}}$ for the $pp$ group is 2.08 across 100 points, reflecting the varied levels of agreement between theory and data.

\begin{figure}[htbp]
	\centering
	\includegraphics[width = 0.9 \textwidth]{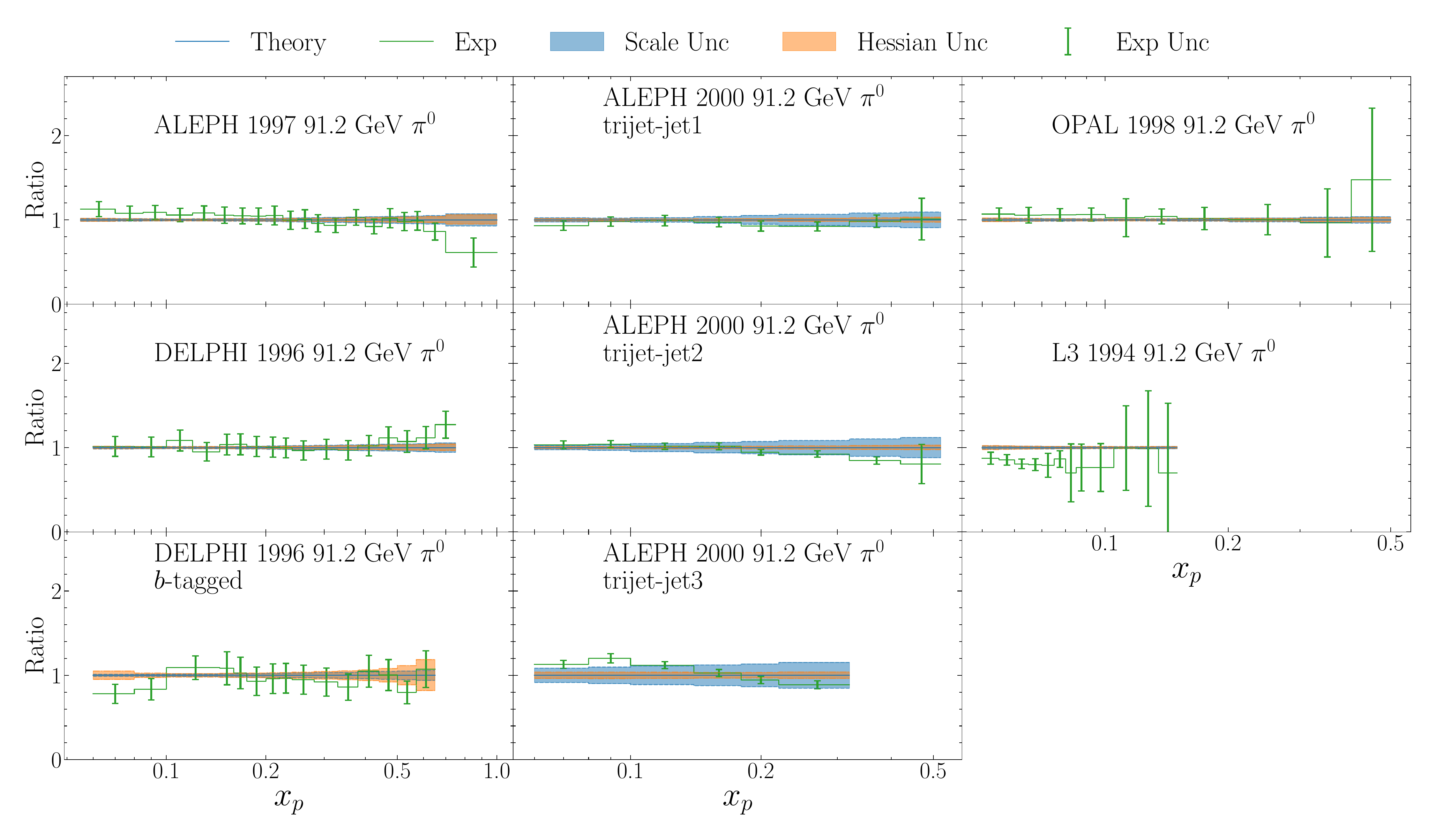}
	\caption{
		Same as \cref{f.SIA-Z_pole-K0S} but for $\pi^0$ production in SIA at $Z$-pole energy.
	}
	\label{f.SIA-Z_pole-pi_0}
\end{figure}

\cref{f.SIA-Z_pole-pi_0} show the comparison between theoretical calculations and SIA measurements at the $Z$-pole energy.
These comparisons encompass inclusive $\pi^0$ production data from four LEP experiments (ALEPH \cite{ALEPH:1996pxg}, DELPHI \cite{DELPHI:1995ase}, L3 \cite{L3:1994gkb} and OPAL \cite{OPAL:1998enc}), as well as the specialized tri-jet measurements from ALEPH \cite{ALEPH:1999udi} and $b$-tagged $\pi^0$ production from DELPHI \cite{DELPHI:1995ase}.
The experimental data from ALEPH, DELPHI and OPAL demonstrate remarkable consistency with theoretical predictions, maintaining ratios close to unity across the measured $x_p$ range.
A notable exception appears in the L3 data, which shows data-theory ratios of $\sim 0.75$, accompanied by substantial experimental uncertainties.
In the context of jet-specific analyses, the scale uncertainties dominate over their Hessian counterparts and show a characteristic increase with $x_p$, as previously discussed in \cref{ss.K0S-fit-quality,ss.Lambda-fit-quality}.
The theoretical predictions align well with ALEPH tri-jet measurements (for the first two jets), with most theoretical values falling within the experimental uncertainty bands.
The exception is on the third jet of the tri-jet events, in which a larger disagreement is found.
However, these differences remain comparable to the size of scale variations in general.
The $b$-tagged measurements from DELPHI show particularly good agreement in the intermediate and high $x_p$ regions ($x_p > 0.1$), though some fluctuations are evident in the low $x_p$ region.
The large Hessian uncertainties in this case reflect the large experimental uncertainties.

\begin{figure}[htbp]
	\centering
	\includegraphics[width = 0.9\textwidth]{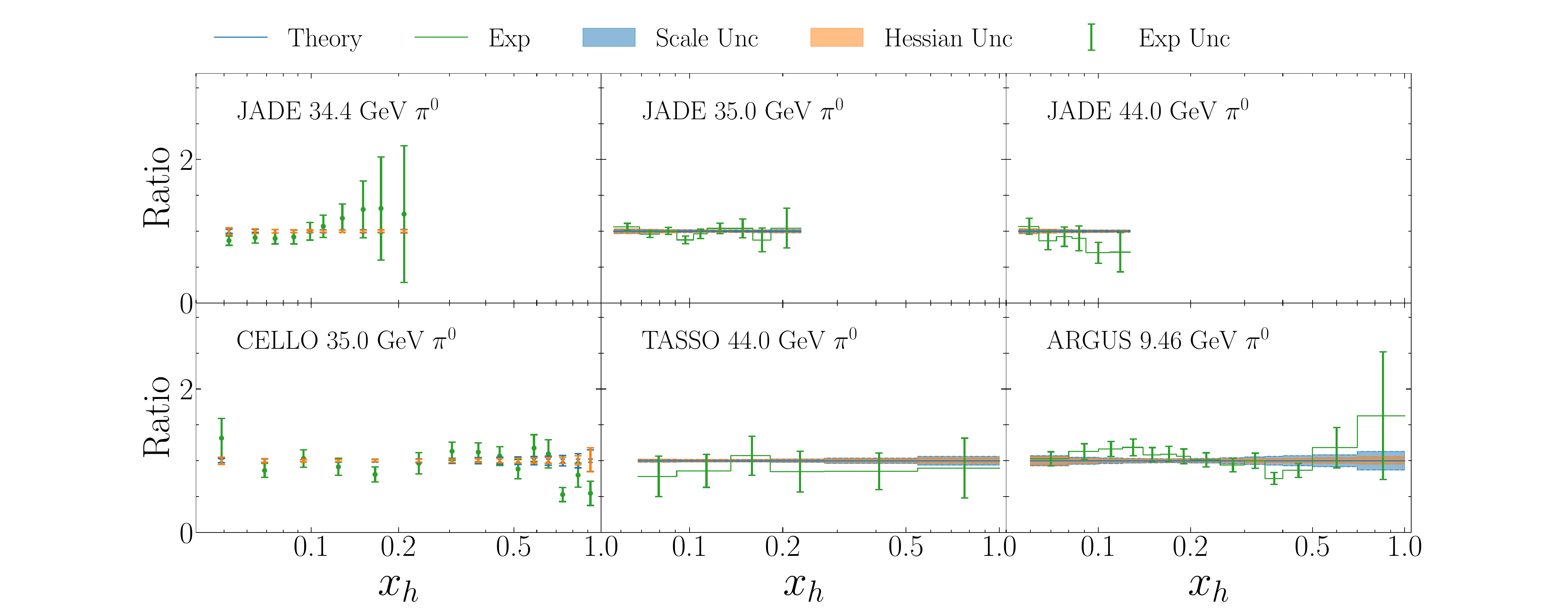}
	\caption{
		Same as \cref{f.SIA-Z_pole-K0S} but for $\pi^0$ production in SIA below $Z$-pole energy.
	}
	\label{f.SIA-below-Z_pole-pi_0}
\end{figure}

\cref{f.SIA-below-Z_pole-pi_0} illustrates the comparison between theoretical predictions and SIA measurements at various low-energy scales, spanning from 9.46 GeV to 44 GeV.
These datasets include ARGUS \cite{ARGUS:1989orf} at 9.49 GeV, JADE at 34.4 GeV \cite{JADE:1985bzp}, 35 and 44 GeV \cite{JADE:1989ewf}, CELLO \cite{Elze:1989gm} at 35 GeV and TASSO \cite{TASSO:1988jma} at 44 GeV.
The JADE 35 GeV data is well described by theoretical predictions, while the JADE 34.4 GeV data is not well described at low $x_p$ values.
The JADE 44 GeV data is well described at low $x_p$, but a large deviation is found at large $x_p$ values.
However, due to the large experimental uncertainties, the $\chi^2 / N_{\mathrm{pt}}$ for JADE experiments are all around 1.
The measurements from CELLO, TASSO and ARGUS, while showing some statistical fluctuations, are generally well described by the theoretical predictions.
Most data points and theoretical values fall within the 1$\sigma$ bands of each other.
The occasional excursions outside these uncertainty bands are consistent with expected statistical variations, suggesting that there are no significant systematic discrepancies between theory and experiment across these energy scales.

\begin{figure}[htbp]
	\centering
	\includegraphics[width = 0.9 \textwidth]{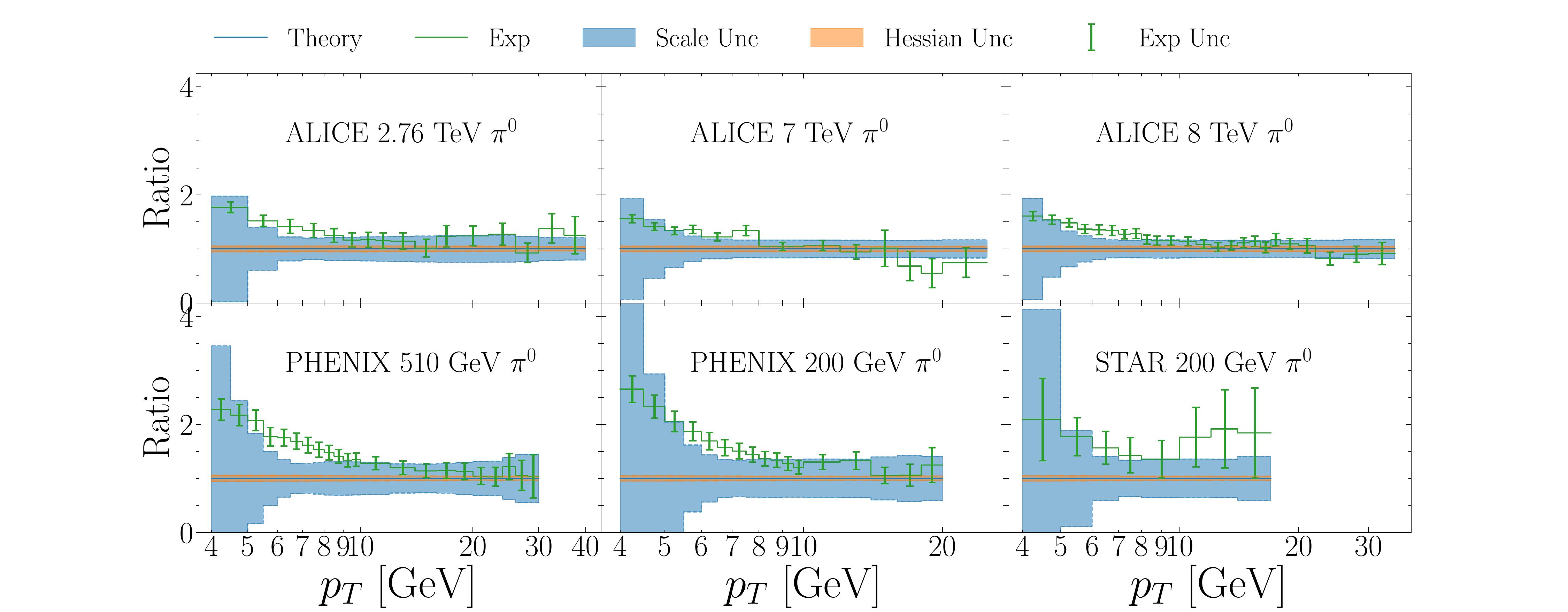}
	\caption{Same as \cref{f.SIA-Z_pole-K0S} but for $\pi^0$ production in $pp$ collisions.}
	\label{f.pp-pi_0}
\end{figure}

\cref{f.pp-pi_0} presents a comparison between theoretical calculations and experimental measurements for $\pi^0$ production in $pp$ collisions.
Each panel displays the ratio between experimental data and theoretical predictions, where the measurements and uncertainties are normalized to the theoretical central values.
The analysis encompasses data from multiple experiments: STAR measurements at 200 GeV \cite{STAR:2009qzv}, PHENIX results at 200 GeV \cite{PHENIX:2007kqm} and 510 GeV \cite{PHENIX:2015fxo}, as well as ALICE measurements at 2.76 TeV \cite{ALICE:2017nce}, 7 TeV \cite{ALICE:2012wos} and 8 TeV \cite{ALICE:2017ryd}.
A striking feature of these comparisons is the substantial scale uncertainty, especially in the low-$p_T$ region.
This occurs because of both the low energy scales and the need to extrapolate the FFs below their parametrization scale during evolution.
In contrast, the Hessian uncertainties exhibit remarkable stability and remain relatively modest across all transverse-momentum ranges.
Most experimental datasets are generally better described in the high $p_T$ region, and are greatly underestimated in the low $p_T$ region.
That is a typical feature as already observed in our previous analysis of charged pions \cite{Gao:2024dbv}.
This behavior suggests a systematic tension that persists across different experimental conditions and collision energies.
The STAR data, on the other hand, are not well described across the entire $p_T$ range.

In summary, for $\pi^0$ we present the predictions using FFs constructed from the NPC fitted $\pi^{\pm}$ FFs.
Although a global fit is not performed, a good agreement between data and prediction is achieved, resulting in a total $\chi^2 / N_{\mathrm{pt}}$ of 1.54.
The individual $\chi^2 / N_{\mathrm{pt}}$ values for SIA and $pp$ datasets are 1.18 and 2.08, respectively.
In general, the SIA datasets are well described by the constructed $\pi^0$ FFs, with exceptions on CELLO at 35 GeV \cite{Elze:1989gm}, ALEPH jet 3 at 91.2 GeV \cite{ALEPH:1999udi} and L3 at 91.2 GeV \cite{L3:1994gkb}.
The $pp$ collision datasets, on the other hand, are only generally well described in the high $p_T$ regions, which is consistent with the findings in \cite{Gao:2024dbv}.

\section{NPC23 delivered FFs}
\label{s.NPC23-FFs}

In this section, we present our FFs obtained from the NLO analysis of global data sets, for $K_S^0$ meson, $\Lambda$ baryon, $\eta$ meson and $\pi^0$ meson.

\begin{figure}[htbp]
	\centering
	\includegraphics[width = 0.95 \textwidth]{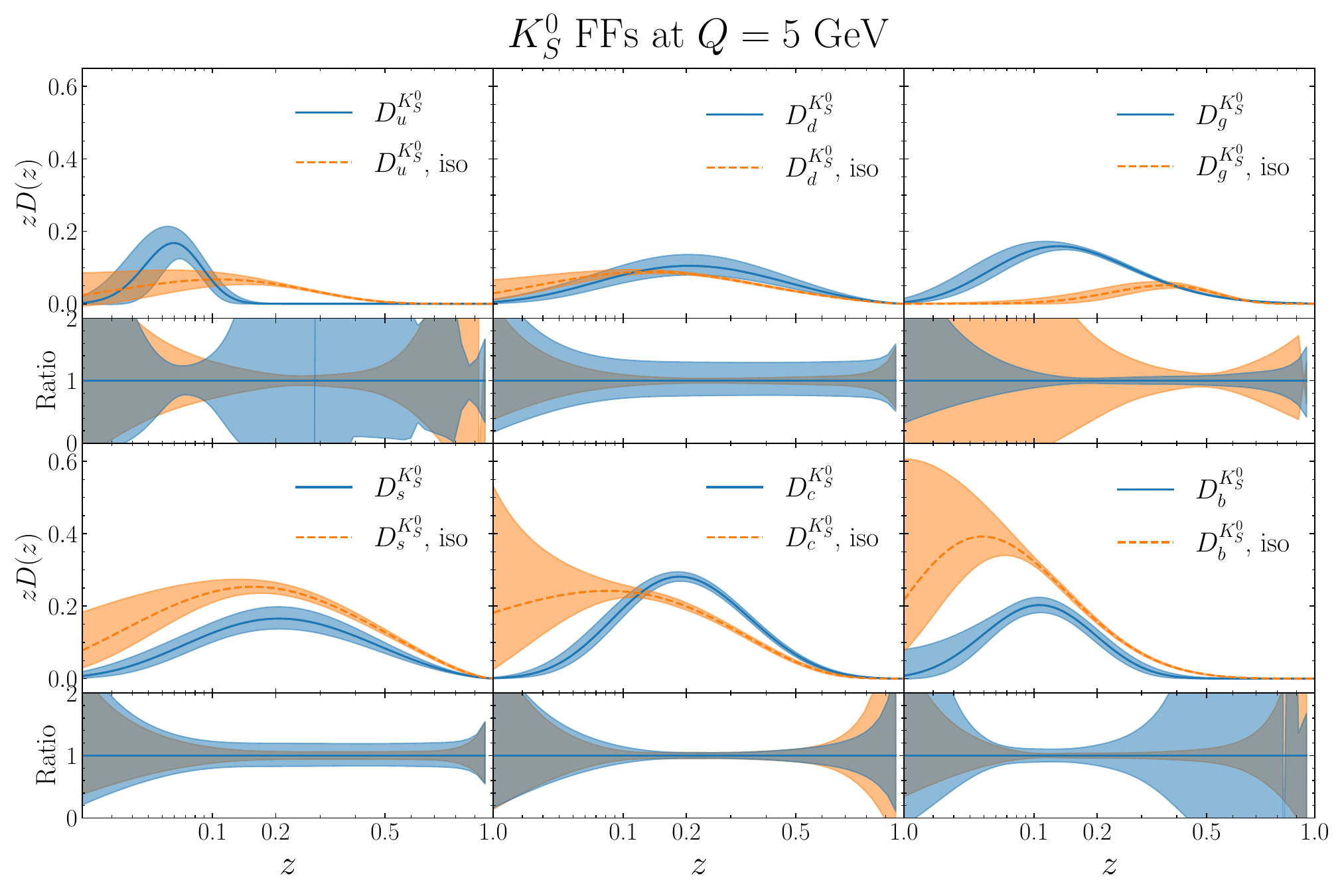}
	\caption{
		The NPC23 fragmentation functions (FFs) for $K_S^0$ at scale $Q = 5~\mathrm{GeV}$.
		The FFs are plotted with respect to $z$, the light-cone momentum fraction of parent parton carried by the hadron.
		The colored bands represent uncertainties estimated from Hessian method with 68\% confidence level, with blue for the fitted $K_S^0$ FFs and orange for the reconstructed $K_S^0$ FFs.
		The relative uncertainties are shown correspondingly in lower panels of each subplots.
		The relative uncertainties are not plotted up to $z = 1$ because the central values vanish towards the end point.
	}
	\label{f.FFs-K0S}
\end{figure}

\begin{figure}
	\centering
	\includegraphics[width = 0.9 \textwidth]{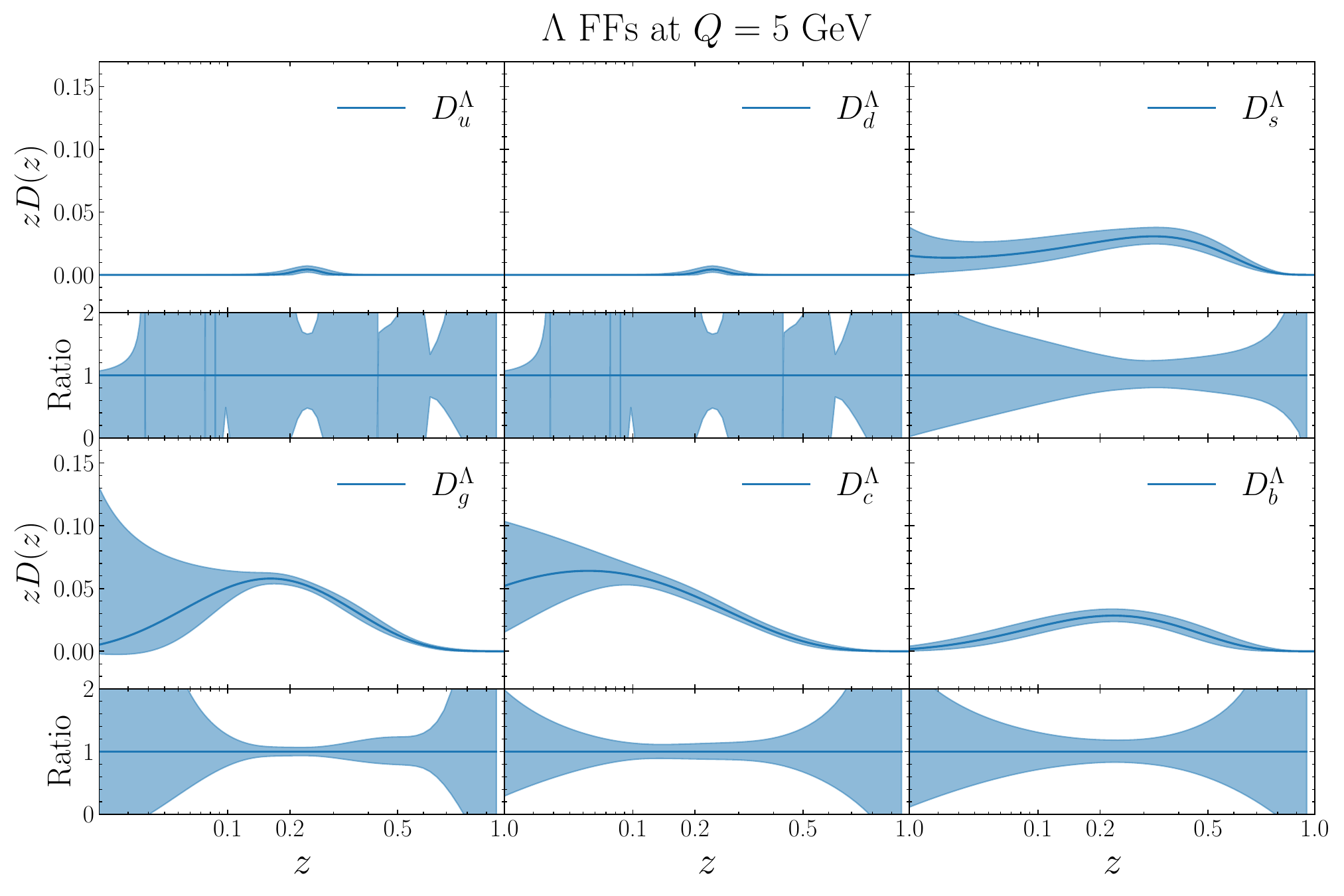}
	\caption{
		Similar to \cref{f.FFs-K0S}, but for $\Lambda$ FFs.
	}
	\label{f.FFs-Lambda}
\end{figure}

\begin{figure}[htbp]
	\centering
	\includegraphics[width = 0.9 \textwidth]{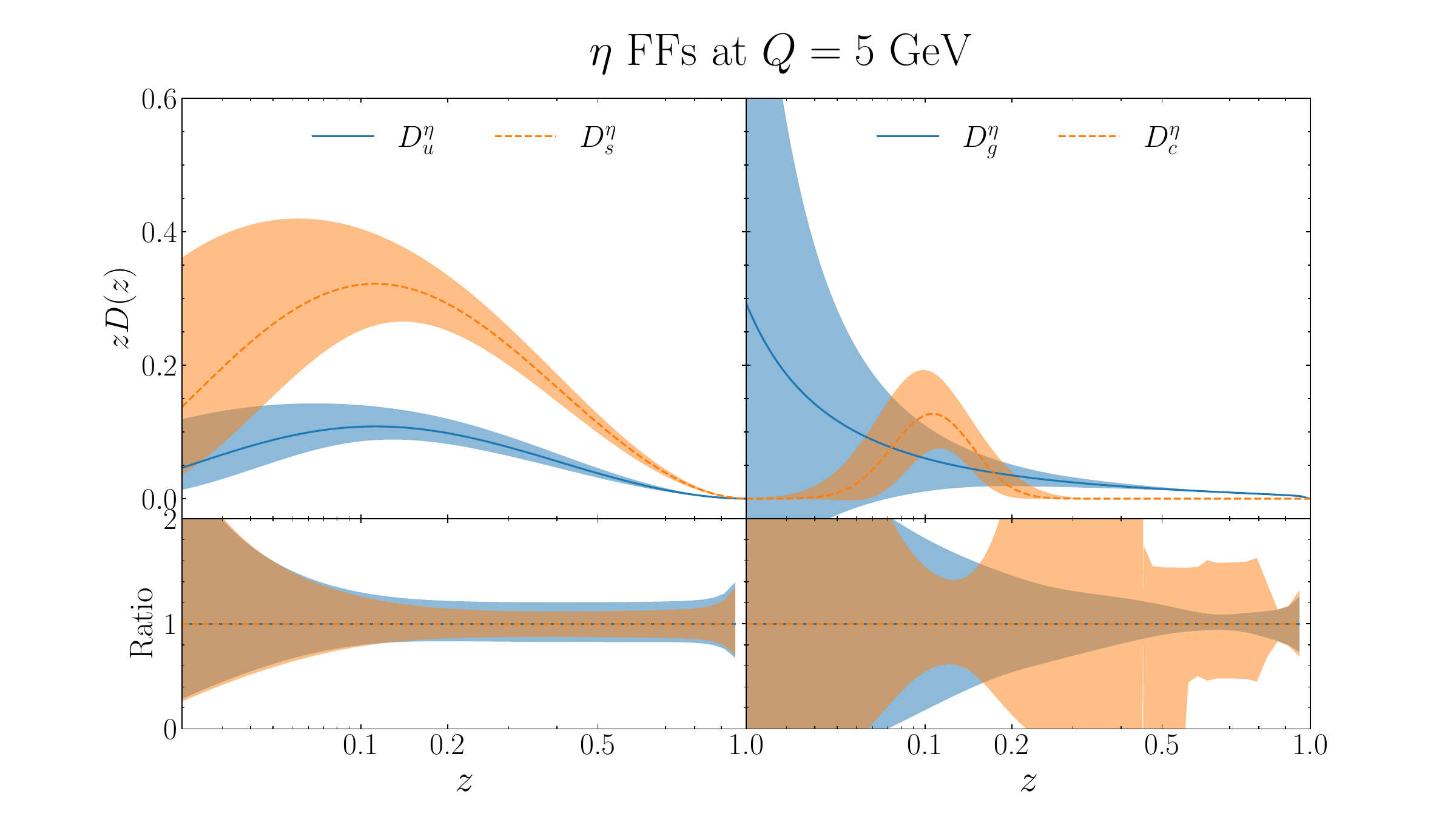}
	\caption{
	Similar to \cref{f.FFs-K0S}, but for $\eta$ FFs.
	We have set $D_u^{\eta} = D_d^{\eta}$ and $D_c^{\eta} = D_b^{\eta}$.
	}
	\label{f.FFs-eta}
\end{figure}

The extracted fragmentation functions for $K_S^0$, $\Lambda$ and $\eta$ are given in \cref{f.FFs-K0S,f.FFs-Lambda,f.FFs-eta} together with their uncertainties.
The FFs are plotted as functions of the light-cone momentum fraction $z$ at an initial scale $Q_0 = 5~\mathrm{GeV}$.
Both the actual values and ratios to the central values are shown in the plots.
We have set $D_q^h = D_{\overline{q}}^h$ for $h = K_S^0$, $\Lambda$ and $\eta$, with $q$ being all the active quark flavors.
The detailed reason for making this choice is given in \cref{ss.parameterization}.
The FFs are in general well constrained within $z \sim \pqty{0.1, 0.5}$.
Otherwise the FFs exhibit large uncertainties due to either precision of the data or the lack of power on flavor separation.
It is especially the case for the FFs of $\eta$ meson from gluon and heavy quarks ($b$ and $c$).
In \cref{f.FFs-K0S}, we show the fitted $K_S^0$ FFs in blue bands for the gluon and all five active quarks.
It should be noted that with SIA data only, one cannot discriminate FFs from $d$ and $s$ quarks.
This limitation arises because both quark types possess identical electric and weak charges, leading to indistinguishable contributions to the observed hadron production in SIA processes.
On the other hand, current data on neutral hadrons from SIDIS and $pp$ collisions are also not sufficient for separating $d$ and $s$ quarks due to limited measurements as well as the large uncertainties.
Thus we have applied a prior constraint on the ratio of FFs from $s$ to $d$ quarks, which is motivated by the relatively larger mass of strange quark, with details given in \cref{ss.parameterization,ss.K0S-fit-quality}.
With the aforementioned prior constraint, the $K_S^0$ FFs are well constrained for all parton flavors in a wide kinematic region.

We also show the $K_S^0$ FFs constructed from $K^{\pm}$ FFs in orange bands, labeled with ``iso'' in \cref{f.FFs-K0S}, assuming isospin symmetry.
The formulae for constructing $K_S^0$ FFs from isospin symmetry read
\begin{equation}
	D_q^{K_S^0}
	=
	\frac{1}{2} \pqty{D_q^{K^0} + D_q^{\overline{K}^0}}
	=
	\frac{1}{2} \pqty{D_{q'}^{K^+} + D_{q'}^{K^-}}
	, \label{e.K0S-construction-u-and-d}
\end{equation}
where $q \pqty{q'} = u \pqty{d}$ or $d \pqty{u}$.
While for flavors $q=s$, $c$, $b$ and gluon, we have:
\begin{equation}
	\label{e.K0S-construction-unfavored}
	\begin{gathered}
		D_q^{K_S^0}
		=
		\frac{1}{2} \pqty{D_q^{K^0} + D_q^{\overline{K}^0}}
		=
		\frac{1}{2} \pqty{D_q^{K^+} + D_q^{K^-}}
		, \\
		D_g^{K_S^0}
		=
		\frac{1}{2} \pqty{D_g^{K^0} + D_g^{\overline{K}^0}}
		=
		\frac{1}{2} \pqty{D_g^{K^+} + D_g^{K^-}}
		. \\
	\end{gathered}
\end{equation}
Here the $K^{\pm}$ FFs are from the NPC23 NLO analysis of charged hadrons~\cite{Gao:2024dbv}.
Both the fitted and constructed $K^0_S$ FFs exhibit a consistent pattern, with the FF originating from the
$s$ quark being slightly larger than that from the $d$ quark. However, it is worth noting that the error bands associated with the FFs of the $s$ quark barely overlap, indicating a degree of uncertainty in this observation.
The FFs from heavy quarks are well determined thanks to the inclusion of SLD measurements with $c$/$b$-tagging~\cite{SLD:1998coh}.
The charm quark FFs are slightly favored over those of the bottom quark, especially at large-$z$, probably due to its preference of decays into the strange quark.
The fitted and constructed FFs from $c$ and $b$ quark show clear differences which are much larger than the quoted uncertainties.
For $u$ quark, we find that the fitted and constructed FFs are significantly different, with the fitted result showing a peak around $z = 0.05$, while the constructed result being relatively flat.
Finally, for the FFs from gluon, the fitted results only agree with the constructed results in large-$z$ region, and is much larger than the constructed result in middle and low $z$ regions.

In general, the fitted and constructed FFs of $K_S^0$ do not coincide with each other particularly for those from gluon and  $u$, $c$, $b$ quarks.
These discrepancies could indicate possible isospin symmetry violation in fragmentation to kaons.
However, that can also be due to the poor quality of data or even tensions between different data sets.
For instance, the smaller gluon FFs to charged kaons are predominantly pulled by the hadron-in-jet data from $pp$ collisions which are absent in the analysis of neutral kaons.
The FFs to neutral kaons from $u$ quark are only directly constrained by a few data points from H1 and ZEUS collaboration with large uncertainties.
Therefore, further measurements, especially data from SIDIS or $pp$ collisions, will be needed for clarifications, as we will elaborate in \cref{ss.K0S-production-prediction}.
Similar observation concerning $K_S^0$ FFs from a direct fit and from a construction with isospin symmetry is also reported in the SAK20 analysis \cite{Soleymaninia:2020ahn}.

The $\Lambda$ FFs are shown in \cref{f.FFs-Lambda}, where a good constraint is found for all quark flavors and gluon in the region $z \gtrsim 0.1$.
We have set the FFs from $u$ and $d$ quarks to be equal due to insufficient experimental data to constrain them independently.
Additionally, it has been argued in previous studies~\cite{Chen:2021hdn, Chen:2021zrr} that isospin symmetry holds well for fragmentation to $\Lambda$ baryon, further supporting the validity of this approach.

Our study shows that strange quark is clearly favored compared to $u$ and $d$ quarks, despite the fact that they are all constituent quarks.
This can be expected because strange quark has a much larger mass than that of $u$ and $d$ quarks, and this makes $\Lambda$ baryon less likely to be produced from $u$ and $d$ quarks.
As for heavy quarks, the $c$ quark FF exhibits a typical behavior in which the probability peaks at low $z$ and decreases monotonically with increasing $z$.
The $b$ quark is clearly disfavored compared to the $c$ quark.
This is probably due to the same fact as explained in the case of $K_S^0$ FFs.
However, we do emphasize that, due to relatively weaker constraints for $\Lambda$ baryon compared to $\pi$ and $K$ mesons, the behavior of $b$ quark FFs has to be further studied when more data are available.
As for the FFs from gluon, one interesting aspect is that, compared to the case of $K_S^0$ meson, the FFs of $\Lambda$ baryon peak at a slightly larger $z$ value.
This can be explained by the fact that a heavier hadron requires gluons to carry larger momentum fractions compared to those in lighter hadrons~\cite{Hirai:2007cx, Sato:2016wqj, Moffat:2021dji}.
The same pattern can also be observed when comparing the FFs from bottom quark to $K_S^0$ and $\Lambda$.

The $\eta$ FFs are shown in \cref{f.FFs-eta}.
According to the symmetries in the $\eta$ wave function  we have set $D_u^{\Lambda} = D_d^{\Lambda}$.
Furthermore, due to lack of direct constraint from the world data with heavy quark tagging, we assume $D_c^{\eta} = D_b^{\eta}$.
The FFs of the strange quark are roughly twice as large as those of the up quark, which is expected given the relatively larger mass of the strange quark and the fact that $s$ is more favored in the $\eta$ wave function $\ket{\eta} = (\ket{u \overline{u}} + \ket{d \overline{d}} - 2 \ket{s \overline{s}})/\sqrt 6$.
We observe that the FFs of $\eta$ meson from heavy quarks are much less than those of the favored quarks, except at $z \sim 0.1$, where the FFs from the charm (bottom) quark are slightly higher than those of the up (down) quarks.
Although relatively good constraints on quarks are obtained, the FFs for gluons remain poorly constrained in the low $z$ region.
The large uncertainties of $\eta$ FFs compared to those of $K_S^0$ and $\Lambda$ clearly indicate that further high-precision measurements of $\eta$ meson production are needed.

\begin{figure}[htbp]
	\centering
	\includegraphics[width = 0.9 \textwidth]{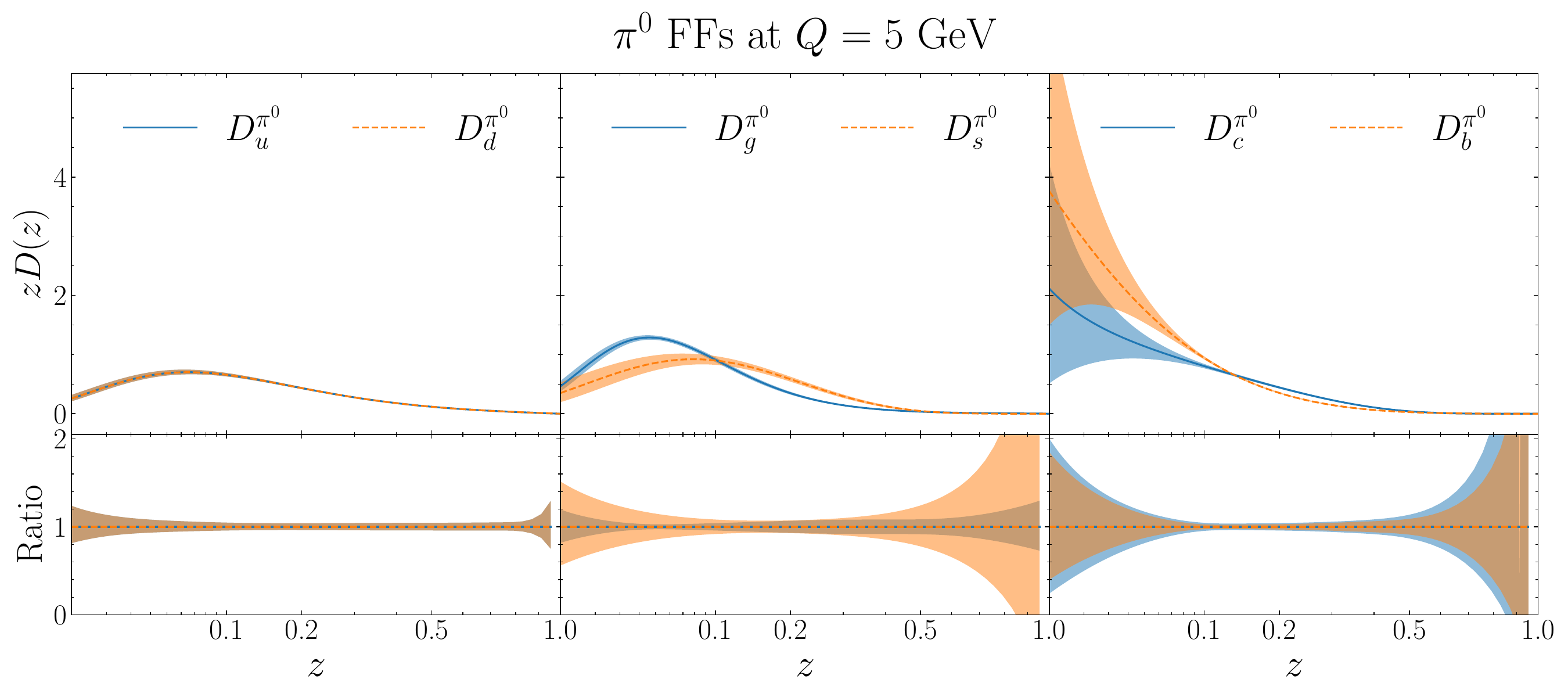}
	\caption{
		Similar to \cref{f.FFs-K0S}, but for $\pi^0$ FFs.
		The FFs are constructed via isospin symmetry from those of charged pions.
	}
	\label{f.FFs-pi_0}
\end{figure}

Finally, we show our $\pi^0$ FFs in \cref{f.FFs-pi_0}, which are computed via isospin symmetry from $\pi^{\pm}$ FFs given in previous NPC23 analyses \cite{Gao:2024dbv}:
\begin{equation}
	D_q^{\pi^0}
	=
	\frac{1}{4} \pqty{D_{q}^{\pi^+} + D_{q}^{\pi^-} + D_{q'}^{\pi^+} + D_{q'}^{\pi^-}}
	, \label{e.pi_0-construction-u-and-d}
\end{equation}
where $q \pqty{q'} = u \pqty{d}$ or $d \pqty{u}$.
While for gluon and the remaining quark flavors, we have:
\begin{equation}
	D_q^{\pi^0}
	=
	\frac{1}{2} \pqty{D_q^{\pi^+} + D_q^{\pi^-}}
	, \quad
	D_g^{\pi^0}
	=
	\frac{1}{2} \pqty{D_g^{\pi^+} + D_g^{\pi^-}}
	. \label{e.pi_0-construction-unfavored}
\end{equation}
By this formalism, $D_d^{\pi^0}=D_{\overline{d}}^{\pi^0}=D_u^{\pi^0}=D_{\overline{u}}^{\pi^0}$.
The $\pi^0$ FFs from the remaining quarks and the gluon are essentially the same as those of the $\pi^{\pm}$.
Importantly, we find that the $\pi^0$ FFs obtained in this way can describe very well the world data on $\pi^0$ production (see \cref{ss.pi_0-cross-section}).
This together with the result from our previous analysis of charged pions indicate that the isospin symmetry is well valid in fragmentation to pions.

\clearpage
\section{The NPC23 predictions}
\label{s.prediction}

In this section, we provide predictions to physical observables/quantities that are of interest.
We first present the truncated moments which are related to the momentum sum rule of FFs.
Then we provide predictions for the in-jet production of neutral hadrons at LHC kinematics.
Finally, the case of kaon production is further studied in \cref{ss.K0S-production-prediction} using fitted $K_S^0$ FFs and constructed $K_S^0$ FFs via \cref{e.K0S-construction-u-and-d,e.K0S-construction-unfavored}, respectively.
The difference in the resulting cross sections is significant, thus it is promising that measurement of such observables will help us further test the isospin symmetry in kaon fragmentation.

\subsection{Moments and sum rules of FFs}

The FFs represent the number density of finding a hadron inside its parent parton, and they satisfy the momentum sum rule.
However, the satisfaction of momentum sum rule has been questioned recently in \cite{Collins:2023cuo}.
In our global analysis, we have the most comprehensive species of FFs, including charged hadrons $\pi^{\pm}$, $K^{\pm}$ and $p/\overline{p}$~\cite{Gao:2024dbv}, as well as neutral hadrons $\pi^0$, $K_S^0$, $\Lambda$ and $\eta$ from this work.
In this section, we provide a test on the momentum sum rule utilizing results on FFs of neutral hadrons.
Since the experimental data only constrain the FFs down to a finite $z_{\min}$, we define the truncated moments as:
\begin{equation}
	\expval{z}_i^h
	=
	\int_{z_{\min}}^1 \dd{z} z D_i^h \pqty{z}
	, \label{e.moments-of-FFs}
\end{equation}
where $i$ is the flavor of the parent parton and $h$ is the final state of the identified hadron.

In \cref{t.moments-of-neutral-FFs}, we present the $\expval{z}_i^h$ for the neutral hadron FFs fitted in this work, \textit{i.e.}, $K_S^0$, $\Lambda$ and $\eta$, as well as for the $\pi^0$ FFs which are constructed from the $\pi^{\pm}$ FFs via isospin symmetry.
The $\expval{z}_i^h$ values are computed with fixed $z_{\min}=0.05$ for different parton flavors.
The momentum fractions for $K_S^0 + K_L^0$ are obtained by multiplying the truncated moments of $K_S^0$ by a factor of two, and similarly, the total momentum fractions for $\Lambda + \overline{\Lambda}$ are twice of those for $\Lambda$.

\begin{table}[ht]
	\setcellgapes{2.5 pt}
	\makegapedcells
	\centering
	\begin{tabular}{|c|c|c|c|c|c|c|}
		\hline
		hadron                       & $g$ ($z_{\min} = 0.05$)      & $d$ ($z_{\min} = 0.05$)      & $u$ ($z_{\min} = 0.05$)      & $s$ ($z_{\min} = 0.05$)      & $c$ ($z_{\min} = 0.05$)      & $b$ ($z_{\min} = 0.05$)      \\
		\hline
		$K_S^0+K_L^0$                & $0.0922^{+0.0051}_{-0.0040}$ & $0.0940^{+0.0278}_{-0.0223}$ & $0.0155^{+0.0046}_{-0.0047}$ & $0.1486^{+0.0280}_{-0.0243}$ & $0.1829^{+0.0112}_{-0.0105}$ & $0.0685^{+0.0092}_{-0.0082}$ \\
		\hline
		$\Lambda+\overline{\Lambda}$ & $0.0385^{+0.0049}_{-0.0037}$ & $0.0006^{+0.0006}_{-0.0004}$ & $0.0006^{+0.0006}_{-0.0004}$ & $0.0296^{+0.0070}_{-0.0059}$ & $0.0321^{+0.0038}_{-0.0034}$ & $0.0228^{+0.0052}_{-0.0044}$ \\
		\hline
		$\eta$                       & $0.0192^{+0.0083}_{-0.0077}$ & $0.0420^{+0.0092}_{-0.0068}$ & $0.0420^{+0.0092}_{-0.0068}$ & $0.1247^{+0.0189}_{-0.0174}$ & $0.0119^{+0.0071}_{-0.0056}$ & $0.0119^{+0.0071}_{-0.0056}$ \\
		\hline
		$\pi^0$                      & $0.1577^{+0.0072}_{-0.0072}$ & $0.1829^{+0.0076}_{-0.0065}$ & $0.1829^{+0.0076}_{-0.0065}$ & $0.1954^{+0.0160}_{-0.0156}$ & $0.1674^{+0.0076}_{-0.0076}$ & $0.1636^{+0.0060}_{-0.0057}$ \\
		\hline
		\textbf{sum}                 & $0.3076^{+0.0131}_{-0.0119}$ & $0.3194^{+0.0303}_{-0.0242}$ & $0.2409^{+0.0128}_{-0.0105}$ & $0.4984^{+0.0380}_{-0.0342}$ & $0.3944^{+0.0157}_{-0.0146}$ & $0.2668^{+0.0141}_{-0.0123}$ \\
		\hline
	\end{tabular}
	\caption{
		The total momentum fractions of parent partons carried by various final-state hadrons, including $K_S^0 + K_L^0$, $\Lambda + \overline{\Lambda}$, $\eta$ and $\pi^0$, are presented.
		The momentum fractions are defined in \cref{e.moments-of-FFs}, with $z_{\min}$ values listed in the first row of this table.
		The cental values and Hessian uncertainties at 68\% confidence level are computed from our nominal fit at $Q = 5~\mathrm{GeV}$.
		The last row lists the sum for all hadrons in this table.
		Since the neutral hadron FFs are not correlated with each other, the uncertainties of the sum are determined by combining the uncertainties of individual hadrons in quadrature.
	}
	\label{t.moments-of-neutral-FFs}
\end{table}

From the results presented in \cref{t.moments-of-neutral-FFs}, we observe that the listed neutral hadrons carry approximately $24\% \sim 39\%$ of the momentum from gluons, $u$ quarks, $d$ quarks, $c$ quarks and $b$ quarks.
A notable feature is the large fraction of momentum carried by the neutral hadrons from the strange quark, which is roughly 49\% of the total $s$ quark momentum.
Of this, both $\pi^0$ and $K_S^0 + K_L^0$ contribute dominantly, suggesting a significant role for these hadrons in the fragmentation of strange quark.

\begin{figure}[hbtp]
	\begin{subfigure}[b]{0.45 \textwidth}
		\includegraphics[width = \textwidth]{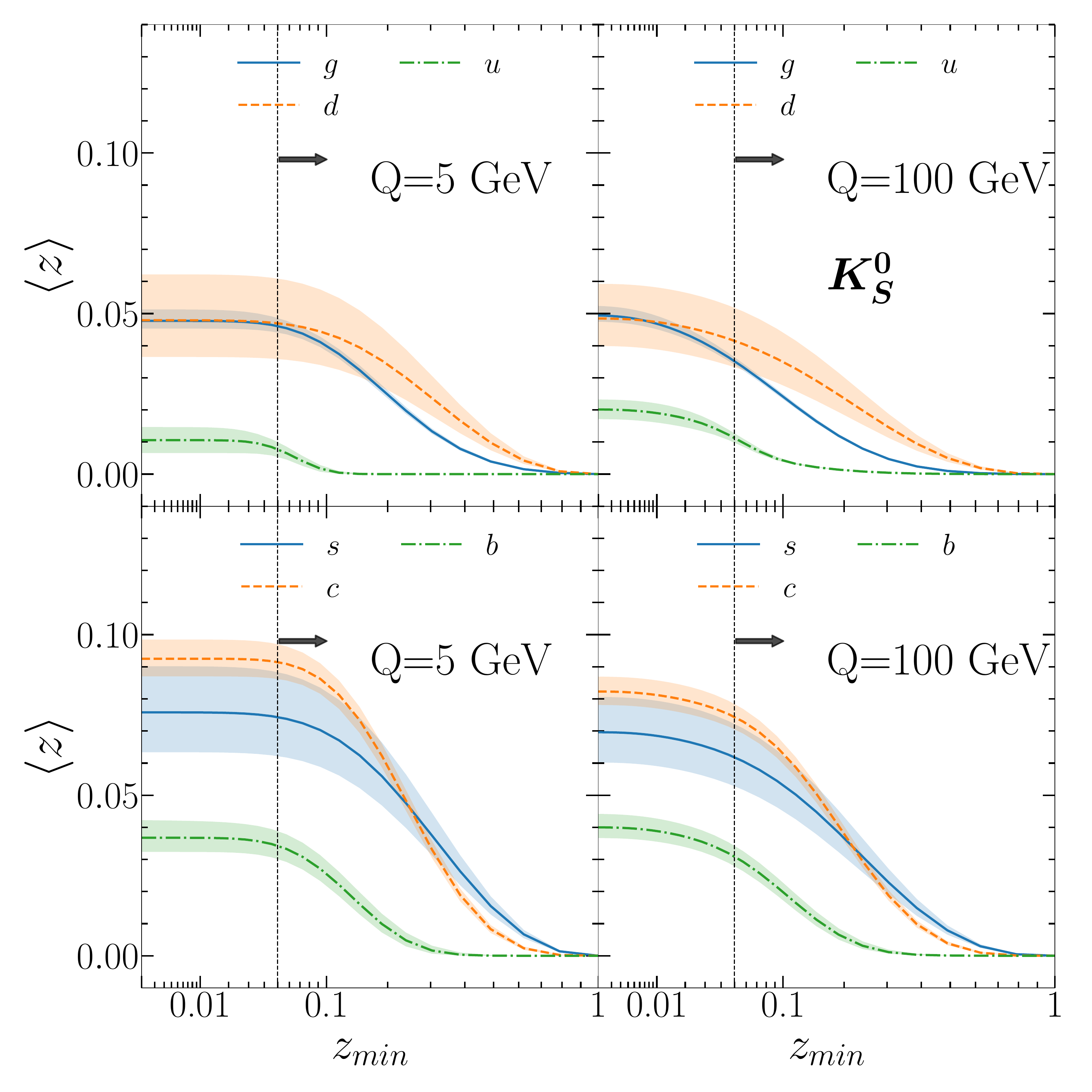}
		\caption{$K_S^0$}
		\label{f.hadron-moments-K0S}
	\end{subfigure}
	\hfil
	\begin{subfigure}[b]{0.45 \textwidth}
		\includegraphics[width = \textwidth]{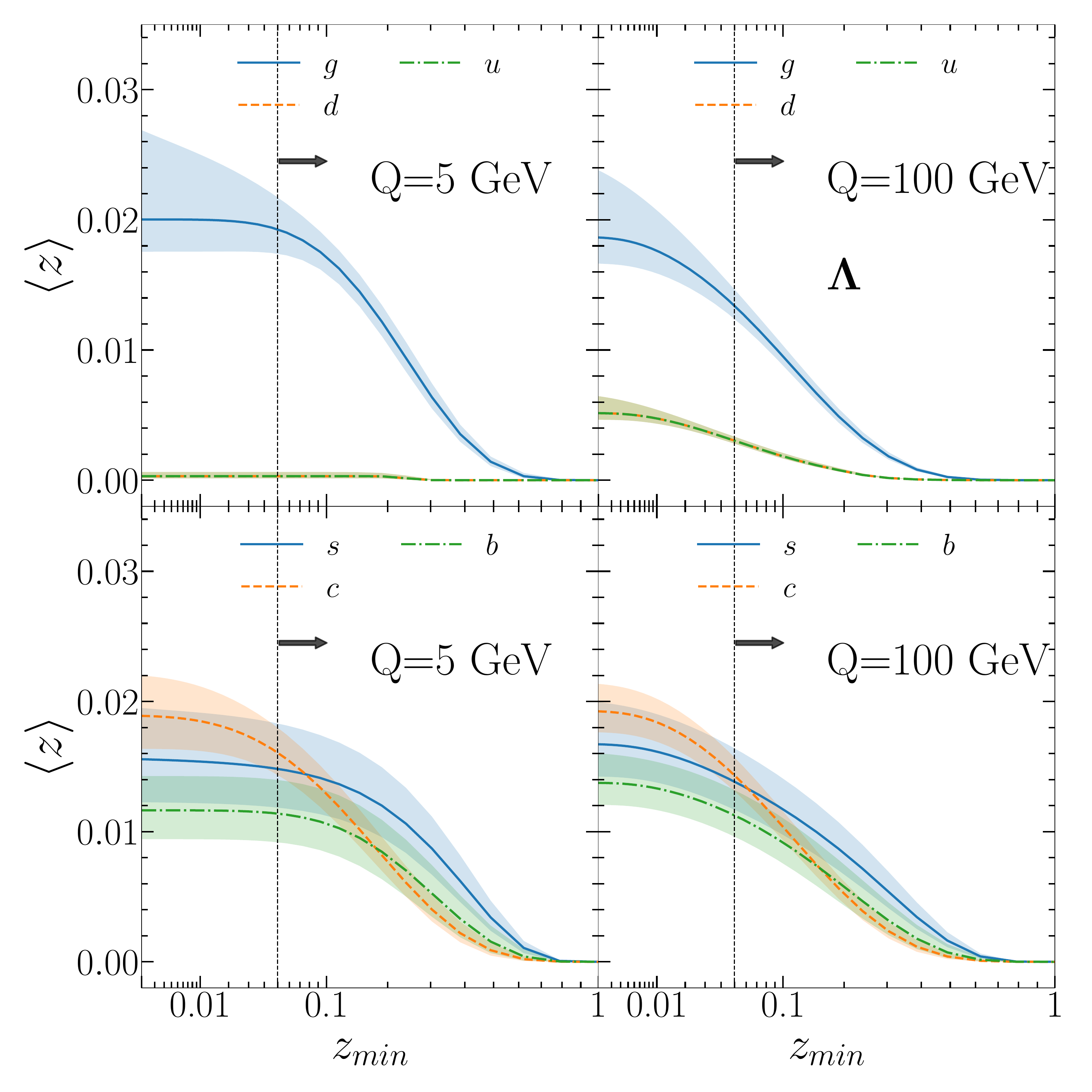}
		\caption{$\Lambda$}
		\label{f.hadron-moments-Lambda}
	\end{subfigure}
	\caption{
		Average momentum fraction carried by different neutral hadrons ($K_S^0$ and $\Lambda$) fragmented from various partons including $u$, $d$, $s$, $c$, $b$ and $g$ as a function of $z_{\min}$.
		Notice in this plot, instead of $K_S^0 + K_L^0$ or $\Lambda + \overline{\Lambda}$, we show the truncated moments for $K_S^0$ and $\Lambda$ alone.
		In order to better visualize both ends of $z_{\min}$ spectrum, we made the plot in $z_{\min}^{1/3}$ scale.
		The central values and uncertainties at 68\% C.L. are calculated from our best-fit and Hessian error FFs at $Q = 5~\mathrm{GeV}$ and 100 GeV.
		The vertical lines indicate the kinematic coverage of experimental data.
	}
	\label{f.hadron-moments-K0S-and-Lambda}
\end{figure}

\begin{figure}[hbtp]
	\begin{subfigure}[b]{0.45 \textwidth}
		\includegraphics[width = \textwidth]{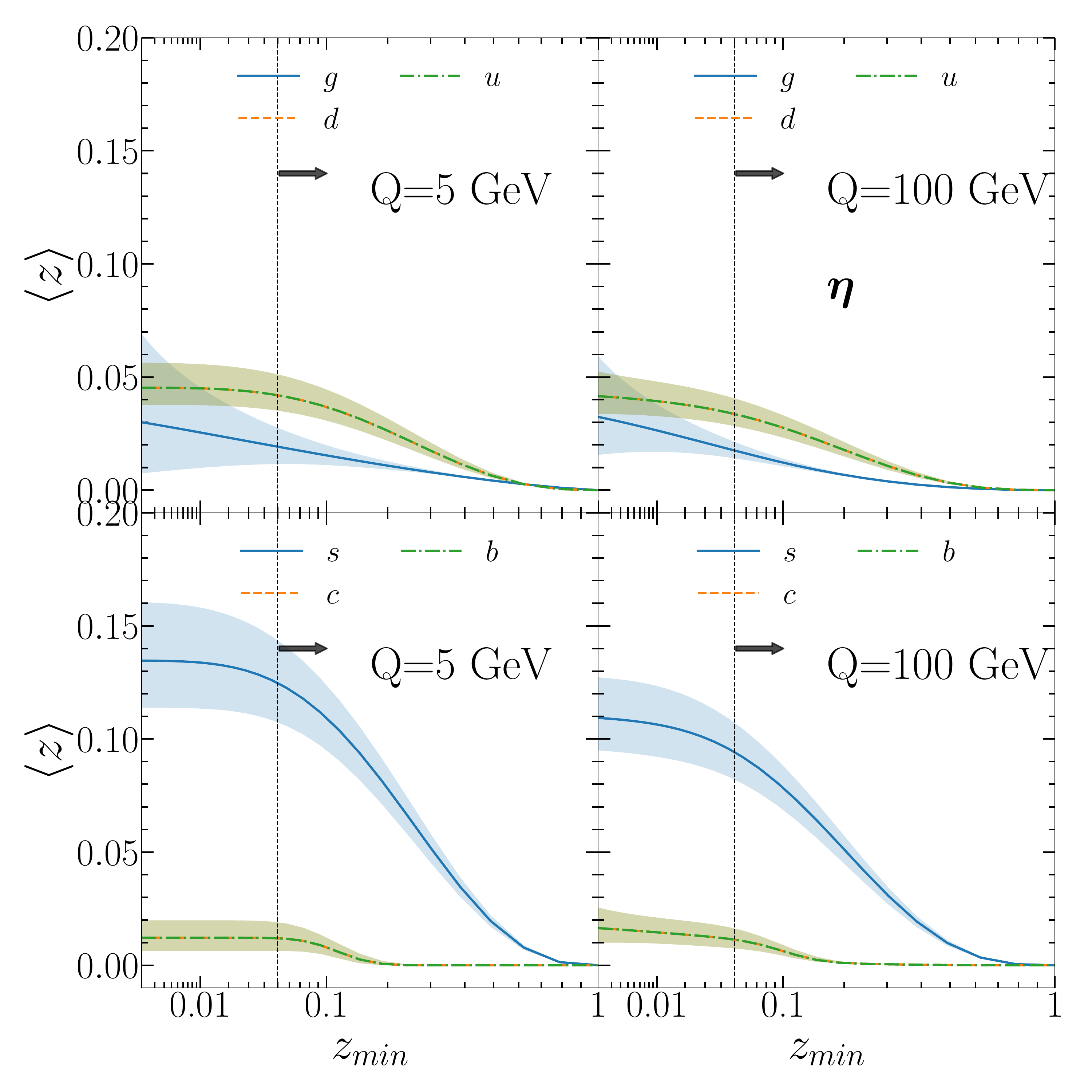}
		\caption{$\eta$}
		\label{f.hadron-moments-eta}
	\end{subfigure}
	\hfil
	\begin{subfigure}[b]{0.45 \textwidth}
		\includegraphics[width = \textwidth]{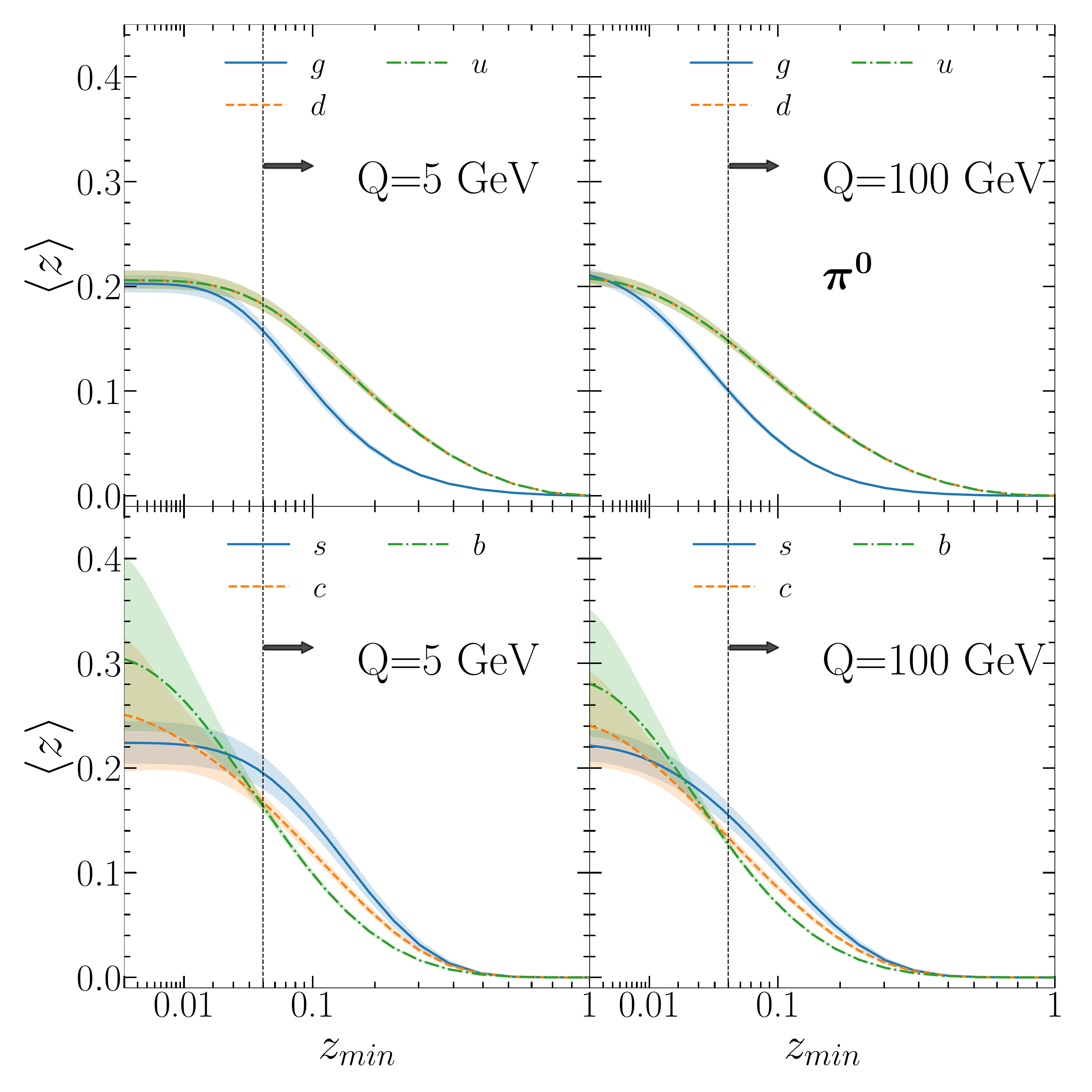}
		\caption{$\pi^0$}
		\label{f.hadron-moments-pi_0}
	\end{subfigure}
	\caption{
	Same as \cref{f.hadron-moments-K0S-and-Lambda} but for $\eta$ and $\pi^0$.
	The spectra of $u$ and $d$, as well as $c$ and $b$ in \cref{f.hadron-moments-eta} completely overlap with each other because we have set $D_u^{\eta} = D_d^{\eta}$ and $D_c^{\eta} = D_b^{\eta}$ in our fits (see \cref{ss.parameterization} for more discussion).
	}
	\label{f.hadron-moments-eta-and-pi_0}
\end{figure}

In \cref{f.hadron-moments-K0S-and-Lambda,f.hadron-moments-eta-and-pi_0}, we present $\expval{z}_i^h$ as a function of $z_{\min}$ for the neutral hadrons $K_S^0$, $\Lambda$, $\eta$, and $\pi^0$.
The plots are organized such that in each subplot, the upper panels show results for the gluon, $d$ quark, and $u$ quark, while the lower panels display the results for the $s$ quark, $c$ quark, and $b$ quark.
In addition, to see the effect of QCD evolution, in each subplot we plot the truncated moments at two different scales.
In the left panel we show results at $Q = 5~\mathrm{GeV}$, and in the right panels for $Q = 100~\mathrm{GeV}$.
We extrapolate $z_{\min}$ down to 0.001, and use the vertical lines to indicate the kinematic limit of data used in the analysis.

As shown in \cref{f.hadron-moments-K0S}, the truncated moments of $K_S^0$ saturate at $z_{\min} \approx 0.06$ at $Q = 5~\mathrm{GeV}$.
However, at $Q = 100~\mathrm{GeV}$, the FFs are pushed toward smaller $z$ and thus the truncated moments flatten at much smaller $z_{\min}$ values.
A similar behavior is also observed for $\Lambda$.
The momentum sharing to $K_S^0$ for $d$ and $s$ quarks, as well as momentum sharing to $\Lambda$ for gluon, $s$ and $b$ quarks are not well constrained by the data, indicating the need for future high-precision measurements.

In \cref{f.hadron-moments-eta-and-pi_0}, the truncated moments for $\eta$ and $\pi^0$ are presented.
Since the $\eta$ FFs from strange quark and gluon are not well constrained by data, they show relatively larger uncertainties at both $Q = 5~\mathrm{GeV}$ and $Q = 100~\mathrm{GeV}$ in \cref{f.hadron-moments-eta}.
Since we have set $D_u^{\eta} = D_d^{\eta}$ and $D_c^{\eta} = D_b^{\eta}$ in our fits, their curves overlap with each other.
The $\eta$ FFs from $u$, $d$, $c$ and $b$ quarks are relatively better constrained, thus their moments show smaller uncertainties.
In the context of $\pi^0$, the truncated moments saturate at $z_{\min} \approx 0.02$ at $Q = 5~\mathrm{GeV}$, except for $c$ and $b$ which are not well constrained at small $z$ values.
At $Q = 100~\mathrm{GeV}$, the $\pi^0$ FFs are pushed toward smaller $z$, and the saturation happens at a much smaller $z_{\min}$.

\begin{figure}[hbtp]
	\begin{subfigure}[b]{0.45 \textwidth}
		\includegraphics[width = \textwidth]{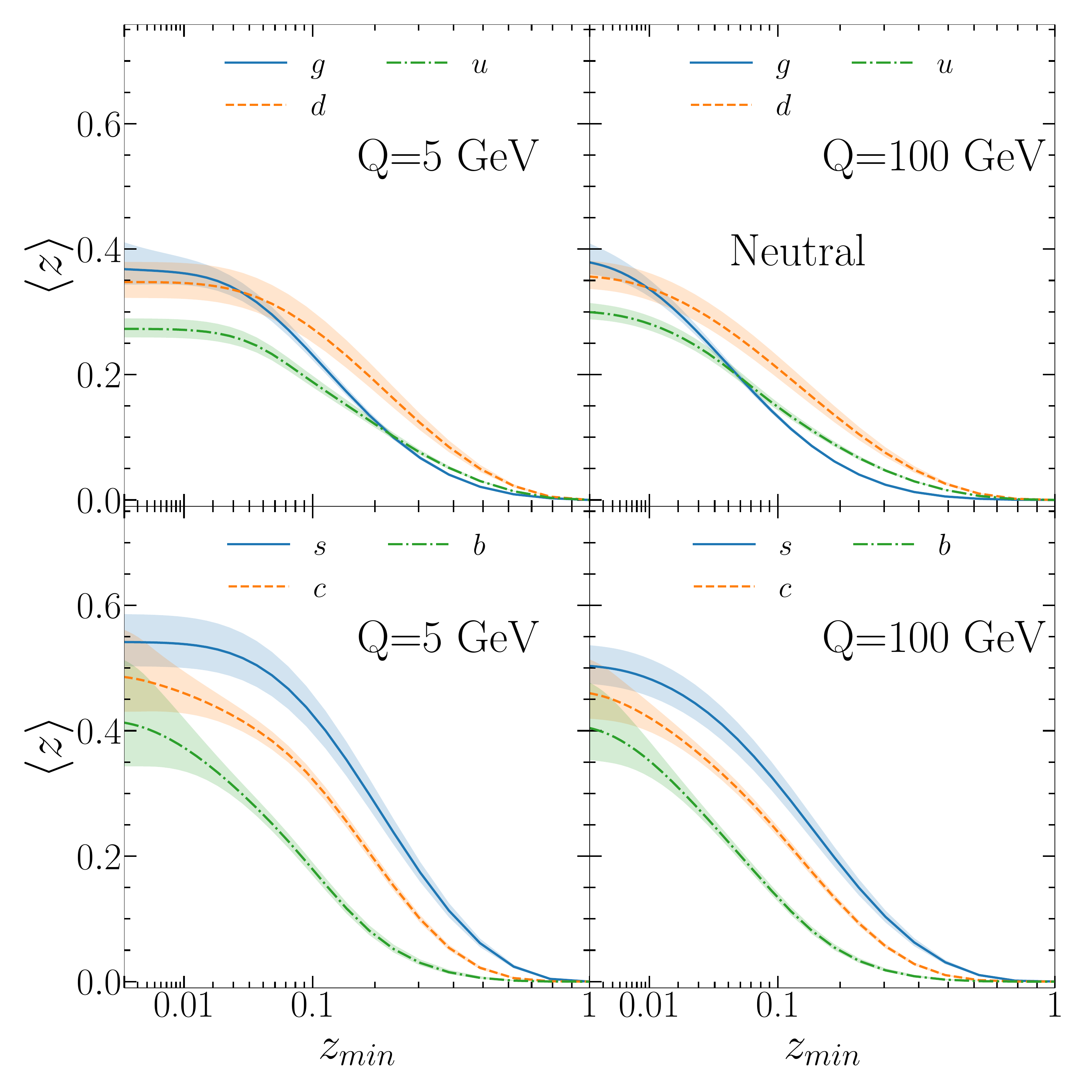}
		\caption{Sum of neutral hadrons}
		\label{f.neutral-hadron-moments}
	\end{subfigure}
	\begin{subfigure}[b]{0.45 \textwidth}
		\includegraphics[width = \textwidth]{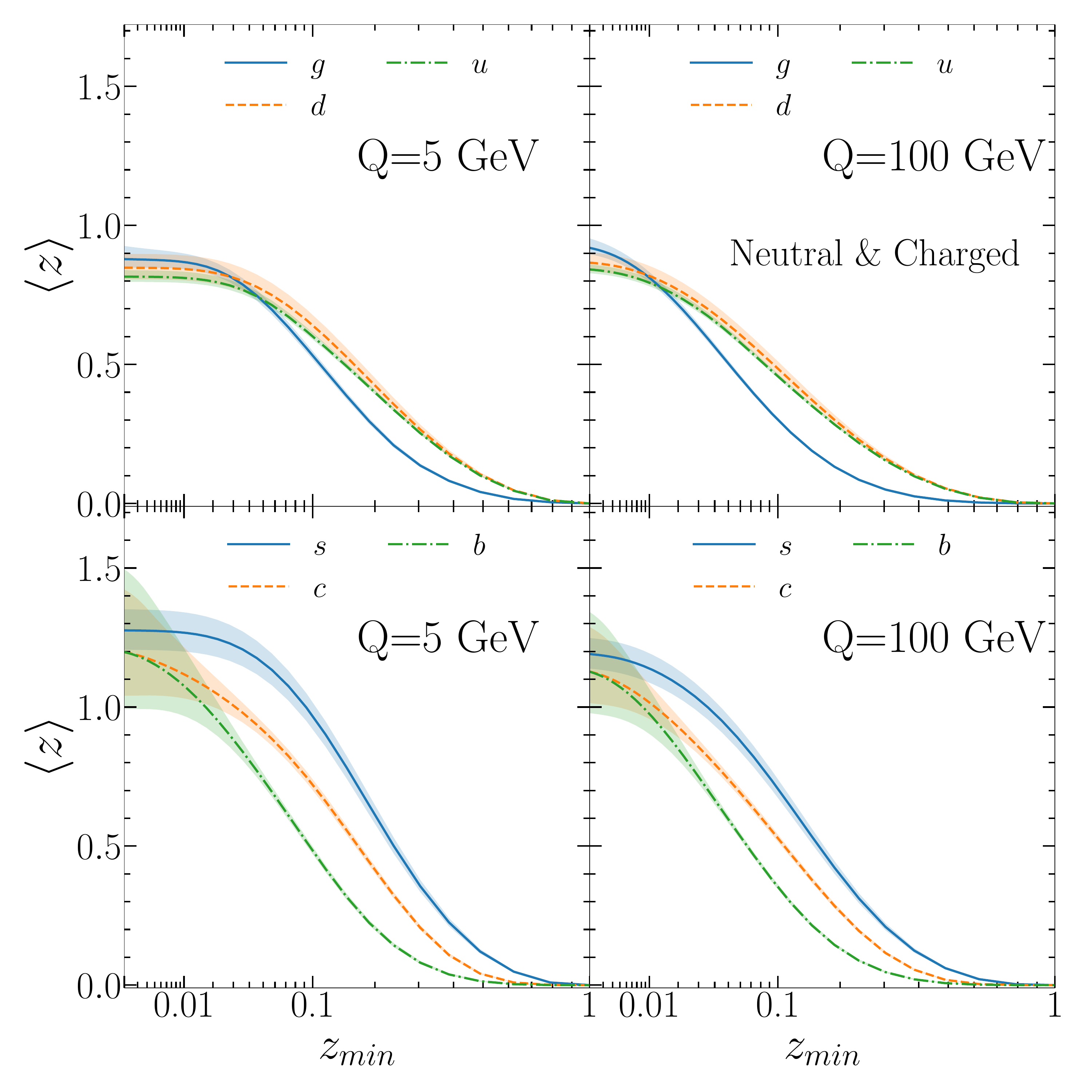}
		\caption{Sum of neutral and charged hadrons}
		\label{f.charged-and-neutral-hadron-moments}
	\end{subfigure}
	\caption{
		Same as \cref{f.hadron-moments-K0S-and-Lambda} but for sums of neutral hadrons, and for sums of charged hadrons from previous work \cite{Gao:2024dbv} and neutral hadrons from this work.
	}
	\label{f.summed-hadron-moments}
\end{figure}

Finally, in \cref{f.summed-hadron-moments} we present the momentum sum of all neutral hadrons from this work, as well as the momentum sum of both charged hadrons and neutral hadrons.
From \cref{f.neutral-hadron-moments}, it can be seen that at the initial scale, the momentum sum of neutral hadrons for gluon and $d$, $u$, $s$ quarks stabilizes at $z_{\min}$ slightly above 0.01.
In contrast, for $c$ and $b$ quarks, the momentum fractions continue to grow at $z_{\min} = 0.01$, although with large uncertainties.
At $Q = 100~\mathrm{GeV}$, again, all curves flatten at much smaller $z_{\min}$ values.
What is much more interesting is the total truncated moment of all available hadrons shown in \cref{f.charged-and-neutral-hadron-moments}.
These include $\pi^{\pm}$, $K^{\pm}$, $p + \overline{p}$, $K_S^0 + K_L^0$, $\Lambda + \overline{\Lambda}$, $\eta$ and $\pi^0$.
We can see that these hadrons take approximately $80\% \sim 90\%$ of momentum from gluon, $u$ and $d$ quarks in the fragmentation.
On the other hand, the total momentum fraction taken by these hadrons can exceed 100\% for $s$, $c$ and $b$ quarks when extrapolated into small $z_{\min}$ values, especially for the strange quark.
This violation of the momentum sum rule could be due to the lack of constraints from the data.
That includes, for example, the poor flavor separation concerning fragmentation from the $d$ and $s$ quarks, and the lack of data as small $z$.
A definitive conclusion cannot be drawn at this point and will require future precision measurements, especially from future electron-ion colliders.
In order to better visualize the numerical values of the truncated moments, in \cref{t.moments-of-all-FFs} we list the sum of truncated moments for the neutral hadron FFs from this work, the charged hadron FFs from the previous work~\cite{Gao:2024dbv}, as well as the total sum for both neutral and charged hadron FFs.

\begin{table}[ht]
	\setcellgapes{2.5 pt}
	\makegapedcells
	\centering
	\begin{tabular}{|c|c|c|c|c|c|c|}
		\hline
		hadron  & $g$ $\pqty{z_{\min} = 0.01}$ & $d$ $\pqty{z_{\min} = 0.01}$ & $u$ $\pqty{z_{\min} = 0.01}$ & $s$ $\pqty{z_{\min} = 0.01}$ & $c$ $\pqty{z_{\min} = 0.01}$ & $b$ $\pqty{z_{\min} = 0.01}$ \\
		\hline
		neutral & $0.3614^{+0.0247}_{-0.0189}$ & $0.3459^{+0.0319}_{-0.0253}$ & $0.2712^{+0.0163}_{-0.0134}$ & $0.5382^{+0.0432}_{-0.0380}$ & $0.4599^{+0.0347}_{-0.0322}$ & $0.3731^{+0.0466}_{-0.0377}$ \\
		\hline
		charged & $0.5068^{+0.0145}_{-0.0133}$ & $0.5066^{+0.0420}_{-0.0367}$ & $0.5306^{+0.0150}_{-0.0135}$ & $0.7293^{+0.0440}_{-0.0416}$ & $0.6581^{+0.0655}_{-0.0572}$ & $0.7033^{+0.0903}_{-0.0723}$ \\
		\hline
		total   & $0.8682^{+0.0310}_{-0.0255}$ & $0.8525^{+0.0590}_{-0.0498}$ & $0.8018^{+0.0206}_{-0.0171}$ & $1.2675^{+0.0733}_{-0.0680}$ & $1.1180^{+0.0973}_{-0.0867}$ & $1.0764^{+0.1350}_{-0.1080}$ \\
		\hline
	\end{tabular}
	\caption{
		Same as \cref{t.moments-of-neutral-FFs}, but for sums of neutral FFs, charged FFs as well as all FFs available from previous \cite{Gao:2024dbv} and current NPC works.
		In particular, the neutral hadrons included are $K_S^0 + K_L^0$, $\Lambda + \overline{\Lambda}$, $\eta$ and $\pi^0$, and the charged hadrons included are $\pi^{\pm}$, $K^{\pm}$ and $p + \overline{p}$.
		The last row lists the sum of neutral and charged hadrons in this table.
		The $\pi^0$ FFs and all charged FFs are correlated with each other, hence their error sets are added together for computation of uncertainties.
	}
	\label{t.moments-of-all-FFs}
\end{table}

\clearpage
\subsection{Neutral hadrons in jet fragmentation}
\label{ss.hadron-in-jet-prediction}

\begin{figure}
	\centering
	\begin{subfigure}[b]{0.45 \textwidth}
		\includegraphics[width = \linewidth]{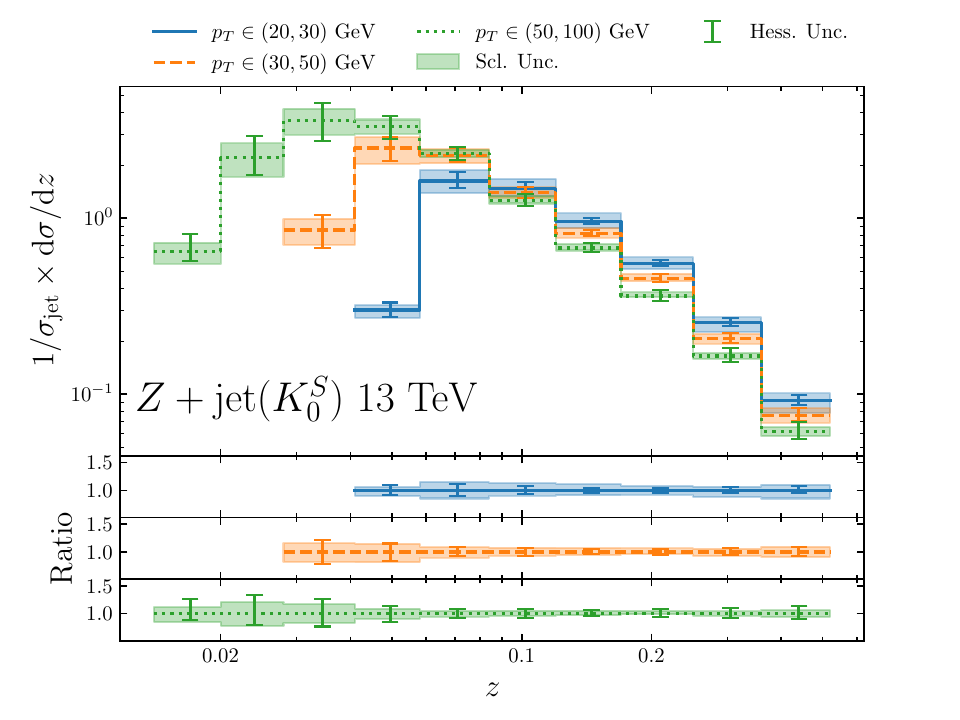}
		\caption{$K_S^0$-in-jet production}
		\label{f.prediction-LHCb-pp-Z_jet-K0S}
	\end{subfigure}
	\begin{subfigure}[b]{0.45 \textwidth}
		\includegraphics[width = \linewidth]{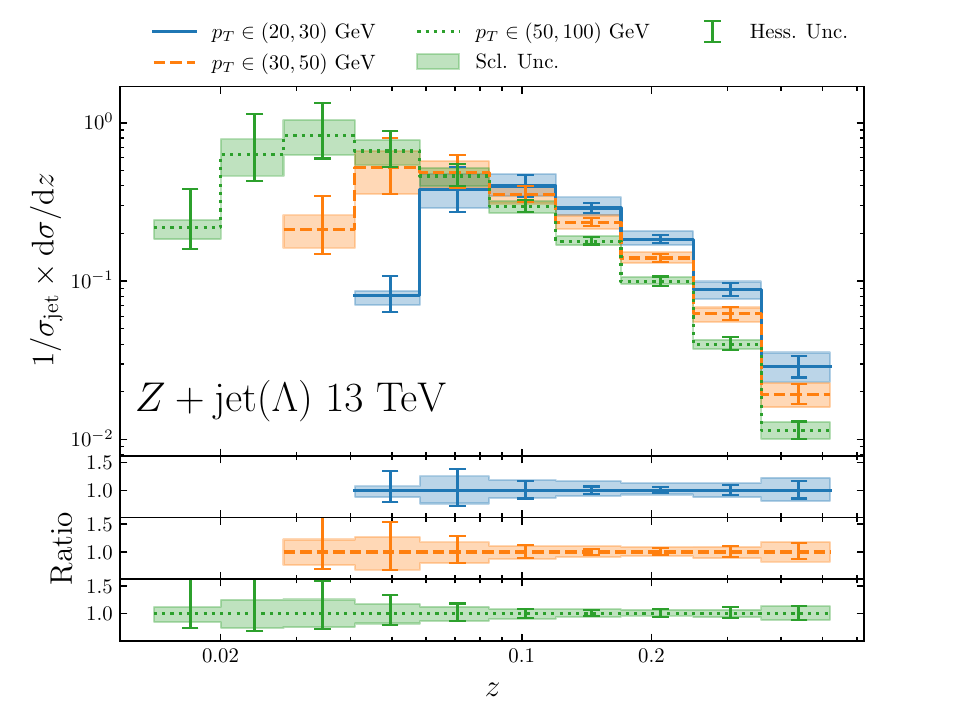}
		\caption{$\Lambda$-in-jet production}
		\label{f.prediction-LHCb-pp-Z_jet-Lambda}
	\end{subfigure}
	\caption{
		Prediction of $K_S^0$-in-jet and $\Lambda$-in-jet production in $pp \to Z + \mathrm{jet}$ at LHCb kinematics with $\sqrt{s} = 13 ~ \mathrm{GeV}$.
		In the $y$-label, the $\sigma_{\mathrm{jet}}$ in the denominator represents the total $pp \to Z + \mathrm{jet}$ cross section.
		The jet and $Z$-boson rapidity ranges are chosen respectively as $y_J \in \pqty{2.5, 4}$ and $y_Z \in \pqty{2, 4.5}$.
		The jet cone parameter is chosen as $R = 0.5$.
		Jets are detected with transverse momentum range $p_T \in \pqty{20, 30} \, \mathrm{GeV}$ (blue), $p_T \in \pqty{30, 50} \, \mathrm{GeV}$ (orange) and $p_T \in \pqty{50, 100} \, \mathrm{GeV}$ (green).
		For all three different $p_T$ ranges, the scale uncertainties are depicted with error bands, while the Hessian uncertainties are shown with error bars.
		In the lower three panels, the distributions, together with the uncertainties, are normalized to their corresponding central values.
		The cross sections and ratios are drawn with respect to the momentum fraction $z$ of jet carried by the fragmented hadron.
	}
	\label{f.prediction-LHCb-pp-Z_jet-K0S-Lambda}
\end{figure}

\begin{figure}
	\centering
	\begin{subfigure}[b]{0.45 \textwidth}
		\includegraphics[width = \linewidth]{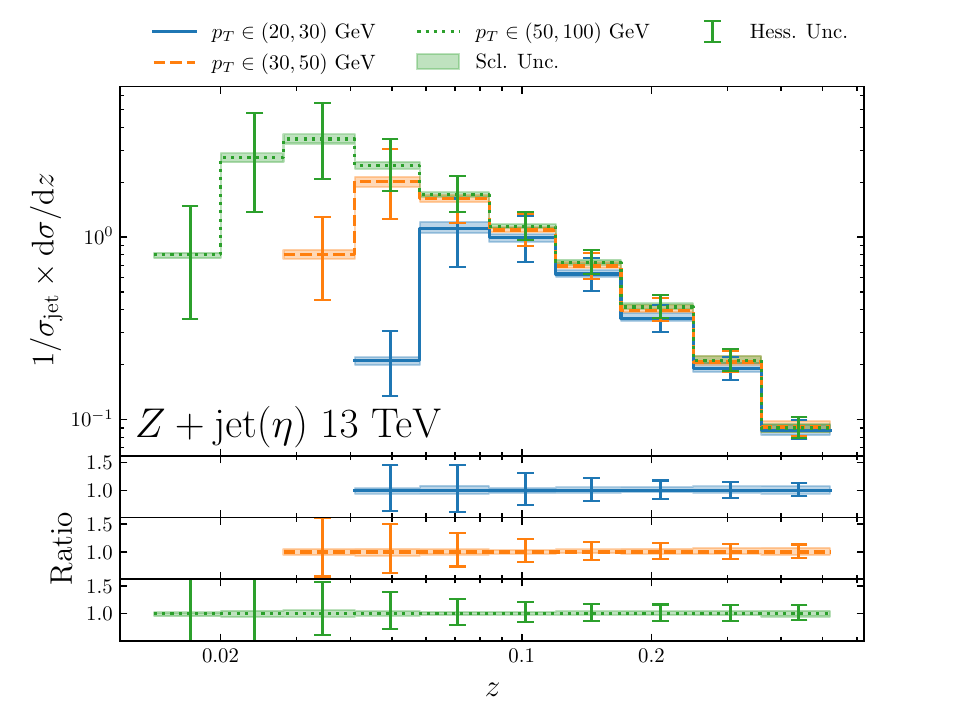}
		\caption{}
		\label{f.prediction-LHCb-pp-Z_jet-eta}
	\end{subfigure}
	\begin{subfigure}[b]{0.45 \textwidth}
		\includegraphics[width = \linewidth]{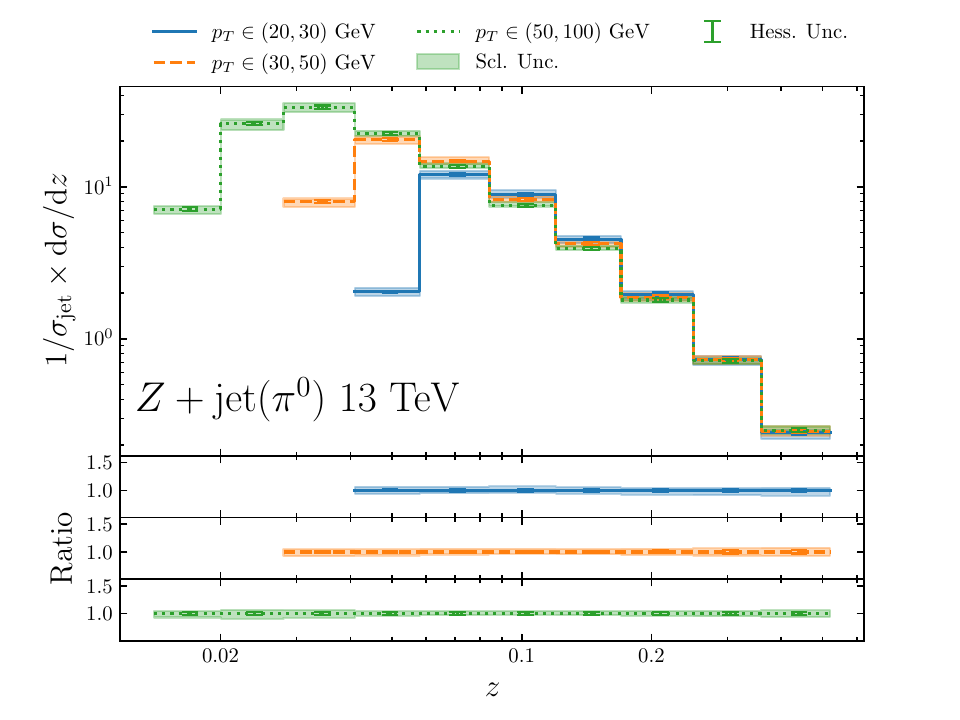}
		\caption{}
		\label{f.prediction-LHCb-pp-Z_jet-pi_0}
	\end{subfigure}
	\caption{
		Same as \cref{f.prediction-LHCb-pp-Z_jet-K0S-Lambda} but for $\eta$-in-jet and $\pi^0$-in-jet production.
	}
	\label{f.prediction-LHCb-pp-Z_jet-eta-pi_0}
\end{figure}

In this section, we provide predictions for hadron-in-jet production for $pp \to Z + \mathrm{jet}$ at the LHCb.
In particular, we follow the kinematics in Ref. \cite{LHCb:2022rky} and set the center-of-mass energy to $\sqrt{s} = 13~\mathrm{TeV}$.
In addition, the rapidity ranges for the jet and $Z$-boson are set to $y_J \in \pqty{2.5, 4}$ and $y_Z \in \pqty{2, 4.5}$, respectively.
The jet cone parameter is chosen as $R = 0.5$.
The central values for the factorization scale and renormalization scale are set to half the sum of the transverse mass of all final state particles.
The central value for the fragmentation scale is set to the transverse momentum of the jet.
Differential cross sections in $z$ are presented for three different jet transverse momentum $p_T$ ranges, namely $\pqty{20, 30} \, \mathrm{GeV}$, $\pqty{30, 50} \, \mathrm{GeV}$ and $\pqty{50, 100} \, \mathrm{GeV}$.
The $Z$-boson transverse momentum $p_{T,Z}$ is chosen to be in the same range as the jet $p_T$, ensuring the back-to-back configuration.
The hadron-in-jet cross sections,
expressed as functions of the momentum fraction of the jet carried by the identified hadron, denoted as $z$,
are predicted and presented in \cref{f.prediction-LHCb-pp-Z_jet-K0S-Lambda,f.prediction-LHCb-pp-Z_jet-eta-pi_0} for all neutral hadrons investigated in this study, \textit{i.e.}, $K_S^0$, $\Lambda$, $\eta$ and $\pi^0$.

In \cref{f.prediction-LHCb-pp-Z_jet-K0S}, $K_S^0$-in-jet production rate is shown for $pp \to Z + \mathrm{jet}$ in a back-to-back configuration.
Since an accurate prediction in the low hadron transverse momentum region requires resummation corrections that are beyond the scope of this work, we cut the hadron transverse momentum $p_{T,h} > 0.7~\mathrm{GeV}$ to exclude such scenarios.
The curves for different $p_T$ ranges exhibit behavior similar to that observed for the charged hadrons in Ref. \cite{LHCb:2022rky}.
Specifically, the production rates in different $p_T$ ranges are nearly identical in the high-$z$ region, with deviations becoming apparent as $z \lesssim 0.06$.
This can be understood as a result of scaling violation, as $z$ becomes lower, the TMD effects will start to play a more important role.
For low $p_T$ jets, the TMD effects appear earlier (at relatively larger $z$) than for jets with higher $p_T$.
Another interesting observation is that the peak positions for $K_S^0$-in-jet production are similar to those for $K^{\pm}$-in-jet production in \cite{LHCb:2022rky}, for each $p_T$ range, respectively.

In \cref{f.prediction-LHCb-pp-Z_jet-Lambda}, $\Lambda$-in-jet production is predicted.
Compared to the case of $K_S^0$ production, we observe that the theoretical uncertainties for $\Lambda$ production are larger in the low $z$ region.
This indicates that the $\Lambda$ FFs are less constrained at small $z$.
Consequently, measurements in this region will contribute to improving the constraints on the $\Lambda$ FFs.

In \cref{f.prediction-LHCb-pp-Z_jet-eta}, $\eta$-in-jet production is predicted.
The Hessian uncertainty in this case is larger in the low $z$ region, which again reflects the relatively less constrained $\eta$ FFs.
In this case, the scale uncertainty is smaller than that predicted for $K_S^0$ and $\Lambda$ in \cref{f.prediction-LHCb-pp-Z_jet-K0S-Lambda}.
The scale uncertainties for $K_S^0$ and $\Lambda$ are relatively larger due to the significant contribution of gluon distributions to the scale evolution.
However, for $\eta$ production, the gluon distribution is smaller than the light quark distributions in the region $z \gtrsim 0.1$, making the scale variation less pronounced.
Additionally, the observable shown in \cref{f.prediction-LHCb-pp-Z_jet-eta} is normalized to the total $pp \to Z + \mathrm{jet}$ cross section, further suppressing the scale uncertainty.

Finally, $\pi^0$-in-jet production is predicted in \cref{f.prediction-LHCb-pp-Z_jet-pi_0}.
Since the $\pi^0$ FFs are well constrained, the calculated observable has small uncertainties.

\clearpage
\subsection{Ratios of neutral and charged kaon production}
\label{ss.K0S-production-prediction}

In this section, we provide predictions for $K_S^0$ over the $K^{\pm}$ production rate:
\begin{equation}
	R \pqty{2 K_S^0 / K^{\pm}}
	\equiv
	\frac{2 \dd{\sigma}_{K_S^0}}{\dd{x}}
	\bigg /
	\frac{\dd{\sigma}_{K^{\pm}}}{\dd{x}}
	\, , \label{e.K0S-K_pm-ratio}
\end{equation}
where $x = x_p$, $z_p$ or $p_T$ for SIA, SIDIS or $pp$ collisions, respectively.
Throughout this section, we adopt the abbreviated notation $R_K \equiv R \pqty{2 K_S^0 / K^{\pm}}$ in the main text for brevity.
Here, the $K^{\pm}$ production rate is calculated by the fitted $K^{\pm}$ fragmentation functions \cite{Gao:2024nkz},
while the $K^0_S$ yield is based on either the $K_S^0$ FFs fitted in this work (with the observable denoted by $R_K^{\mathrm{fit}}$) or the $K_S^0$ FFs constructed via isospin symmetry by \cref{e.K0S-construction-u-and-d,e.K0S-construction-unfavored} (with the observable denoted by $R_K^{\mathrm{iso}}$).
In the following, we will illustrate the discrimination power of $R_K$ in clarifying if the isospin symmetry in kaon FFs is violated, by comparing theoretical predictions $R_K^{\mathrm{fit}}$ with $R_K^{\mathrm{iso}}$ in SIA, SIDIS or $pp$ collisions.

\begin{figure}
	\centering
	\includegraphics[width = 0.8 \linewidth]{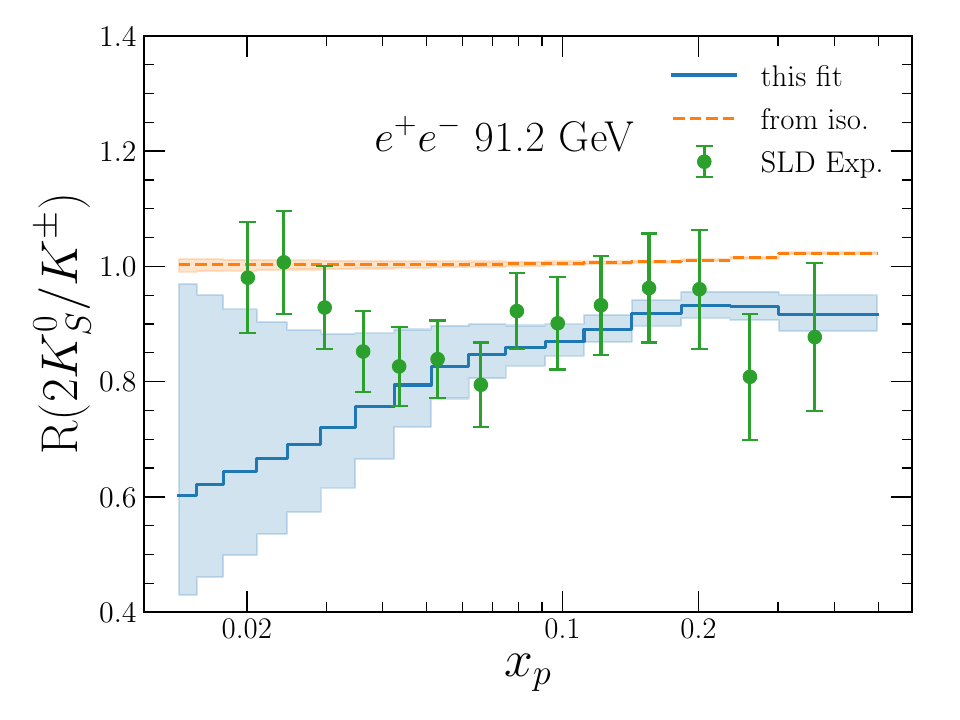}
	\caption{
	Theoretical predictions of $R_K^{\mathrm{fit}}$ (blue) and $R_K^{\mathrm{iso}}$ (orange)
	in $e^+ e^-$ collisions at $\sqrt{s} = 91.2~\mathrm{GeV}$.
	The green data points and error bars are provided by SLD measurement \cite{SLD:1998coh}.
	}
	\label{f.K0S-over-K_pm-SLD}
\end{figure}

\begin{figure}
	\centering
	\includegraphics[width = 0.8 \linewidth]{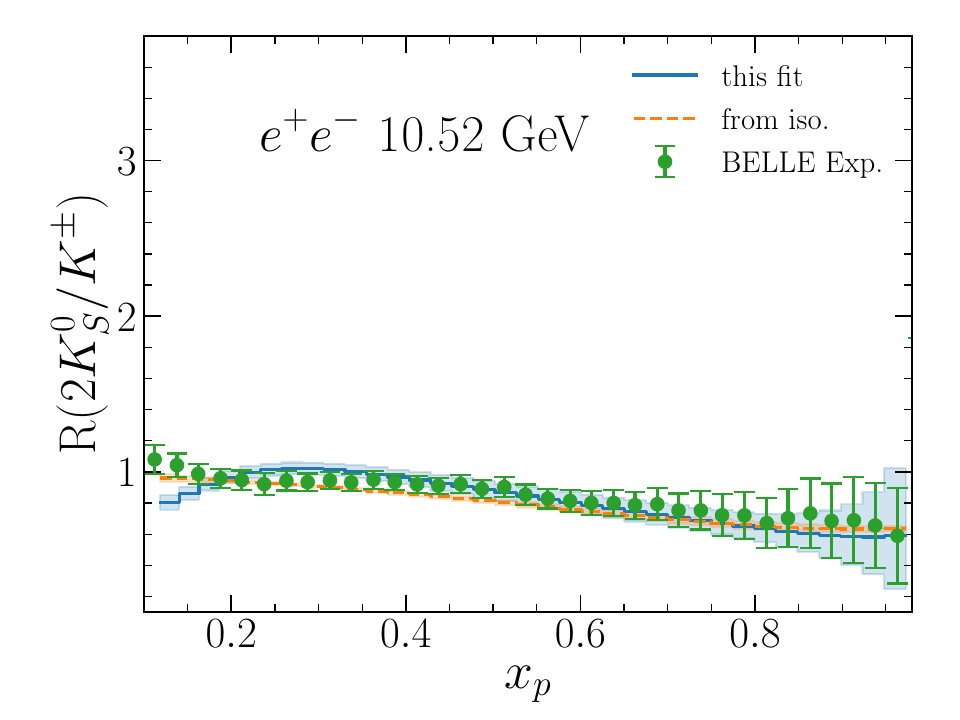}
	\caption{
		Same as \cref{f.K0S-over-K_pm-SLD} but at Belle kinematics with $\sqrt{s} = 10.52~\mathrm{GeV}$.
		The data points are taken from Ref. \cite{Belle:2024vua}.
	}
	\label{f.K0S-over-K_pm-Belle}
\end{figure}

In \cref{f.K0S-over-K_pm-SLD}, we provide the predictions of $R_K^{\mathrm{fit}}$ and $R_K^{\mathrm{iso}}$ as functions of $x_p$ for SIA at the $Z$-pole, together with the results from the SLD measurement \cite{SLD:1998coh}.
The $K_S^0$ FFs extracted in this work correctly reproduce the ratio in both high and intermediate $x_p$ regions.
In the low $x_p$ region, although the theory prediction and data are consistent within uncertainties, their central values deviate substantially from each other.
On the other hand, the $K_S^0$ FFs constructed via isospin symmetry predict a ratio $R_K^{\mathrm{iso}}$ that is close to unity across the entire $x_p$ range.
To comprehend this phenomenon, we begin by recognizing that the constructed $2 \times K_S^0$ FFs and $K^{\pm}$ FFs differ solely by an interchange of the $u$ and $d$ quark distributions.
Additionally, the parton-level production rates of $u$ and $d$ quarks are similar at the $Z$-pole.
These factors result in $R_K^{\mathrm{iso}}$ being very close to unity.
Although moderate tension is observed between $R_K^{\mathrm{iso}}$ and the SLD measurement around $x_p \sim 0.05$, it is not sufficient to conclude the violation of the isospin symmetry in kaon fragmentation.
In \cref{f.K0S-over-K_pm-Belle}, we present the predictions for $R_K^{\mathrm{fit}}$ and $R_K^{\mathrm{iso}}$ at $\sqrt{s} = 10.52~\mathrm{GeV}$, together with the Belle measurement at the same energy scale \cite{Belle:2024vua}.
Apart from the tension in the small $x_p$ region, the $R_K^{\mathrm{fit}}$ and $R_K^{\mathrm{iso}}$ roughly describe the Belle data equally well in the region $0.2 < x_p < 0.8$.

\begin{figure}
	\centering
	\includegraphics[width = 0.9 \linewidth]{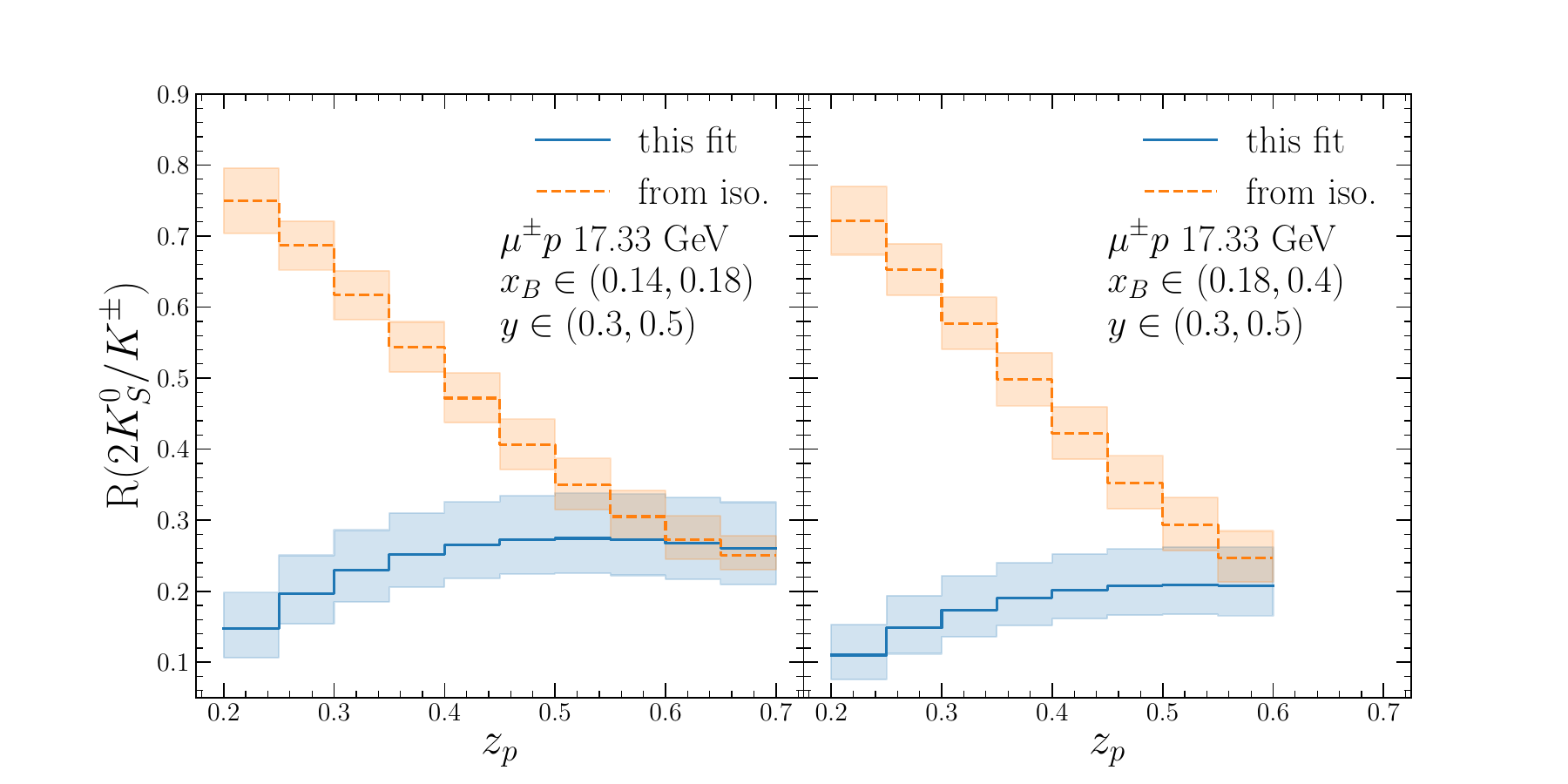}
	\caption{
		Same as \cref{f.K0S-over-K_pm-SLD} but for SIDIS off a proton target at COMPASS kinematics with $\sqrt{s} = 17.33~\mathrm{GeV}$.
		The variables $x_B$ and $y$ denote the Bjorken-$x$ variable and event inelasticity, respectively.
		The cross section is differential in $z_p$.
		The variables $x_B$, $y$ and $z_p$ are all defined in \cref{ss.K0S-data}.
	}
	\label{f.K0S-over-K_pm-COMPASS}
\end{figure}

\begin{figure}
	\centering
	\includegraphics[width = 0.9 \linewidth]{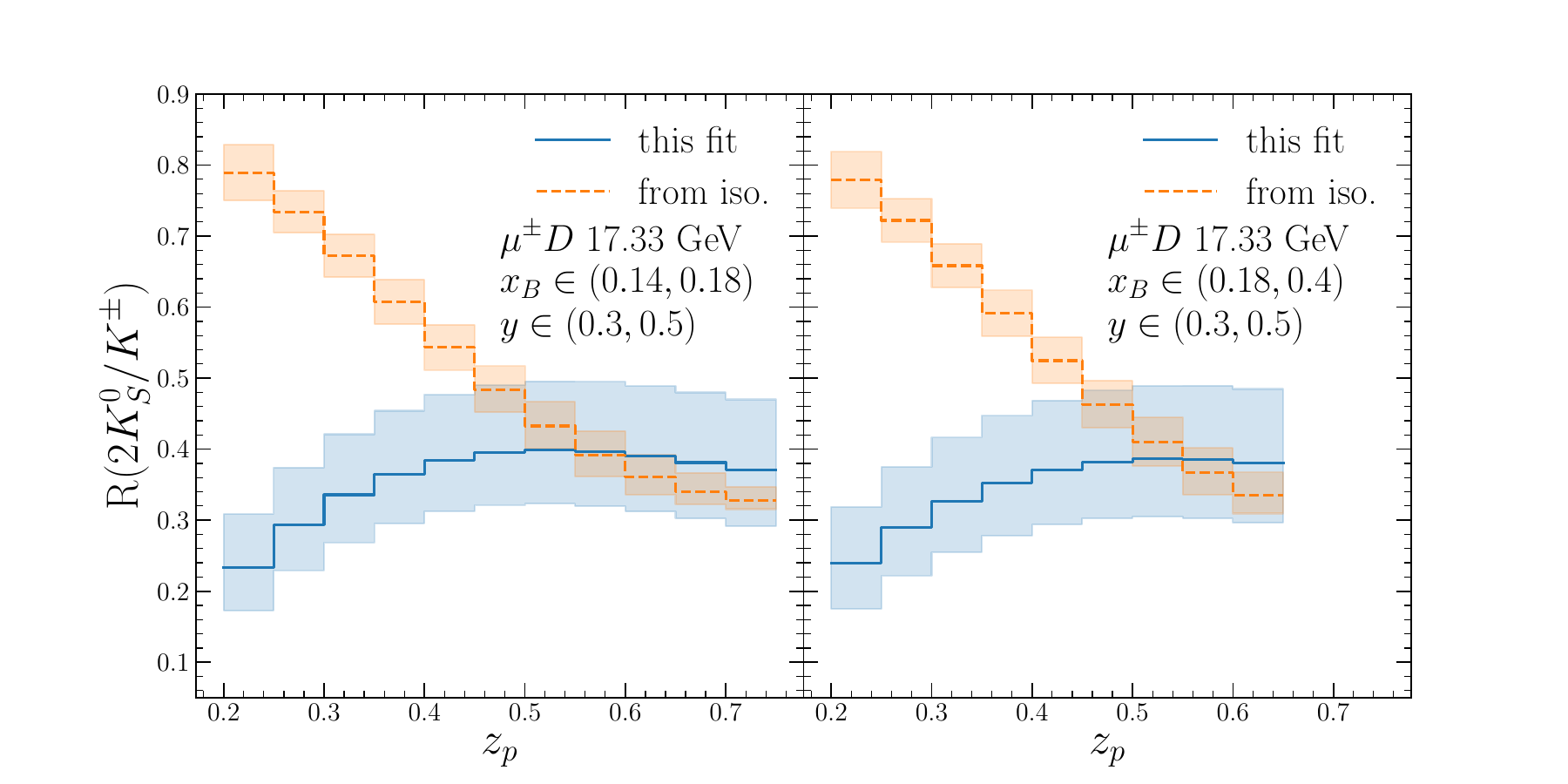}
	\caption{
		Same as \cref{f.K0S-over-K_pm-COMPASS} but for SIDIS off a deuteron target.
	}
	\label{f.K0S-over-K_pm-COMPASS-deutron}
\end{figure}

In \cref{f.K0S-over-K_pm-COMPASS}, we present the predicted $R_K^{\mathrm{fit}}$ and $R_K^{\mathrm{iso}}$, as a function of $z_p$, for SIDIS off a proton target with COMPASS kinematics.
In particular, the center-of-mass energy is set to $\sqrt{s} = 17.33~\mathrm{GeV}$, the $x_B$ ranges are set to $\pqty{0.14, 0.18}$ and $\pqty{0.18, 0.4}$ for the left and right panels, respectively.
Finally, the event inelasticity $y$ range is $\pqty{0.3, 0.5}$ for both panels.
Although the prediction is still at relatively low $\sqrt{s}$, the two curves (blue and orange) are well separated from each other, unlike what is observed in \cref{f.K0S-over-K_pm-Belle}.
This is because the SIDIS process has the advantage for flavor separation, which contributes to the discrimination between the two scenarios.
Similar behavior is observed in \cref{f.K0S-over-K_pm-COMPASS-deutron} for SIDIS off a deuteron target in the same kinematic region.
Future SIDIS measurements on $K^0_S$ production, especially in the middle and small $z_p$ regions,
can be essential to test the isospin symmetry in kaon FFs.

\begin{figure}
	\centering
	\includegraphics[width = 0.8 \linewidth]{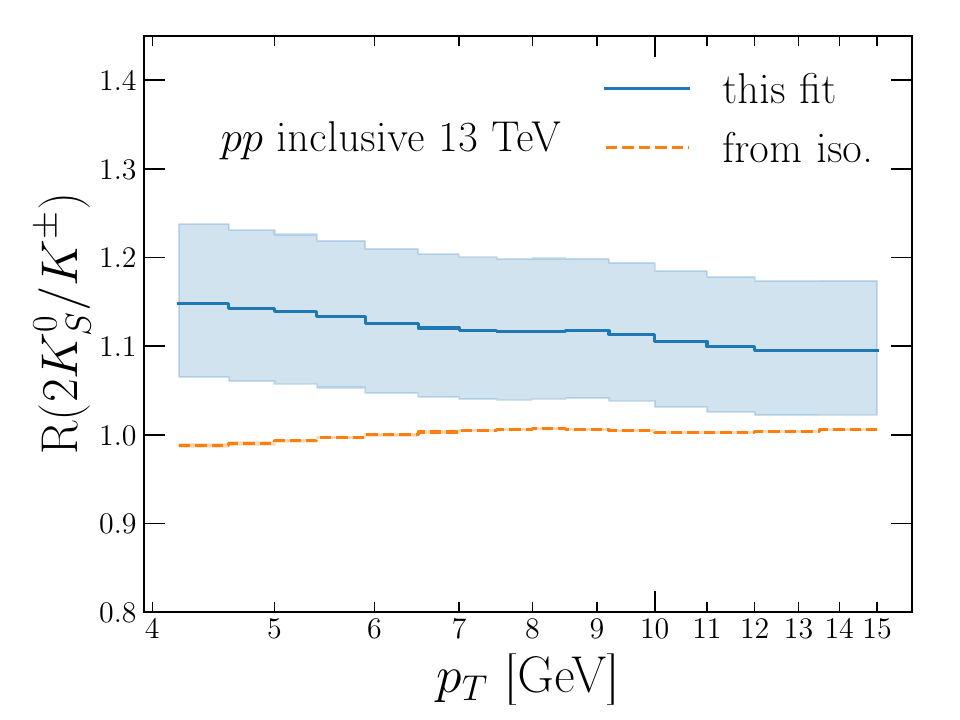}
	\caption{
		Same as \cref{f.K0S-over-K_pm-SLD} but for inclusive hadron production in $pp$ collisions at ALICE kinematics with $\sqrt{s} = 13~\mathrm{TeV}$.
		The ratios are drawn as functions of hadron transverse momentum $p_T$.
	}
	\label{f.K0S-over-K_pm-ALICE}
\end{figure}

In \cref{f.K0S-over-K_pm-ALICE}, the ratios $R_K^{\mathrm{fit}}$ and $R_K^{\mathrm{iso}}$ are presented as functions of the transverse momentum of the identified hadron for inclusive hadron production in $pp$ collisions.
Following ALICE kinematics \cite{ALICE:2020jsh}, we set $\sqrt{s} = 13~\mathrm{TeV}$ and the hadron rapidity range as $\pqty{-0.5, 0.5}$.
The orange curve, which is plotted from the $K_S^0$ FFs constructed by isospin symmetry, remains close to unity. In contrast, the theoretical predictions based on the fitted $K_S^0$ FFs (blue) deviate slightly from unity and exhibit larger uncertainties.
Nevertheless, both $R_K^{\mathrm{fit}}$ and $R_K^{\mathrm{iso}}$ are relatively flat.
This occurs because inclusive hadron production measurements primarily probe the integral of FFs, making them insensitive to subtle differences like shapes of FFs.

\begin{figure}
	\centering
	\includegraphics[width = 0.98 \textwidth]{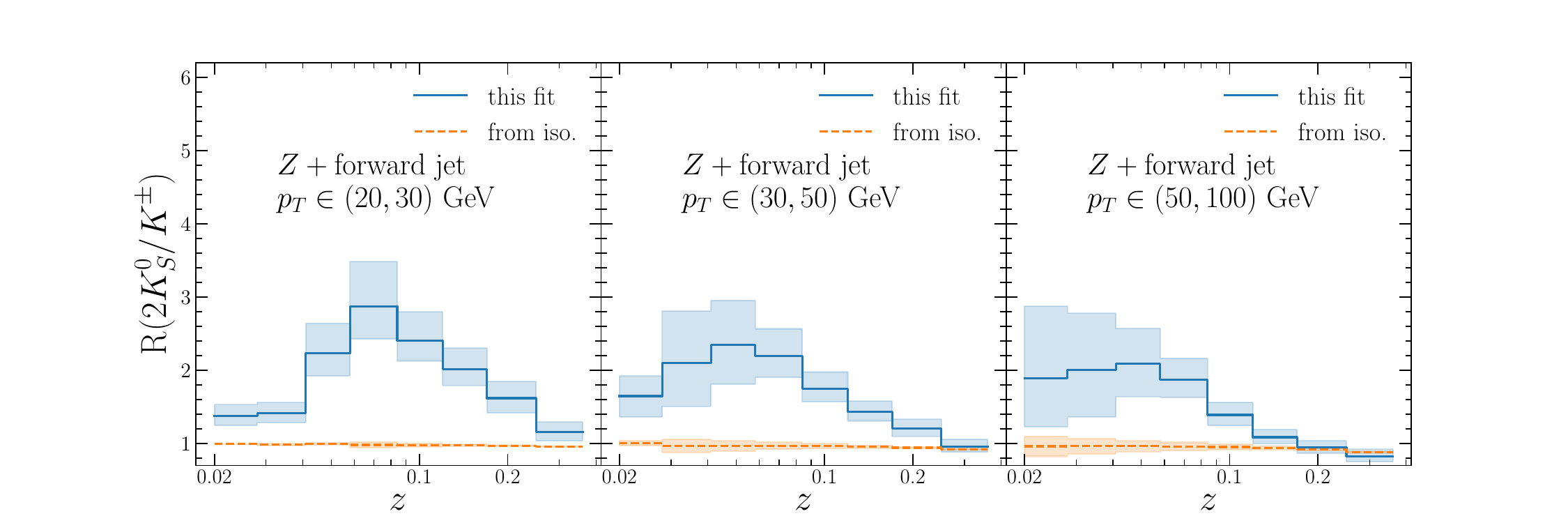}
	\caption{
		Same as \cref{f.K0S-over-K_pm-SLD} but for hadron-in-jet production in $pp \to Z + \mathrm{jet}$ at LHCb kinematics with $\sqrt{s} = 13~\mathrm{TeV}$.
		The kinematics used in this figure are the same as those used in \cref{f.prediction-LHCb-pp-Z_jet-K0S-Lambda}.
	}
	\label{f.K0S-over-K_pm-LHCb}
\end{figure}

While the ratios $R_K^{\mathrm{fit}}$ and $R_K^{\mathrm{iso}}$ for inclusive hadron production are only directly sensitive to the integrated FFs, we supplement these with hadron-in-jet production measurements that directly probe FFs as a function of $z$.
Specifically, we analyze the processes $pp \to Z + \mathrm{jet} \pqty{K_S^0}$ and $pp \to Z + \mathrm{jet} \pqty{K^{\pm}}$, as shown in \cref{f.K0S-over-K_pm-LHCb}.
We follow \cref{ss.hadron-in-jet-prediction} and set the kinematics according to the LHCb measurement~\cite{LHCb:2022rky} at 13 TeV.
The predictions for $R_K^{\mathrm{fit}}$ and $R_K^{\mathrm{iso}}$ are provided in three jet $p_T$ bins, \textit{i.e.}, $p_T \in \pqty{20, 30} \, \mathrm{GeV}$, $p_T \in \pqty{30, 50} \, \mathrm{GeV}$ and $p_T \in \pqty{50, 100} \, \mathrm{GeV}$.
Here again, the $Z$-boson $p_{T,Z}$ ranges and the jet $p_T$ ranges are set to be identical in order to maintain the back-to-back configuration.
While $R_K^{\mathrm{iso}}$ remains consistent with unity as expected, $R_K^{\mathrm{fit}}$ exhibits a significant deviation from 1.
This behavior is mostly driven by the difference $u$ quark FFs, where the fitted $K_S^0$ distribution develops more prominent peaks compared to that of $K^{\pm}$.
The peak positions for the observable shown in \cref{f.K0S-over-K_pm-LHCb} (between $z \in \pqty{0.06, 0.07}$) also roughly align with that of the $u$ quark distribution shown in \cref{f.FFs-K0S}.
In conclusion, this ratio observable clearly serves a better role in distinguishing the $K_S^0$ FFs fitted from global datasets (blue) and the $K_S^0$ FFs constructed from isospin symmetry (orange).

In summary, in this section predictions for the ratio of production rates of $K_S^0$ and $K^{\pm}$ in SIA, SIDIS and $pp$ collisions are presented.
We find that future measurements of the proposed observable, especially for SIDIS or for hadron-in-jet production, demonstrate significant potential for testing isospin symmetry in kaon fragmentations.

\clearpage

\section{Discussion and Conclusions}
\label{s.conclusion}

In this paper, we have presented the extraction of fragmentation functions (FFs) for light neutral hadrons including $K_S^0$ meson, $\Lambda$ baryon and $\eta$ meson from global analysis of world data.
In addition, we provide the $\pi^0$ meson FFs constructed from those of $\pi^{\pm}$ via isospin symmetry.
In the region $z \sim \pqty{0.1, 0.5}$, where precise data are available, the extracted FFs are in general well constrained.
As for the quality of the fit, good $\chi^2$ values are found for most datasets from various processes, including SIA, SIDIS and $pp$ collisions.
With the most comprehensive datasets included in the fit to neutral hadrons, we are able to deliver the state-of-art neutral hadron FFs.
In particular, with the $K_S^0$-in-jet data incorporated, the extracted gluon FFs get better constrained.
In addition, thanks to the FMNLO framework, we are able to provide the $\Lambda$ FFs and $\eta$ FFs with uncertainties that are not available in previous global analyses.
In identifying the direction of future measurements, uncertainties of FFs will be of important reference, since larger uncertainty implies areas of valuable future exploration.
Furthermore, with the uncertainties now available for $\Lambda$ and $\eta$ FFs, future predictions of related observables can include theoretical uncertainties as well.
Finally, in order to analyze the constraints from different experiments, we also perform alternative fits with a particular group of datasets subtracted during the fit.
As applications of the extracted FFs, we provide several predictions to physical quantities that are of interest.
We first calculate the fraction of momentum carried by the fragmented hadrons from the final-state partons, for each hadrons analyzed in this paper as well as all the charged and neutral hadrons from this and previous NPC analyses.
The neutral hadrons are found to carry approximately $25\% \sim 38\%$ of the momentum from gluon and $u$, $d$, $c$ and $b$ quarks, while about 51\% of the momentum from the $s$ quark.
Moreover, the total momentum carried by neutral and charged hadrons generally obeys the momentum sum rule.
Though moderate violation is observed for the $s$ quark, a definitive conclusion will require future high precision data.
Furthermore, we predict the hadron-in-jet production rates for $pp \to Z + \mathrm{jet}$ process with LHCb kinematics.
Since the hadron-in-jet observable is differential in the momentum fraction $z$, future measurements of such observable will greatly help in further constraining the FFs of neutral hadrons.
Lastly, we provide predictions on ratios of $K_S^0$ and $K^{\pm}$ production rates dedicated to current and future measurements at SLD, Belle, COMPASS, ALICE and LHCb experiments.
Such ratios are of great importance in studying the isospin symmetry in kaon fragmentation.
In our analysis, we find that certain distributions are not very well constrained by current data.
For $K_S^0$ FFs, the strange and down quark distributions can only be separated with certain prior assumptions.
The $\eta$ FFs from the gluon also show large uncertainties, especially in the low $z$ region.
Such limitations highlight the necessity of future high precision measurements, with emphasis on flavor separation.
For example, the FFs of $K_S^0$ from the strange and down quark can be better constrained by future measurements of the SIDIS process, and the FFs of $\eta$ from the gluon can benefit from future measurements of hadron-in-jet productions, both in SIA and $pp$ collisions.
In addition, the $c$ and $b$-tagged SIA data can help improve the determination of $\eta$ FFs.
This work is a continuation of the previous NPC analysis in which the light charged hadron FFs are determined.
As a collective endeavor into providing comprehensive and accurate fragmentation functions, our work has made an important step forward into the less constrained realm of light neutral hadron FFs.
In addition, we provide comparisons of our FFs with the determination of other groups in \cref{s.FFs-comparison}.
In \cref{s.alternative-fits}, alternative fits with subtraction of existing global data as well as addition of recent released measurements are provided.
The FFs presented in this work are publicly available in \texttt{LHAPDF} format with the details given in \cref{s.LHAPDF-grid}.

\noindent
\textbf{Acknowledgments.}

The work of J.G. is supported by the National Natural Science Foundation of China (NSFC) under Grant No. 12275173, Shanghai Municipal Education Commission under Grant No. 2024AIZD007, and open fund of Key Laboratory of Atomic and Subatomic Structure and Quantum Control (Ministry of Education).
H.X. is supported by the NSFC under Grants No. 12475139 and Guangdong Major Project of Basic and Applied Basic Research Nos. 2020B0301030008.
Y. Zhao is supported by the NSFC under Grant No. U2032105 and the CAS Project for Young Scientists in Basic Research No. YSBR-117.
X.S. is supported by the Helmholtz-OCPC Postdoctoral Exchange Program under Grant No. ZD2022004.
Y. Zhou is supported by the European Union ``Next Generation EU'' program through the Italian PRIN 2022 grant n. 20225ZHA7W.

\appendix

\section{Comparison to other groups}
\label{s.FFs-comparison}

In this section, we present the comparison between our fitted FFs and the FFs obtained by other groups.
The comparison is conducted at two energy scales: $Q = 5~\mathrm{GeV}$ and $Q = 91.2~\mathrm{GeV}$.

\begin{figure}[H]
	\centering
	\includegraphics[width = 0.8 \textwidth]{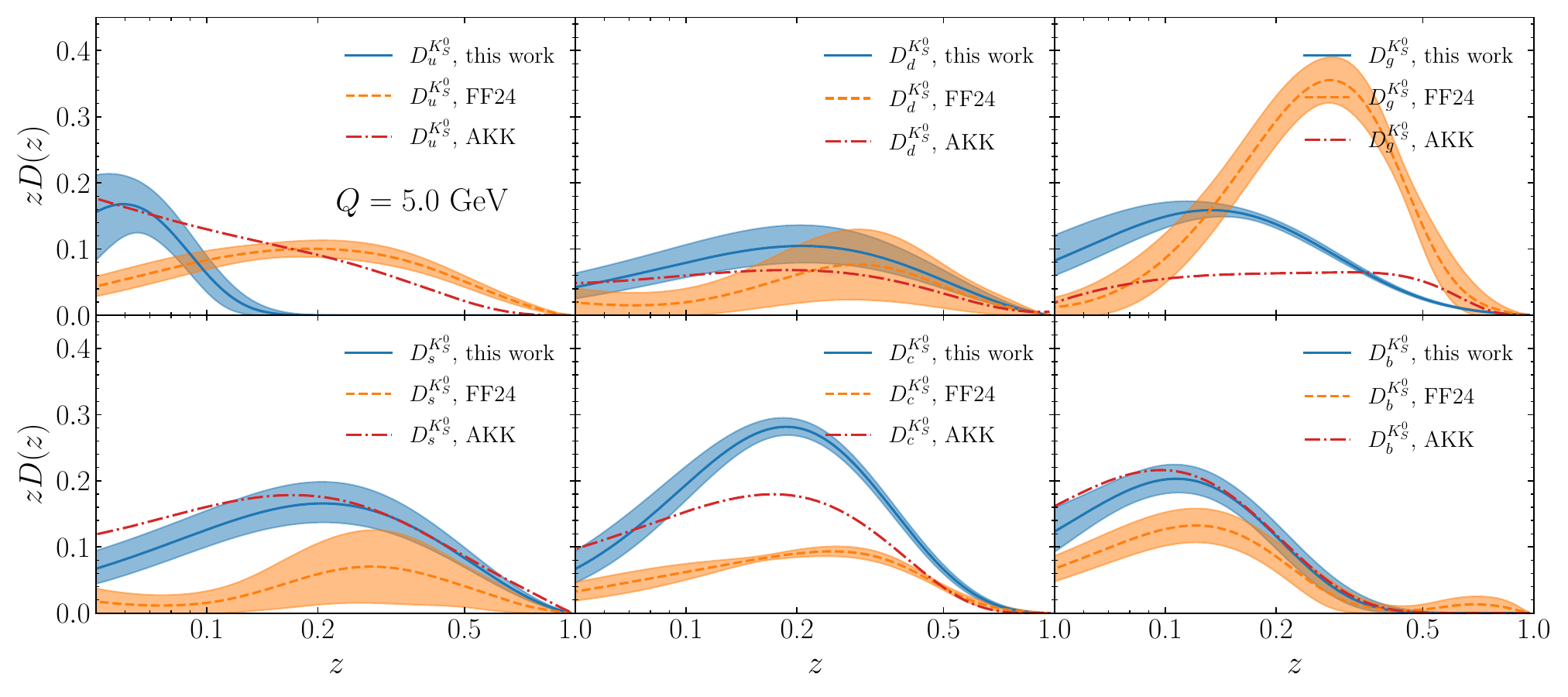}
	\caption{
		Comparison of our $K_S^0$ FFs at $Q = 5~\mathrm{GeV}$ with AKK08 \cite{Albino:2008fy} and FF24 \cite{Soleymaninia:2024jam}.
	}
	\label{f.comparison-K0S}
\end{figure}

\begin{figure}[H]
	\centering
	\includegraphics[width = 0.8 \textwidth]{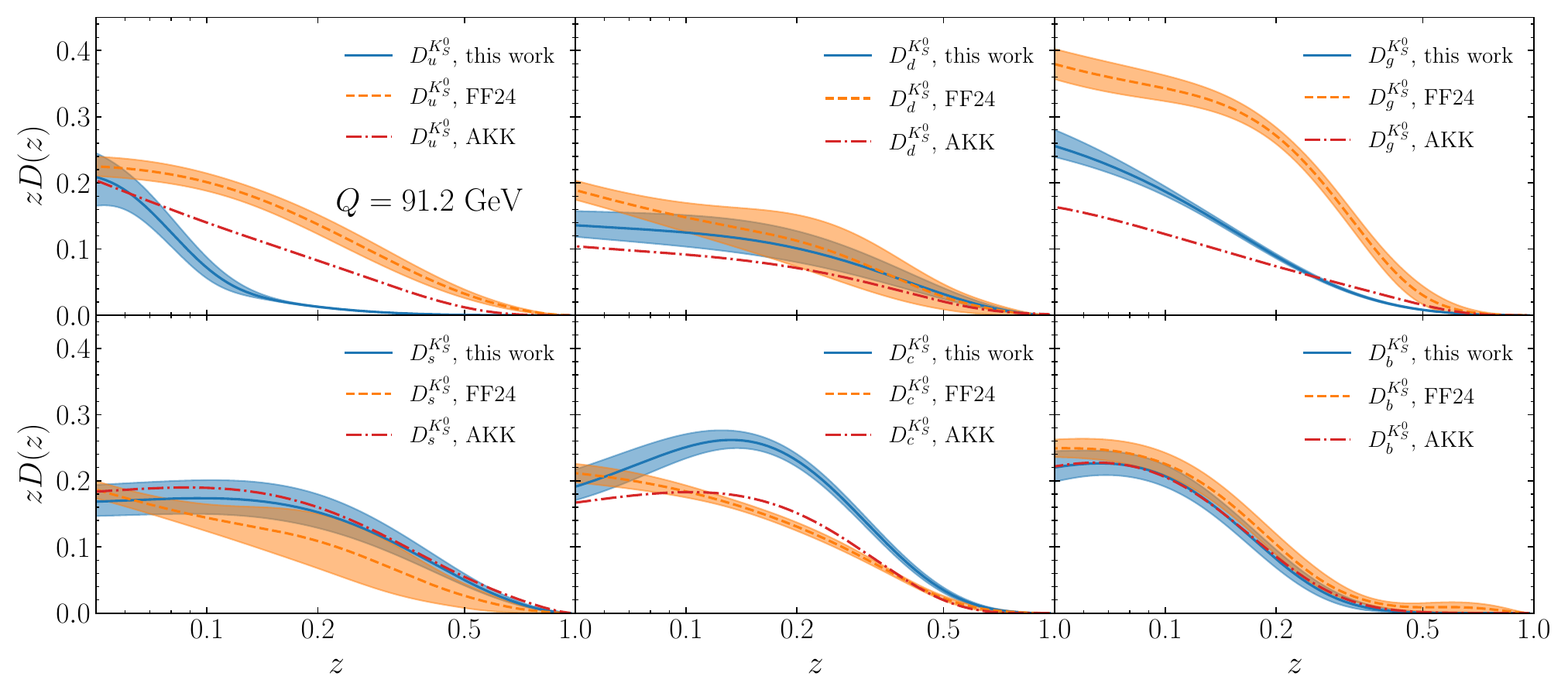}
	\caption{Same as \cref{f.comparison-K0S} but at $Q = 91.2~\mathrm{GeV}$.}
	\label{f.comparison-K0S-mZ}
\end{figure}

The comparison for $K_S^0$ with AKK08 \cite{Albino:2008fy} and FF24 \cite{Soleymaninia:2024jam} is given in \cref{f.comparison-K0S}, at $Q = 5~\mathrm{GeV}$.
The $u$ quark distribution derived in this study exhibits a distinct peak at $z \approx 0.06$ and rapidly declines to zero as $z$ increases.
In contrast, the FF24 $u$ distribution is more uniform, while the $u$ quark distribution from AKK08 remains comparatively larger over the entire $z$ range.
The observed discrepancy between the $u$ quark distribution in this work and FF24 likely stems from FF24's inclusion of the SIA data from BESIII \cite{BESIII:2022zit} at very low center-of-mass energies.
The difference between the $u$-quark FF in this analysis and AKK08 likely stems from differences in the selection of experimental datasets.
For the $d$ and $b$ distributions, we observe that our results as well as those from AKK08 and FF24 are generally consistent with each other.
The gluon distribution obtained in this work exhibits a peak around $z = 0.15$, whereas FF24 shows a peak around $z = 0.3$, and AKK08 remains relatively flat until $z \approx 0.5$.
The dramatic difference between our gluon FF and those from AKK08 and FF24 can likely be attributed to their exclusion of the $K_S^0$-in-jet production data from SIA.
As demonstrated in \cref{ss.K0S-subtraction}, these measurements provide the strongest constraints on the gluon distribution in our analysis.
Regarding the distribution of $s$ quark, our result is in close agreement with those of AKK08, while the result from FF24 shows discernible difference in the low $z$ region.
This difference can be likely attributed to FF24's employment of the neural network parameterization of the FFs.
The $c$ quark distribution, on the other hand, shows notable differences between our result and those from AKK08 and FF24.
In the AKK08 analysis, the relatively smaller $c$ quark distribution compensates for their correspondingly larger $u$ quark distribution compared to ours.
For FF24, their even smaller $c$ quark distribution serves to balance their significantly larger gluon distribution relative to ours.
Finally, as illustrated in \cref{f.comparison-K0S-mZ}, when evolved to $Q = 91.2~\mathrm{GeV}$, the shapes and magnitudes of the aforementioned distributions undergo some changes, but the relative behavior among them remains largely consistent.

\begin{figure}[H]
	\centering
	\includegraphics[width = 0.8 \textwidth]{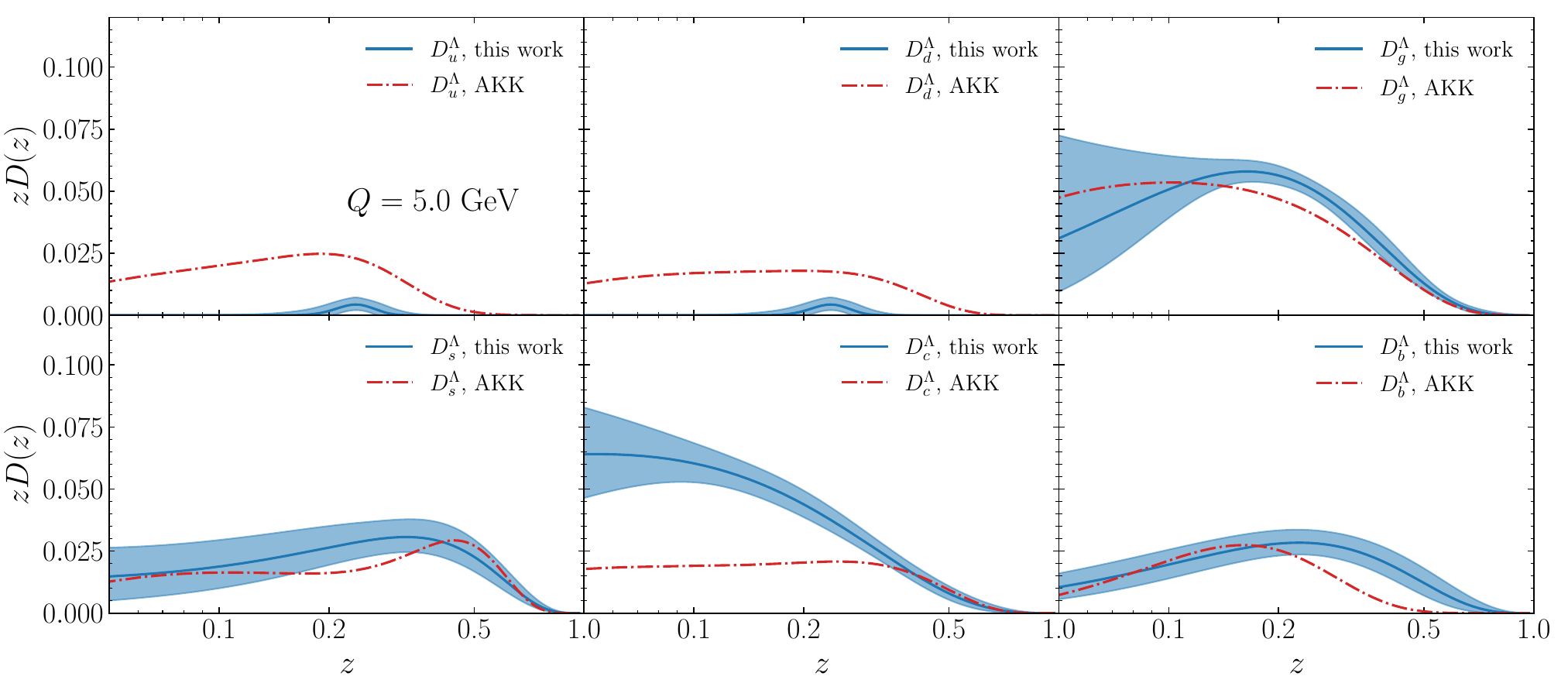}
	\caption{
		Comparison of our $\Lambda$ FFs at $Q = 5~\mathrm{GeV}$ with AKK08 \cite{Albino:2008fy}.
	}
	\label{f.comparison-Lambda}
\end{figure}
\begin{figure}[H]
	\centering
	\includegraphics[width = 0.8 \textwidth]{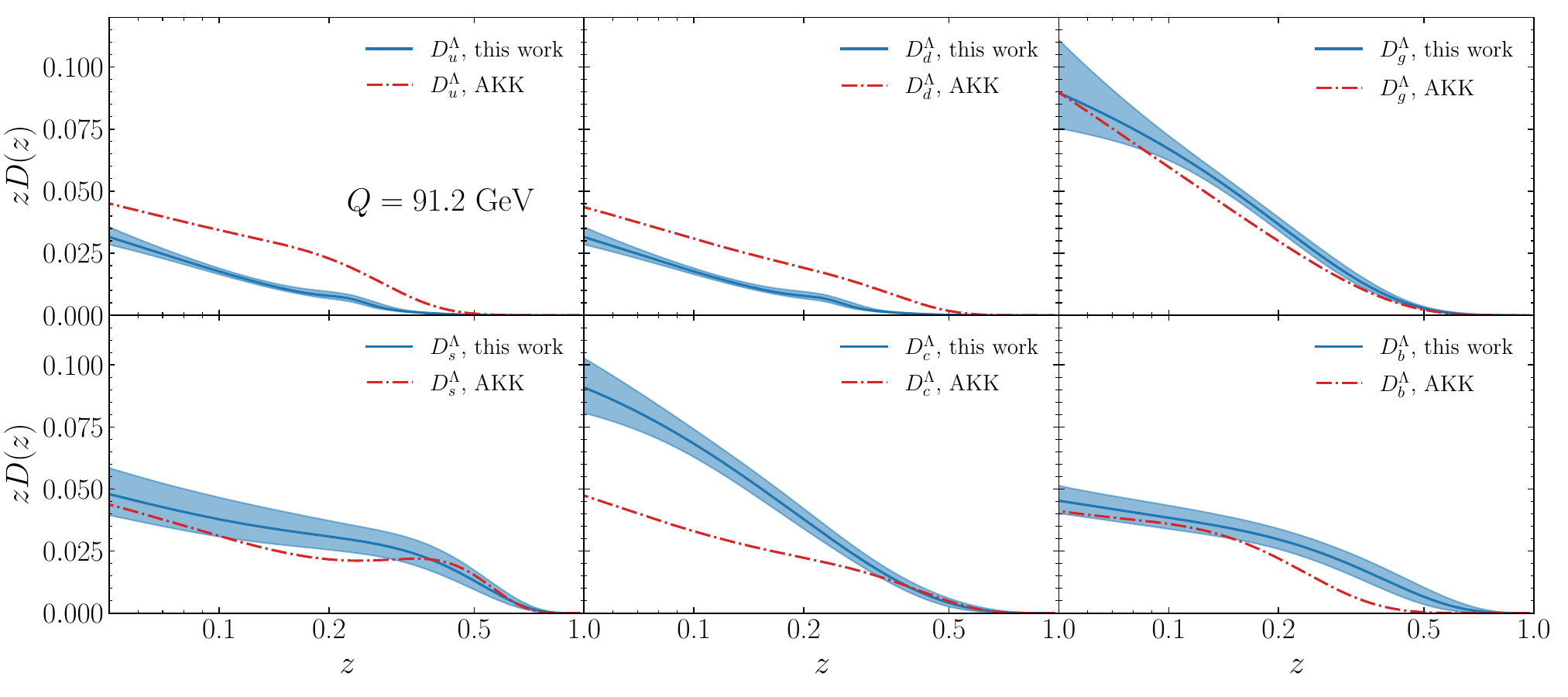}
	\caption{Same as \cref{f.comparison-Lambda} but at $Q = 91.2~\mathrm{GeV}$.}
	\label{f.comparison-Lambda-mZ}
\end{figure}

For $\Lambda$, the $u$ and $d$ quark distributions from AKK08 \cite{Albino:2008fy} are notably larger than those in this work, across the entire $z$ range.
As for the distributions of gluon, $s$ and $b$ quarks, the results from this work are in good agreement with those from AKK08, with some exception in the intermediate $z$ region.
Regarding the $c$ quark distribution, a significant difference between our result and the AKK08 result is observed at low and intermediate $z$ regions.
However, some agreement is found in the large $z$ region is observed for the $c$ distribution.
Our extracted $u$ and $d$ quark fragmentation functions (FFs) are substantially smaller than those of AKK08, while the $c$ quark FF compensates by being significantly larger.
This behavior reflects limitations in the current $\Lambda$ production data, which lack the precision required for robust flavor separation in a global analysis.
Particularly, the $c$ quark distribution at $z \lesssim 0.1$ is only directly constrained by two data points from the SLD $c$-tagged experiment \cite{SLD:1998coh}.
Finally, as demonstrated in \cref{f.comparison-Lambda-mZ}, when evolved to $Q = 91.2~\mathrm{GeV}$, the shapes and magnitudes of the $\Lambda$ FFs exhibit changes, but the relative behavior among them stays the same in general.

\begin{figure}[H]
	\centering
	\includegraphics[width = 0.9 \textwidth]{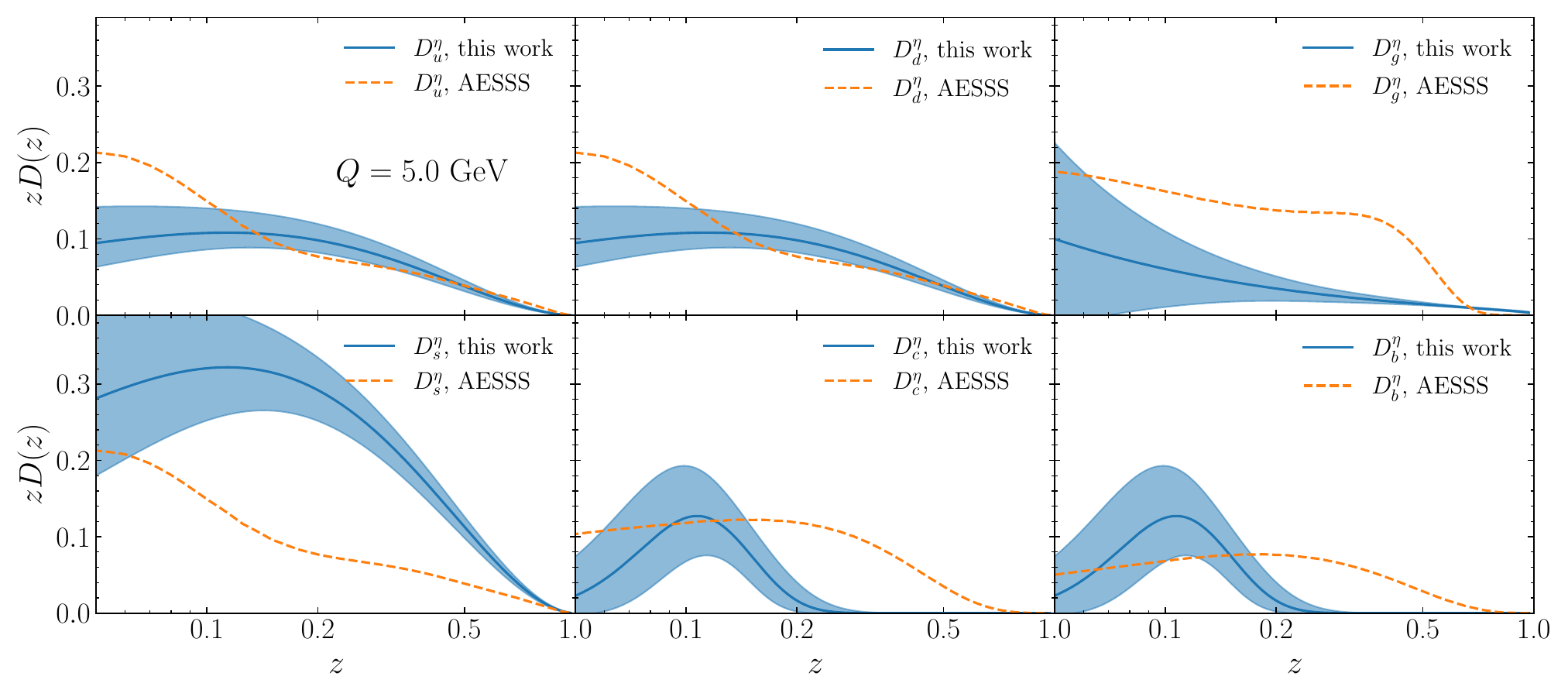}
	\caption{
		Comparison of our $\eta$ FFs at $Q = 5~\mathrm{GeV}$ with AESSS \cite{Aidala:2010bn}.
	}
	\label{f.comparison-eta}
\end{figure}
\begin{figure}[H]
	\centering
	\includegraphics[width = 0.9 \textwidth]{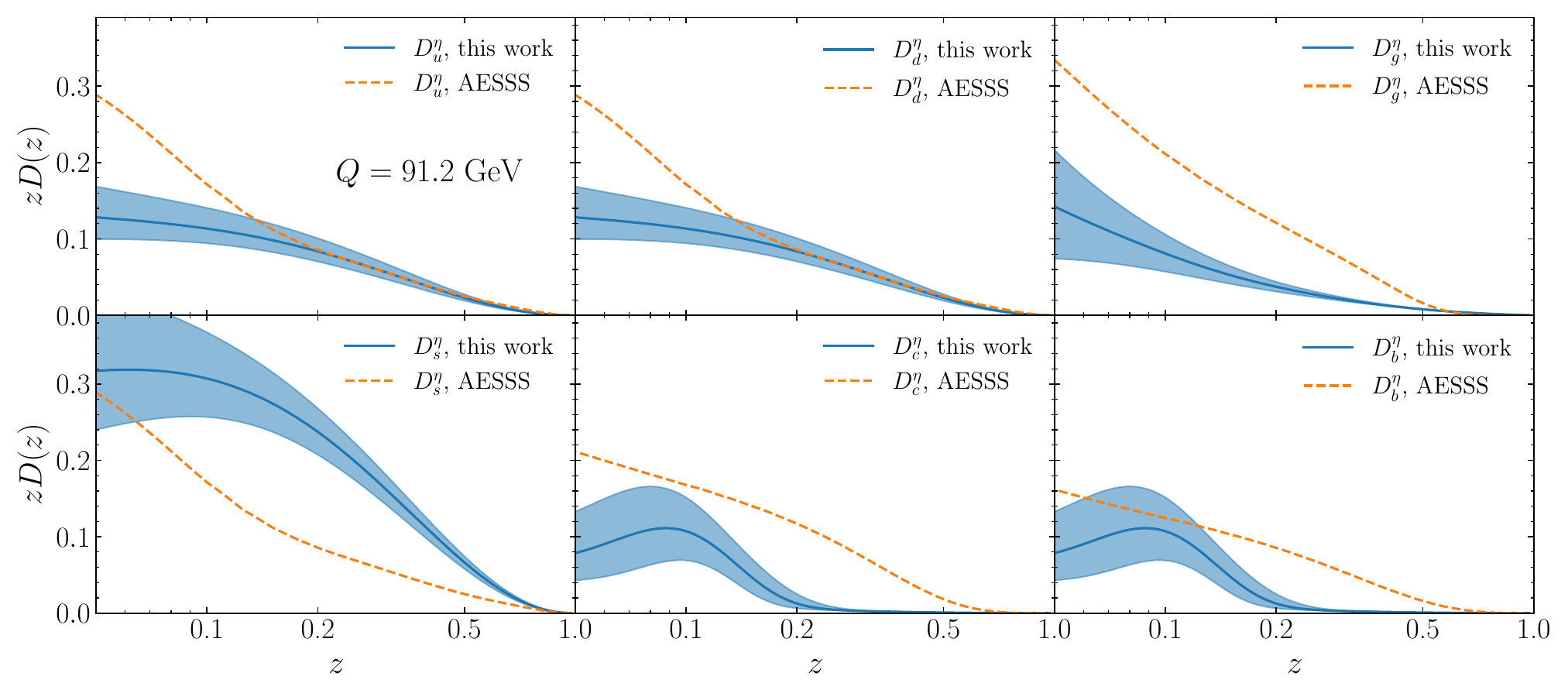}
	\caption{Same as \cref{f.comparison-eta} but at $Q = 91.2~\mathrm{GeV}$.}
	\label{f.comparison-eta-mZ}
\end{figure}

For the $\eta$ FFs, the $u$ and $d$ quark distributions fitted in this work are in good agreement with those from AESSS \cite{Aidala:2010bn} at $z \gtrsim 0.2$.
In the lower $z$ region, however, the $u$ and $d$ distributions obtained from AESSS are discernibly larger than those obtained in this work.
On the contrary, the strange distribution obtained in this work is generally larger than that from AESSS, compensating the relative smallness of the $u$ and $d$ distributions fitted in this analysis.
This difference arises because the AESSS analysis sets the $u$, $d$ and $s$ quark distributions to be equal, whereas our analysis allows the $s$ quark distribution to have an independent normalization.
As for the gluon distribution, we observe that the AESSS result is generally larger than that obtained in this work.
The observed difference in the gluon distribution likely originates from the distinct datasets used: while AESSS incorporated only low-energy $pp$ collision data from PHENIX, our analysis additionally includes higher-energy $pp$ data from ALICE.
Furthermore, this study identifies the $c$ and $b$ quark distributions with a prominent peak at $z \approx 0.1$, whereas the corresponding AESSS distributions remain generally flat across the most of the $z$ region before decreasing at $z \approx 0.3$.
This difference occurs because while our analysis fits $c$ and $b$ quark distributions with complete shape and normalization parameters, AESSS constrains their shapes to those of charged hadron residuals and fits only normalization parameters.
Finally, as shown in \cref{f.comparison-eta-mZ}, when evolved to $Q = 91.2~\mathrm{GeV}$, the shapes and magnitudes of the $\eta$ FFs exhibit changes, but the relative behavior among them generally stays the same.

\section{Alternative fits}
\label{s.alternative-fits}

In this section, we provide alternative fits that are varied based on the nominal NPC23 fits.
In \cref{ss.K0S-subtraction,ss.Lambda-subtraction}, variations are performed to study in detail the constraints coming from each dataset.
This is done by systematically subtracting different experiments from the nominal fit (baseline), while keeping other settings, including the kinematic cuts and treatment of experimental uncertainties, intact.
By doing this, we can learn the impact of a certain experiment on the final extracted FFs.
The global datasets are categorized into five distinct groups based on the types of reaction: SIA measurements at the $Z$-pole (excluding hadron-in-jet productions), SIA hadron-in-jet measurements, SIA measurements below the $Z$-pole, SIDIS measurements and $pp$ collisions.
This study is performed for the FFs of $K_S^0$ and $\Lambda$.
As for $\eta$ FFs, due to the limited constraint provided by existing world data, we plan to defer this analysis to a future time when more high-precision data on $\eta$ production become available.

We would like to emphasize that when analyzing the subtraction diagrams, one needs to realize that when the subtracted result prefers a relatively larger distribution than that of the baseline result, it indicates that the subtracted datasets prefer a smaller distribution and vice versa.

Furthermore, in \cref{ss.Belle-impact} we study the impact of recent released measurements on neutral hadron production from the Belle experiment.
And finally we study the impact of recent measurements on the $\eta/\pi^0$ ratio from LHCb in \cref{ss.LHCb-impact}.

\subsection{$K_S^0$ production}
\label{ss.K0S-subtraction}

\begin{figure}
	\centering
	\includegraphics[width = 0.9 \textwidth]{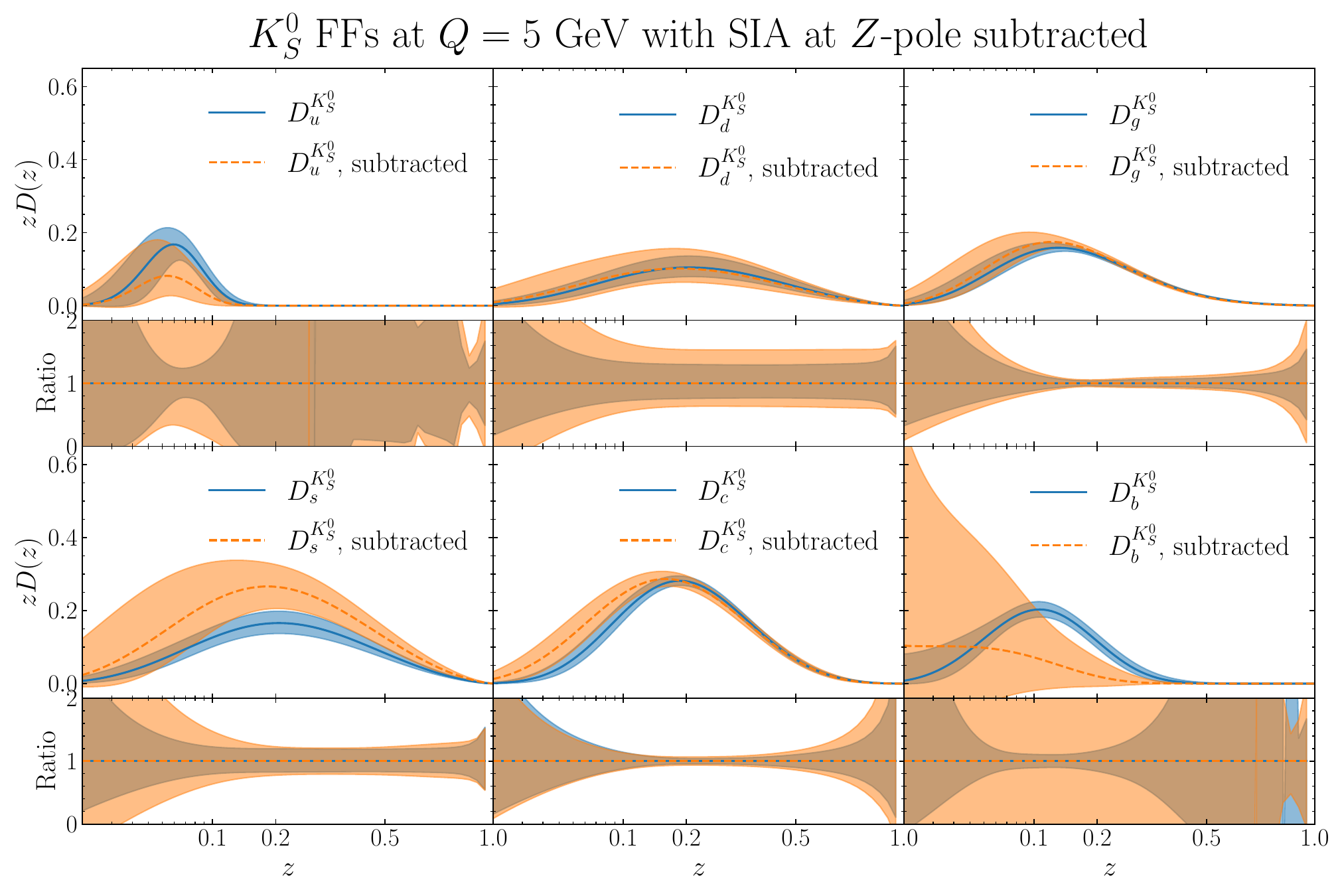}
	\caption{
		Comparison of the baseline fit (blue) and the alternative fit (orange) that subtracts the SIA data at $Z$-pole energy (the $K_S^0$-in-jet data are not subtracted).
		FFs and their uncertainties for different partons are plotted in each panel.
		Relative errors are plotted in the lower panel of each parton.
		The FFs are plotted at $Q^2 = 25~\mathrm{GeV}^2$ with respect to the momentum fraction $z$.
		In order to better display the behaviour of FFs at both low and high $z$ regions, we made the plot in $z^{1/3}$ scale.
		The uncertainty bands represent Hessian uncertainties at confidence level of 68\%.
	}
	\label{f.subtraction-SIA-Z_pole-K0S}
\end{figure}

\begin{figure}[htb]
	\centering
	\includegraphics[width = 0.9 \textwidth]{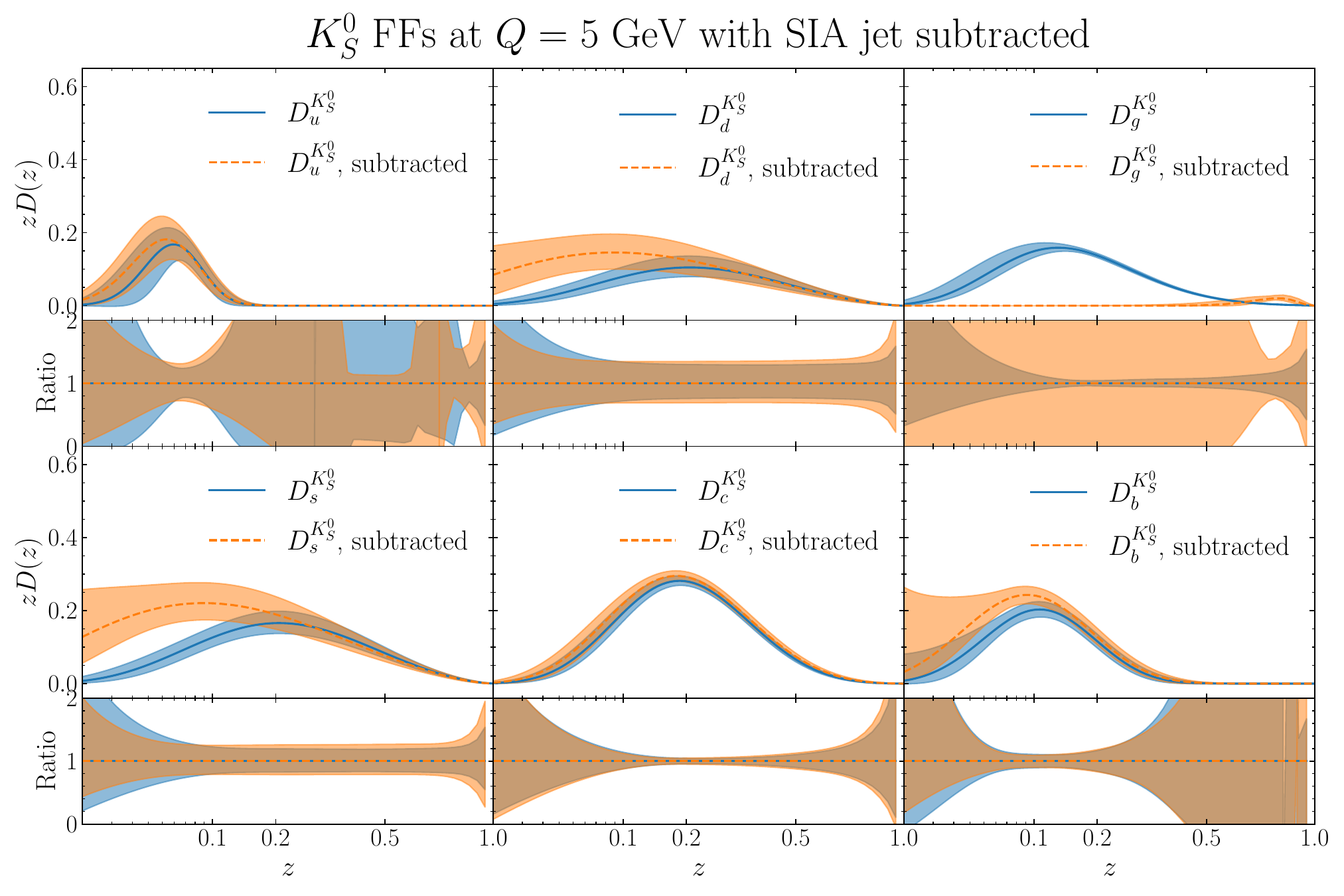}
	\caption{
		Same as \cref{f.subtraction-SIA-Z_pole-K0S} but for subtraction of $K_S^0$-in-jet production in SIA at $Z$-pole energy.
	}
	\label{f.subtraction-SIA-jet-K0S}
\end{figure}

\begin{figure}
	\centering
	\includegraphics[width = 0.9 \textwidth]{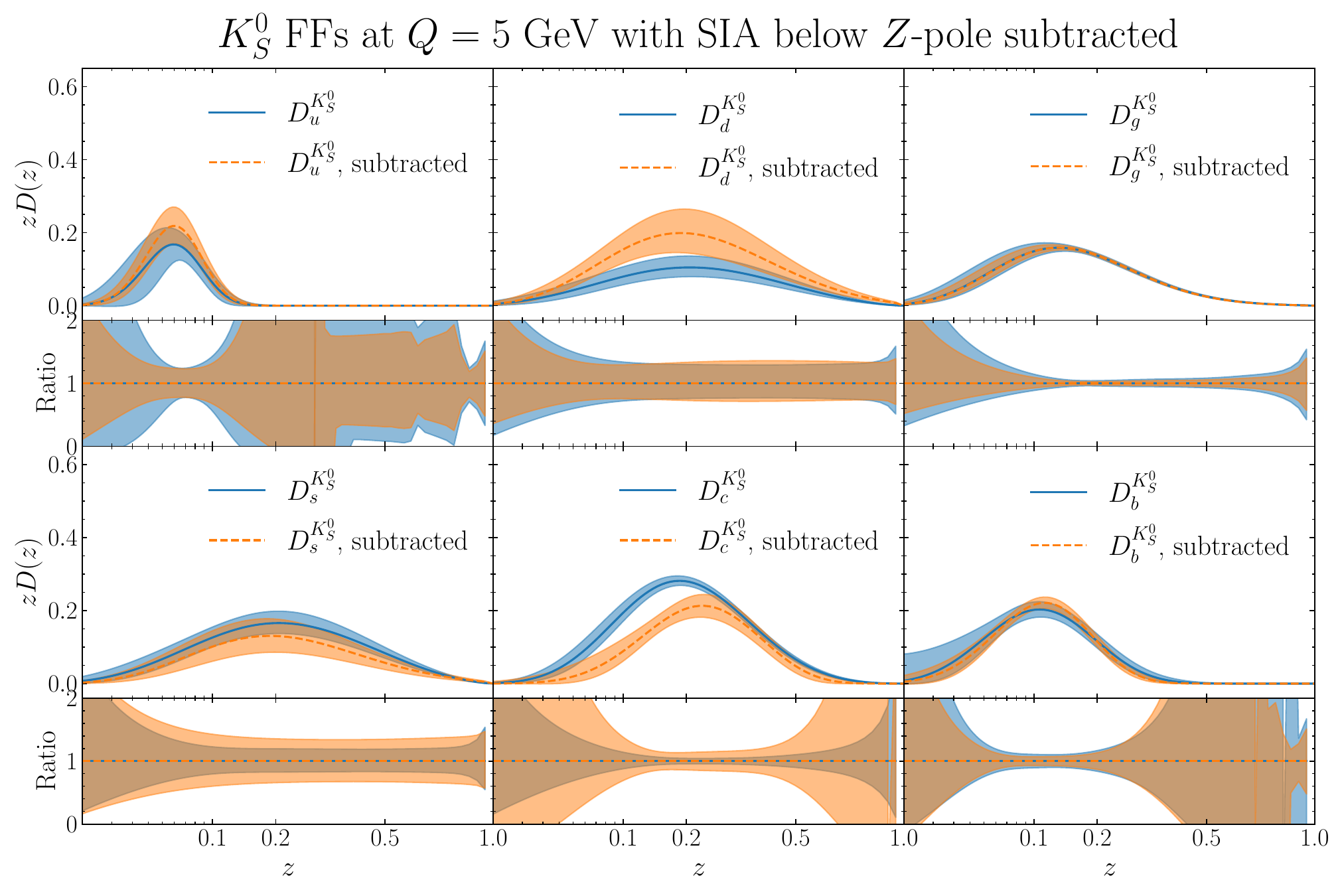}
	\caption{
		Same as \cref{f.subtraction-SIA-Z_pole-K0S} but for subtraction of $K_S^0$ production in SIA below $Z$-pole energy.
	}
	\label{f.subtraction-SIA-below-Z_pole-K0S}
\end{figure}

\begin{figure}
	\centering
	\includegraphics[width = 0.9 \textwidth]{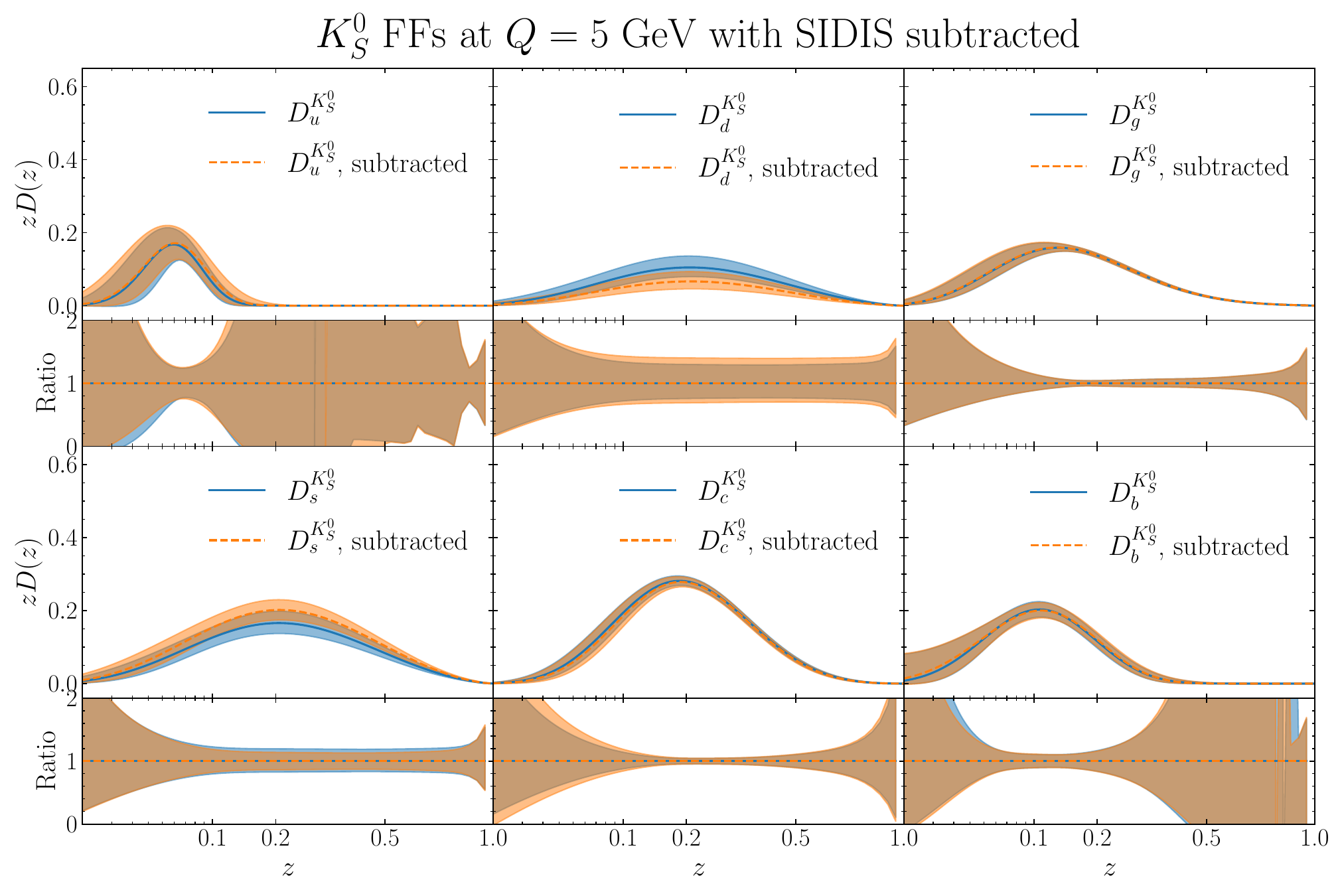}
	\caption{
		Same as \cref{f.subtraction-SIA-Z_pole-K0S} but for subtraction of $K_S^0$ production in SIDIS.
	}
	\label{f.subtraction-SIDIS-K0S}
\end{figure}

\begin{figure}[htb]
	\centering
	\includegraphics[width = 0.9 \textwidth]{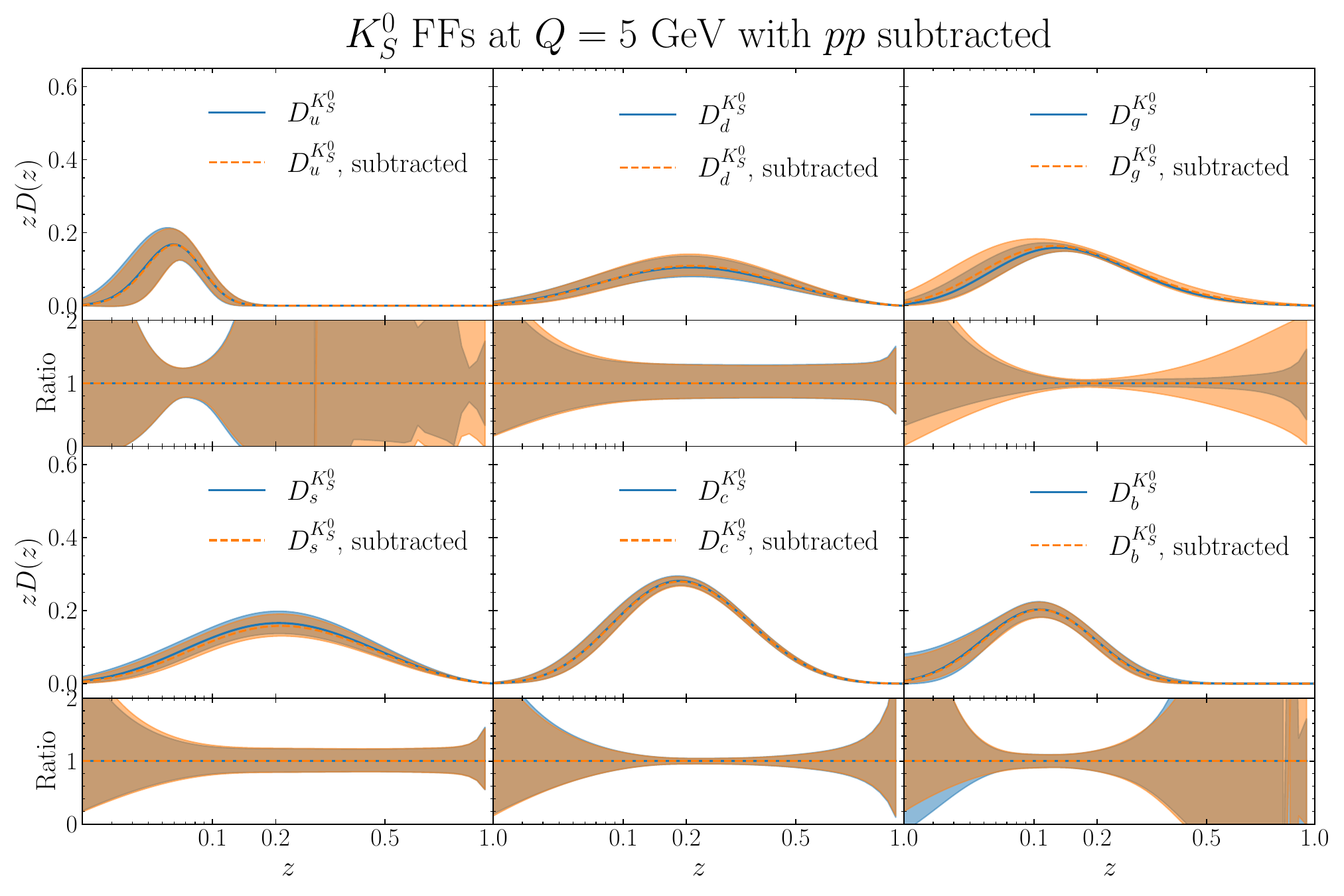}
	\caption{
		Same as \cref{f.subtraction-SIA-Z_pole-K0S} but for subtraction of $K_S^0$ production in $pp$ collisions.
	}
	\label{f.subtraction-pp-K0S}
\end{figure}

In this section, we present the data subtraction analysis for the $K_S^0$ production.
The comparison between each alternative fit and the baseline fit is shown in \cref{f.subtraction-SIA-Z_pole-K0S,f.subtraction-SIA-jet-K0S,f.subtraction-SIA-below-Z_pole-K0S,f.subtraction-SIDIS-K0S,f.subtraction-pp-K0S}.
The figures show the comparison of FFs between the baseline fit and subtracted fit for each parton flavor in separate panels, with FFs plotted at scale $Q^2 = 25~\mathrm{GeV}^2$.
For each flavor we plot the actual $z D \pqty{z}$ values along with their uncertainties normalized to the corresponding central values in baseline and subtracted fits.
The baseline results are plotted in blue, while the subtracted results are plotted in orange.
The bands represent Hessian uncertainties at a confidence level of 68\%.
The impact of removing each subgroup can be visualized as deviations from the baseline results.

The results after the removal of the SIA data at $Z$-pole energy are presented in \cref{f.subtraction-SIA-Z_pole-K0S}.
The distributions that undergo significant change are those of $u$ and $b$ quarks.
The $d$, $s$, $c$ and gluon distributions also experience an enlargement of uncertainties;
however, the shape and central values are generally within 1$\sigma$ from those of the baseline fit.
Clearly, the SIA data provides the most constraints to the $u$ quark distribution, putting both its shape and the uncertainty under control.
For the $b$ quark distribution, a significant increase in uncertainty is observed.
This is caused by the subtraction of SLD $b$-tagged data \cite{SLD:1998coh}.
However, the $c$ quark distribution remains largely well-constrained. This is likely because the remaining Belle data \cite{Belle:2024vua}, which lies below the $b$ quark production threshold, still provides constraining power for the $c$ quark FF.

The impact of removing SIA $K_S^0$-in-jet data at $Z$-pole energy is displayed in \cref{f.subtraction-SIA-jet-K0S}.
Significant effect is observed primarily in the gluon FFs, this is expected as the third jet in the tri-jet event is usually a gluon-initiated jet.
This further highlights the crucial role of the SIA $K_S^0$-in-jet data in constraining the gluon distribution.
Regarding the quark distributions, we notice that the changes in the $u$ and $c$ quark FFs are marginal and within 1$\sigma$ of the baseline fit.
However, the $d$ and $s$ distributions exhibit noticeable changes in their central values, along with a broadening of uncertainty bands in the low $z$ region.
The upward shift of the central values in $d$ and $s$ distributions compensates the downward shift observed in gluon distribution.
Finally, while the $b$ quark distribution shows increased uncertainties, the changes in its central values remain relatively modest.

The results after the removal of the SIA data below $Z$-pole energy are illustrated in \cref{f.subtraction-SIA-below-Z_pole-K0S}.
The $u$ quark distribution after subtraction is still consistent with the baseline result within uncertainties, so are the gluon and $b$ quark distributions.
However, for the $s$ and $d$ quark distributions, we notice that they are both dramatically different from those of the baseline fit.
This indicates that the SIA data below $Z$-pole energy are crucial in constraining the favored quark distributions in $K_S^0$.
Apart from the constituent quarks, the $c$ quark distribution also shifts downward after subtraction, especially in the mid-$z$ region.

The impact of removing SIDIS measurements is presented in \cref{f.subtraction-SIDIS-K0S}.
Our analysis shows that both quark and gluon distributions remain largely stable upon removal of the SIDIS data. This stability primarily stems from the constraining power of Belle data \cite{Belle:2024vua} collected below the $b$ quark production threshold, which effectively anchors the $c$ quark distribution.
Consequently, all other quark flavors maintain their stability through constraints provided by the remaining SIA data, both at and below $Z$-pole.
Notably, the $d$ and $s$ also exhibit resilience to the SIDIS data subtraction, despite the theoretical expectations that SIDIS measurements would help distinguish between these flavor components.
This again urges further precision SIDIS measurement for $K_S^0$ production, so that we can achieve flavor separation between the two constituent quark flavors of $K_S^0$.
Lastly, the gluon and $b$ quark distribution are fairly stable after removal of SIDIS datasets, and this is expected since gluon contribution starts at NLO and bottom quark is not copiously produced in SIDIS.

\cref{f.subtraction-pp-K0S} demonstrates that excluding the $pp$ data has minimal impact on both quark and gluon distributions.
This behavior is expected since: (1) the quark distributions are already tightly constrained by SIA data, including $c$/$b$-tagged $K_S^0$ production measurements, and (2) the gluon distribution is primarily determined by $K_S^0$-in-jet data.
Consequently, the $pp$ collision data can only provide limited additional constraints.

\clearpage
\subsection{$\Lambda$ production}
\label{ss.Lambda-subtraction}

\begin{figure}[htb]
	\centering
	\includegraphics[width = 0.9 \textwidth]{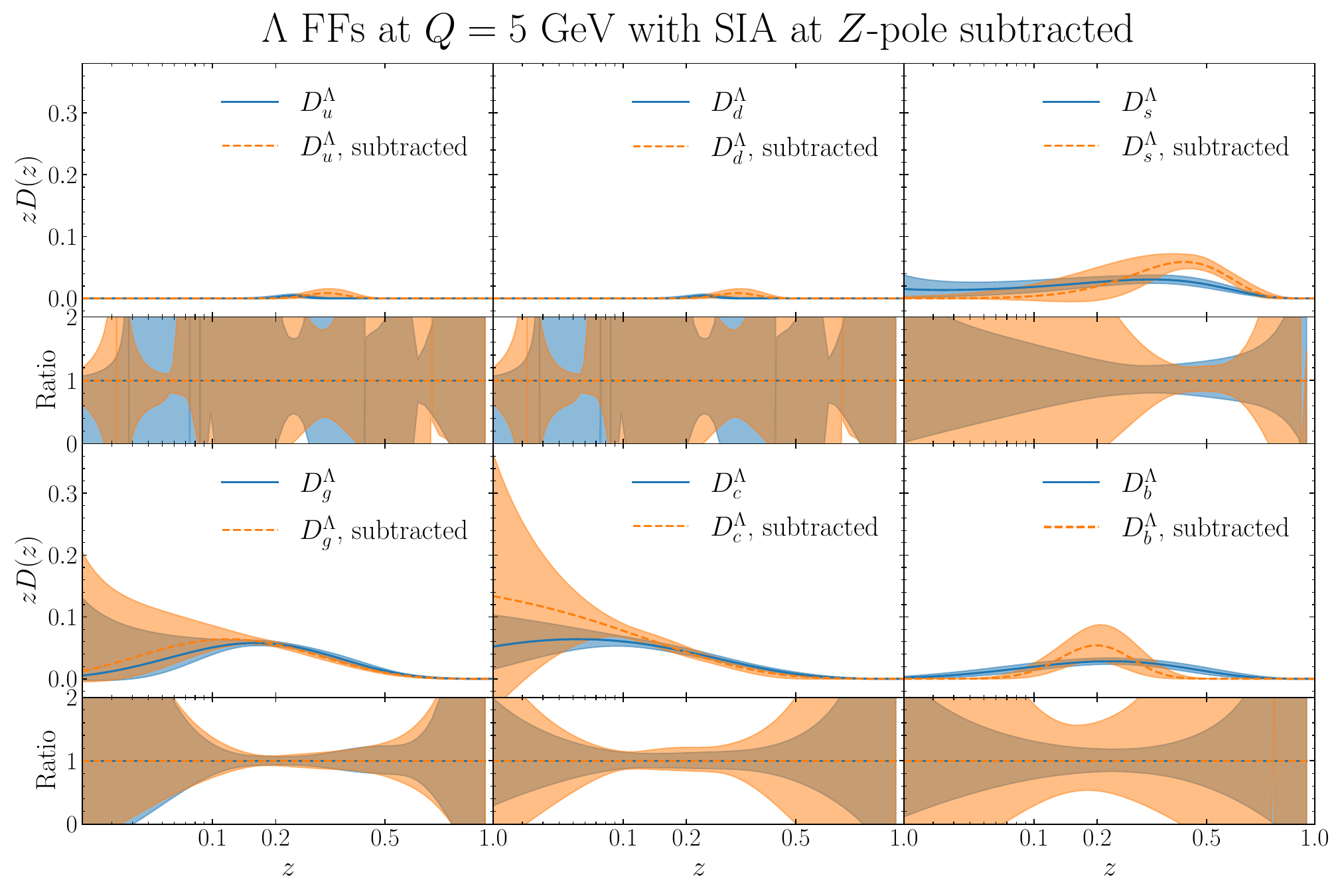}
	\caption{
		Same as \cref{f.subtraction-SIA-Z_pole-K0S} but for subtraction of $\Lambda$ production in SIA at $Z$-pole energy.
		(The $\Lambda$-in-jet data are not subtracted.)
	}
	\label{f.subtraction-SIA-Z_pole-Lambda}
\end{figure}

\begin{figure}[htb]
	\centering
	\includegraphics[width = 0.9 \textwidth]{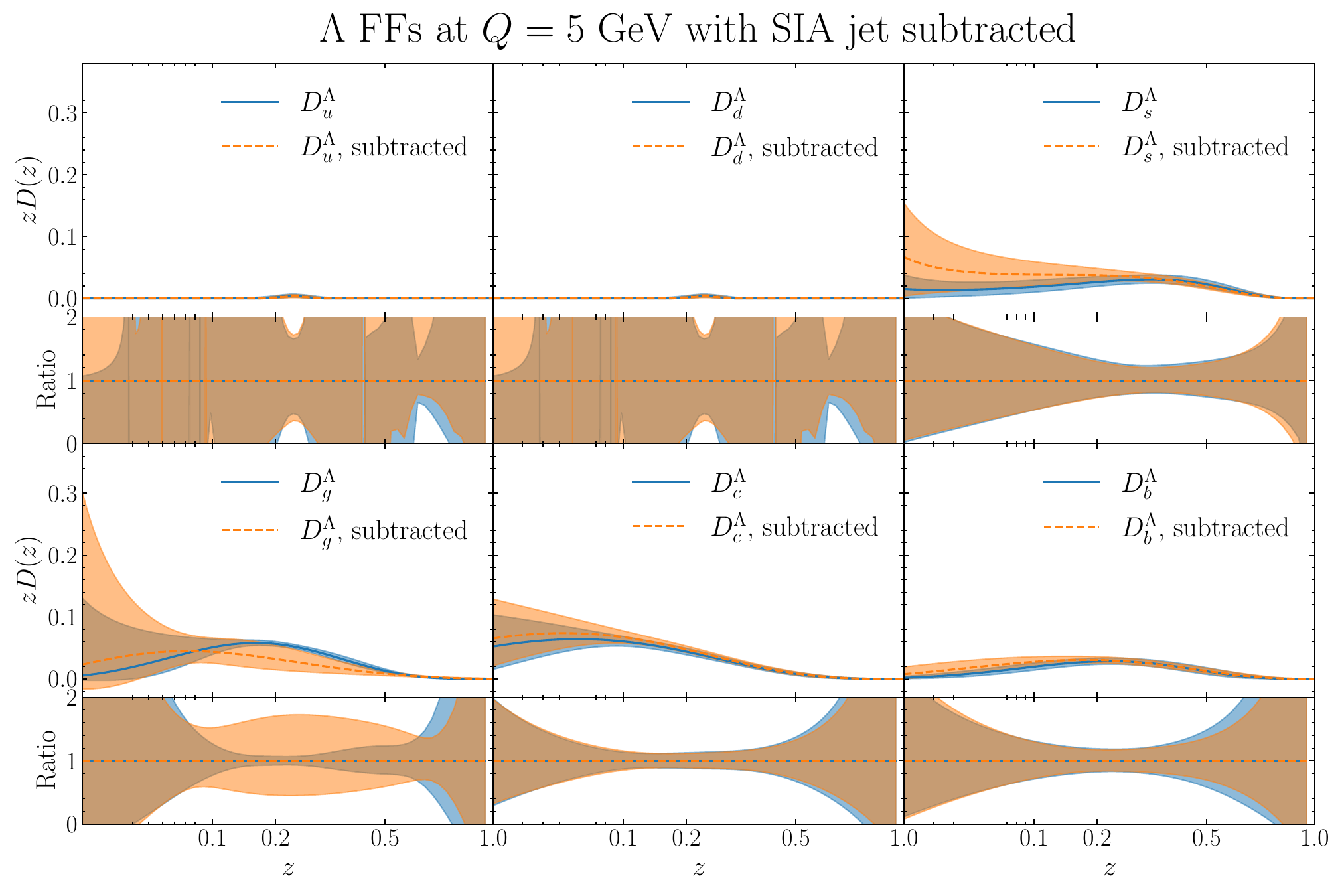}
	\caption{
		Same as \cref{f.subtraction-SIA-Z_pole-K0S} but for subtraction of $\Lambda$-in-jet production in SIA at $Z$-pole energy.
	}
	\label{f.subtraction-SIA-jet-Lambda}
\end{figure}

\begin{figure}[htb]
	\centering
	\includegraphics[width = 0.9 \textwidth]{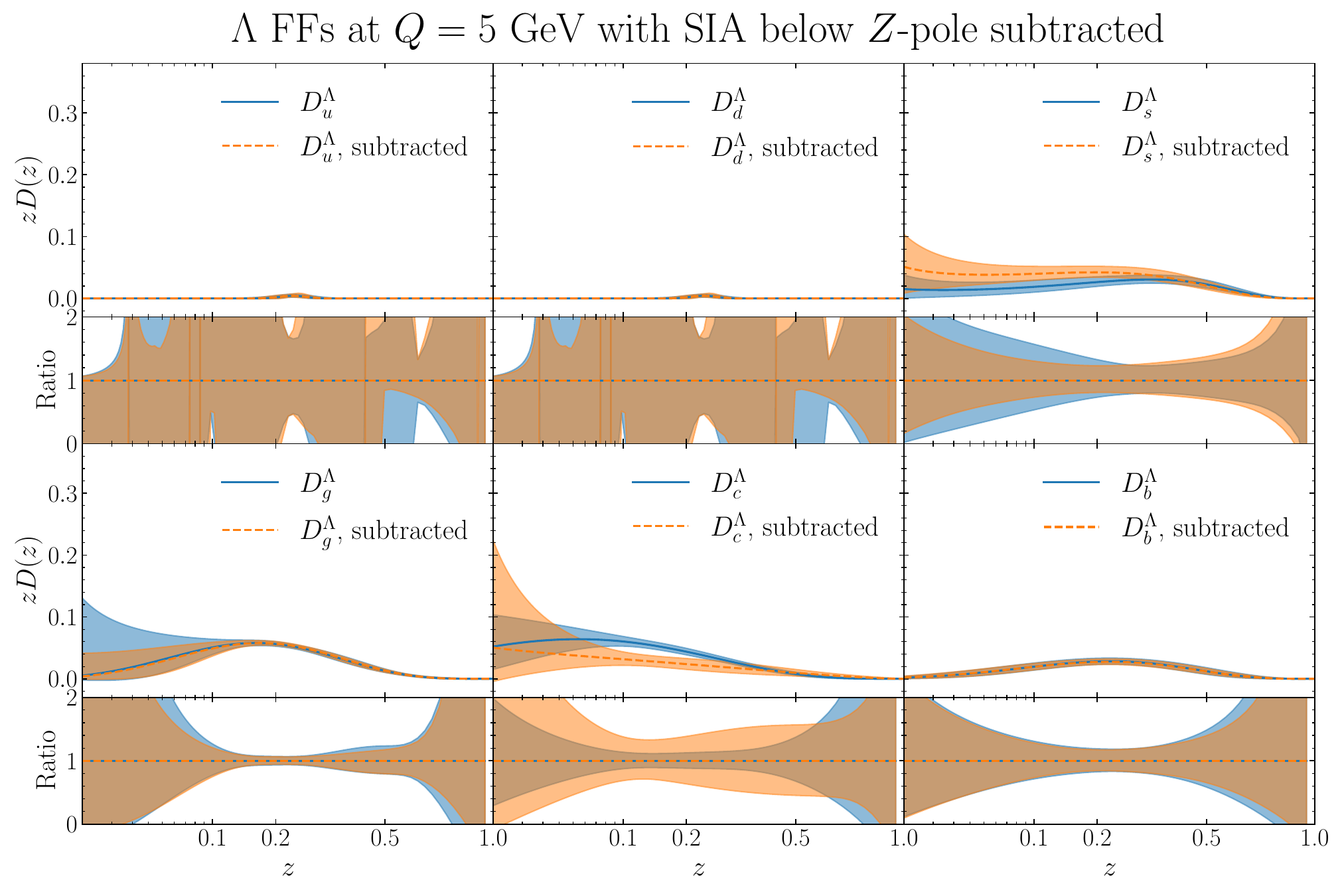}
	\caption{
		Same as \cref{f.subtraction-SIA-Z_pole-K0S} but for subtraction of $\Lambda$ production in SIA below $Z$-pole energy.
	}
	\label{f.subtraction-SIA-below-Z_pole-Lambda}
\end{figure}

In this section, we present the data subtraction analysis for $\Lambda$ production.
The comparison between each alternative fit and the baseline fit is shown in \cref{f.subtraction-SIA-Z_pole-Lambda,f.subtraction-SIA-jet-Lambda,f.subtraction-SIA-below-Z_pole-Lambda}.
The labels and meaning of the colored bands are the same as in the case of $K_S^0$ subtraction, and are given in \cref{ss.K0S-subtraction}.

In \cref{f.subtraction-SIA-Z_pole-Lambda}, the $\Lambda$ production in SIA at $Z$-pole energy is eliminated from the global datasets, and the FFs fitted with these modified datasets are compared to the baseline FFs.
Notably, the $\Lambda$-in-jet data remain unchanged in this context. Taking into account $u$ and $d$ FFs, it is evident that the SIA data at the $Z$-pole exert a notable influence, favoring smaller (almost negligible) distributions of $u$ and $d$.
For the $s$ distribution, a significant impact is also observed from the SIA data at the $Z$-pole, with a similar preference for a reduced distribution in the intermediate $z$ region.
Evidently, The gluon distribution is more effectively constrained at low $z$ by SIA data at the $Z$-pole, owing to their elevated center-of-mass energies.
Moreover, the SIA data at the $Z$-pole impose tighter constraints on the $c$ and $b$ distributions derived from SLD $c$/$b$-tagged data \cite{SLD:1998coh}.

The subtraction of SIA $\Lambda$-in-jet data at $Z$-pole energy is demonstrated in \cref{f.subtraction-SIA-jet-Lambda}.
First, we observe that the $u$ and $d$ distributions exhibit minimal change, suggesting that an effective constraint is provided by the remaining global data.
The strange distribution, while aligning with the baseline outcome in the large $z$ region, has its uncertainty potentially reducible with the inclusion of jet data.
The constraint on the gluon distribution is also enhanced by incorporating jet data, which aligns with the expectation that the third jet in a three-jet event is typically a gluon jet.
Lastly, both $c$ and $b$ quark distributions remain stable upon the subtraction of $\Lambda$-in-jet data as they are mainly constrained by the SLD $c$/$b$-tagged data \cite{SLD:1998coh}.

In \cref{f.subtraction-SIA-below-Z_pole-Lambda}, we present the results from subtraction of the SIA data below the $Z$-pole energy.
The SIA data below $Z$-pole energy has almost no impact on the $u$, $d$ and $b$ distributions, however, they do provide significant constraints for the strange and charm production rates at low and intermediate $z$ regions.

The subtraction of SIDIS and $pp$ data has almost no impact compared to the baseline results.
Figures are omitted for the sake of brevity.
This indicates that the $\Lambda$ FFs are already well determined by the SIA data, while the SIDIS and $pp$ data, with relatively larger experimental uncertainties, do not provide further constraints.

\clearpage
\subsection{Impact of the Belle measurements}
\label{ss.Belle-impact}

Recently, the Belle collaboration released a new set of measurements \cite{Belle:2024vua} for light and charmed mesons, including $K_S^0$ and $\eta$, which are of particular interest to us.
While the $K_S^0$ production data has been included in our nominal result, here we perform an alternative fit to investigate the impact of the $\eta$ production data from Belle.

The experiment was conducted at center-of-mass energies of 10.58 GeV and 10.52 GeV, just below the threshold for $b$-quark production.
The dataset features numerous data points with reduced uncertainties, making it a valuable resource for enhancing constraints and enabling more precise flavor separation.

In order to study the impact of the Belle $\eta$ production data, we performed an additional fit under the same conditions as our baseline analysis, this time including the Belle data \cite{Belle:2024vua} alongside the existing dataset.
After the inclusion of the Belle data, we found that the $\chi^2$ values deteriorated for some datasets, particularly for the SIA data measured below the $Z$-pole energy, as listed in \cref{t.chi_2-eta-with-Belle}.
From the table, it is evident that the $\chi^2/N_{\mathrm{pt}}$ values have increased significantly for ARGUS \cite{ARGUS:1989orf}, HRS \cite{HRS:1987aky} and JADE \cite{JADE:1985bzp} compared to those in \cref{t.chi_2-eta}.
The Belle data, however, shows a good $\chi^2/N_{\mathrm{pt}}$ of 0.4.

\begin{table}[]
	\setcellgapes{2.5 pt}
	\makegapedcells
	\begin{tabular}{|c|c|c|c|c|c|}
		\hline
		collaboration              & year & $\sqrt{s}~\bqty{\mathrm{GeV}}$ & $\chi^2$       & $N_{\mathrm{pt}}$ & $\chi^2/N_{\mathrm{pt}}$ \\
		\hline
		ARGUS \cite{ARGUS:1989orf} & 1990 & 9.46                           & 21.79          & 6                 & 3.63                     \\
		HRS \cite{HRS:1987aky}     & 1988 & 29                             & 34.35          & 13                & 2.64                     \\
		JADE \cite{JADE:1985bzp}   & 1985 & 34.4                           & 7.13           & 2                 & 3.57                     \\
		JADE \cite{JADE:1989ewf}   & 1990 & 35                             & 0.57           & 3                 & 0.19                     \\
		CELLO \cite{CELLO:1989byk} & 1990 & 35                             & 2.17           & 5                 & 0.43                     \\
		Belle \cite{Belle:2024vua} & 2024 & 10.52                          & 12.65          & 32                & 0.40                     \\
		\hline
		\textbf{sum}               &      &                                & \textbf{78.66} & \textbf{61}       & \textbf{1.29}            \\
		\hline
	\end{tabular}
	\caption{
		The $\chi^2$, number of data points ($N_{\mathrm{pt}}$) and $\chi^2 / N_{\mathrm{pt}}$ for the SIA datasets of $\eta$ production with Belle dataset \cite{Belle:2024vua} added.
		Additional details, including information on collaboration, the year of publication, and center-of-mass energy, are also provided.
		Sum of $\chi^2$ is also given for all the datasets listed in this table.
	}
	\label{t.chi_2-eta-with-Belle}
\end{table}

\begin{figure}[b]
	\centering
	\includegraphics[width = 0.9 \linewidth]{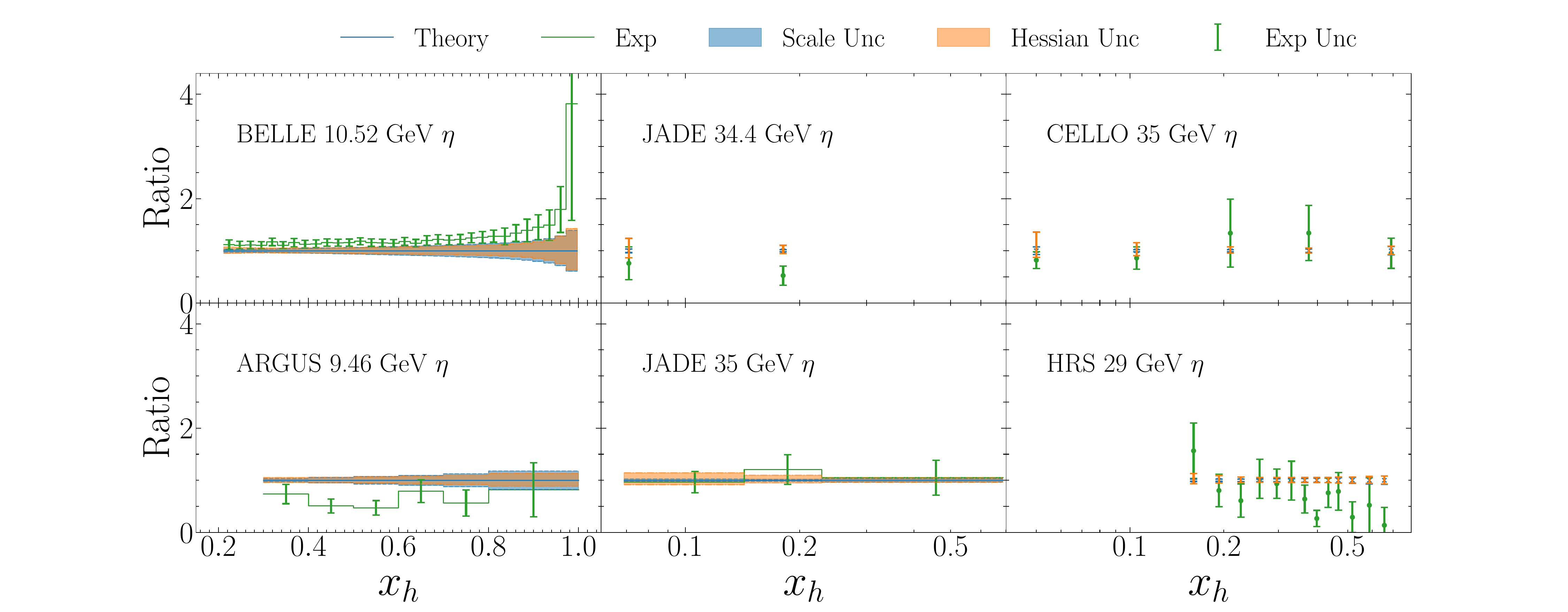}
	\caption{
		Same as \cref{f.SIA-below-Z_pole-eta} but with inclusion of the Belle data \cite{Belle:2024vua}.
	}
	\label{f.SIA-below-Z_pole-eta-with-Belle}
\end{figure}

In \cref{f.SIA-below-Z_pole-eta-with-Belle}, we present the experimental data normalized to theoretical predictions for the datasets listed in \cref{t.chi_2-eta-with-Belle}.
The legends used in this figure are the same as those in \cref{f.SIA-below-Z_pole-eta}.
From the figure, it can be seen that for most experiments (ARGUS at 9.46 GeV \cite{ARGUS:1989orf}, HRS at 29 GeV \cite{HRS:1987aky} and JADE at 34.4 GeV \cite{JADE:1985bzp}), the data points tends to lie below the theoretical predictions.
In contrast, the Belle data points are generally above the theoretical predictions, with a deviation typically exceeding 10\%.
This indicates a significant tension between Belle and other low-energy SIA measurements.

\begin{figure}
	\centering
	\includegraphics[width = 0.9 \linewidth]{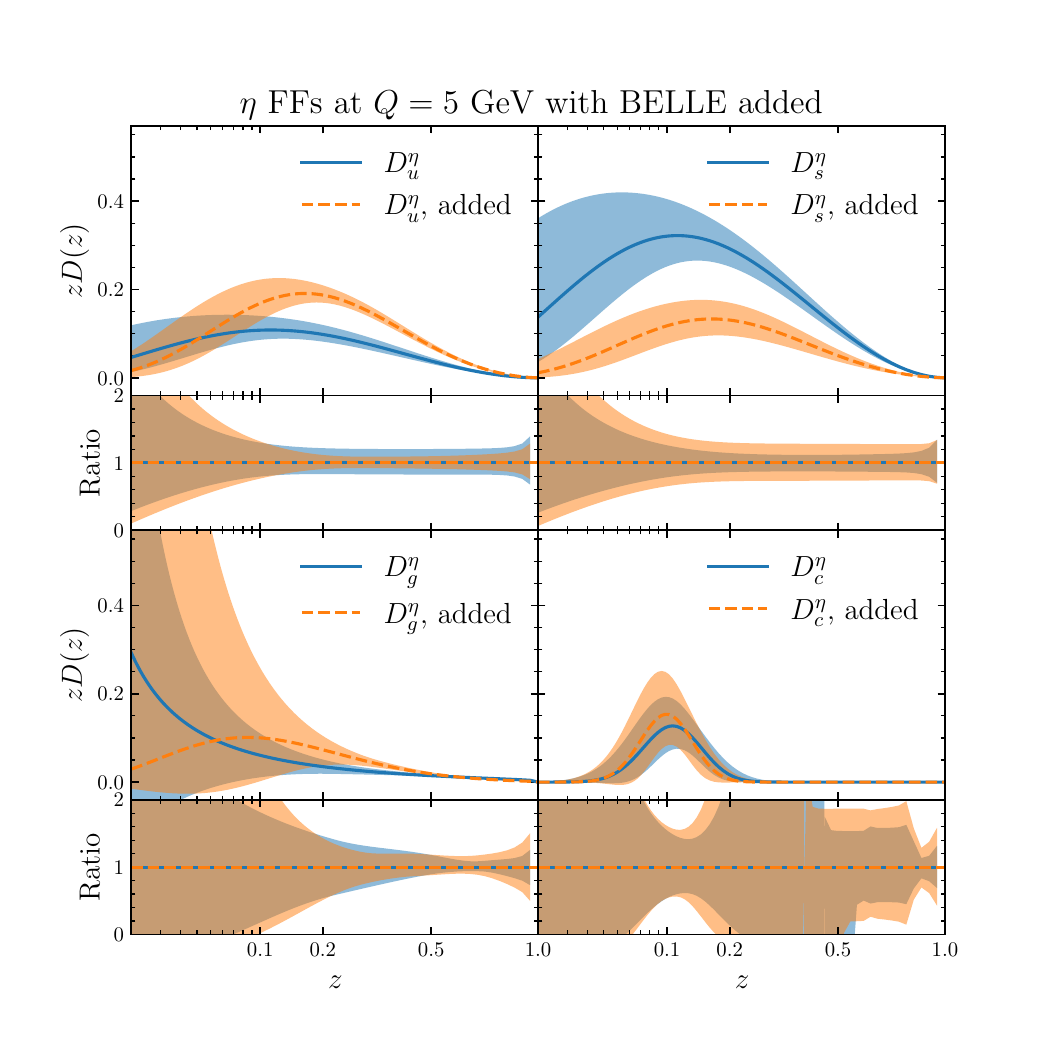}
	\caption{
		Comparison of our nominal $\eta$ FFs (blue) and the alternative fit (orange) with Belle data \cite{Belle:2024vua} added, at $Q = 5~\mathrm{GeV}$.
		The setup of the plot follows the same as \cref{f.FFs-K0S}.
	}
	\label{f.FFs-eta-with-Belle}
\end{figure}

Additionally, we compare the $\eta$ FFs from our baseline fit with those from the alternative fit with Belle data included in \cref{f.FFs-eta-with-Belle}.
The Belle data appears to favor a larger $u$ quark fragmentation while suppressing the $s$ quark distribution compared to the baseline fit.
For gluon fragmentation, although the central values exhibit suppression in the low $z$ region and enhancement in the high $z$ region, the two results are generally consistent within uncertainties.
Finally, the $c$ quark distribution remains generally unchanged even with inclusion of the Belle data.

In summary, due to the tension found between the $\eta$ production data from Belle \cite{Belle:2024vua} and the other world data including ARGUS \cite{ARGUS:1989orf}, HRS \cite{HRS:1987aky} and JADE \cite{JADE:1985bzp}, we decide not to include the Belle measurement in our nominal fit for $\eta$ FFs.
In order to maintain consistency with the $\eta$ data selection and considering that the improvement in FFs is not substantial, we decided not to include the Belle $K_S^0$ data in our nominal results either.

\clearpage
\subsection{Impact of the LHCb measurements}
\label{ss.LHCb-impact}

In this section, we examine the influence of the LHCb measurement for the $\eta/\pi^0$ production ratio $R \pqty{\eta / \pi^0}$ at 5.02 TeV and 13 TeV \cite{LHCb:2023iyw}.
Here $R \pqty{\eta / \pi^0}$ is defined as:
\begin{equation}
	R \pqty{\eta / \pi^0}
	\equiv
	\frac{\dd{\sigma_{\eta}}}{\dd{p_T}}
	\bigg /
	\frac{\dd{\sigma_{\pi^0}}}{\dd{p_T}}
	\, ,
\end{equation}
First, we provide theoretical predictions for the LHCb data using the nominal NPC23 $\eta$ and $\pi^0$ FFs.
Then we assess the impact of this measurement by including it in an alternative fit, in which the nominal NPC23 $\pi^0$ FFs provided in this work are used for determining the denominator of $\eta/\pi^0$ production ratios.
Additionally, the Hessian uncertainties of the constructed $\pi^0$ FFs are very small and can be omitted as indicated in \cref{ss.eta-data}.

\begin{figure}[htbp]
	\centering
	\includegraphics[width = 0.45 \textwidth]{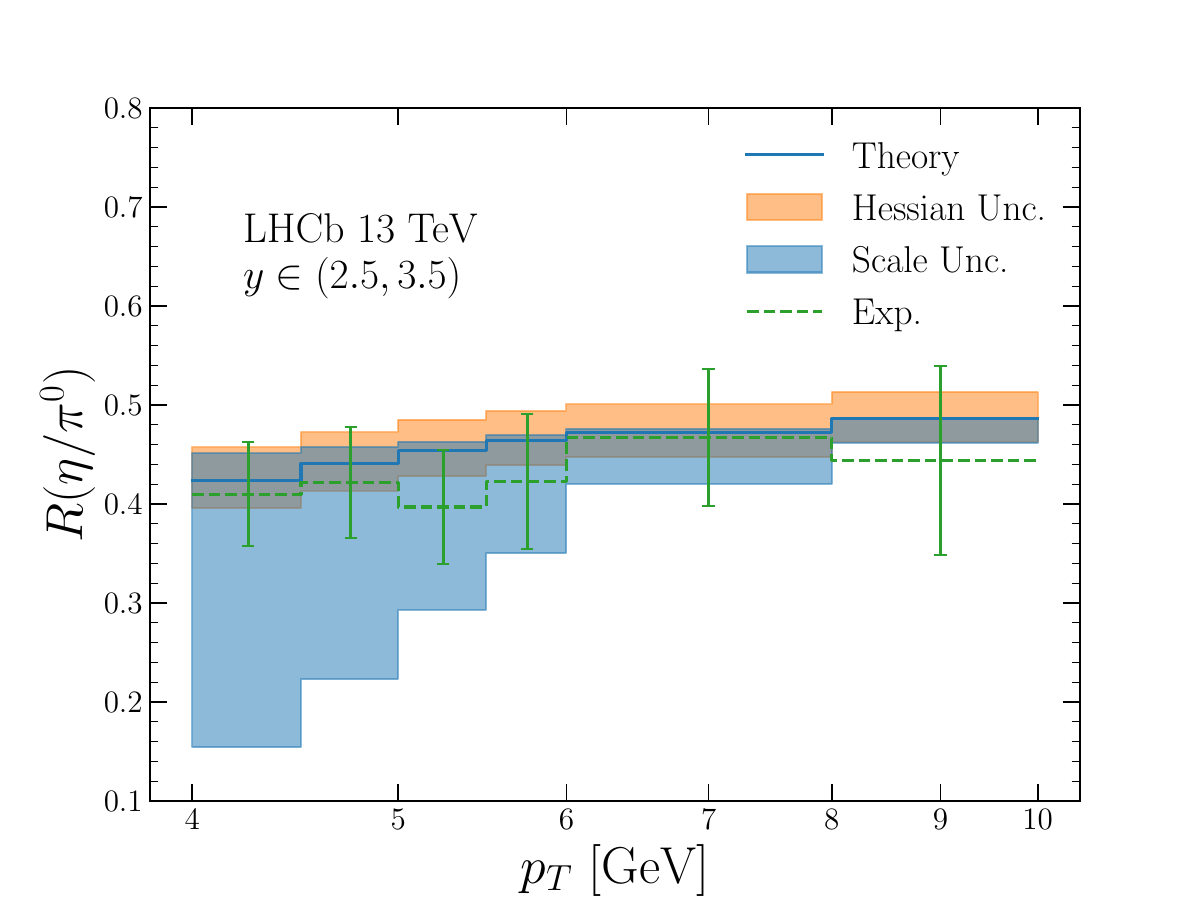}
	\includegraphics[width = 0.45 \textwidth]{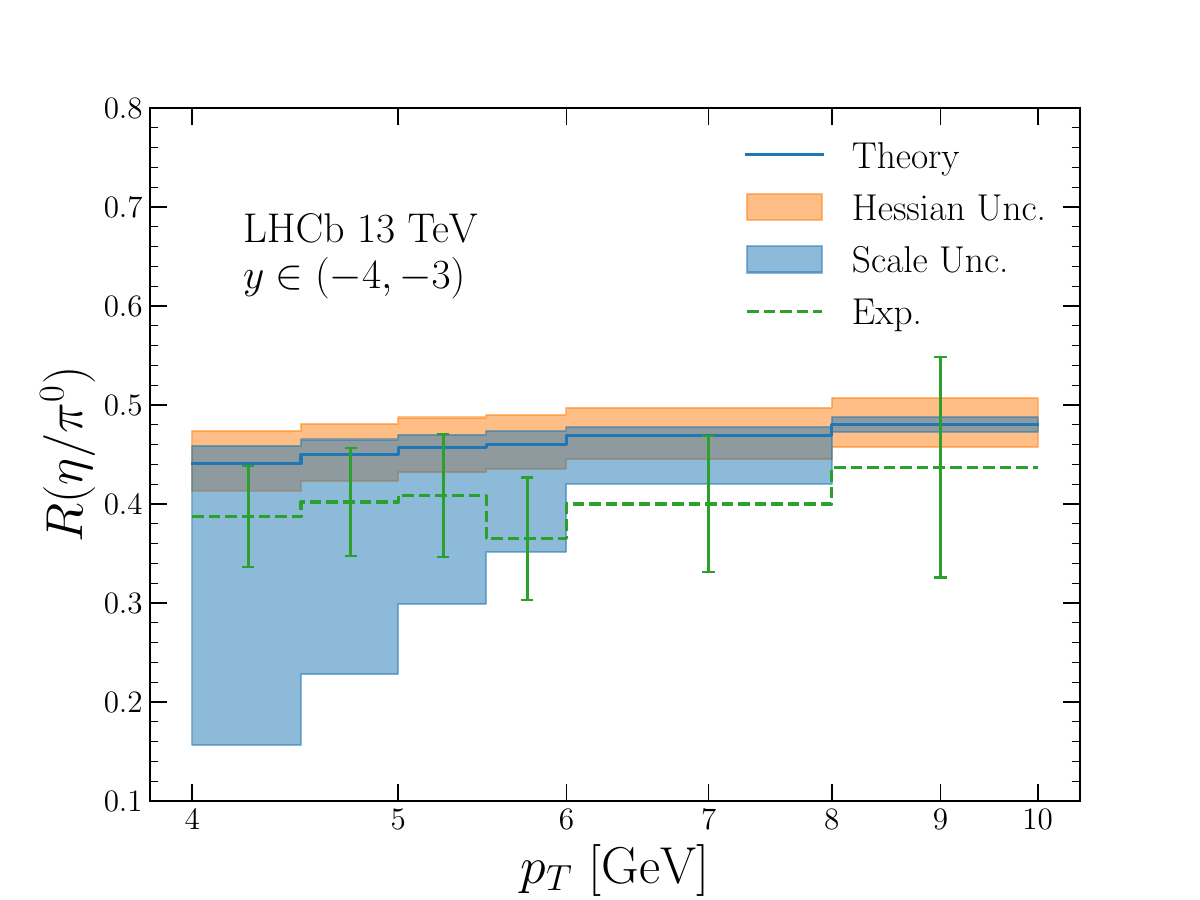}
	\caption{
		Comparison between the theoretical predictions generated with the NPC nominal FFs and the experimental data from LHCb \cite{LHCb:2023iyw} at a center-of-mass energy of 13 TeV.
		The left and right panels correspond the rapidity regions $y \in \pqty{2.5, 3.5}$ and $y \in \pqty{-4, -3}$, respectively.
		The data and experimental uncertainties are shown in green error bars, while the scale and Hessian uncertainties are shown in blue and orange bands, respectively.
	}
	\label{f.LHCb-eta_over_pi0_13T_scl2}
\end{figure}

\begin{figure}[htbp]
	\centering
	\includegraphics[width = 0.45 \textwidth]{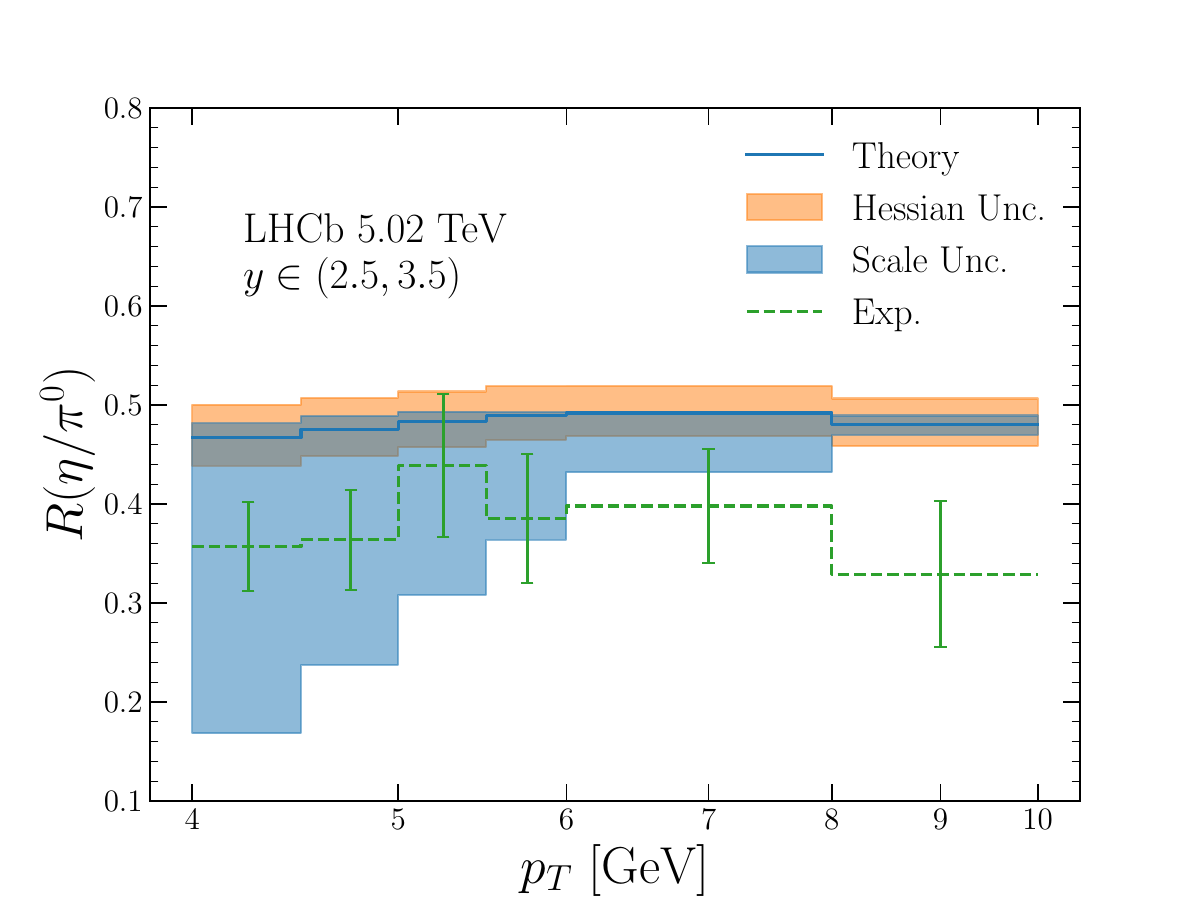}
	\includegraphics[width = 0.45 \textwidth]{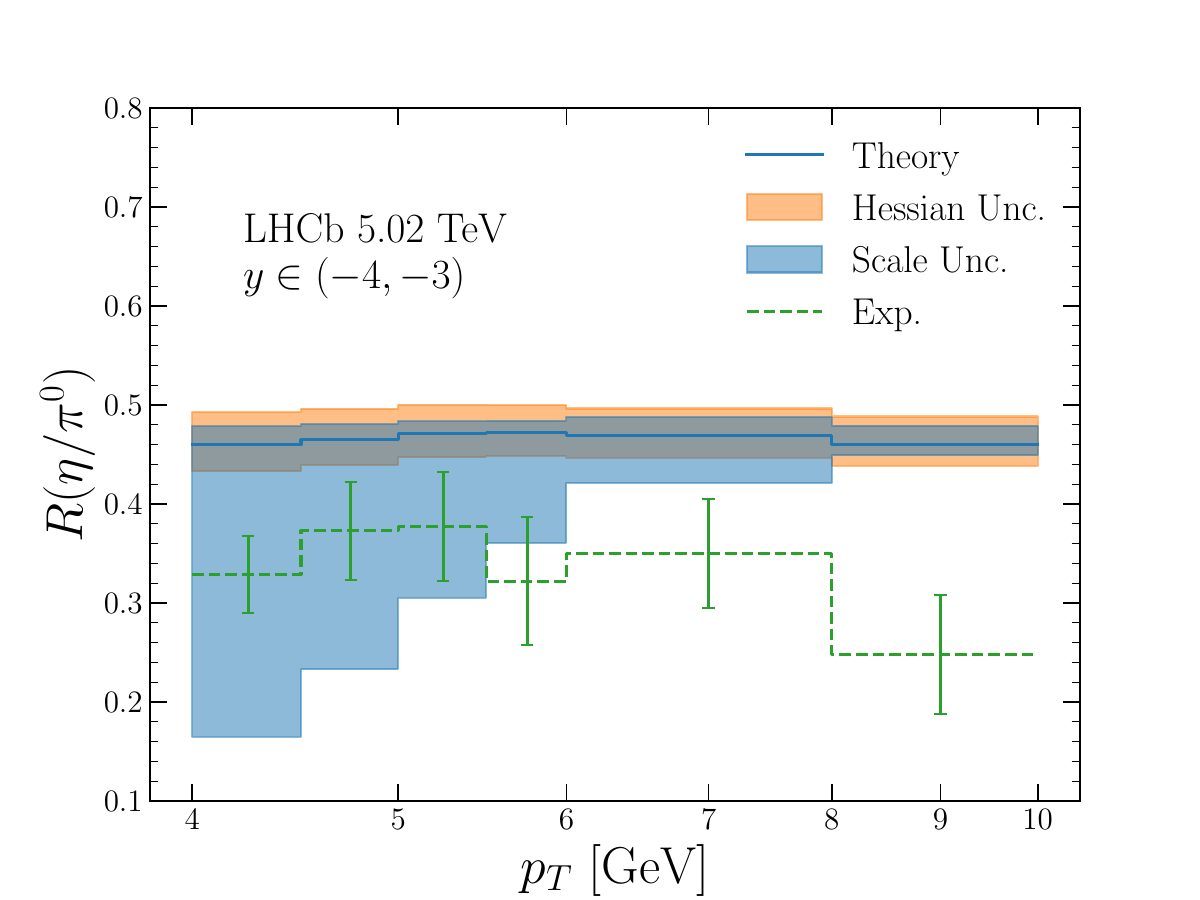}
	\caption{
		Same as \cref{f.LHCb-eta_over_pi0_13T_scl2} but at $\sqrt{s} = 5.02~\mathrm{TeV}$.
	}
	\label{f.LHCb-eta_over_pi0_5T_scl2}
\end{figure}

\cref{f.LHCb-eta_over_pi0_5T_scl2,f.LHCb-eta_over_pi0_13T_scl2} display the comparison between theoretical predictions and experimental data for center-of-mass energies of 5.02 TeV and 13 TeV, respectively.
The theoretical predictions are drawn with solid blue lines, along with their corresponding uncertainty bands.
The Hessian uncertainties at the 68\% confidence level are represented by orange bands, while the scale uncertainties are shown as blue bands.
The scale uncertainty is calculated following the approach outlined in \cref{ss.theory-computation-method}.
The experimental values and uncertainties are shown as green error bars, which incorporates both statistical and systematic uncertainties.
In the FMNLO grid generation, the renormalization and factorization scales are chosen as $\mu_R = \mu_F = p_T$, where $p_T$ denotes the transverse momentum of the identified hadron ($\eta$ or $\pi^0$ in this case).
This choice leads to enhanced scale uncertainties in the low-$p_T$ region ($p_T < 6~\mathrm{GeV}$), where the more pronounced running of the strong coupling constant at lower energy scales results in larger variations.

A systematic trend is observed across all rapidity regions and collision energies, with theoretical predictions consistently overestimating the $\eta/\pi^0$ ratio compared to the experimental data.
This discrepancy can be attributed to the sensitivity of the theoretical calculations to the choice of factorization, renormalization and fragmentation scales, as reflected by the large scale uncertainties in \cref{f.LHCb-eta_over_pi0_5T_scl2,f.LHCb-eta_over_pi0_13T_scl2}.

At the center-of-mass energy of 13 TeV (\cref{f.LHCb-eta_over_pi0_13T_scl2}), the theoretical predictions demonstrate good agreement with the experimental data across the entire $p_T$ range.
The theoretical calculations consistently fall within the experimental uncertainties, indicating robust compatibility between theory and experiment at $\sqrt{s} = 13~\mathrm{TeV}$.
Additionally, the experimental data reveals no significant rapidity dependence in the $\eta/\pi^0$ ratios at $\sqrt{s} = 13~\mathrm{TeV}$.

At the center-of-mass energy of 5.02 TeV (\cref{f.LHCb-eta_over_pi0_5T_scl2}), the theoretical predictions remain stable across both rapidity intervals.
However, they consistently overestimate the experimental values by approximately 30\%.
The agreement is better in the intermediate $p_T$ region compared to both low and high $p_T$ ends.
This behavior is observed in both rapidity ranges that are considered in \cref{f.LHCb-eta_over_pi0_5T_scl2}.

\begin{figure}[htbp]
	\centering
	\includegraphics[width = 0.45 \textwidth]{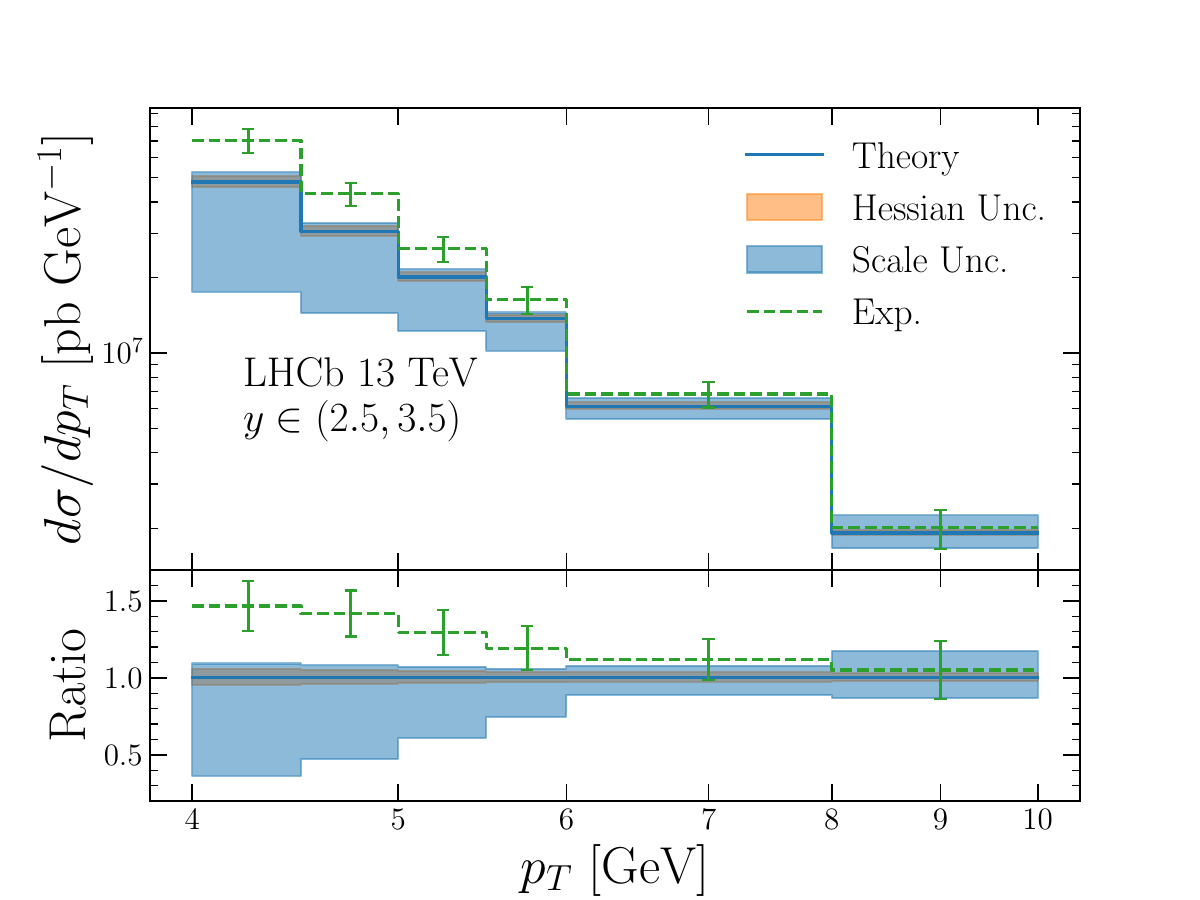}
	\includegraphics[width = 0.45 \textwidth]{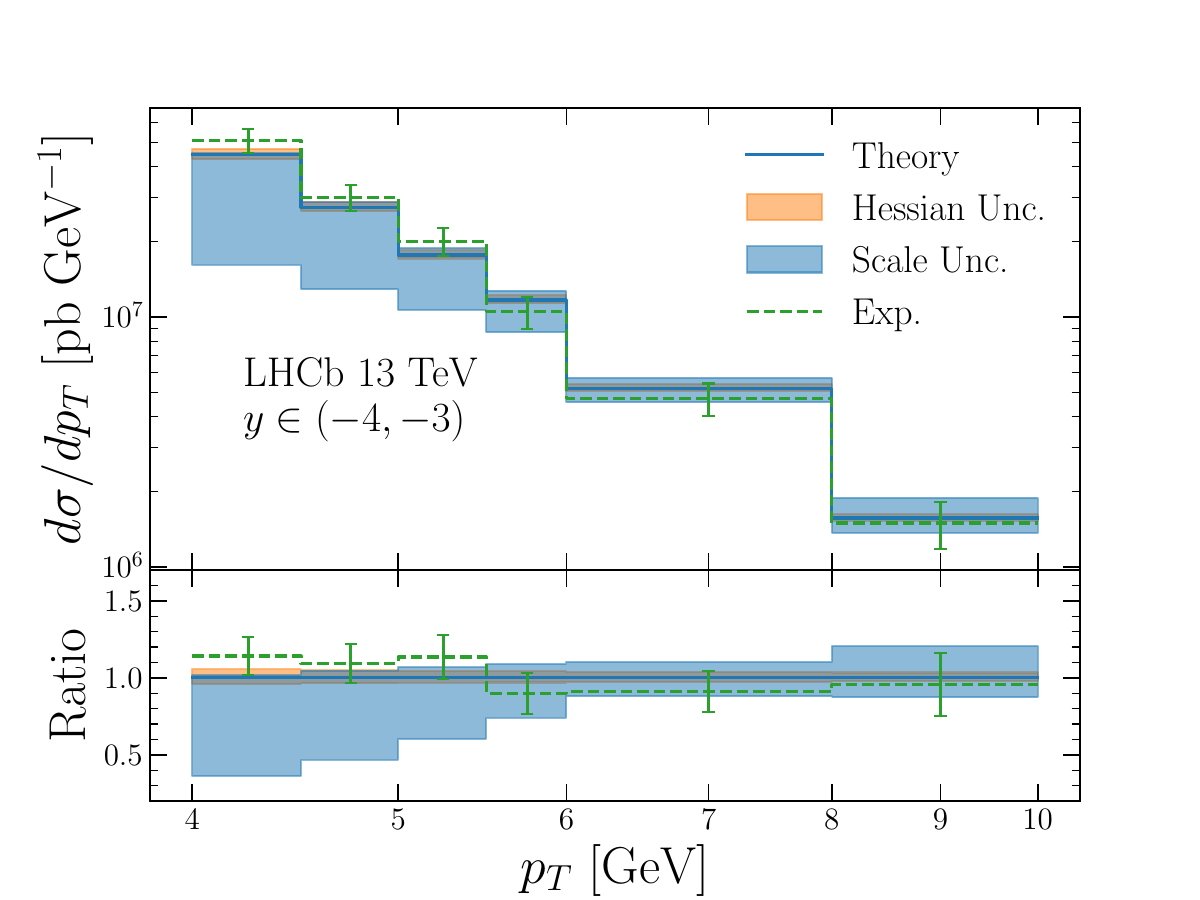}
	\caption{
		Comparison of the differential cross section for $\eta$ production as a function of transverse momentum $p_T$ between theoretical predictions and experimental measurements at a center-of-mass energy of 13 TeV.
		The left and right panels correspond the rapidity regions $y \in \pqty{2.5, 3.5}$ and $y \in \pqty{-4, -3}$, respectively.
		The lower panels display the experimental data normalized to the central value of the theoretical predictions.
	}
	\label{f.LHCb_eta_xsection_13T}
\end{figure}

\begin{figure}[htbp]
	\centering
	\includegraphics[width = 0.45 \textwidth]{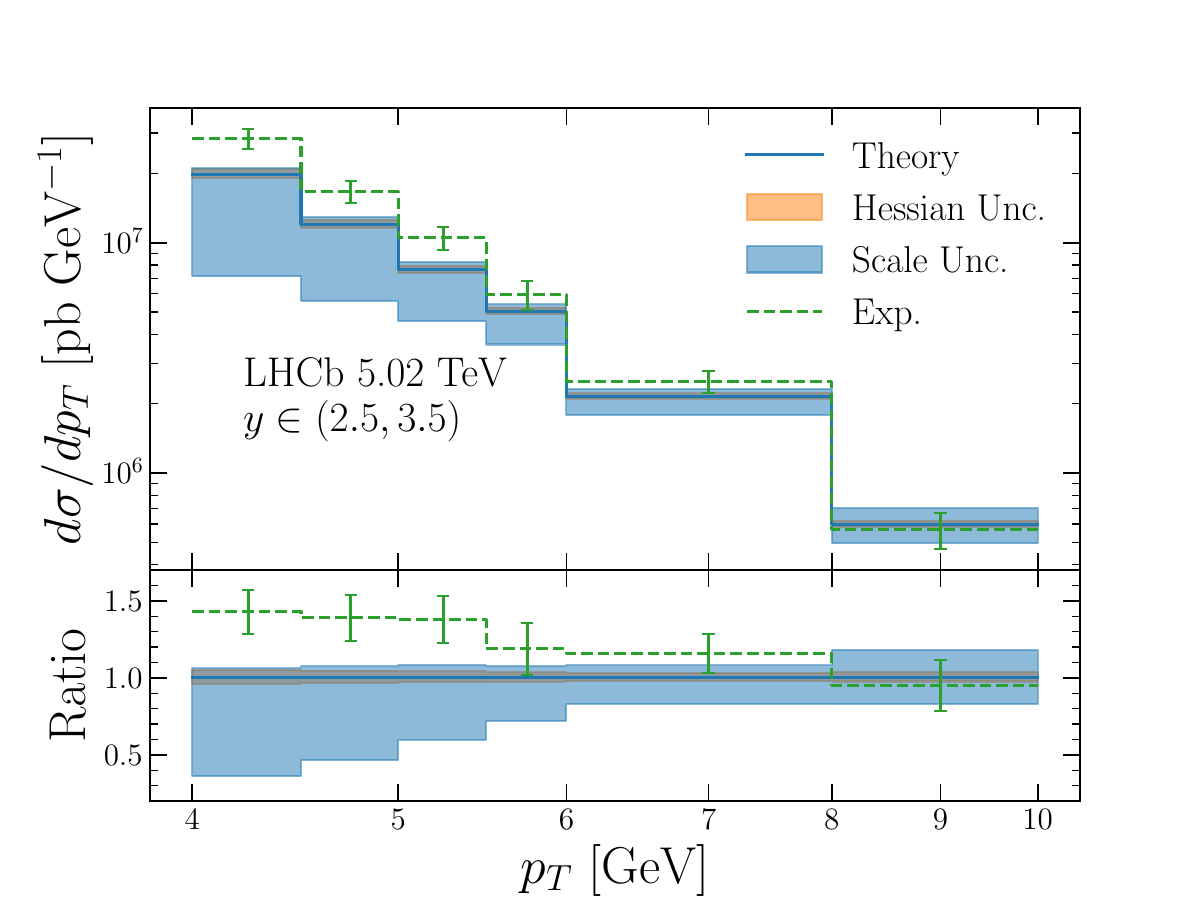}
	\includegraphics[width = 0.45 \textwidth]{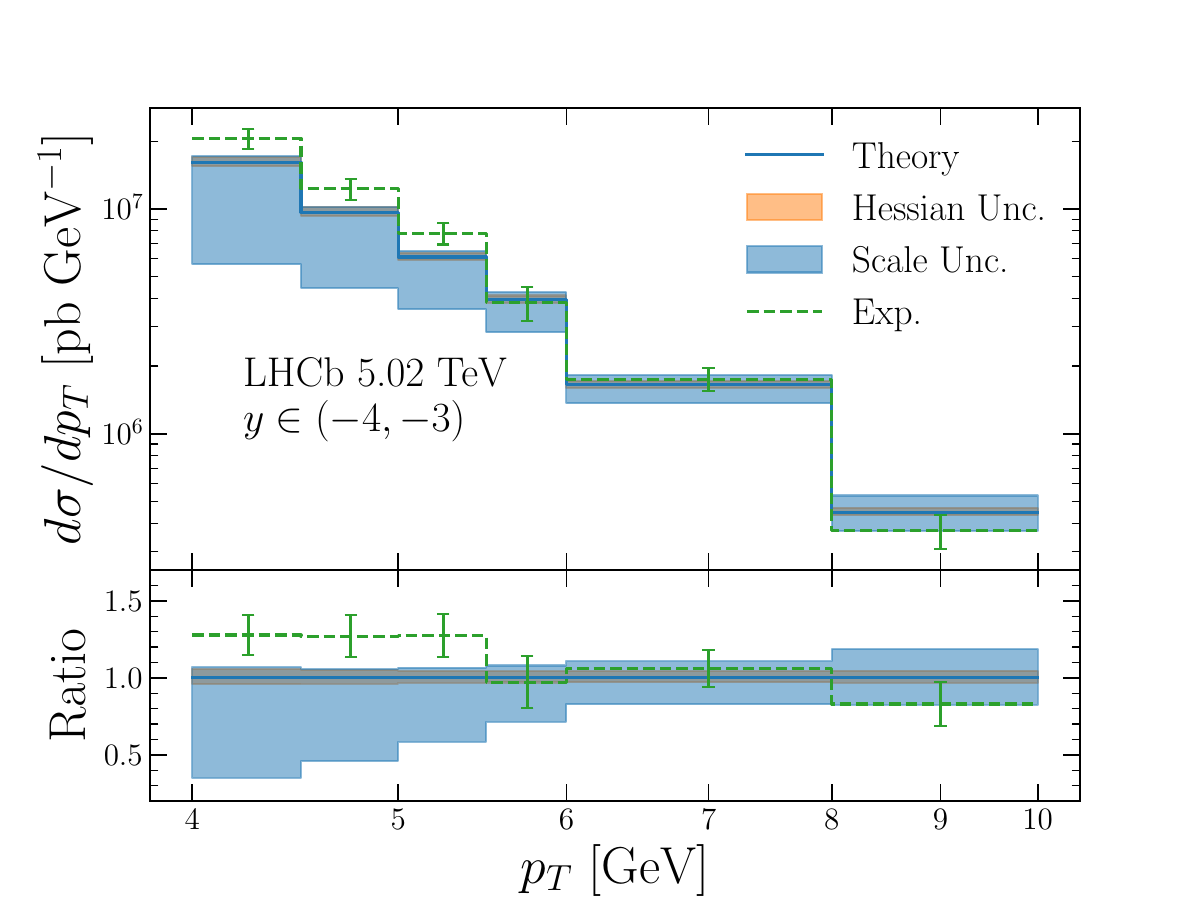}
	\caption{
		Same as \cref{f.LHCb_eta_xsection_13T} but at $\sqrt{s} = 5.02~\mathrm{TeV}$.
	}
	\label{f.LHCb_eta_xsection_5T}
\end{figure}

We further present a detailed comparison of the differential cross section for $\eta$ production as a function of transverse momentum $p_T$ between theoretical predictions and experimental measurements at center-of-mass energies of 5.02 TeV and 13 TeV, covering rapidity regions $y \in \pqty{2.5, 3.5}$ and $y \in \pqty{-4, -3}$.
The results are presented in \cref{f.LHCb_eta_xsection_5T,f.LHCb_eta_xsection_13T}.

The upper panels of these figures display the differential cross sections, while the lower panels depict the data normalized to the central value of the theoretical predictions.
At both collision energies, the theoretical curves agree well with experimental data in the high-$p_T$ region.
However, discernible deviations appear at low-$p_T$ region, particularly in the rapidity region $y \in \pqty{2.5, 3.5}$, where the theoretical predictions overestimate the experimental data by approximately $40\% \sim 50\%$ for both collision energies.
This discrepancy is accompanied by significant scale uncertainties in this kinematic region, which further highlights the sensitivity of the theoretical calculations to the choice of factorization and renormalization scales.

In contrast, better agreement at low-$p_T$ region is observed for the rapidity region $y \in \pqty{-4, -3}$.
At 5.02 TeV, the theoretical predictions overestimate the experimental data by approximately $30\% \sim 40\%$ in the low-$p_T$ bins.
At 7 TeV, the agreement further improves, where experimental measurements and theoretical calculations overlap within their uncertainty bands across the entire $p_T$ region.

These findings demonstrate the robustness of the theoretical framework in describing $\eta$ production, particularly at high $p_T$ and larger collision energies.
However, the persistent discrepancies at lower $p_T$ values suggest areas for further refinement, particularly in addressing scale uncertainties and optimizing the treatment of low-$p_T$ dynamics.

\begin{table}[]
	\setcellgapes{2.5 pt}
	\makegapedcells
	\begin{tabular}{|c|c|c|c|c|c|}
		\hline
		collaboration                     & year & $\sqrt{s}~\bqty{\mathrm{GeV}}$ & $\chi^2$        & $N_{\mathrm{pt}}$ & $\chi^2/N_{\mathrm{pt}}$ \\
		\hline
		ARGUS \cite{ARGUS:1989orf}        & 1990 & 9.46                           & 3.42            & 6                 & 0.57                     \\
		HRS \cite{HRS:1987aky}            & 1988 & 29                             & 15.58           & 13                & 1.20                     \\
		JADE \cite{JADE:1985bzp}          & 1985 & 34.4                           & 1.55            & 2                 & 0.77                     \\
		JADE \cite{JADE:1989ewf}          & 1990 & 35                             & 4.65            & 3                 & 1.55                     \\
		CELLO \cite{CELLO:1989byk}        & 1990 & 35                             & 4.12            & 5                 & 0.82                     \\
		L3 \cite{L3:1992pbe}              & 1992 & 91.2                           & 5.56            & 4                 & 1.39                     \\
		L3 \cite{L3:1994gkb}              & 1994 & 91.2                           & 9.89            & 10                & 0.99                     \\
		ALEPH \cite{ALEPH:1992zhm}        & 1992 & 91.2                           & 1.50            & 8                 & 0.19                     \\
		ALEPH \cite{ALEPH:1999udi}        & 2000 & 91.2                           & 19.68           & 18                & 1.09                     \\
		ALEPH jet 1 \cite{ALEPH:1999udi}  & 2000 & 91.2                           & 11.50           & 7                 & 1.64                     \\
		ALEPH jet 2 \cite{ALEPH:1999udi}  & 2000 & 91.2                           & 2.44            & 6                 & 0.41                     \\
		ALEPH jet 3 \cite{ALEPH:1999udi}  & 2000 & 91.2                           & 13.83           & 4                 & 3.46                     \\
		ALEPH \cite{ALEPH:2001tfk}        & 2002 & 91.2                           & 16.68           & 5                 & 3.34                     \\
		OPAL \cite{OPAL:1998enc}          & 1998 & 91.2                           & 6.48            & 11                & 0.59                     \\
		\hline
		SIA sum                           &      &                                & 116.87          & 102               & 1.15                     \\
		\hline
		PHENIX \cite{PHENIX:2010hvs}      & 2011 & 200                            & 10.89           & 14                & 0.78                     \\
		ALICE \cite{ALICE:2017nce}        & 2017 & 2760                           & 5.08            & 6                 & 0.85                     \\
		ALICE \cite{ALICE:2012wos}        & 2012 & 7000                           & 1.38            & 4                 & 0.345                    \\
		ALICE \cite{ALICE:2017ryd}        & 2018 & 8000                           & 11.67           & 13                & 0.90                     \\
		ALICE \cite{ALICE:2024vgi}        & 2024 & 13000                          & 10.68           & 14                & 0.76                     \\
		\hline
		$pp$ sum                          &      &                                & 39.71           & 51                & 0.87                     \\
		\hline
		LHCb forward \cite{LHCb:2023iyw}  & 2024 & 5020                           & 5.06            & 6                 & 0.84                     \\
		LHCb forward \cite{LHCb:2023iyw}  & 2024 & 13000                          & 0.60            & 6                 & 0.10                     \\
		LHCb backward \cite{LHCb:2023iyw} & 2024 & 5020                           & 13.08           & 6                 & 2.18                     \\
		LHCb backward \cite{LHCb:2023iyw} & 2024 & 13000                          & 1.42            & 6                 & 0.24                     \\
		\hline
		\textbf{sum}                      &      &                                & \textbf{20.16}  & \textbf{24}       & \textbf{0.84}            \\
		\hline
		\textbf{total sum}                &      &                                & \textbf{176.73} & \textbf{177}      & \textbf{1.00}            \\
		\hline
	\end{tabular}
	\caption{
		The $\chi^2$, number of data points ($N_{\mathrm{pt}}$) and $\chi^2 / N_{\mathrm{pt}}$ for $\eta/\pi^0$ production data with LHCb dataset \cite{LHCb:2023iyw} added.
		The terms ``forward'' and ``backward'' denote rapidity bins $y \in \pqty{2.5, 3.5}$ and $y \in \pqty{-4, -3}$, respectively.
		Additional details, including information on collaboration, the year of publication, and center-of-mass energy, are also provided.
		Sum of $\chi^2$ is also given for all the datasets listed in this table.
	}
	\label{t.chi_2-eta-with-lhcb}
\end{table}

To further investigate the influence of the LHCb data on our analysis, we performed an alternative fit under identical conditions as our baseline analysis, incorporating only the LHCb data \cite{LHCb:2023iyw} in addition.
The inclusion of the LHCb data resulted in minimal changes to the overall $\chi^2$ values.
The $\chi^2$ values for newly added sets are provided in \cref{t.chi_2-eta-with-lhcb}.

As shown in \cref{t.chi_2-eta-with-lhcb}, a good quality fit is achieved for the LHCb datasets, with a total $\chi^2/N_{\mathrm{pt}} = 0.84$.
However, the LHCb data measured in the backward rapidity region ($\eta \in \pqty{-4, -3}$) exhibits a slightly worse agreement, with $\chi^2/N_{\mathrm{pt}} = 2.18$.
In addition, the inclusion of the LHCb datasets only slightly deteriorates the fit quality of the global datasets, resulting in an increase of $\chi^2$ by 5.22 units.

\begin{figure}
	\centering
	\includegraphics[width=0.8\linewidth]{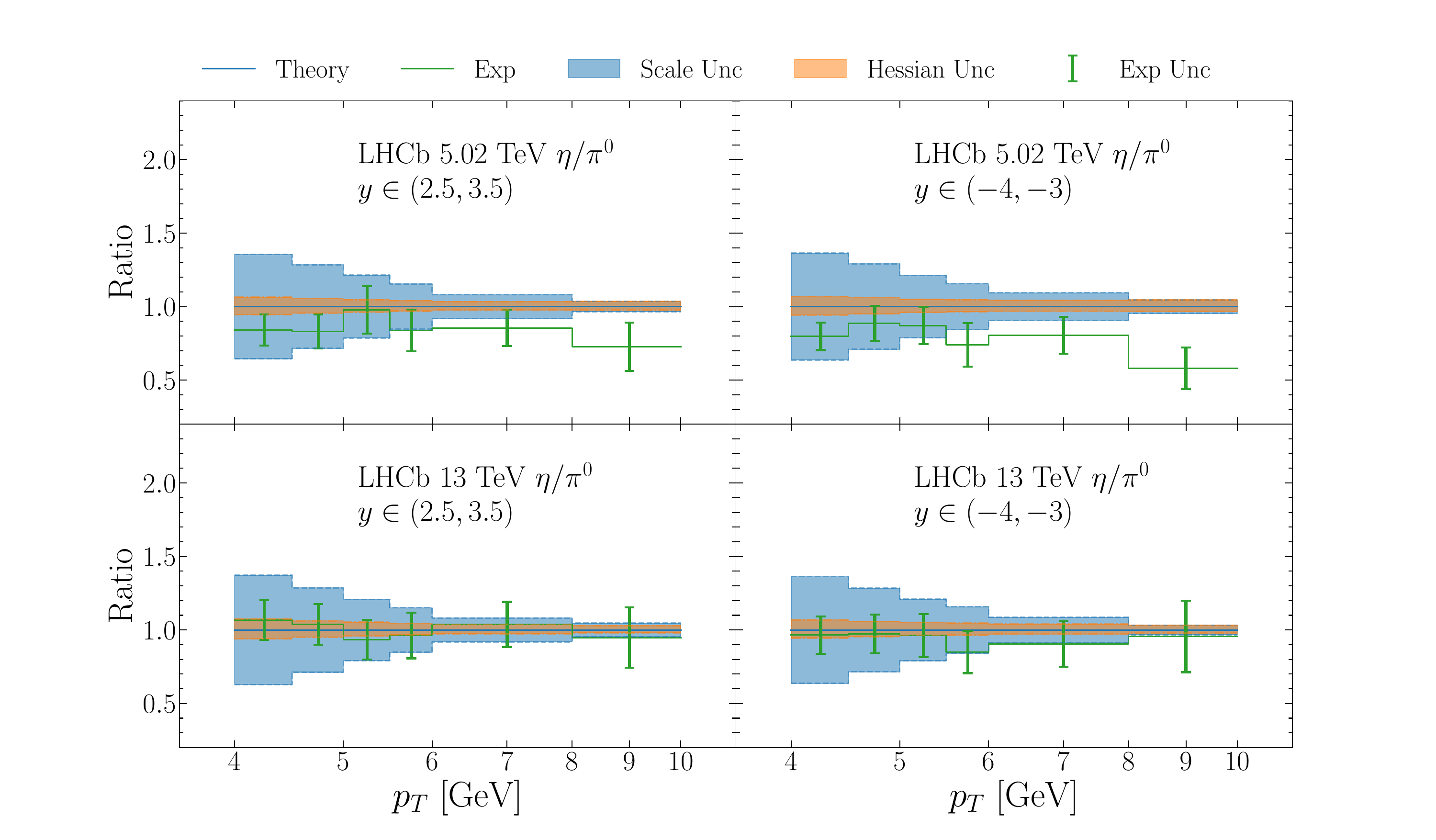}
	\caption{
		Same as \cref{f.pp-eta} but for LHCb data on production ratio of $\eta / \pi^0$ with inclusion of the LHCb data \cite{LHCb:2023iyw} on top of the baseline fit.
	}
	\label{f.eta_lhcb_theory_exp}
\end{figure}

In \cref{f.eta_lhcb_theory_exp}, we present the experimental data normalized to the central values of the theoretical predictions for the LHCb datasets listed in \cref{t.chi_2-eta-with-lhcb}.
The plot compares the measured and theoretically computed $\eta/\pi^0$ production ratios at center-of-mass energies 5.02 TeV and 13 TeV in both forward and backward rapidity regions.
The legends used in this figure are consistent with those in \cref{f.pp-eta}.

By comparing with \cref{f.LHCb-eta_over_pi0_5T_scl2,f.LHCb-eta_over_pi0_13T_scl2}, it is clear that the agreement between theory and experimental data improves significantly after incorporating the LHCb data into the fit.
The 13 TeV measurements now show excellent consistency, while discrepancies at 5 TeV are substantially reduced.
Nevertheless, in the high $p_T$ region, deviation between the theoretical predictions and the experimental data is still observed at $\sqrt{s} = 5.02~\mathrm{GeV}$.

\begin{figure}[htbp]
	\centering
	\includegraphics[width = 0.8 \textwidth]{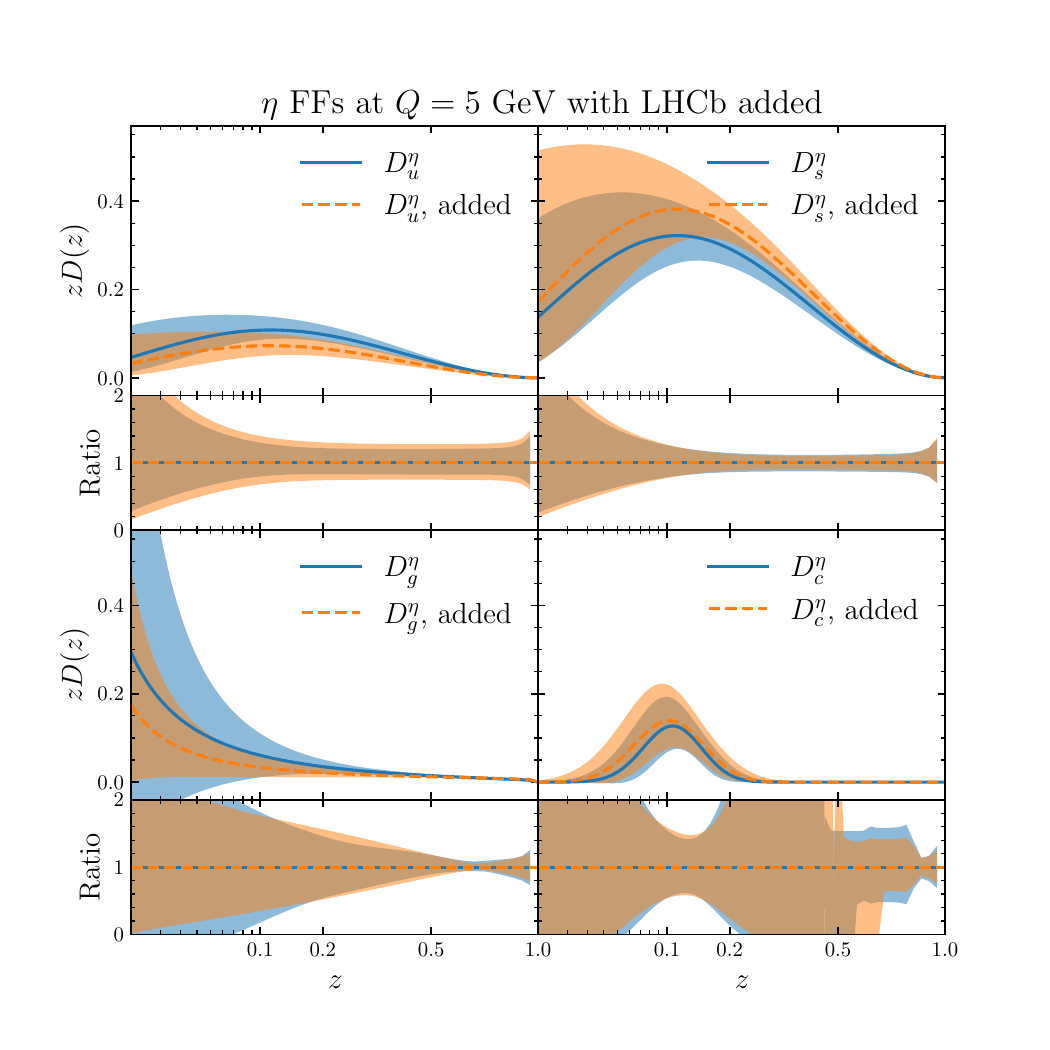}
	\caption{
		Comparison of our nominal $\eta$ FFs (blue) and the alternative fit (orange) with LHCb data \cite{LHCb:2023iyw} added, at $Q = 5~\mathrm{GeV}$.
		The setup of the plot follows the same as \cref{f.FFs-K0S}.
	}
	\label{f.etaff_with_lhcb}
\end{figure}

Furthermore, in \cref{f.etaff_with_lhcb}, we compare the $\eta$ FFs obtained from our baseline fit with those derived from the alternative fit that includes the LHCb data.
The inclusion of the LHCb data only resulted in minor impact on the FFs.
Specifically, the $u$-quark and gluon fragmentation functions are slightly suppressed compared to the baseline fit, while the $s$-quark and $c$-quark distributions are moderately enhanced.
Importantly, all variations remain within the original uncertainty bands.
A notable observation is the reduction in uncertainties for the gluon distribution in the low $z$ region, suggesting that the LHCb measurements provide valuable constraints on gluon FFs.

In summary, the addition of the LHCb data yields results that are consistent with our baseline fit.
Although the overall improvements are modest, the enhanced constraints on the gluon fragmentation function, particularly in the low $z$ region, underscore the importance of LHCb measurements in probing gluon dynamics.

\clearpage
\section{LHAPDF6 grid}
\label{s.LHAPDF-grid}

In this section, we provide a brief summary of the files containing our fragmentation functions.
Interpolation tables of the NPC23 neutral hadron FFs follow the \texttt{LHAGRID1} format \cite{Buckley_2015}, which is the same format employed in PDF grid files.
Access to the FFs is facilitated through the unified interface of \texttt{LHAPDF6}, available via Fortran, C++ and Python code.
To use these sets, one can extract them into the LHAPDF data directory, accessible through the \texttt{lhapdf --datadir} command or direct download directly from the website~\footnote{\url{https://lhapdf.hepforge.org/pdfsets.html}}.
The FFs fitted (or constructed via isospin symmetry) in this work correspond to partons fragmenting into various hadrons including $K^0_S$, $\Lambda$, $\eta$ and $\pi^0$.
They are named as \verb|NPC23_KA0S_nlo|, \verb|NPC23_Lambda_nlo|, \verb|NPC23_Eta_nlo| and \verb|NPC23_PI0_nlo|, respectively.
In addition, we provide the $K_S^0$ FFs constructed from $K^{\pm}$ FFs via isospin symmetry, as given in \cref{e.K0S-construction-u-and-d,e.K0S-construction-unfavored}, under the name \verb|NPC23_KA0S_nlo_iso|. %% 127 subsets

The FF sets \verb|NPC23_KA0S_nlo|, \verb|NPC23_Lambda_nlo|, \verb|NPC23_Eta_nlo|, \verb|NPC23_PI0_nlo| and \verb|NPC23_KA0S_nlo_iso| consist of 41, 37, 25, 127 and 127 subsets, respectively.
The zeroth subset is the central set, while the remaining subsets correspond to the Hessian error sets.
For the estimation of FFs uncertainties of any observable $X$, the following formula for asymmetric errors is employed \cite{Nadolsky:2001yg}:
\begin{equation}
	\label{e.Hessian-error}
	\begin{aligned}
		\delta^+ X
		 & =
		\sqrt{\sum_{i=1}^{N_d} \bqty{\max \pqty{X_{2 i-1}-X_0, X_{2 i}-X_0, 0}}^2}
		,    \\
		\delta^- X
		 & =
		\sqrt{\sum_{i=1}^{N_d} \bqty{\max \pqty{X_0-X_{2 i-1}, X_0-X_{2 i}, 0}}^2}
		.
	\end{aligned}
\end{equation}
Here, $X_0$ represents the prediction obtained with the central set of FFs, and $X_{2i-1} \pqty{X_{2i}}$ represents the predictions obtained with the error set for the $i$-th eigenvector in positive (negative) direction.
When dealing with observables involving FFs of different hadrons, it is crucial to consider their correlations.
In particular, the $\pi^0$ FFs and $K_S^0$ FFs constructed via isospin symmetry, namely \verb|NPC23_PI0_nlo| and \verb|NPC23_KA0S_nlo_iso|, are correlated with each other, as well as with the charged hadron FFs fitted in the previous NPC work \cite{Gao:2024dbv}.
However, the $K^0_S$, $\Lambda$ and $\eta$ FFs are not inter-correlated, nor are they correlated with the other FFs from the current and previous NPC analyses.
As an example, when calculating ratios between $\pi^0$ FFs and $K^{\pm}$ FFs, it is essential to first compute the ratios using consistent error sets for $\pi^0$ and $K^{\pm}$.
Afterward, the uncertainties can be estimated by applying \cref{e.Hessian-error}. This ensures that the correlations between the $\pi^0$ and $K^{\pm}$ FFs are properly accounted for in the uncertainty estimation.

Our FFs grids are provided as 2-dimensional interpolation tables evaluated at different momentum fractions $z$ and different fragmentation scales $Q$.
The $z$ values range from 0.003 to 1 with 99 nodes, while $Q$ spans from 4 to 4000 GeV with 32 nodes in total.
Our grids also include numerical values of $\alpha_S$ with $\alpha_S \pqty{m_Z} = 0.118$ and two-loop running.
The number of active quark flavors is fixed to 5.
The interpolator is configured to use the default log-cubic method, and the FFs are frozen when exceeding the specified ranges of $z$ or $Q$ mentioned above.

\bibliography{refs}

\begin{thebibliography}{152}
\expandafter\ifx\csname natexlab\endcsname\relax\def\natexlab#1{#1}\fi
\expandafter\ifx\csname bibnamefont\endcsname\relax
  \def\bibnamefont#1{#1}\fi
\expandafter\ifx\csname bibfnamefont\endcsname\relax
  \def\bibfnamefont#1{#1}\fi
\expandafter\ifx\csname citenamefont\endcsname\relax
  \def\citenamefont#1{#1}\fi
\expandafter\ifx\csname url\endcsname\relax
  \def\url#1{\texttt{#1}}\fi
\expandafter\ifx\csname urlprefix\endcsname\relax\def\urlprefix{URL }\fi
\providecommand{\bibinfo}[2]{#2}
\providecommand{\eprint}[2][]{\url{#2}}

\bibitem[{\citenamefont{Berman et~al.}(1971)\citenamefont{Berman, Bjorken, and
  Kogut}}]{Berman:1971xz}
\bibinfo{author}{\bibfnamefont{S.~M.} \bibnamefont{Berman}},
  \bibinfo{author}{\bibfnamefont{J.~D.} \bibnamefont{Bjorken}},
  \bibnamefont{and} \bibinfo{author}{\bibfnamefont{J.~B.} \bibnamefont{Kogut}},
  \bibinfo{journal}{Phys. Rev. D} \textbf{\bibinfo{volume}{4}},
  \bibinfo{pages}{3388} (\bibinfo{year}{1971}).

\bibitem[{\citenamefont{Field and Feynman}(1978)}]{Field:1977fa}
\bibinfo{author}{\bibfnamefont{R.~D.} \bibnamefont{Field}} \bibnamefont{and}
  \bibinfo{author}{\bibfnamefont{R.~P.} \bibnamefont{Feynman}},
  \bibinfo{journal}{Nucl. Phys. B} \textbf{\bibinfo{volume}{136}},
  \bibinfo{pages}{1} (\bibinfo{year}{1978}).

\bibitem[{\citenamefont{Feynman et~al.}(1978)\citenamefont{Feynman, Field, and
  Fox}}]{Feynman:1978dt}
\bibinfo{author}{\bibfnamefont{R.~P.} \bibnamefont{Feynman}},
  \bibinfo{author}{\bibfnamefont{R.~D.} \bibnamefont{Field}}, \bibnamefont{and}
  \bibinfo{author}{\bibfnamefont{G.~C.} \bibnamefont{Fox}},
  \bibinfo{journal}{Phys. Rev. D} \textbf{\bibinfo{volume}{18}},
  \bibinfo{pages}{3320} (\bibinfo{year}{1978}).

\bibitem[{\citenamefont{Metz and Vossen}(2016)}]{Metz:2016swz}
\bibinfo{author}{\bibfnamefont{A.}~\bibnamefont{Metz}} \bibnamefont{and}
  \bibinfo{author}{\bibfnamefont{A.}~\bibnamefont{Vossen}},
  \bibinfo{journal}{Prog. Part. Nucl. Phys.} \textbf{\bibinfo{volume}{91}},
  \bibinfo{pages}{136} (\bibinfo{year}{2016}), \eprint{1607.02521}.

\bibitem[{\citenamefont{Collins et~al.}(1989)\citenamefont{Collins, Soper, and
  Sterman}}]{Collins:1989gx}
\bibinfo{author}{\bibfnamefont{J.~C.} \bibnamefont{Collins}},
  \bibinfo{author}{\bibfnamefont{D.~E.} \bibnamefont{Soper}}, \bibnamefont{and}
  \bibinfo{author}{\bibfnamefont{G.~F.} \bibnamefont{Sterman}},
  \bibinfo{journal}{Adv. Ser. Direct. High Energy Phys.}
  \textbf{\bibinfo{volume}{5}}, \bibinfo{pages}{1} (\bibinfo{year}{1989}),
  \eprint{hep-ph/0409313}.

\bibitem[{\citenamefont{Collins}(2023)}]{Collins:2011zzd}
\bibinfo{author}{\bibfnamefont{J.}~\bibnamefont{Collins}},
  \emph{\bibinfo{title}{{Foundations of Perturbative QCD}}},
  vol.~\bibinfo{volume}{32} (\bibinfo{publisher}{Cambridge University Press},
  \bibinfo{year}{2023}), ISBN \bibinfo{isbn}{978-1-009-40184-5,
  978-1-009-40183-8, 978-1-009-40182-1}.

\bibitem[{\citenamefont{Li et~al.}(2020)\citenamefont{Li, Vitev, and
  Zhu}}]{Li:2020bub}
\bibinfo{author}{\bibfnamefont{H.~T.} \bibnamefont{Li}},
  \bibinfo{author}{\bibfnamefont{I.}~\bibnamefont{Vitev}}, \bibnamefont{and}
  \bibinfo{author}{\bibfnamefont{Y.~J.} \bibnamefont{Zhu}},
  \bibinfo{journal}{JHEP} \textbf{\bibinfo{volume}{11}}, \bibinfo{pages}{051}
  (\bibinfo{year}{2020}), \eprint{2006.02437}.

\bibitem[{\citenamefont{Kang et~al.}(2024{\natexlab{a}})\citenamefont{Kang,
  Penttala, Zhao, and Zhou}}]{Kang:2023oqj}
\bibinfo{author}{\bibfnamefont{Z.-B.} \bibnamefont{Kang}},
  \bibinfo{author}{\bibfnamefont{J.}~\bibnamefont{Penttala}},
  \bibinfo{author}{\bibfnamefont{F.}~\bibnamefont{Zhao}}, \bibnamefont{and}
  \bibinfo{author}{\bibfnamefont{Y.}~\bibnamefont{Zhou}},
  \bibinfo{journal}{Phys. Rev. D} \textbf{\bibinfo{volume}{109}},
  \bibinfo{pages}{094012} (\bibinfo{year}{2024}{\natexlab{a}}),
  \eprint{2311.17142}.

\bibitem[{\citenamefont{Kang et~al.}(2024{\natexlab{b}})\citenamefont{Kang,
  Lee, Penttala, Zhao, and Zhou}}]{Kang:2024otf}
\bibinfo{author}{\bibfnamefont{Z.-B.} \bibnamefont{Kang}},
  \bibinfo{author}{\bibfnamefont{S.}~\bibnamefont{Lee}},
  \bibinfo{author}{\bibfnamefont{J.}~\bibnamefont{Penttala}},
  \bibinfo{author}{\bibfnamefont{F.}~\bibnamefont{Zhao}}, \bibnamefont{and}
  \bibinfo{author}{\bibfnamefont{Y.}~\bibnamefont{Zhou}}
  (\bibinfo{year}{2024}{\natexlab{b}}), \eprint{2410.02747}.

\bibitem[{\citenamefont{Sassot et~al.}(2010)\citenamefont{Sassot, Stratmann,
  and Zurita}}]{Sassot:2009sh}
\bibinfo{author}{\bibfnamefont{R.}~\bibnamefont{Sassot}},
  \bibinfo{author}{\bibfnamefont{M.}~\bibnamefont{Stratmann}},
  \bibnamefont{and} \bibinfo{author}{\bibfnamefont{P.}~\bibnamefont{Zurita}},
  \bibinfo{journal}{Phys. Rev. D} \textbf{\bibinfo{volume}{81}},
  \bibinfo{pages}{054001} (\bibinfo{year}{2010}), \eprint{0912.1311}.

\bibitem[{\citenamefont{Zurita}(2021)}]{Zurita:2021kli}
\bibinfo{author}{\bibfnamefont{P.}~\bibnamefont{Zurita}}
  (\bibinfo{year}{2021}), \eprint{2101.01088}.

\bibitem[{\citenamefont{Soleymaninia
  et~al.}(2024{\natexlab{a}})\citenamefont{Soleymaninia, Hashamipour, Khanpour,
  Shoeibi, and Mohamaditabar}}]{Soleymaninia:2023dds}
\bibinfo{author}{\bibfnamefont{M.}~\bibnamefont{Soleymaninia}},
  \bibinfo{author}{\bibfnamefont{H.}~\bibnamefont{Hashamipour}},
  \bibinfo{author}{\bibfnamefont{H.}~\bibnamefont{Khanpour}},
  \bibinfo{author}{\bibfnamefont{S.}~\bibnamefont{Shoeibi}}, \bibnamefont{and}
  \bibinfo{author}{\bibfnamefont{A.}~\bibnamefont{Mohamaditabar}},
  \bibinfo{journal}{Eur. Phys. J. Plus} \textbf{\bibinfo{volume}{139}},
  \bibinfo{pages}{794} (\bibinfo{year}{2024}{\natexlab{a}}),
  \eprint{2305.02664}.

\bibitem[{\citenamefont{Doradau et~al.}(2025)\citenamefont{Doradau, Martinez,
  Sassot, and Stratmann}}]{Doradau:2024wli}
\bibinfo{author}{\bibfnamefont{M.}~\bibnamefont{Doradau}},
  \bibinfo{author}{\bibfnamefont{R.~T.} \bibnamefont{Martinez}},
  \bibinfo{author}{\bibfnamefont{R.}~\bibnamefont{Sassot}}, \bibnamefont{and}
  \bibinfo{author}{\bibfnamefont{M.}~\bibnamefont{Stratmann}},
  \bibinfo{journal}{Phys. Rev. D} \textbf{\bibinfo{volume}{111}},
  \bibinfo{pages}{034045} (\bibinfo{year}{2025}), \eprint{2411.08222}.

\bibitem[{\citenamefont{Kang et~al.}(2022)\citenamefont{Kang, Terry, Vossen,
  Xu, and Zhang}}]{Kang:2021kpt}
\bibinfo{author}{\bibfnamefont{Z.-B.} \bibnamefont{Kang}},
  \bibinfo{author}{\bibfnamefont{J.}~\bibnamefont{Terry}},
  \bibinfo{author}{\bibfnamefont{A.}~\bibnamefont{Vossen}},
  \bibinfo{author}{\bibfnamefont{Q.}~\bibnamefont{Xu}}, \bibnamefont{and}
  \bibinfo{author}{\bibfnamefont{J.}~\bibnamefont{Zhang}},
  \bibinfo{journal}{Phys. Rev. D} \textbf{\bibinfo{volume}{105}},
  \bibinfo{pages}{094033} (\bibinfo{year}{2022}), \eprint{2108.05383}.

\bibitem[{\citenamefont{Li et~al.}(2024{\natexlab{a}})\citenamefont{Li, Xing,
  and Zhang}}]{Li:2024nod}
\bibinfo{author}{\bibfnamefont{T.}~\bibnamefont{Li}},
  \bibinfo{author}{\bibfnamefont{H.}~\bibnamefont{Xing}}, \bibnamefont{and}
  \bibinfo{author}{\bibfnamefont{D.-B.} \bibnamefont{Zhang}}
  (\bibinfo{year}{2024}{\natexlab{a}}), \eprint{2406.05683}.

\bibitem[{\citenamefont{Grieninger and Zahed}(2024)}]{Grieninger:2024axp}
\bibinfo{author}{\bibfnamefont{S.}~\bibnamefont{Grieninger}} \bibnamefont{and}
  \bibinfo{author}{\bibfnamefont{I.}~\bibnamefont{Zahed}}
  (\bibinfo{year}{2024}), \eprint{2406.01891}.

\bibitem[{\citenamefont{Rijken and van Neerven}(1996)}]{Rijken:1996vr}
\bibinfo{author}{\bibfnamefont{P.~J.} \bibnamefont{Rijken}} \bibnamefont{and}
  \bibinfo{author}{\bibfnamefont{W.~L.} \bibnamefont{van Neerven}},
  \bibinfo{journal}{Phys. Lett. B} \textbf{\bibinfo{volume}{386}},
  \bibinfo{pages}{422} (\bibinfo{year}{1996}), \eprint{hep-ph/9604436}.

\bibitem[{\citenamefont{Rijken and van Neerven}(1997)}]{Rijken:1996ns}
\bibinfo{author}{\bibfnamefont{P.~J.} \bibnamefont{Rijken}} \bibnamefont{and}
  \bibinfo{author}{\bibfnamefont{W.~L.} \bibnamefont{van Neerven}},
  \bibinfo{journal}{Nucl. Phys. B} \textbf{\bibinfo{volume}{487}},
  \bibinfo{pages}{233} (\bibinfo{year}{1997}), \eprint{hep-ph/9609377}.

\bibitem[{\citenamefont{Mitov and Moch}(2006)}]{Mitov:2006wy}
\bibinfo{author}{\bibfnamefont{A.}~\bibnamefont{Mitov}} \bibnamefont{and}
  \bibinfo{author}{\bibfnamefont{S.-O.} \bibnamefont{Moch}},
  \bibinfo{journal}{Nucl. Phys. B} \textbf{\bibinfo{volume}{751}},
  \bibinfo{pages}{18} (\bibinfo{year}{2006}), \eprint{hep-ph/0604160}.

\bibitem[{\citenamefont{Blumlein and Ravindran}(2006)}]{Blumlein:2006rr}
\bibinfo{author}{\bibfnamefont{J.}~\bibnamefont{Blumlein}} \bibnamefont{and}
  \bibinfo{author}{\bibfnamefont{V.}~\bibnamefont{Ravindran}},
  \bibinfo{journal}{Nucl. Phys. B} \textbf{\bibinfo{volume}{749}},
  \bibinfo{pages}{1} (\bibinfo{year}{2006}), \eprint{hep-ph/0604019}.

\bibitem[{\citenamefont{Cacciari and Catani}(2001)}]{Cacciari:2001cw}
\bibinfo{author}{\bibfnamefont{M.}~\bibnamefont{Cacciari}} \bibnamefont{and}
  \bibinfo{author}{\bibfnamefont{S.}~\bibnamefont{Catani}},
  \bibinfo{journal}{Nucl. Phys. B} \textbf{\bibinfo{volume}{617}},
  \bibinfo{pages}{253} (\bibinfo{year}{2001}), \eprint{hep-ph/0107138}.

\bibitem[{\citenamefont{Moch and Vogt}(2009)}]{Moch:2009my}
\bibinfo{author}{\bibfnamefont{S.}~\bibnamefont{Moch}} \bibnamefont{and}
  \bibinfo{author}{\bibfnamefont{A.}~\bibnamefont{Vogt}},
  \bibinfo{journal}{Phys. Lett. B} \textbf{\bibinfo{volume}{680}},
  \bibinfo{pages}{239} (\bibinfo{year}{2009}), \eprint{0908.2746}.

\bibitem[{\citenamefont{Xu and Zhu}(2024)}]{Xu:2024rbt}
\bibinfo{author}{\bibfnamefont{Z.}~\bibnamefont{Xu}} \bibnamefont{and}
  \bibinfo{author}{\bibfnamefont{H.~X.} \bibnamefont{Zhu}}
  (\bibinfo{year}{2024}), \eprint{2411.11595}.

\bibitem[{\citenamefont{Altarelli et~al.}(1979)\citenamefont{Altarelli, Ellis,
  Martinelli, and Pi}}]{Altarelli:1979kv}
\bibinfo{author}{\bibfnamefont{G.}~\bibnamefont{Altarelli}},
  \bibinfo{author}{\bibfnamefont{R.~K.} \bibnamefont{Ellis}},
  \bibinfo{author}{\bibfnamefont{G.}~\bibnamefont{Martinelli}},
  \bibnamefont{and} \bibinfo{author}{\bibfnamefont{S.-Y.} \bibnamefont{Pi}},
  \bibinfo{journal}{Nucl. Phys. B} \textbf{\bibinfo{volume}{160}},
  \bibinfo{pages}{301} (\bibinfo{year}{1979}).

\bibitem[{\citenamefont{Nason and Webber}(1994)}]{Nason:1993xx}
\bibinfo{author}{\bibfnamefont{P.}~\bibnamefont{Nason}} \bibnamefont{and}
  \bibinfo{author}{\bibfnamefont{B.~R.} \bibnamefont{Webber}},
  \bibinfo{journal}{Nucl. Phys. B} \textbf{\bibinfo{volume}{421}},
  \bibinfo{pages}{473} (\bibinfo{year}{1994}), \bibinfo{note}{[Erratum:
  Nucl.Phys.B 480, 755 (1996)]}.

\bibitem[{\citenamefont{Furmanski and Petronzio}(1982)}]{Furmanski:1981cw}
\bibinfo{author}{\bibfnamefont{W.}~\bibnamefont{Furmanski}} \bibnamefont{and}
  \bibinfo{author}{\bibfnamefont{R.}~\bibnamefont{Petronzio}},
  \bibinfo{journal}{Z. Phys. C} \textbf{\bibinfo{volume}{11}},
  \bibinfo{pages}{293} (\bibinfo{year}{1982}).

\bibitem[{\citenamefont{Graudenz}(1994)}]{Graudenz:1994dq}
\bibinfo{author}{\bibfnamefont{D.}~\bibnamefont{Graudenz}},
  \bibinfo{journal}{Nucl. Phys. B} \textbf{\bibinfo{volume}{432}},
  \bibinfo{pages}{351} (\bibinfo{year}{1994}), \eprint{hep-ph/9406274}.

\bibitem[{\citenamefont{de~Florian et~al.}(1998)\citenamefont{de~Florian,
  Stratmann, and Vogelsang}}]{deFlorian:1997zj}
\bibinfo{author}{\bibfnamefont{D.}~\bibnamefont{de~Florian}},
  \bibinfo{author}{\bibfnamefont{M.}~\bibnamefont{Stratmann}},
  \bibnamefont{and}
  \bibinfo{author}{\bibfnamefont{W.}~\bibnamefont{Vogelsang}},
  \bibinfo{journal}{Phys. Rev. D} \textbf{\bibinfo{volume}{57}},
  \bibinfo{pages}{5811} (\bibinfo{year}{1998}), \eprint{hep-ph/9711387}.

\bibitem[{\citenamefont{de~Florian and
  Rotstein~Habarnau}(2013)}]{deFlorian:2012wk}
\bibinfo{author}{\bibfnamefont{D.}~\bibnamefont{de~Florian}} \bibnamefont{and}
  \bibinfo{author}{\bibfnamefont{Y.}~\bibnamefont{Rotstein~Habarnau}},
  \bibinfo{journal}{Eur. Phys. J. C} \textbf{\bibinfo{volume}{73}},
  \bibinfo{pages}{2356} (\bibinfo{year}{2013}), \eprint{1210.7203}.

\bibitem[{\citenamefont{Goyal et~al.}(2024)\citenamefont{Goyal, Moch, Pathak,
  Rana, and Ravindran}}]{Goyal:2023zdi}
\bibinfo{author}{\bibfnamefont{S.}~\bibnamefont{Goyal}},
  \bibinfo{author}{\bibfnamefont{S.-O.} \bibnamefont{Moch}},
  \bibinfo{author}{\bibfnamefont{V.}~\bibnamefont{Pathak}},
  \bibinfo{author}{\bibfnamefont{N.}~\bibnamefont{Rana}}, \bibnamefont{and}
  \bibinfo{author}{\bibfnamefont{V.}~\bibnamefont{Ravindran}},
  \bibinfo{journal}{Phys. Rev. Lett.} \textbf{\bibinfo{volume}{132}},
  \bibinfo{pages}{251902} (\bibinfo{year}{2024}), \eprint{2312.17711}.

\bibitem[{\citenamefont{Bonino et~al.}(2024{\natexlab{a}})\citenamefont{Bonino,
  Gehrmann, and Stagnitto}}]{Bonino:2024qbh}
\bibinfo{author}{\bibfnamefont{L.}~\bibnamefont{Bonino}},
  \bibinfo{author}{\bibfnamefont{T.}~\bibnamefont{Gehrmann}}, \bibnamefont{and}
  \bibinfo{author}{\bibfnamefont{G.}~\bibnamefont{Stagnitto}},
  \bibinfo{journal}{Phys. Rev. Lett.} \textbf{\bibinfo{volume}{132}},
  \bibinfo{pages}{251901} (\bibinfo{year}{2024}{\natexlab{a}}),
  \eprint{2401.16281}.

\bibitem[{\citenamefont{Abele et~al.}(2021)\citenamefont{Abele, de~Florian, and
  Vogelsang}}]{Abele:2021nyo}
\bibinfo{author}{\bibfnamefont{M.}~\bibnamefont{Abele}},
  \bibinfo{author}{\bibfnamefont{D.}~\bibnamefont{de~Florian}},
  \bibnamefont{and}
  \bibinfo{author}{\bibfnamefont{W.}~\bibnamefont{Vogelsang}},
  \bibinfo{journal}{Phys. Rev. D} \textbf{\bibinfo{volume}{104}},
  \bibinfo{pages}{094046} (\bibinfo{year}{2021}), \eprint{2109.00847}.

\bibitem[{\citenamefont{Abele et~al.}(2022)\citenamefont{Abele, de~Florian, and
  Vogelsang}}]{Abele:2022wuy}
\bibinfo{author}{\bibfnamefont{M.}~\bibnamefont{Abele}},
  \bibinfo{author}{\bibfnamefont{D.}~\bibnamefont{de~Florian}},
  \bibnamefont{and}
  \bibinfo{author}{\bibfnamefont{W.}~\bibnamefont{Vogelsang}},
  \bibinfo{journal}{Phys. Rev. D} \textbf{\bibinfo{volume}{106}},
  \bibinfo{pages}{014015} (\bibinfo{year}{2022}), \eprint{2203.07928}.

\bibitem[{\citenamefont{Aversa et~al.}(1989)\citenamefont{Aversa, Chiappetta,
  Greco, and Guillet}}]{Aversa:1988vb}
\bibinfo{author}{\bibfnamefont{F.}~\bibnamefont{Aversa}},
  \bibinfo{author}{\bibfnamefont{P.}~\bibnamefont{Chiappetta}},
  \bibinfo{author}{\bibfnamefont{M.}~\bibnamefont{Greco}}, \bibnamefont{and}
  \bibinfo{author}{\bibfnamefont{J.~P.} \bibnamefont{Guillet}},
  \bibinfo{journal}{Nucl. Phys. B} \textbf{\bibinfo{volume}{327}},
  \bibinfo{pages}{105} (\bibinfo{year}{1989}).

\bibitem[{\citenamefont{de~Florian}(2003)}]{deFlorian:2002az}
\bibinfo{author}{\bibfnamefont{D.}~\bibnamefont{de~Florian}},
  \bibinfo{journal}{Phys. Rev. D} \textbf{\bibinfo{volume}{67}},
  \bibinfo{pages}{054004} (\bibinfo{year}{2003}), \eprint{hep-ph/0210442}.

\bibitem[{\citenamefont{Jager et~al.}(2003)\citenamefont{Jager, Schafer,
  Stratmann, and Vogelsang}}]{Jager:2002xm}
\bibinfo{author}{\bibfnamefont{B.}~\bibnamefont{Jager}},
  \bibinfo{author}{\bibfnamefont{A.}~\bibnamefont{Schafer}},
  \bibinfo{author}{\bibfnamefont{M.}~\bibnamefont{Stratmann}},
  \bibnamefont{and}
  \bibinfo{author}{\bibfnamefont{W.}~\bibnamefont{Vogelsang}},
  \bibinfo{journal}{Phys. Rev. D} \textbf{\bibinfo{volume}{67}},
  \bibinfo{pages}{054005} (\bibinfo{year}{2003}), \eprint{hep-ph/0211007}.

\bibitem[{\citenamefont{Czakon et~al.}(2025)\citenamefont{Czakon, Generet,
  Mitov, and Poncelet}}]{Czakon:2025yti}
\bibinfo{author}{\bibfnamefont{M.}~\bibnamefont{Czakon}},
  \bibinfo{author}{\bibfnamefont{T.}~\bibnamefont{Generet}},
  \bibinfo{author}{\bibfnamefont{A.}~\bibnamefont{Mitov}}, \bibnamefont{and}
  \bibinfo{author}{\bibfnamefont{R.}~\bibnamefont{Poncelet}}
  (\bibinfo{year}{2025}), \eprint{2503.11489}.

\bibitem[{\citenamefont{Kang et~al.}(2016)\citenamefont{Kang, Ringer, and
  Vitev}}]{Kang:2016ehg}
\bibinfo{author}{\bibfnamefont{Z.-B.} \bibnamefont{Kang}},
  \bibinfo{author}{\bibfnamefont{F.}~\bibnamefont{Ringer}}, \bibnamefont{and}
  \bibinfo{author}{\bibfnamefont{I.}~\bibnamefont{Vitev}},
  \bibinfo{journal}{JHEP} \textbf{\bibinfo{volume}{11}}, \bibinfo{pages}{155}
  (\bibinfo{year}{2016}), \eprint{1606.07063}.

\bibitem[{\citenamefont{Kaufmann et~al.}(2020)\citenamefont{Kaufmann, Liu,
  Mukherjee, Ringer, and Vogelsang}}]{Kaufmann:2019ksh}
\bibinfo{author}{\bibfnamefont{T.}~\bibnamefont{Kaufmann}},
  \bibinfo{author}{\bibfnamefont{X.}~\bibnamefont{Liu}},
  \bibinfo{author}{\bibfnamefont{A.}~\bibnamefont{Mukherjee}},
  \bibinfo{author}{\bibfnamefont{F.}~\bibnamefont{Ringer}}, \bibnamefont{and}
  \bibinfo{author}{\bibfnamefont{W.}~\bibnamefont{Vogelsang}},
  \bibinfo{journal}{JHEP} \textbf{\bibinfo{volume}{02}}, \bibinfo{pages}{040}
  (\bibinfo{year}{2020}), \eprint{1910.11746}.

\bibitem[{\citenamefont{Kang et~al.}(2019)\citenamefont{Kang, Lee, Terry, and
  Xing}}]{Kang:2019ahe}
\bibinfo{author}{\bibfnamefont{Z.-B.} \bibnamefont{Kang}},
  \bibinfo{author}{\bibfnamefont{K.}~\bibnamefont{Lee}},
  \bibinfo{author}{\bibfnamefont{J.}~\bibnamefont{Terry}}, \bibnamefont{and}
  \bibinfo{author}{\bibfnamefont{H.}~\bibnamefont{Xing}},
  \bibinfo{journal}{Phys. Lett. B} \textbf{\bibinfo{volume}{798}},
  \bibinfo{pages}{134978} (\bibinfo{year}{2019}), \eprint{1906.07187}.

\bibitem[{\citenamefont{Wang et~al.}(2021)\citenamefont{Wang, Kang, Xing, and
  Zhang}}]{Wang:2020kar}
\bibinfo{author}{\bibfnamefont{L.}~\bibnamefont{Wang}},
  \bibinfo{author}{\bibfnamefont{Z.-B.} \bibnamefont{Kang}},
  \bibinfo{author}{\bibfnamefont{H.}~\bibnamefont{Xing}}, \bibnamefont{and}
  \bibinfo{author}{\bibfnamefont{B.-W.} \bibnamefont{Zhang}},
  \bibinfo{journal}{Phys. Rev. D} \textbf{\bibinfo{volume}{103}},
  \bibinfo{pages}{054043} (\bibinfo{year}{2021}), \eprint{2003.03796}.

\bibitem[{\citenamefont{Anderle et~al.}(2017)\citenamefont{Anderle, Kaufmann,
  Stratmann, Ringer, and Vitev}}]{Anderle:2017cgl}
\bibinfo{author}{\bibfnamefont{D.~P.} \bibnamefont{Anderle}},
  \bibinfo{author}{\bibfnamefont{T.}~\bibnamefont{Kaufmann}},
  \bibinfo{author}{\bibfnamefont{M.}~\bibnamefont{Stratmann}},
  \bibinfo{author}{\bibfnamefont{F.}~\bibnamefont{Ringer}}, \bibnamefont{and}
  \bibinfo{author}{\bibfnamefont{I.}~\bibnamefont{Vitev}},
  \bibinfo{journal}{Phys. Rev. D} \textbf{\bibinfo{volume}{96}},
  \bibinfo{pages}{034028} (\bibinfo{year}{2017}), \eprint{1706.09857}.

\bibitem[{\citenamefont{Gao et~al.}(2024{\natexlab{a}})\citenamefont{Gao, Liu,
  Shen, Xing, and Zhao}}]{Gao:2024nkz}
\bibinfo{author}{\bibfnamefont{J.}~\bibnamefont{Gao}},
  \bibinfo{author}{\bibfnamefont{C.}~\bibnamefont{Liu}},
  \bibinfo{author}{\bibfnamefont{X.}~\bibnamefont{Shen}},
  \bibinfo{author}{\bibfnamefont{H.}~\bibnamefont{Xing}}, \bibnamefont{and}
  \bibinfo{author}{\bibfnamefont{Y.}~\bibnamefont{Zhao}},
  \bibinfo{journal}{Phys. Rev. Lett.} \textbf{\bibinfo{volume}{132}},
  \bibinfo{pages}{261903} (\bibinfo{year}{2024}{\natexlab{a}}),
  \eprint{2401.02781}.

\bibitem[{\citenamefont{Gao et~al.}(2024{\natexlab{b}})\citenamefont{Gao, Liu,
  Shen, Xing, and Zhao}}]{Gao:2024dbv}
\bibinfo{author}{\bibfnamefont{J.}~\bibnamefont{Gao}},
  \bibinfo{author}{\bibfnamefont{C.}~\bibnamefont{Liu}},
  \bibinfo{author}{\bibfnamefont{X.}~\bibnamefont{Shen}},
  \bibinfo{author}{\bibfnamefont{H.}~\bibnamefont{Xing}}, \bibnamefont{and}
  \bibinfo{author}{\bibfnamefont{Y.}~\bibnamefont{Zhao}}
  (\bibinfo{year}{2024}{\natexlab{b}}), \eprint{2407.04422}.

\bibitem[{\citenamefont{Arleo et~al.}(2014)\citenamefont{Arleo, Fontannaz,
  Guillet, and Nguyen}}]{Arleo:2013tya}
\bibinfo{author}{\bibfnamefont{F.}~\bibnamefont{Arleo}},
  \bibinfo{author}{\bibfnamefont{M.}~\bibnamefont{Fontannaz}},
  \bibinfo{author}{\bibfnamefont{J.-P.} \bibnamefont{Guillet}},
  \bibnamefont{and} \bibinfo{author}{\bibfnamefont{C.~L.}
  \bibnamefont{Nguyen}}, \bibinfo{journal}{JHEP} \textbf{\bibinfo{volume}{04}},
  \bibinfo{pages}{147} (\bibinfo{year}{2014}), \eprint{1311.7356}.

\bibitem[{\citenamefont{Liu et~al.}(2023)\citenamefont{Liu, Shen, Zhou, and
  Gao}}]{Liu:2023fsq}
\bibinfo{author}{\bibfnamefont{C.}~\bibnamefont{Liu}},
  \bibinfo{author}{\bibfnamefont{X.}~\bibnamefont{Shen}},
  \bibinfo{author}{\bibfnamefont{B.}~\bibnamefont{Zhou}}, \bibnamefont{and}
  \bibinfo{author}{\bibfnamefont{J.}~\bibnamefont{Gao}},
  \bibinfo{journal}{JHEP} \textbf{\bibinfo{volume}{09}}, \bibinfo{pages}{108}
  (\bibinfo{year}{2023}), \eprint{2305.14620}.

\bibitem[{\citenamefont{Zidi et~al.}(2024)\citenamefont{Zidi, Guillet,
  Schienbein, and Zaraket}}]{Zidi:2024lid}
\bibinfo{author}{\bibfnamefont{M.~S.} \bibnamefont{Zidi}},
  \bibinfo{author}{\bibfnamefont{J.~P.} \bibnamefont{Guillet}},
  \bibinfo{author}{\bibfnamefont{I.}~\bibnamefont{Schienbein}},
  \bibnamefont{and} \bibinfo{author}{\bibfnamefont{H.}~\bibnamefont{Zaraket}},
  \bibinfo{journal}{Eur. Phys. J. C} \textbf{\bibinfo{volume}{84}},
  \bibinfo{pages}{611} (\bibinfo{year}{2024}), \eprint{2403.14574}.

\bibitem[{\citenamefont{Caletti et~al.}(2024)\citenamefont{Caletti,
  Gehrmann-De~Ridder, Huss, Garcia, and Stagnitto}}]{Caletti:2024xaw}
\bibinfo{author}{\bibfnamefont{S.}~\bibnamefont{Caletti}},
  \bibinfo{author}{\bibfnamefont{A.}~\bibnamefont{Gehrmann-De~Ridder}},
  \bibinfo{author}{\bibfnamefont{A.}~\bibnamefont{Huss}},
  \bibinfo{author}{\bibfnamefont{A.~R.} \bibnamefont{Garcia}},
  \bibnamefont{and}
  \bibinfo{author}{\bibfnamefont{G.}~\bibnamefont{Stagnitto}},
  \bibinfo{journal}{JHEP} \textbf{\bibinfo{volume}{10}}, \bibinfo{pages}{027}
  (\bibinfo{year}{2024}), \eprint{2405.17540}.

\bibitem[{\citenamefont{Bonino et~al.}(2024{\natexlab{b}})\citenamefont{Bonino,
  Gehrmann, Marcoli, Sch\"urmann, and Stagnitto}}]{Bonino:2024adk}
\bibinfo{author}{\bibfnamefont{L.}~\bibnamefont{Bonino}},
  \bibinfo{author}{\bibfnamefont{T.}~\bibnamefont{Gehrmann}},
  \bibinfo{author}{\bibfnamefont{M.}~\bibnamefont{Marcoli}},
  \bibinfo{author}{\bibfnamefont{R.}~\bibnamefont{Sch\"urmann}},
  \bibnamefont{and}
  \bibinfo{author}{\bibfnamefont{G.}~\bibnamefont{Stagnitto}},
  \bibinfo{journal}{JHEP} \textbf{\bibinfo{volume}{08}}, \bibinfo{pages}{073}
  (\bibinfo{year}{2024}{\natexlab{b}}), \eprint{2406.09925}.

\bibitem[{\citenamefont{Mitov et~al.}(2006)\citenamefont{Mitov, Moch, and
  Vogt}}]{Mitov:2006ic}
\bibinfo{author}{\bibfnamefont{A.}~\bibnamefont{Mitov}},
  \bibinfo{author}{\bibfnamefont{S.}~\bibnamefont{Moch}}, \bibnamefont{and}
  \bibinfo{author}{\bibfnamefont{A.}~\bibnamefont{Vogt}},
  \bibinfo{journal}{Phys. Lett. B} \textbf{\bibinfo{volume}{638}},
  \bibinfo{pages}{61} (\bibinfo{year}{2006}), \eprint{hep-ph/0604053}.

\bibitem[{\citenamefont{Moch and Vogt}(2008)}]{Moch:2007tx}
\bibinfo{author}{\bibfnamefont{S.}~\bibnamefont{Moch}} \bibnamefont{and}
  \bibinfo{author}{\bibfnamefont{A.}~\bibnamefont{Vogt}},
  \bibinfo{journal}{Phys. Lett. B} \textbf{\bibinfo{volume}{659}},
  \bibinfo{pages}{290} (\bibinfo{year}{2008}), \eprint{0709.3899}.

\bibitem[{\citenamefont{Almasy et~al.}(2012)\citenamefont{Almasy, Moch, and
  Vogt}}]{Almasy:2011eq}
\bibinfo{author}{\bibfnamefont{A.~A.} \bibnamefont{Almasy}},
  \bibinfo{author}{\bibfnamefont{S.}~\bibnamefont{Moch}}, \bibnamefont{and}
  \bibinfo{author}{\bibfnamefont{A.}~\bibnamefont{Vogt}},
  \bibinfo{journal}{Nucl. Phys. B} \textbf{\bibinfo{volume}{854}},
  \bibinfo{pages}{133} (\bibinfo{year}{2012}), \eprint{1107.2263}.

\bibitem[{\citenamefont{Chen et~al.}(2021{\natexlab{a}})\citenamefont{Chen,
  Yang, Zhu, and Zhu}}]{Chen:2020uvt}
\bibinfo{author}{\bibfnamefont{H.}~\bibnamefont{Chen}},
  \bibinfo{author}{\bibfnamefont{T.-Z.} \bibnamefont{Yang}},
  \bibinfo{author}{\bibfnamefont{H.~X.} \bibnamefont{Zhu}}, \bibnamefont{and}
  \bibinfo{author}{\bibfnamefont{Y.~J.} \bibnamefont{Zhu}},
  \bibinfo{journal}{Chin. Phys. C} \textbf{\bibinfo{volume}{45}},
  \bibinfo{pages}{043101} (\bibinfo{year}{2021}{\natexlab{a}}),
  \eprint{2006.10534}.

\bibitem[{\citenamefont{Ebert et~al.}(2021)\citenamefont{Ebert, Mistlberger,
  and Vita}}]{Ebert:2020qef}
\bibinfo{author}{\bibfnamefont{M.~A.} \bibnamefont{Ebert}},
  \bibinfo{author}{\bibfnamefont{B.}~\bibnamefont{Mistlberger}},
  \bibnamefont{and} \bibinfo{author}{\bibfnamefont{G.}~\bibnamefont{Vita}},
  \bibinfo{journal}{JHEP} \textbf{\bibinfo{volume}{07}}, \bibinfo{pages}{121}
  (\bibinfo{year}{2021}), \eprint{2012.07853}.

\bibitem[{\citenamefont{Luo et~al.}(2021)\citenamefont{Luo, Yang, Zhu, and
  Zhu}}]{Luo:2020epw}
\bibinfo{author}{\bibfnamefont{M.-x.} \bibnamefont{Luo}},
  \bibinfo{author}{\bibfnamefont{T.-Z.} \bibnamefont{Yang}},
  \bibinfo{author}{\bibfnamefont{H.~X.} \bibnamefont{Zhu}}, \bibnamefont{and}
  \bibinfo{author}{\bibfnamefont{Y.~J.} \bibnamefont{Zhu}},
  \bibinfo{journal}{JHEP} \textbf{\bibinfo{volume}{06}}, \bibinfo{pages}{115}
  (\bibinfo{year}{2021}), \eprint{2012.03256}.

\bibitem[{\citenamefont{Binnewies
  et~al.}(1995{\natexlab{a}})\citenamefont{Binnewies, Kniehl, and
  Kramer}}]{Binnewies:1994ju}
\bibinfo{author}{\bibfnamefont{J.}~\bibnamefont{Binnewies}},
  \bibinfo{author}{\bibfnamefont{B.~A.} \bibnamefont{Kniehl}},
  \bibnamefont{and} \bibinfo{author}{\bibfnamefont{G.}~\bibnamefont{Kramer}},
  \bibinfo{journal}{Z. Phys. C} \textbf{\bibinfo{volume}{65}},
  \bibinfo{pages}{471} (\bibinfo{year}{1995}{\natexlab{a}}),
  \eprint{hep-ph/9407347}.

\bibitem[{\citenamefont{Binnewies
  et~al.}(1995{\natexlab{b}})\citenamefont{Binnewies, Kniehl, and
  Kramer}}]{Binnewies:1995pt}
\bibinfo{author}{\bibfnamefont{J.}~\bibnamefont{Binnewies}},
  \bibinfo{author}{\bibfnamefont{B.~A.} \bibnamefont{Kniehl}},
  \bibnamefont{and} \bibinfo{author}{\bibfnamefont{G.}~\bibnamefont{Kramer}},
  \bibinfo{journal}{Phys. Rev. D} \textbf{\bibinfo{volume}{52}},
  \bibinfo{pages}{4947} (\bibinfo{year}{1995}{\natexlab{b}}),
  \eprint{hep-ph/9503464}.

\bibitem[{\citenamefont{Kniehl et~al.}(2000)\citenamefont{Kniehl, Kramer, and
  Potter}}]{Kniehl:2000fe}
\bibinfo{author}{\bibfnamefont{B.~A.} \bibnamefont{Kniehl}},
  \bibinfo{author}{\bibfnamefont{G.}~\bibnamefont{Kramer}}, \bibnamefont{and}
  \bibinfo{author}{\bibfnamefont{B.}~\bibnamefont{Potter}},
  \bibinfo{journal}{Nucl. Phys. B} \textbf{\bibinfo{volume}{582}},
  \bibinfo{pages}{514} (\bibinfo{year}{2000}), \eprint{hep-ph/0010289}.

\bibitem[{\citenamefont{Bourhis et~al.}(2001)\citenamefont{Bourhis, Fontannaz,
  Guillet, and Werlen}}]{Bourhis:2000gs}
\bibinfo{author}{\bibfnamefont{L.}~\bibnamefont{Bourhis}},
  \bibinfo{author}{\bibfnamefont{M.}~\bibnamefont{Fontannaz}},
  \bibinfo{author}{\bibfnamefont{J.~P.} \bibnamefont{Guillet}},
  \bibnamefont{and} \bibinfo{author}{\bibfnamefont{M.}~\bibnamefont{Werlen}},
  \bibinfo{journal}{Eur. Phys. J. C} \textbf{\bibinfo{volume}{19}},
  \bibinfo{pages}{89} (\bibinfo{year}{2001}), \eprint{hep-ph/0009101}.

\bibitem[{\citenamefont{Kretzer}(2000)}]{Kretzer:2000yf}
\bibinfo{author}{\bibfnamefont{S.}~\bibnamefont{Kretzer}},
  \bibinfo{journal}{Phys. Rev. D} \textbf{\bibinfo{volume}{62}},
  \bibinfo{pages}{054001} (\bibinfo{year}{2000}), \eprint{hep-ph/0003177}.

\bibitem[{\citenamefont{Kretzer et~al.}(2001)\citenamefont{Kretzer, Leader, and
  Christova}}]{Kretzer:2001pz}
\bibinfo{author}{\bibfnamefont{S.}~\bibnamefont{Kretzer}},
  \bibinfo{author}{\bibfnamefont{E.}~\bibnamefont{Leader}}, \bibnamefont{and}
  \bibinfo{author}{\bibfnamefont{E.}~\bibnamefont{Christova}},
  \bibinfo{journal}{Eur. Phys. J. C} \textbf{\bibinfo{volume}{22}},
  \bibinfo{pages}{269} (\bibinfo{year}{2001}), \eprint{hep-ph/0108055}.

\bibitem[{\citenamefont{de~Florian et~al.}(2007)\citenamefont{de~Florian,
  Sassot, and Stratmann}}]{deFlorian:2007ekg}
\bibinfo{author}{\bibfnamefont{D.}~\bibnamefont{de~Florian}},
  \bibinfo{author}{\bibfnamefont{R.}~\bibnamefont{Sassot}}, \bibnamefont{and}
  \bibinfo{author}{\bibfnamefont{M.}~\bibnamefont{Stratmann}},
  \bibinfo{journal}{Phys. Rev. D} \textbf{\bibinfo{volume}{76}},
  \bibinfo{pages}{074033} (\bibinfo{year}{2007}), \eprint{0707.1506}.

\bibitem[{\citenamefont{de~Florian et~al.}(2015)\citenamefont{de~Florian,
  Sassot, Epele, Hern\'andez-Pinto, and Stratmann}}]{deFlorian:2014xna}
\bibinfo{author}{\bibfnamefont{D.}~\bibnamefont{de~Florian}},
  \bibinfo{author}{\bibfnamefont{R.}~\bibnamefont{Sassot}},
  \bibinfo{author}{\bibfnamefont{M.}~\bibnamefont{Epele}},
  \bibinfo{author}{\bibfnamefont{R.~J.} \bibnamefont{Hern\'andez-Pinto}},
  \bibnamefont{and}
  \bibinfo{author}{\bibfnamefont{M.}~\bibnamefont{Stratmann}},
  \bibinfo{journal}{Phys. Rev. D} \textbf{\bibinfo{volume}{91}},
  \bibinfo{pages}{014035} (\bibinfo{year}{2015}), \eprint{1410.6027}.

\bibitem[{\citenamefont{de~Florian et~al.}(2017)\citenamefont{de~Florian,
  Epele, Hernandez-Pinto, Sassot, and Stratmann}}]{deFlorian:2017lwf}
\bibinfo{author}{\bibfnamefont{D.}~\bibnamefont{de~Florian}},
  \bibinfo{author}{\bibfnamefont{M.}~\bibnamefont{Epele}},
  \bibinfo{author}{\bibfnamefont{R.~J.} \bibnamefont{Hernandez-Pinto}},
  \bibinfo{author}{\bibfnamefont{R.}~\bibnamefont{Sassot}}, \bibnamefont{and}
  \bibinfo{author}{\bibfnamefont{M.}~\bibnamefont{Stratmann}},
  \bibinfo{journal}{Phys. Rev. D} \textbf{\bibinfo{volume}{95}},
  \bibinfo{pages}{094019} (\bibinfo{year}{2017}), \eprint{1702.06353}.

\bibitem[{\citenamefont{Borsa et~al.}(2022{\natexlab{a}})\citenamefont{Borsa,
  de~Florian, Sassot, and Stratmann}}]{Borsa:2021ran}
\bibinfo{author}{\bibfnamefont{I.}~\bibnamefont{Borsa}},
  \bibinfo{author}{\bibfnamefont{D.}~\bibnamefont{de~Florian}},
  \bibinfo{author}{\bibfnamefont{R.}~\bibnamefont{Sassot}}, \bibnamefont{and}
  \bibinfo{author}{\bibfnamefont{M.}~\bibnamefont{Stratmann}},
  \bibinfo{journal}{Phys. Rev. D} \textbf{\bibinfo{volume}{105}},
  \bibinfo{pages}{L031502} (\bibinfo{year}{2022}{\natexlab{a}}),
  \eprint{2110.14015}.

\bibitem[{\citenamefont{Hirai et~al.}(2007)\citenamefont{Hirai, Kumano, Nagai,
  and Sudoh}}]{Hirai:2007cx}
\bibinfo{author}{\bibfnamefont{M.}~\bibnamefont{Hirai}},
  \bibinfo{author}{\bibfnamefont{S.}~\bibnamefont{Kumano}},
  \bibinfo{author}{\bibfnamefont{T.~H.} \bibnamefont{Nagai}}, \bibnamefont{and}
  \bibinfo{author}{\bibfnamefont{K.}~\bibnamefont{Sudoh}},
  \bibinfo{journal}{Phys. Rev. D} \textbf{\bibinfo{volume}{75}},
  \bibinfo{pages}{094009} (\bibinfo{year}{2007}), \eprint{hep-ph/0702250}.

\bibitem[{\citenamefont{Albino et~al.}(2008)\citenamefont{Albino, Kniehl, and
  Kramer}}]{Albino:2008fy}
\bibinfo{author}{\bibfnamefont{S.}~\bibnamefont{Albino}},
  \bibinfo{author}{\bibfnamefont{B.~A.} \bibnamefont{Kniehl}},
  \bibnamefont{and} \bibinfo{author}{\bibfnamefont{G.}~\bibnamefont{Kramer}},
  \bibinfo{journal}{Nucl. Phys. B} \textbf{\bibinfo{volume}{803}},
  \bibinfo{pages}{42} (\bibinfo{year}{2008}), \eprint{0803.2768}.

\bibitem[{\citenamefont{Bertone et~al.}(2018)\citenamefont{Bertone, Hartland,
  Nocera, Rojo, and Rottoli}}]{Bertone:2018ecm}
\bibinfo{author}{\bibfnamefont{V.}~\bibnamefont{Bertone}},
  \bibinfo{author}{\bibfnamefont{N.~P.} \bibnamefont{Hartland}},
  \bibinfo{author}{\bibfnamefont{E.~R.} \bibnamefont{Nocera}},
  \bibinfo{author}{\bibfnamefont{J.}~\bibnamefont{Rojo}}, \bibnamefont{and}
  \bibinfo{author}{\bibfnamefont{L.}~\bibnamefont{Rottoli}}
  (\bibinfo{collaboration}{NNPDF}), \bibinfo{journal}{Eur. Phys. J. C}
  \textbf{\bibinfo{volume}{78}}, \bibinfo{pages}{651} (\bibinfo{year}{2018}),
  \eprint{1807.03310}.

\bibitem[{\citenamefont{Khalek et~al.}(2021)\citenamefont{Khalek, Bertone, and
  Nocera}}]{Khalek:2021gxf}
\bibinfo{author}{\bibfnamefont{R.~A.} \bibnamefont{Khalek}},
  \bibinfo{author}{\bibfnamefont{V.}~\bibnamefont{Bertone}}, \bibnamefont{and}
  \bibinfo{author}{\bibfnamefont{E.~R.} \bibnamefont{Nocera}}
  (\bibinfo{collaboration}{MAP (Multi-dimensional Analyses of Partonic
  distributions)}), \bibinfo{journal}{Phys. Rev. D}
  \textbf{\bibinfo{volume}{104}}, \bibinfo{pages}{034007}
  (\bibinfo{year}{2021}), \eprint{2105.08725}.

\bibitem[{\citenamefont{Moffat et~al.}(2021)\citenamefont{Moffat, Melnitchouk,
  Rogers, and Sato}}]{Moffat:2021dji}
\bibinfo{author}{\bibfnamefont{E.}~\bibnamefont{Moffat}},
  \bibinfo{author}{\bibfnamefont{W.}~\bibnamefont{Melnitchouk}},
  \bibinfo{author}{\bibfnamefont{T.~C.} \bibnamefont{Rogers}},
  \bibnamefont{and} \bibinfo{author}{\bibfnamefont{N.}~\bibnamefont{Sato}}
  (\bibinfo{collaboration}{Jefferson Lab Angular Momentum (JAM)}),
  \bibinfo{journal}{Phys. Rev. D} \textbf{\bibinfo{volume}{104}},
  \bibinfo{pages}{016015} (\bibinfo{year}{2021}), \eprint{2101.04664}.

\bibitem[{\citenamefont{Gao et~al.}(2025)\citenamefont{Gao, Shen, Xing, Zhao,
  and Zhou}}]{Gao:2025hlm}
\bibinfo{author}{\bibfnamefont{J.}~\bibnamefont{Gao}},
  \bibinfo{author}{\bibfnamefont{X.}~\bibnamefont{Shen}},
  \bibinfo{author}{\bibfnamefont{H.}~\bibnamefont{Xing}},
  \bibinfo{author}{\bibfnamefont{Y.}~\bibnamefont{Zhao}}, \bibnamefont{and}
  \bibinfo{author}{\bibfnamefont{B.}~\bibnamefont{Zhou}}
  (\bibinfo{year}{2025}), \eprint{2502.17837}.

\bibitem[{\citenamefont{Anderle et~al.}(2015)\citenamefont{Anderle, Ringer, and
  Stratmann}}]{Anderle:2015lqa}
\bibinfo{author}{\bibfnamefont{D.~P.} \bibnamefont{Anderle}},
  \bibinfo{author}{\bibfnamefont{F.}~\bibnamefont{Ringer}}, \bibnamefont{and}
  \bibinfo{author}{\bibfnamefont{M.}~\bibnamefont{Stratmann}},
  \bibinfo{journal}{Phys. Rev. D} \textbf{\bibinfo{volume}{92}},
  \bibinfo{pages}{114017} (\bibinfo{year}{2015}), \eprint{1510.05845}.

\bibitem[{\citenamefont{Bertone et~al.}(2017)\citenamefont{Bertone, Carrazza,
  Hartland, Nocera, and Rojo}}]{Bertone:2017tyb}
\bibinfo{author}{\bibfnamefont{V.}~\bibnamefont{Bertone}},
  \bibinfo{author}{\bibfnamefont{S.}~\bibnamefont{Carrazza}},
  \bibinfo{author}{\bibfnamefont{N.~P.} \bibnamefont{Hartland}},
  \bibinfo{author}{\bibfnamefont{E.~R.} \bibnamefont{Nocera}},
  \bibnamefont{and} \bibinfo{author}{\bibfnamefont{J.}~\bibnamefont{Rojo}}
  (\bibinfo{collaboration}{NNPDF}), \bibinfo{journal}{Eur. Phys. J. C}
  \textbf{\bibinfo{volume}{77}}, \bibinfo{pages}{516} (\bibinfo{year}{2017}),
  \eprint{1706.07049}.

\bibitem[{\citenamefont{Soleymaninia et~al.}(2018)\citenamefont{Soleymaninia,
  Goharipour, and Khanpour}}]{Soleymaninia:2018uiv}
\bibinfo{author}{\bibfnamefont{M.}~\bibnamefont{Soleymaninia}},
  \bibinfo{author}{\bibfnamefont{M.}~\bibnamefont{Goharipour}},
  \bibnamefont{and} \bibinfo{author}{\bibfnamefont{H.}~\bibnamefont{Khanpour}},
  \bibinfo{journal}{Phys. Rev. D} \textbf{\bibinfo{volume}{98}},
  \bibinfo{pages}{074002} (\bibinfo{year}{2018}), \eprint{1805.04847}.

\bibitem[{\citenamefont{Soleymaninia et~al.}(2021)\citenamefont{Soleymaninia,
  Goharipour, Khanpour, and Spiesberger}}]{Soleymaninia:2020bsq}
\bibinfo{author}{\bibfnamefont{M.}~\bibnamefont{Soleymaninia}},
  \bibinfo{author}{\bibfnamefont{M.}~\bibnamefont{Goharipour}},
  \bibinfo{author}{\bibfnamefont{H.}~\bibnamefont{Khanpour}}, \bibnamefont{and}
  \bibinfo{author}{\bibfnamefont{H.}~\bibnamefont{Spiesberger}},
  \bibinfo{journal}{Phys. Rev. D} \textbf{\bibinfo{volume}{103}},
  \bibinfo{pages}{054045} (\bibinfo{year}{2021}), \eprint{2008.05342}.

\bibitem[{\citenamefont{Abdolmaleki et~al.}(2021)\citenamefont{Abdolmaleki,
  Soleymaninia, Khanpour, Amoroso, Giuli, Glazov, Luszczak, Olness, and
  Zenaiev}}]{Abdolmaleki:2021yjf}
\bibinfo{author}{\bibfnamefont{H.}~\bibnamefont{Abdolmaleki}},
  \bibinfo{author}{\bibfnamefont{M.}~\bibnamefont{Soleymaninia}},
  \bibinfo{author}{\bibfnamefont{H.}~\bibnamefont{Khanpour}},
  \bibinfo{author}{\bibfnamefont{S.}~\bibnamefont{Amoroso}},
  \bibinfo{author}{\bibfnamefont{F.}~\bibnamefont{Giuli}},
  \bibinfo{author}{\bibfnamefont{A.}~\bibnamefont{Glazov}},
  \bibinfo{author}{\bibfnamefont{A.}~\bibnamefont{Luszczak}},
  \bibinfo{author}{\bibfnamefont{F.}~\bibnamefont{Olness}}, \bibnamefont{and}
  \bibinfo{author}{\bibfnamefont{O.}~\bibnamefont{Zenaiev}}
  (\bibinfo{collaboration}{xfitter Developers\textquoteright{} Team}),
  \bibinfo{journal}{Phys. Rev. D} \textbf{\bibinfo{volume}{104}},
  \bibinfo{pages}{056019} (\bibinfo{year}{2021}), \eprint{2105.11306}.

\bibitem[{\citenamefont{Borsa et~al.}(2022{\natexlab{b}})\citenamefont{Borsa,
  Sassot, de~Florian, Stratmann, and Vogelsang}}]{Borsa:2022vvp}
\bibinfo{author}{\bibfnamefont{I.}~\bibnamefont{Borsa}},
  \bibinfo{author}{\bibfnamefont{R.}~\bibnamefont{Sassot}},
  \bibinfo{author}{\bibfnamefont{D.}~\bibnamefont{de~Florian}},
  \bibinfo{author}{\bibfnamefont{M.}~\bibnamefont{Stratmann}},
  \bibnamefont{and}
  \bibinfo{author}{\bibfnamefont{W.}~\bibnamefont{Vogelsang}},
  \bibinfo{journal}{Phys. Rev. Lett.} \textbf{\bibinfo{volume}{129}},
  \bibinfo{pages}{012002} (\bibinfo{year}{2022}{\natexlab{b}}),
  \eprint{2202.05060}.

\bibitem[{\citenamefont{Abdul~Khalek et~al.}(2022)\citenamefont{Abdul~Khalek,
  Bertone, Khoudli, and Nocera}}]{AbdulKhalek:2022laj}
\bibinfo{author}{\bibfnamefont{R.}~\bibnamefont{Abdul~Khalek}},
  \bibinfo{author}{\bibfnamefont{V.}~\bibnamefont{Bertone}},
  \bibinfo{author}{\bibfnamefont{A.}~\bibnamefont{Khoudli}}, \bibnamefont{and}
  \bibinfo{author}{\bibfnamefont{E.~R.} \bibnamefont{Nocera}}
  (\bibinfo{collaboration}{MAP (Multi-dimensional Analyses of Partonic
  distributions)}), \bibinfo{journal}{Phys. Lett. B}
  \textbf{\bibinfo{volume}{834}}, \bibinfo{pages}{137456}
  (\bibinfo{year}{2022}), \eprint{2204.10331}.

\bibitem[{\citenamefont{Binnewies et~al.}(1996)\citenamefont{Binnewies, Kniehl,
  and Kramer}}]{Binnewies:1995kg}
\bibinfo{author}{\bibfnamefont{J.}~\bibnamefont{Binnewies}},
  \bibinfo{author}{\bibfnamefont{B.~A.} \bibnamefont{Kniehl}},
  \bibnamefont{and} \bibinfo{author}{\bibfnamefont{G.}~\bibnamefont{Kramer}},
  \bibinfo{journal}{Phys. Rev. D} \textbf{\bibinfo{volume}{53}},
  \bibinfo{pages}{3573} (\bibinfo{year}{1996}), \eprint{hep-ph/9506437}.

\bibitem[{\citenamefont{Albino et~al.}(2006)\citenamefont{Albino, Kniehl, and
  Kramer}}]{Albino:2005mv}
\bibinfo{author}{\bibfnamefont{S.}~\bibnamefont{Albino}},
  \bibinfo{author}{\bibfnamefont{B.~A.} \bibnamefont{Kniehl}},
  \bibnamefont{and} \bibinfo{author}{\bibfnamefont{G.}~\bibnamefont{Kramer}},
  \bibinfo{journal}{Nucl. Phys. B} \textbf{\bibinfo{volume}{734}},
  \bibinfo{pages}{50} (\bibinfo{year}{2006}), \eprint{hep-ph/0510173}.

\bibitem[{\citenamefont{Soleymaninia et~al.}(2020)\citenamefont{Soleymaninia,
  Abdolmaleki, and Khanpour}}]{Soleymaninia:2020ahn}
\bibinfo{author}{\bibfnamefont{M.}~\bibnamefont{Soleymaninia}},
  \bibinfo{author}{\bibfnamefont{H.}~\bibnamefont{Abdolmaleki}},
  \bibnamefont{and} \bibinfo{author}{\bibfnamefont{H.}~\bibnamefont{Khanpour}},
  \bibinfo{journal}{Phys. Rev. D} \textbf{\bibinfo{volume}{102}},
  \bibinfo{pages}{114029} (\bibinfo{year}{2020}), \eprint{2009.08139}.

\bibitem[{\citenamefont{Li et~al.}(2024{\natexlab{b}})\citenamefont{Li,
  Anderle, Xing, and Zhao}}]{Li:2024etc}
\bibinfo{author}{\bibfnamefont{M.}~\bibnamefont{Li}},
  \bibinfo{author}{\bibfnamefont{D.~P.} \bibnamefont{Anderle}},
  \bibinfo{author}{\bibfnamefont{H.}~\bibnamefont{Xing}}, \bibnamefont{and}
  \bibinfo{author}{\bibfnamefont{Y.}~\bibnamefont{Zhao}}
  (\bibinfo{year}{2024}{\natexlab{b}}), \eprint{2404.11527}.

\bibitem[{\citenamefont{Soleymaninia
  et~al.}(2024{\natexlab{b}})\citenamefont{Soleymaninia, Hashamipour,
  Salajegheh, Khanpour, Spiesberger, and Mei{\ss}ner}}]{Soleymaninia:2024jam}
\bibinfo{author}{\bibfnamefont{M.}~\bibnamefont{Soleymaninia}},
  \bibinfo{author}{\bibfnamefont{H.}~\bibnamefont{Hashamipour}},
  \bibinfo{author}{\bibfnamefont{M.}~\bibnamefont{Salajegheh}},
  \bibinfo{author}{\bibfnamefont{H.}~\bibnamefont{Khanpour}},
  \bibinfo{author}{\bibfnamefont{H.}~\bibnamefont{Spiesberger}},
  \bibnamefont{and} \bibinfo{author}{\bibfnamefont{U.-G.}
  \bibnamefont{Mei{\ss}ner}}, \bibinfo{journal}{Phys. Rev. D}
  \textbf{\bibinfo{volume}{110}}, \bibinfo{pages}{014019}
  (\bibinfo{year}{2024}{\natexlab{b}}), \eprint{2404.07334}.

\bibitem[{\citenamefont{Aidala et~al.}(2011)\citenamefont{Aidala, Ellinghaus,
  Sassot, Seele, and Stratmann}}]{Aidala:2010bn}
\bibinfo{author}{\bibfnamefont{C.~A.} \bibnamefont{Aidala}},
  \bibinfo{author}{\bibfnamefont{F.}~\bibnamefont{Ellinghaus}},
  \bibinfo{author}{\bibfnamefont{R.}~\bibnamefont{Sassot}},
  \bibinfo{author}{\bibfnamefont{J.~P.} \bibnamefont{Seele}}, \bibnamefont{and}
  \bibinfo{author}{\bibfnamefont{M.}~\bibnamefont{Stratmann}},
  \bibinfo{journal}{Phys. Rev. D} \textbf{\bibinfo{volume}{83}},
  \bibinfo{pages}{034002} (\bibinfo{year}{2011}), \eprint{1009.6145}.

\bibitem[{\citenamefont{Abramowicz et~al.}(2012)}]{ZEUS:2011cdi}
\bibinfo{author}{\bibfnamefont{H.}~\bibnamefont{Abramowicz}}
  \bibnamefont{et~al.} (\bibinfo{collaboration}{ZEUS}), \bibinfo{journal}{JHEP}
  \textbf{\bibinfo{volume}{03}}, \bibinfo{pages}{020} (\bibinfo{year}{2012}),
  \eprint{1111.3526}.

\bibitem[{\citenamefont{Zhou and Gao}(2024)}]{Zhou:2024cyk}
\bibinfo{author}{\bibfnamefont{B.}~\bibnamefont{Zhou}} \bibnamefont{and}
  \bibinfo{author}{\bibfnamefont{J.}~\bibnamefont{Gao}} (\bibinfo{year}{2024}),
  \eprint{2407.10059}.

\bibitem[{\citenamefont{Alexander et~al.}(1991)}]{OPAL:1991ixp}
\bibinfo{author}{\bibfnamefont{G.}~\bibnamefont{Alexander}}
  \bibnamefont{et~al.} (\bibinfo{collaboration}{OPAL}), \bibinfo{journal}{Phys.
  Lett. B} \textbf{\bibinfo{volume}{264}}, \bibinfo{pages}{467}
  (\bibinfo{year}{1991}).

\bibitem[{\citenamefont{Akers et~al.}(1995)}]{OPAL:1995ebr}
\bibinfo{author}{\bibfnamefont{R.}~\bibnamefont{Akers}} \bibnamefont{et~al.}
  (\bibinfo{collaboration}{OPAL}), \bibinfo{journal}{Z. Phys. C}
  \textbf{\bibinfo{volume}{67}}, \bibinfo{pages}{389} (\bibinfo{year}{1995}).

\bibitem[{\citenamefont{Abbiendi et~al.}(2000{\natexlab{a}})}]{OPAL:2000dkf}
\bibinfo{author}{\bibfnamefont{G.}~\bibnamefont{Abbiendi}} \bibnamefont{et~al.}
  (\bibinfo{collaboration}{OPAL}), \bibinfo{journal}{Eur. Phys. J. C}
  \textbf{\bibinfo{volume}{17}}, \bibinfo{pages}{373}
  (\bibinfo{year}{2000}{\natexlab{a}}), \eprint{hep-ex/0007017}.

\bibitem[{\citenamefont{Barate et~al.}(1998)}]{ALEPH:1996oqp}
\bibinfo{author}{\bibfnamefont{R.}~\bibnamefont{Barate}} \bibnamefont{et~al.}
  (\bibinfo{collaboration}{ALEPH}), \bibinfo{journal}{Phys. Rept.}
  \textbf{\bibinfo{volume}{294}}, \bibinfo{pages}{1} (\bibinfo{year}{1998}).

\bibitem[{\citenamefont{Abreu et~al.}(1995)}]{DELPHI:1994qgk}
\bibinfo{author}{\bibfnamefont{P.}~\bibnamefont{Abreu}} \bibnamefont{et~al.}
  (\bibinfo{collaboration}{DELPHI}), \bibinfo{journal}{Z. Phys. C}
  \textbf{\bibinfo{volume}{65}}, \bibinfo{pages}{587} (\bibinfo{year}{1995}).

\bibitem[{\citenamefont{Abe et~al.}(1999)}]{SLD:1998coh}
\bibinfo{author}{\bibfnamefont{K.}~\bibnamefont{Abe}} \bibnamefont{et~al.}
  (\bibinfo{collaboration}{SLD}), \bibinfo{journal}{Phys. Rev. D}
  \textbf{\bibinfo{volume}{59}}, \bibinfo{pages}{052001}
  (\bibinfo{year}{1999}), \eprint{hep-ex/9805029}.

\bibitem[{\citenamefont{Aihara et~al.}(1984)}]{TPCTwoGamma:1984eoj}
\bibinfo{author}{\bibfnamefont{H.}~\bibnamefont{Aihara}} \bibnamefont{et~al.}
  (\bibinfo{collaboration}{TPC/Two Gamma}), \bibinfo{journal}{Phys. Rev. Lett.}
  \textbf{\bibinfo{volume}{53}}, \bibinfo{pages}{2378} (\bibinfo{year}{1984}).

\bibitem[{\citenamefont{Schellman et~al.}(1985)}]{Schellman:1984yz}
\bibinfo{author}{\bibfnamefont{H.}~\bibnamefont{Schellman}}
  \bibnamefont{et~al.}, \bibinfo{journal}{Phys. Rev. D}
  \textbf{\bibinfo{volume}{31}}, \bibinfo{pages}{3013} (\bibinfo{year}{1985}).

\bibitem[{\citenamefont{Althoff et~al.}(1985)}]{TASSO:1984nda}
\bibinfo{author}{\bibfnamefont{M.}~\bibnamefont{Althoff}} \bibnamefont{et~al.}
  (\bibinfo{collaboration}{TASSO}), \bibinfo{journal}{Z. Phys. C}
  \textbf{\bibinfo{volume}{27}}, \bibinfo{pages}{27} (\bibinfo{year}{1985}).

\bibitem[{\citenamefont{Braunschweig et~al.}(1990)}]{TASSO:1989jyt}
\bibinfo{author}{\bibfnamefont{W.}~\bibnamefont{Braunschweig}}
  \bibnamefont{et~al.} (\bibinfo{collaboration}{TASSO}), \bibinfo{journal}{Z.
  Phys. C} \textbf{\bibinfo{volume}{47}}, \bibinfo{pages}{167}
  (\bibinfo{year}{1990}).

\bibitem[{\citenamefont{Derrick et~al.}(1987)}]{Derrick:1985wd}
\bibinfo{author}{\bibfnamefont{M.}~\bibnamefont{Derrick}} \bibnamefont{et~al.},
  \bibinfo{journal}{Phys. Rev. D} \textbf{\bibinfo{volume}{35}},
  \bibinfo{pages}{2639} (\bibinfo{year}{1987}).

\bibitem[{\citenamefont{Behrend et~al.}(1990{\natexlab{a}})}]{CELLO:1989adw}
\bibinfo{author}{\bibfnamefont{H.~J.} \bibnamefont{Behrend}}
  \bibnamefont{et~al.} (\bibinfo{collaboration}{CELLO}), \bibinfo{journal}{Z.
  Phys. C} \textbf{\bibinfo{volume}{46}}, \bibinfo{pages}{397}
  (\bibinfo{year}{1990}{\natexlab{a}}).

\bibitem[{\citenamefont{Itoh et~al.}(1995)}]{TOPAZ:1994voc}
\bibinfo{author}{\bibfnamefont{R.}~\bibnamefont{Itoh}} \bibnamefont{et~al.}
  (\bibinfo{collaboration}{TOPAZ}), \bibinfo{journal}{Phys. Lett. B}
  \textbf{\bibinfo{volume}{345}}, \bibinfo{pages}{335} (\bibinfo{year}{1995}),
  \eprint{hep-ex/9412015}.

\bibitem[{\citenamefont{Seidl et~al.}(2024)}]{Belle:2024vua}
\bibinfo{author}{\bibfnamefont{R.}~\bibnamefont{Seidl}} \bibnamefont{et~al.}
  (\bibinfo{collaboration}{Belle}) (\bibinfo{year}{2024}), \eprint{2411.12216}.

\bibitem[{\citenamefont{Barate et~al.}(2000)}]{ALEPH:1999udi}
\bibinfo{author}{\bibfnamefont{R.}~\bibnamefont{Barate}} \bibnamefont{et~al.}
  (\bibinfo{collaboration}{ALEPH}), \bibinfo{journal}{Eur. Phys. J. C}
  \textbf{\bibinfo{volume}{16}}, \bibinfo{pages}{613} (\bibinfo{year}{2000}).

\bibitem[{\citenamefont{Abreu et~al.}(2000)}]{DELPHI:2000ahn}
\bibinfo{author}{\bibfnamefont{P.}~\bibnamefont{Abreu}} \bibnamefont{et~al.}
  (\bibinfo{collaboration}{DELPHI}), \bibinfo{journal}{Eur. Phys. J. C}
  \textbf{\bibinfo{volume}{18}}, \bibinfo{pages}{203} (\bibinfo{year}{2000}),
  \bibinfo{note}{[Erratum: Eur.Phys.J.C 25, 493 (2002)]},
  \eprint{hep-ex/0103031}.

\bibitem[{\citenamefont{Abbiendi et~al.}(2000{\natexlab{b}})}]{OPAL:1999zfe}
\bibinfo{author}{\bibfnamefont{G.}~\bibnamefont{Abbiendi}} \bibnamefont{et~al.}
  (\bibinfo{collaboration}{OPAL}), \bibinfo{journal}{Eur. Phys. J. C}
  \textbf{\bibinfo{volume}{16}}, \bibinfo{pages}{407}
  (\bibinfo{year}{2000}{\natexlab{b}}), \eprint{hep-ex/0001054}.

\bibitem[{\citenamefont{Ablikim et~al.}(2023)}]{BESIII:2022zit}
\bibinfo{author}{\bibfnamefont{M.}~\bibnamefont{Ablikim}} \bibnamefont{et~al.}
  (\bibinfo{collaboration}{BESIII}), \bibinfo{journal}{Phys. Rev. Lett.}
  \textbf{\bibinfo{volume}{130}}, \bibinfo{pages}{231901}
  (\bibinfo{year}{2023}), \eprint{2211.11253}.

\bibitem[{\citenamefont{Ablikim et~al.}(2025)}]{BESIII:2025mbc}
\bibinfo{author}{\bibfnamefont{M.}~\bibnamefont{Ablikim}} \bibnamefont{et~al.}
  (\bibinfo{collaboration}{BESIII}) (\bibinfo{year}{2025}),
  \eprint{2502.16084}.

\bibitem[{\citenamefont{Albrecht et~al.}(1990)}]{ARGUS:1989orf}
\bibinfo{author}{\bibfnamefont{H.}~\bibnamefont{Albrecht}} \bibnamefont{et~al.}
  (\bibinfo{collaboration}{ARGUS}), \bibinfo{journal}{Z. Phys. C}
  \textbf{\bibinfo{volume}{46}}, \bibinfo{pages}{15} (\bibinfo{year}{1990}).

\bibitem[{\citenamefont{Abachi et~al.}(1988)}]{HRS:1987aky}
\bibinfo{author}{\bibfnamefont{S.}~\bibnamefont{Abachi}} \bibnamefont{et~al.}
  (\bibinfo{collaboration}{HRS}), \bibinfo{journal}{Phys. Lett. B}
  \textbf{\bibinfo{volume}{205}}, \bibinfo{pages}{111} (\bibinfo{year}{1988}).

\bibitem[{\citenamefont{de~la Vaissiere et~al.}(1985)}]{delaVaissiere:1984xg}
\bibinfo{author}{\bibfnamefont{C.}~\bibnamefont{de~la Vaissiere}}
  \bibnamefont{et~al.}, \bibinfo{journal}{Phys. Rev. Lett.}
  \textbf{\bibinfo{volume}{54}}, \bibinfo{pages}{2071} (\bibinfo{year}{1985}),
  \bibinfo{note}{[Erratum: Phys.Rev.Lett. 55, 263 (1985)]}.

\bibitem[{\citenamefont{Brandelik et~al.}(1981)}]{TASSO:1981uqa}
\bibinfo{author}{\bibfnamefont{R.}~\bibnamefont{Brandelik}}
  \bibnamefont{et~al.} (\bibinfo{collaboration}{TASSO}),
  \bibinfo{journal}{Phys. Lett. B} \textbf{\bibinfo{volume}{105}},
  \bibinfo{pages}{75} (\bibinfo{year}{1981}).

\bibitem[{\citenamefont{Bartel et~al.}(1985)}]{JADE:1985bzp}
\bibinfo{author}{\bibfnamefont{W.}~\bibnamefont{Bartel}} \bibnamefont{et~al.}
  (\bibinfo{collaboration}{JADE}), \bibinfo{journal}{Z. Phys. C}
  \textbf{\bibinfo{volume}{28}}, \bibinfo{pages}{343} (\bibinfo{year}{1985}).

\bibitem[{\citenamefont{Braunschweig
  et~al.}(1989{\natexlab{a}})}]{TASSO:1988qlu}
\bibinfo{author}{\bibfnamefont{W.}~\bibnamefont{Braunschweig}}
  \bibnamefont{et~al.} (\bibinfo{collaboration}{TASSO}), \bibinfo{journal}{Z.
  Phys. C} \textbf{\bibinfo{volume}{45}}, \bibinfo{pages}{209}
  (\bibinfo{year}{1989}{\natexlab{a}}).

\bibitem[{\citenamefont{Pitzl et~al.}(1990)}]{JADE:1989ewf}
\bibinfo{author}{\bibfnamefont{D.~D.} \bibnamefont{Pitzl}} \bibnamefont{et~al.}
  (\bibinfo{collaboration}{JADE}), \bibinfo{journal}{Z. Phys. C}
  \textbf{\bibinfo{volume}{46}}, \bibinfo{pages}{1} (\bibinfo{year}{1990}),
  \bibinfo{note}{[Erratum: Z.Phys.C 47, 676 (1990)]}.

\bibitem[{\citenamefont{Behrend et~al.}(1990{\natexlab{b}})}]{CELLO:1989byk}
\bibinfo{author}{\bibfnamefont{H.~J.} \bibnamefont{Behrend}}
  \bibnamefont{et~al.} (\bibinfo{collaboration}{CELLO}), \bibinfo{journal}{Z.
  Phys. C} \textbf{\bibinfo{volume}{47}}, \bibinfo{pages}{1}
  (\bibinfo{year}{1990}{\natexlab{b}}).

\bibitem[{\citenamefont{Braunschweig
  et~al.}(1989{\natexlab{b}})}]{TASSO:1988jma}
\bibinfo{author}{\bibfnamefont{W.}~\bibnamefont{Braunschweig}}
  \bibnamefont{et~al.} (\bibinfo{collaboration}{TASSO}), \bibinfo{journal}{Z.
  Phys. C} \textbf{\bibinfo{volume}{42}}, \bibinfo{pages}{189}
  (\bibinfo{year}{1989}{\natexlab{b}}).

\bibitem[{\citenamefont{Alexander et~al.}(1997)}]{OPAL:1996gsw}
\bibinfo{author}{\bibfnamefont{G.}~\bibnamefont{Alexander}}
  \bibnamefont{et~al.} (\bibinfo{collaboration}{OPAL}), \bibinfo{journal}{Z.
  Phys. C} \textbf{\bibinfo{volume}{73}}, \bibinfo{pages}{569}
  (\bibinfo{year}{1997}).

\bibitem[{\citenamefont{Ackerstaff et~al.}(1998)}]{OPAL:1998enc}
\bibinfo{author}{\bibfnamefont{K.}~\bibnamefont{Ackerstaff}}
  \bibnamefont{et~al.} (\bibinfo{collaboration}{OPAL}), \bibinfo{journal}{Eur.
  Phys. J. C} \textbf{\bibinfo{volume}{5}}, \bibinfo{pages}{411}
  (\bibinfo{year}{1998}), \eprint{hep-ex/9805011}.

\bibitem[{\citenamefont{Buskulic et~al.}(1992)}]{ALEPH:1992zhm}
\bibinfo{author}{\bibfnamefont{D.}~\bibnamefont{Buskulic}} \bibnamefont{et~al.}
  (\bibinfo{collaboration}{ALEPH}), \bibinfo{journal}{Phys. Lett. B}
  \textbf{\bibinfo{volume}{292}}, \bibinfo{pages}{210} (\bibinfo{year}{1992}).

\bibitem[{\citenamefont{Buskulic et~al.}(1994)}]{ALEPH:1994fts}
\bibinfo{author}{\bibfnamefont{D.}~\bibnamefont{Buskulic}} \bibnamefont{et~al.}
  (\bibinfo{collaboration}{ALEPH}), \bibinfo{journal}{Z. Phys. C}
  \textbf{\bibinfo{volume}{64}}, \bibinfo{pages}{361} (\bibinfo{year}{1994}).

\bibitem[{\citenamefont{Barate et~al.}(1997)}]{ALEPH:1996pxg}
\bibinfo{author}{\bibfnamefont{R.}~\bibnamefont{Barate}} \bibnamefont{et~al.}
  (\bibinfo{collaboration}{ALEPH}), \bibinfo{journal}{Z. Phys. C}
  \textbf{\bibinfo{volume}{74}}, \bibinfo{pages}{451} (\bibinfo{year}{1997}).

\bibitem[{\citenamefont{Heister et~al.}(2002)}]{ALEPH:2001tfk}
\bibinfo{author}{\bibfnamefont{A.}~\bibnamefont{Heister}} \bibnamefont{et~al.}
  (\bibinfo{collaboration}{ALEPH}), \bibinfo{journal}{Phys. Lett. B}
  \textbf{\bibinfo{volume}{528}}, \bibinfo{pages}{19} (\bibinfo{year}{2002}),
  \eprint{hep-ex/0201012}.

\bibitem[{\citenamefont{Abreu et~al.}(1993)}]{DELPHI:1993vpj}
\bibinfo{author}{\bibfnamefont{P.}~\bibnamefont{Abreu}} \bibnamefont{et~al.}
  (\bibinfo{collaboration}{DELPHI}), \bibinfo{journal}{Phys. Lett. B}
  \textbf{\bibinfo{volume}{318}}, \bibinfo{pages}{249} (\bibinfo{year}{1993}).

\bibitem[{\citenamefont{Adam et~al.}(1996)}]{DELPHI:1995ase}
\bibinfo{author}{\bibfnamefont{W.}~\bibnamefont{Adam}} \bibnamefont{et~al.}
  (\bibinfo{collaboration}{DELPHI}), \bibinfo{journal}{Z. Phys. C}
  \textbf{\bibinfo{volume}{69}}, \bibinfo{pages}{561} (\bibinfo{year}{1996}).

\bibitem[{\citenamefont{Adriani et~al.}(1992)}]{L3:1992pbe}
\bibinfo{author}{\bibfnamefont{O.}~\bibnamefont{Adriani}} \bibnamefont{et~al.}
  (\bibinfo{collaboration}{L3}), \bibinfo{journal}{Phys. Lett. B}
  \textbf{\bibinfo{volume}{286}}, \bibinfo{pages}{403} (\bibinfo{year}{1992}).

\bibitem[{\citenamefont{Acciarri et~al.}(1994)}]{L3:1994gkb}
\bibinfo{author}{\bibfnamefont{M.}~\bibnamefont{Acciarri}} \bibnamefont{et~al.}
  (\bibinfo{collaboration}{L3}), \bibinfo{journal}{Phys. Lett. B}
  \textbf{\bibinfo{volume}{328}}, \bibinfo{pages}{223} (\bibinfo{year}{1994}).

\bibitem[{\citenamefont{Khachatryan et~al.}(2011)}]{CMS:2011jlm}
\bibinfo{author}{\bibfnamefont{V.}~\bibnamefont{Khachatryan}}
  \bibnamefont{et~al.} (\bibinfo{collaboration}{CMS}), \bibinfo{journal}{JHEP}
  \textbf{\bibinfo{volume}{05}}, \bibinfo{pages}{064} (\bibinfo{year}{2011}),
  \eprint{1102.4282}.

\bibitem[{\citenamefont{Abelev et~al.}(2010)}]{STAR:2009qzv}
\bibinfo{author}{\bibfnamefont{B.~I.} \bibnamefont{Abelev}}
  \bibnamefont{et~al.} (\bibinfo{collaboration}{STAR}), \bibinfo{journal}{Phys.
  Rev. C} \textbf{\bibinfo{volume}{81}}, \bibinfo{pages}{064904}
  (\bibinfo{year}{2010}), \eprint{0912.3838}.

\bibitem[{\citenamefont{Adare et~al.}(2011)}]{PHENIX:2010hvs}
\bibinfo{author}{\bibfnamefont{A.}~\bibnamefont{Adare}} \bibnamefont{et~al.}
  (\bibinfo{collaboration}{PHENIX}), \bibinfo{journal}{Phys. Rev. D}
  \textbf{\bibinfo{volume}{83}}, \bibinfo{pages}{032001}
  (\bibinfo{year}{2011}), \eprint{1009.6224}.

\bibitem[{\citenamefont{Adare et~al.}(2007)}]{PHENIX:2007kqm}
\bibinfo{author}{\bibfnamefont{A.}~\bibnamefont{Adare}} \bibnamefont{et~al.}
  (\bibinfo{collaboration}{PHENIX}), \bibinfo{journal}{Phys. Rev. D}
  \textbf{\bibinfo{volume}{76}}, \bibinfo{pages}{051106}
  (\bibinfo{year}{2007}), \eprint{0704.3599}.

\bibitem[{\citenamefont{Adare et~al.}(2016)}]{PHENIX:2015fxo}
\bibinfo{author}{\bibfnamefont{A.}~\bibnamefont{Adare}} \bibnamefont{et~al.}
  (\bibinfo{collaboration}{PHENIX}), \bibinfo{journal}{Phys. Rev. D}
  \textbf{\bibinfo{volume}{93}}, \bibinfo{pages}{011501}
  (\bibinfo{year}{2016}), \eprint{1510.02317}.

\bibitem[{\citenamefont{Acharya et~al.}(2017)}]{ALICE:2017nce}
\bibinfo{author}{\bibfnamefont{S.}~\bibnamefont{Acharya}} \bibnamefont{et~al.}
  (\bibinfo{collaboration}{ALICE}), \bibinfo{journal}{Eur. Phys. J. C}
  \textbf{\bibinfo{volume}{77}}, \bibinfo{pages}{339} (\bibinfo{year}{2017}),
  \eprint{1702.00917}.

\bibitem[{\citenamefont{Abelev et~al.}(2012)}]{ALICE:2012wos}
\bibinfo{author}{\bibfnamefont{B.}~\bibnamefont{Abelev}} \bibnamefont{et~al.}
  (\bibinfo{collaboration}{ALICE}), \bibinfo{journal}{Phys. Lett. B}
  \textbf{\bibinfo{volume}{717}}, \bibinfo{pages}{162} (\bibinfo{year}{2012}),
  \eprint{1205.5724}.

\bibitem[{\citenamefont{Acharya et~al.}(2018)}]{ALICE:2017ryd}
\bibinfo{author}{\bibfnamefont{S.}~\bibnamefont{Acharya}} \bibnamefont{et~al.}
  (\bibinfo{collaboration}{ALICE}), \bibinfo{journal}{Eur. Phys. J. C}
  \textbf{\bibinfo{volume}{78}}, \bibinfo{pages}{263} (\bibinfo{year}{2018}),
  \eprint{1708.08745}.

\bibitem[{\citenamefont{Acharya et~al.}(2021)}]{ALICE:2020jsh}
\bibinfo{author}{\bibfnamefont{S.}~\bibnamefont{Acharya}} \bibnamefont{et~al.}
  (\bibinfo{collaboration}{ALICE}), \bibinfo{journal}{Eur. Phys. J. C}
  \textbf{\bibinfo{volume}{81}}, \bibinfo{pages}{256} (\bibinfo{year}{2021}),
  \eprint{2005.11120}.

\bibitem[{\citenamefont{Acharya et~al.}(2024)}]{ALICE:2024vgi}
\bibinfo{author}{\bibfnamefont{S.}~\bibnamefont{Acharya}} \bibnamefont{et~al.}
  (\bibinfo{collaboration}{ALICE}) (\bibinfo{year}{2024}), \eprint{2411.09560}.

\bibitem[{\citenamefont{Aid et~al.}(1996)}]{H1:1996kfw}
\bibinfo{author}{\bibfnamefont{S.}~\bibnamefont{Aid}} \bibnamefont{et~al.}
  (\bibinfo{collaboration}{H1}), \bibinfo{journal}{Nucl. Phys. B}
  \textbf{\bibinfo{volume}{480}}, \bibinfo{pages}{3} (\bibinfo{year}{1996}),
  \eprint{hep-ex/9607010}.

\bibitem[{\citenamefont{Acton et~al.}(1992)}]{OPAL:1992asw}
\bibinfo{author}{\bibfnamefont{P.~D.} \bibnamefont{Acton}} \bibnamefont{et~al.}
  (\bibinfo{collaboration}{OPAL}), \bibinfo{journal}{Phys. Lett. B}
  \textbf{\bibinfo{volume}{291}}, \bibinfo{pages}{503} (\bibinfo{year}{1992}).

\bibitem[{\citenamefont{Elze}(1990)}]{Elze:1989gm}
\bibinfo{author}{\bibfnamefont{H.-T.} \bibnamefont{Elze}}, \bibinfo{journal}{Z.
  Phys. C} \textbf{\bibinfo{volume}{47}}, \bibinfo{pages}{647}
  (\bibinfo{year}{1990}).

\bibitem[{\citenamefont{Stratmann and Vogelsang}(1997)}]{Stratmann:1996hn}
\bibinfo{author}{\bibfnamefont{M.}~\bibnamefont{Stratmann}} \bibnamefont{and}
  \bibinfo{author}{\bibfnamefont{W.}~\bibnamefont{Vogelsang}},
  \bibinfo{journal}{Nucl. Phys. B} \textbf{\bibinfo{volume}{496}},
  \bibinfo{pages}{41} (\bibinfo{year}{1997}), \eprint{hep-ph/9612250}.

\bibitem[{\citenamefont{Salam and Rojo}(2009)}]{Salam:2008qg}
\bibinfo{author}{\bibfnamefont{G.~P.} \bibnamefont{Salam}} \bibnamefont{and}
  \bibinfo{author}{\bibfnamefont{J.}~\bibnamefont{Rojo}},
  \bibinfo{journal}{Comput. Phys. Commun.} \textbf{\bibinfo{volume}{180}},
  \bibinfo{pages}{120} (\bibinfo{year}{2009}), \eprint{0804.3755}.

\bibitem[{\citenamefont{Salam and Rojo}(2008)}]{Salam:2008sz}
\bibinfo{author}{\bibfnamefont{G.}~\bibnamefont{Salam}} \bibnamefont{and}
  \bibinfo{author}{\bibfnamefont{J.}~\bibnamefont{Rojo}}, in
  \emph{\bibinfo{booktitle}{{16th International Workshop on Deep Inelastic
  Scattering and Related Subjects}}} (\bibinfo{year}{2008}),
  p.~\bibinfo{pages}{42}, \eprint{0807.0198}.

\bibitem[{\citenamefont{Gao}(2024)}]{FMNLO}
\bibinfo{author}{\bibfnamefont{J.}~\bibnamefont{Gao}},
  \emph{\bibinfo{title}{\textsc{FMNLO}}},
  \bibinfo{howpublished}{\url{https://fmnlo.sjtu.edu.cn/~fmnlo/}}
  (\bibinfo{year}{2024}).

\bibitem[{\citenamefont{Dulat et~al.}(2016)\citenamefont{Dulat, Hou, Gao,
  Guzzi, Huston, Nadolsky, Pumplin, Schmidt, Stump, and Yuan}}]{Dulat:2015mca}
\bibinfo{author}{\bibfnamefont{S.}~\bibnamefont{Dulat}},
  \bibinfo{author}{\bibfnamefont{T.-J.} \bibnamefont{Hou}},
  \bibinfo{author}{\bibfnamefont{J.}~\bibnamefont{Gao}},
  \bibinfo{author}{\bibfnamefont{M.}~\bibnamefont{Guzzi}},
  \bibinfo{author}{\bibfnamefont{J.}~\bibnamefont{Huston}},
  \bibinfo{author}{\bibfnamefont{P.}~\bibnamefont{Nadolsky}},
  \bibinfo{author}{\bibfnamefont{J.}~\bibnamefont{Pumplin}},
  \bibinfo{author}{\bibfnamefont{C.}~\bibnamefont{Schmidt}},
  \bibinfo{author}{\bibfnamefont{D.}~\bibnamefont{Stump}}, \bibnamefont{and}
  \bibinfo{author}{\bibfnamefont{C.~P.} \bibnamefont{Yuan}},
  \bibinfo{journal}{Phys. Rev. D} \textbf{\bibinfo{volume}{93}},
  \bibinfo{pages}{033006} (\bibinfo{year}{2016}), \eprint{1506.07443}.

\bibitem[{\citenamefont{Stump et~al.}(2001)\citenamefont{Stump, Pumplin, Brock,
  Casey, Huston, Kalk, Lai, and Tung}}]{Stump:2001gu}
\bibinfo{author}{\bibfnamefont{D.}~\bibnamefont{Stump}},
  \bibinfo{author}{\bibfnamefont{J.}~\bibnamefont{Pumplin}},
  \bibinfo{author}{\bibfnamefont{R.}~\bibnamefont{Brock}},
  \bibinfo{author}{\bibfnamefont{D.}~\bibnamefont{Casey}},
  \bibinfo{author}{\bibfnamefont{J.}~\bibnamefont{Huston}},
  \bibinfo{author}{\bibfnamefont{J.}~\bibnamefont{Kalk}},
  \bibinfo{author}{\bibfnamefont{H.~L.} \bibnamefont{Lai}}, \bibnamefont{and}
  \bibinfo{author}{\bibfnamefont{W.~K.} \bibnamefont{Tung}},
  \bibinfo{journal}{Phys. Rev. D} \textbf{\bibinfo{volume}{65}},
  \bibinfo{pages}{014012} (\bibinfo{year}{2001}), \eprint{hep-ph/0101051}.

\bibitem[{\citenamefont{Pumplin et~al.}(2002)\citenamefont{Pumplin, Stump,
  Huston, Lai, Nadolsky, and Tung}}]{Pumplin:2002vw}
\bibinfo{author}{\bibfnamefont{J.}~\bibnamefont{Pumplin}},
  \bibinfo{author}{\bibfnamefont{D.~R.} \bibnamefont{Stump}},
  \bibinfo{author}{\bibfnamefont{J.}~\bibnamefont{Huston}},
  \bibinfo{author}{\bibfnamefont{H.~L.} \bibnamefont{Lai}},
  \bibinfo{author}{\bibfnamefont{P.~M.} \bibnamefont{Nadolsky}},
  \bibnamefont{and} \bibinfo{author}{\bibfnamefont{W.~K.} \bibnamefont{Tung}},
  \bibinfo{journal}{JHEP} \textbf{\bibinfo{volume}{07}}, \bibinfo{pages}{012}
  (\bibinfo{year}{2002}), \eprint{hep-ph/0201195}.

\bibitem[{\citenamefont{Chen et~al.}(2021{\natexlab{b}})\citenamefont{Chen,
  Liang, Pan, Song, and Wei}}]{Chen:2021hdn}
\bibinfo{author}{\bibfnamefont{K.-B.} \bibnamefont{Chen}},
  \bibinfo{author}{\bibfnamefont{Z.-T.} \bibnamefont{Liang}},
  \bibinfo{author}{\bibfnamefont{Y.-L.} \bibnamefont{Pan}},
  \bibinfo{author}{\bibfnamefont{Y.-K.} \bibnamefont{Song}}, \bibnamefont{and}
  \bibinfo{author}{\bibfnamefont{S.-Y.} \bibnamefont{Wei}},
  \bibinfo{journal}{Phys. Lett. B} \textbf{\bibinfo{volume}{816}},
  \bibinfo{pages}{136217} (\bibinfo{year}{2021}{\natexlab{b}}),
  \eprint{2102.00658}.

\bibitem[{\citenamefont{Chen et~al.}(2022)\citenamefont{Chen, Liang, Song, and
  Wei}}]{Chen:2021zrr}
\bibinfo{author}{\bibfnamefont{K.-b.} \bibnamefont{Chen}},
  \bibinfo{author}{\bibfnamefont{Z.-t.} \bibnamefont{Liang}},
  \bibinfo{author}{\bibfnamefont{Y.-k.} \bibnamefont{Song}}, \bibnamefont{and}
  \bibinfo{author}{\bibfnamefont{S.-y.} \bibnamefont{Wei}},
  \bibinfo{journal}{Phys. Rev. D} \textbf{\bibinfo{volume}{105}},
  \bibinfo{pages}{034027} (\bibinfo{year}{2022}), \eprint{2108.07740}.

\bibitem[{\citenamefont{Sato et~al.}(2016)\citenamefont{Sato, Ethier,
  Melnitchouk, Hirai, Kumano, and Accardi}}]{Sato:2016wqj}
\bibinfo{author}{\bibfnamefont{N.}~\bibnamefont{Sato}},
  \bibinfo{author}{\bibfnamefont{J.~J.} \bibnamefont{Ethier}},
  \bibinfo{author}{\bibfnamefont{W.}~\bibnamefont{Melnitchouk}},
  \bibinfo{author}{\bibfnamefont{M.}~\bibnamefont{Hirai}},
  \bibinfo{author}{\bibfnamefont{S.}~\bibnamefont{Kumano}}, \bibnamefont{and}
  \bibinfo{author}{\bibfnamefont{A.}~\bibnamefont{Accardi}},
  \bibinfo{journal}{Phys. Rev. D} \textbf{\bibinfo{volume}{94}},
  \bibinfo{pages}{114004} (\bibinfo{year}{2016}), \eprint{1609.00899}.

\bibitem[{\citenamefont{Collins and Rogers}(2024)}]{Collins:2023cuo}
\bibinfo{author}{\bibfnamefont{J.}~\bibnamefont{Collins}} \bibnamefont{and}
  \bibinfo{author}{\bibfnamefont{T.~C.} \bibnamefont{Rogers}},
  \bibinfo{journal}{Phys. Rev. D} \textbf{\bibinfo{volume}{109}},
  \bibinfo{pages}{016006} (\bibinfo{year}{2024}), \eprint{2309.03346}.

\bibitem[{\citenamefont{Aaij et~al.}(2023)}]{LHCb:2022rky}
\bibinfo{author}{\bibfnamefont{R.}~\bibnamefont{Aaij}} \bibnamefont{et~al.}
  (\bibinfo{collaboration}{LHCb}), \bibinfo{journal}{Phys. Rev. D}
  \textbf{\bibinfo{volume}{108}}, \bibinfo{pages}{L031103}
  (\bibinfo{year}{2023}), \eprint{2208.11691}.

\bibitem[{\citenamefont{Aaij et~al.}(2024)}]{LHCb:2023iyw}
\bibinfo{author}{\bibfnamefont{R.}~\bibnamefont{Aaij}} \bibnamefont{et~al.}
  (\bibinfo{collaboration}{LHCb}), \bibinfo{journal}{Phys. Rev. C}
  \textbf{\bibinfo{volume}{109}}, \bibinfo{pages}{024907}
  (\bibinfo{year}{2024}), \eprint{2310.17326}.

\bibitem[{\citenamefont{Buckley et~al.}(2015)\citenamefont{Buckley, Ferrando,
  Lloyd, Nordström, Page, Rüfenacht, Schönherr, and Watt}}]{Buckley_2015}
\bibinfo{author}{\bibfnamefont{A.}~\bibnamefont{Buckley}},
  \bibinfo{author}{\bibfnamefont{J.}~\bibnamefont{Ferrando}},
  \bibinfo{author}{\bibfnamefont{S.}~\bibnamefont{Lloyd}},
  \bibinfo{author}{\bibfnamefont{K.}~\bibnamefont{Nordström}},
  \bibinfo{author}{\bibfnamefont{B.}~\bibnamefont{Page}},
  \bibinfo{author}{\bibfnamefont{M.}~\bibnamefont{Rüfenacht}},
  \bibinfo{author}{\bibfnamefont{M.}~\bibnamefont{Schönherr}},
  \bibnamefont{and} \bibinfo{author}{\bibfnamefont{G.}~\bibnamefont{Watt}},
  \bibinfo{journal}{The European Physical Journal C}
  \textbf{\bibinfo{volume}{75}} (\bibinfo{year}{2015}), ISSN
  \bibinfo{issn}{1434-6052}.

\bibitem[{\citenamefont{Nadolsky and Sullivan}(2001)}]{Nadolsky:2001yg}
\bibinfo{author}{\bibfnamefont{P.~M.} \bibnamefont{Nadolsky}} \bibnamefont{and}
  \bibinfo{author}{\bibfnamefont{Z.}~\bibnamefont{Sullivan}},
  \bibinfo{journal}{eConf} \textbf{\bibinfo{volume}{C010630}},
  \bibinfo{pages}{P510} (\bibinfo{year}{2001}), \eprint{hep-ph/0110378}.

\end{thebibliography}

\end{document}